\documentclass[manuscript=article]{achemso}
\setkeys{acs}{articletitle = true}
\usepackage[T1]{fontenc}
\usepackage[utf8]{inputenc}
\usepackage{graphicx}
\usepackage{amsmath}
\usepackage{color}
\usepackage{comment}
\usepackage{hyperref}
\usepackage{multirow}
\usepackage{subfig}
\usepackage{bm}

\usepackage{verbatim}

\title{Projection-based Density Matrix  Renormalization Group in Density Functional Theory Embedding}

\author{Pavel Beran}
\affiliation{J. Heyrovsk\'{y} Institute of Physical Chemistry, Academy of Sciences of the Czech \mbox{Republic, v.v.i.}, Dolej\v{s}kova 3, 18223 Prague 8, Czech Republic}
\alsoaffiliation{Faculty of Mathematics and Physics, Charles University, Prague, Czech Republic}

\author{Katarzyna Pernal}
\affiliation{Institute of Physics, Lodz University of Technology, \mbox{ul.\ Wolczanska 217/221, 93-005 Lodz, Poland}}

\author{Fabijan Pavosevic}
\affiliation{Center for Computational Quantum Physics, Flatiron Institute, 162 5th Ave., New York, 10010  NY,  USA}
\email{fpavosevic@gmail.com}

\author{Libor Veis}
\email{libor.veis@jh-inst.cas.cz}
\affiliation{J. Heyrovsk\'{y} Institute of Physical Chemistry, Academy of Sciences of the Czech \mbox{Republic, v.v.i.}, Dolej\v{s}kova 3, 18223 Prague 8, Czech Republic}

\begin{document}

\begin{abstract}
\textbf{Abstract:} The density matrix renormalization group (DMRG) method has already proved itself as a very efficient and accurate computational method, which can treat large active spaces and capture the major part of strong correlation. Its application on larger molecules is, however, limited by its own computational scaling as well as 
demands of methods for treatment of the missing dynamical electron correlation. In this work, we present the first step in the direction of combining DMRG with density functional theory (DFT), one of the most employed quantum chemical methods with favourable scaling, by means of the projection-based wave function (WF)-in-DFT embedding. On the two proof-of-concept but important molecular examples, we demonstrate that the developed DMRG-in-DFT approach provides a very accurate description of molecules with a strongly correlated fragment.

\end{abstract}

\maketitle



Strong correlation plays a crucial role in many aspects of chemistry, such as bond breaking processes, open-shell systems, excited electronic states, as well as in catalysis.~\cite{lyakh2012multireference, Szalay2011} Accurate and efficient description of strongly correlated molecules, however, belongs to long-standing challenges of quantum chemistry. In principle, it can be accounted for by the exact full configuration interaction (FCI) method, but it is prohibitively expensive due to its exponential scaling. In order to bypass the limitations of FCI, several approximate polynomially scaling wave function (WF) methods were developed over the years, which can be systematically improved towards FCI. In case of molecules with weakly correlated electrons, such as organic molecules composed from the main elements and at equilibrium geometries, the most prominent example is undoubtedly the coupled cluster method~\cite{Bartlett2007}, whereas the concept of the complete active space (CAS)~\cite{roos1987complete} can be considered as a standard tool for strongly correlated molecules, such as transition metal complexes and bond breaking processes. The last two cases are also the focus of this work.

The complete active space self-consistent field (CASSCF) method~\cite{Roos1980}, which couples FCI in a small active space with orbital optimization,
is usually the starting point of multireference (MR) calculations. The missing dynamical electron correlation is then taken into account by post-SCF methods, such as the complete active space second-order perturbation theory (CASPT2)~\cite{Andersson1992}, the second-order $n$-electron valence state perturbation theory (NEVPT2)~\cite{Angeli2001}, or the multireference configuration interaction (MRCI)~\cite{Szalay2011}. The common hurdle of all these methods is the limited CAS size to less than 20 orbitals, due to the FCI exponential scaling.

Since many molecules, such as transition metal complexes, require larger CAS than FCI can handle, several approximate FCI solvers have been developed, one of them being the density matrix renormalization group (DMRG) method.~\cite{White1992} After its introduction in the quantum chemistry~\cite{White1999}, it has established itself as a powerful technique suitable for generic strongly correlated molecules with a few dozens of active orbitals~\cite{chan_review,Szalay2015, reiher_perspective}. This sparked interest in development of many post-DMRG methods for treatment of the missing (out-of-CAS) dynamical correlation are available~\cite{Cheng2022}. However, these WF-based methods are still too costly for large systems of particular interest. Their alternative, the density functional theory (DFT) represent a cost-effective approach applicable to very large molecules, which however, has its own limitations. The major shortcomings of DFT are undoubtedly the approximate form of the exchange-correlation functional as well as the single reference character, which makes it unsuitable for strongly correlated problems.~\cite{burke2012}

One way of extending the range of applicability of accurate (single or multireference) WF-based methods can be achieved by means of the quantum embedding~\cite{Jones2020}. This approach relies on locality of chemical interactions and splits the whole system into the active subsystem that is treated at a high level, and the environment subsystem that is treated at a lower level of theory.~\cite{sun2016quantum,Jones2020} Previously, Neugebauer, Reiher, and co-workers presented the first and to the best of our knowledge the only attempt to embed DMRG calculations in DFT environment by means of the frozen density embedding approach~\cite{Dresselhaus2015} for treatment of strongly correlated systems. However, due to the approximate form of the non-additive kinetic potential (NAKP), their proof-of-principle applications were restricted to systems in which the active subsystem is not covalently bonded to the environment. 

The projection-based DFT (PB-DFT) embedding~\cite{Manby2012} method is free of the NAKP problem, due to the orthogonality of occupied orbitals of both subsystems, which is achieved by the level shift projection operator~\cite{Manby2012}. This additionally ensures that the sum of energies of the active system and the environment effects is equal to the energy of the full system if both fragments are treated at the same level of theory. Encouraged by an impressive performance of the projection-based embedding for various chemical systems such as, transition metal catalysis, enzyme reactivity, or battery electrolyte decomposition~\cite{Lee2019,puccembedding}, as well as by robustness of the DMRG method, herein we develop and implement the DMRG-in-DFT projection-based embedding method. 
As demonstrated in the remainder of this letter, this approach has a tremendous potential for applications to large strongly correlated systems.



The DMRG method is a variational procedure for approximating the exact FCI wave function with the so called matrix product state (MPS)~\cite{Schollwock2011}. The FCI wave function in the occupation basis representation reads as

\begin{equation}
  | \Psi_{\text{FCI}} \rangle = \sum_{\{\alpha\}} c^{\alpha_1 \alpha_2 \ldots \alpha_n} | \alpha_1 \alpha_2 \cdots \alpha_n \rangle,
\end{equation}

\noindent
where occupation of each orbital corresponds to $\alpha_i \in \{ | 0 \rangle, | \downarrow \rangle, | \uparrow \rangle, | \downarrow \uparrow \rangle \}$ and the expansion coefficients $c^{\alpha_1 \ldots \alpha_n}$ form the FCI tensor. By successive applications of the singular value decomposition (SVD), the FCI tensor can be factorized to the MPS form~\cite{Schollwock2011}

\begin{equation}
  \label{mps_factorization}
  c^{\alpha_1 \ldots \alpha_n} = \sum_{i_1 \ldots i_{n-1}} A[1]_{i_1}^{\alpha_1} A[2]_{i_1 i_2}^{\alpha_2} A[3]_{i_2 i_3}^{\alpha_3} \cdots A[n]_{i_{n-1}}^{\alpha_n},
\end{equation}

\noindent
where $\mathbf{A}[j]^{\alpha_j}$ are the MPS matrices specific to each orbital and the newly introduced auxiliary indices $i_j$ are contracted over. If the MPS factorization is exact, the dimensions of the MPS matrices grow in a similar fashion as the size of the original FCI tensor, i.e. exponentially (with an increasing system size). In DMRG, the dimensions of auxiliary indices are bounded. These dimensions are called bond dimensions and are usually denoted with $M$.

A practical version of DMRG is the two-site algorithm, which provides the wave function in the two-site MPS form

\begin{equation}
  \label{eq:MPS_2site}
  | \Psi_{\text{MPS}} \rangle = \sum_{\{\alpha\}} \mathbf{A}^{\alpha_1} \cdots \mathbf{W}^{\alpha_i \alpha_{i+1}} \cdots \mathbf{A}^{\alpha_n}| \alpha_1 \cdots \alpha_n \rangle.
\end{equation}

\noindent
For a given pair of adjacent indices $[i, (i+1)]$, $\mathbf{W}$ is a four-index tensor, which corresponds to the eigenfunction of the second-quantized electronic Hamiltonian 

\begin{equation}
  \hat{H} = \sum_{\sigma} \sum_{pq} h_{pq} a_{p_{\sigma}}^{\dagger} a_{q_{\sigma}} +
    \frac{1}{2} \sum_{\sigma \sigma^{\prime}}\sum_{pqrs} \langle pq | rs \rangle a_{p_{\sigma}}^{\dagger} a_{q_{\sigma^{\prime}}}^{\dagger} a_{s_{\sigma^{\prime}}} a_{r_{\sigma}},
  \label{ham_sec_quant}
\end{equation}

\noindent
expanded in the tensor product space of four tensor spaces. The tensor spaces are defined on an ordered orbital chain, so called left block ($M_l$ dimensional tensor space), left site (four dimensional tensor space of $i^{\text{th}}$ orbital), right site (four dimensional tensor space of $(i+1)^{\text{th}}$ orbital), and right block ($M_r$ dimensional tensor space).
In Eq.~\ref{ham_sec_quant}, $h_{pq}$ and $\langle pq | rs \rangle$ denote standard one and two-electron integrals in the molecular orbital basis, and $\sigma$ and $\sigma^{\prime}$ denote spin. The MPS matrices $\mathbf{A}$ are obtained by successive application of SVD with truncation on $\mathbf{W}$'s and iterative optimization by going through the ordered orbital chain from left to right and then sweeping back and forth~\cite{Szalay2015}. The maximum bond dimension ($M_{\text{max}}$) which is required for a given accuracy,
can be regarded as a function of the level of entanglement in the studied system~\cite{Legeza2003}. 



In the following, we will briefly describe the projection-based embedding WF-in-DFT technique. The WF-in-DFT embedding procedure starts with an initial DFT calculation of the whole system. Based on some criteria for associating the molecular orbitals to the active and environment subsystems, the corresponding density matrix $\gamma$ is partitioned into the active subsystem A and the environment subsystem B, $\gamma_A$ and $\gamma_B$, respectively. Originally, this was achieved by means of the occupied orbitals localization and Mulliken population analysis~\cite{Manby2012}, though alternative more robust approaches have also been developed~\cite{Claudino2019a, Waldrop2021}. In case of the DFT-in-DFT embedded calculation, the total energy can be expressed as~\cite{Lee2019}
\begin{eqnarray}
    E_{\text{DFT-in-DFT}}[\bm{\gamma}_{\text{emb}}^{\text{A}}; \bm{\gamma}^{\text{A}}, \bm{\gamma}^{\text{B}}] =  E_{\text{DFT}}[\bm{\gamma}_{\text{emb}}^{\text{A}}] + E_{\text{DFT}}[\bm{\gamma}^{\text{A}} + \bm{\gamma}^{\text{B}}] - E_{\text{DFT}}[\bm{\gamma}^{\text{A}}] \\
    + \text{tr}[(\bm{\gamma}_{\text{emb}}^{\text{A}} - \bm{\gamma}^{\text{A}}) \textbf{v}_{\text{emb}}[\bm{\gamma}^{\text{A}}, \bm{\gamma}^{\text{B}}]] + \mu \text{tr}[\bm{\gamma}_{\text{emb}}^{\text{A}} \textbf{P}^{\text{B}}], \nonumber
    \label{dftindft}
\end{eqnarray}

\noindent
where $E_{\text{DFT}}$ denotes the DFT energy evaluated using the bracketed density matrix, $\bm{\gamma}_{\text{emb}}^{\text{A}}$ is the embedded subsystem A density matrix, and $\textbf{P}^{\text{B}}$ is a projection operator enforcing mutual orthogonalization, $\textbf{P}^{\text{B}} = \textbf{S} \bm{\gamma}^{\text{B}} \textbf{S}$. $\textbf{S}$ denotes the atomic orbital overlap matrix. In the limit where the level shift parameter $\mu \rightarrow \infty$, the A and B orbitals are exactly orthogonal, but $\mu$ is for practical purposes taken to be $10^6$, causing negligible error~\cite{Manby2012}. The embedding potential $\textbf{v}_{\text{emb}}$ contains all interactions between subsystems A and B
\begin{equation}
  \textbf{v}_{\text{emb}}[\bm{\gamma}^{\text{A}}, \bm{\gamma}^{\text{B}}] = \textbf{g}[\bm{\gamma}^{\text{A}} + \bm{\gamma}^{\text{B}}] - \textbf{g}[\bm{\gamma}^{\text{A}}].
\end{equation}

\noindent
The matrix $\textbf{g}$ groups all the two-electron contributions (Coulomb, exchange, and exchange-correlation). Because, the projection-based embedding approach is free from non-additive kinetic energy problem~\cite{Manby2012} it is formally exact, i.e. when the active part was treated with the same exchange-correlation functional as the environment, it would be equivalent to the Kohn-Sham solution of the entire system. 

The Fock matrix of subsystem A embedded in B has the following form~\cite{Lee2019}
\begin{equation}
  \textbf{F}^{\text{A}} = \textbf{h} + \textbf{g}[\bm{\gamma}_{\text{emb}}^{\text{A}}] + \textbf{v}_{\text{emb}}[\bm{\gamma}^{\text{A}}, \bm{\gamma}^{\text{B}}] + \mu \textbf{P}^{\text{B}},
\end{equation}
\noindent
where $\textbf{h}$ is the core Hamiltonian matrix and it is self-consistently optimized with respect to $\bm{\gamma}_{\text{emb}}^{\text{A}}$.
In case of single reference WF-in-DFT calculations, HF-in-DFT with the following effective core Hamiltonian
\begin{equation}
  \textbf{h}^{\text{A-in-B}}[\bm{\gamma}^{\text{A}}, \bm{\gamma}^{\text{B}}] = \textbf{h} + \textbf{v}_{\text{emb}}[\bm{\gamma}^{\text{A}}, \bm{\gamma}^{\text{B}}] + \mu \textbf{P}^{\text{B}}
  \label{hinb}
\end{equation}
\noindent
precedes the WF calculation. For MR problems, CASSCF-in-DFT can be performed~\cite{deLimaBatista2017}. 
However, since we employ the accurate DMRG which approaches the FCI solution of the active subsystem, we are free to use HF-in-DFT for the MR problems. 

Most importantly, the DFT-in-DFT method can be straightforwardly employed for a WF-in-DFT embedding where the active subsystem is treated with the DMRG method and the environment subsystem is described with the DFT method. Then the DMRG-in-DFT energy is simply obtained by substituting the DFT energy of the active subsystem A with the DMRG energy as
\begin{eqnarray}
    E_{\text{DMRG-in-DFT}}[\Psi^{\text{A}}_{\text{MPS}}; \bm{\gamma}^{\text{A}}, \bm{\gamma}^{\text{B}}] = E_{\text{DMRG}}[\Psi^{\text{A}}_{\text{MPS}}] + E_{\text{DFT}}[\bm{\gamma}^{\text{A}} + \bm{\gamma}^{\text{B}}] - E_{\text{DFT}}[\bm{\gamma}^{\text{A}}] \\
    + \text{tr}[(\bm{\gamma}_{\text{emb}}^{\text{A}} - \bm{\gamma}^{\text{A}}) \textbf{v}_{\text{emb}}[\bm{\gamma}^{\text{A}}, \bm{\gamma}^{\text{B}}]] + \mu \text{tr}[\bm{\gamma}_{\text{emb}}^{\text{A}} \textbf{P}^{\text{B}}]. \nonumber
    \label{dftindft}
\end{eqnarray}

\noindent
In this equation, $E_{\text{DMRG}}[\Psi^{\text{A}}_{\text{MPS}}]$ is the DMRG energy of the active subsystem corresponding to the MPS wave function $|\Psi^{\text{A}}_{\text{MPS}}\rangle$, which minimizes the active subsystem Hamiltonian (\ref{ham_sec_quant}) with the one-electron part replaced by the effective core Hamiltonian from Eq. \ref{hinb}.

\begin{figure}[!ht]
  \subfloat[\label{bond_stretch}]{%
    \includegraphics[height=3cm]{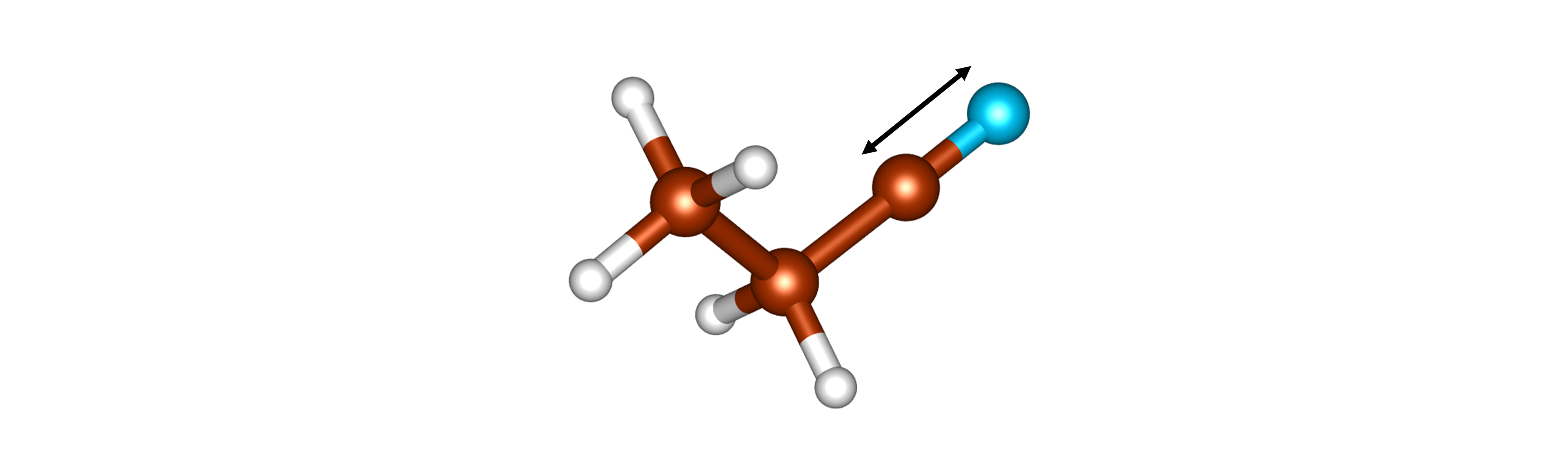}
  }
  \hfill
  \subfloat[\label{reaction}]{%
    \includegraphics[height=4cm]{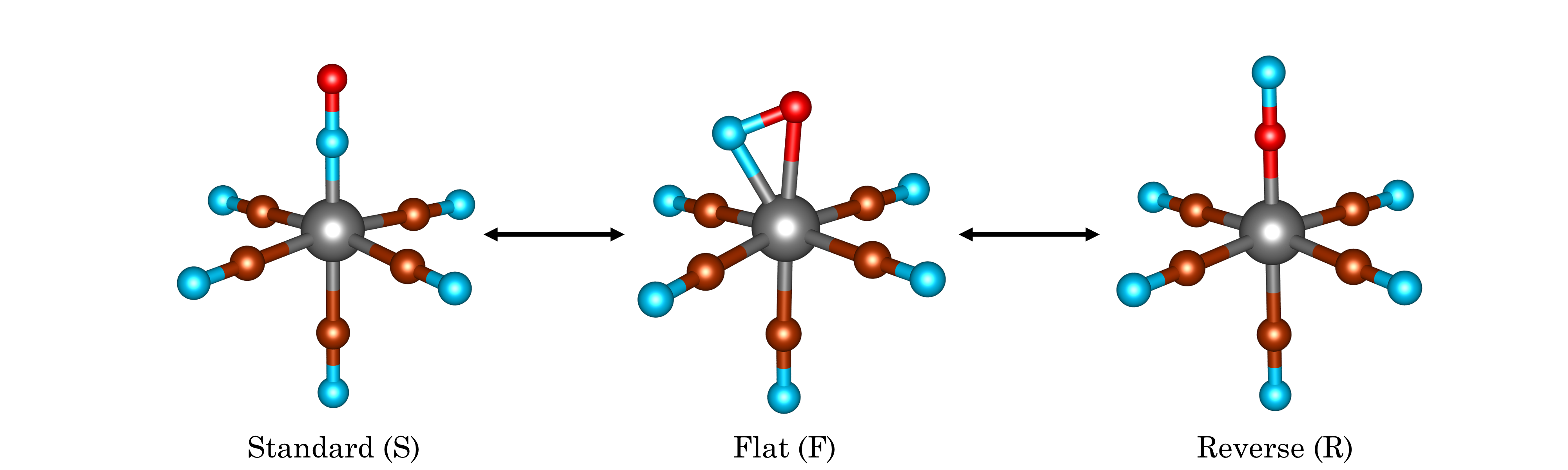}
  }
  \caption{Benchmark problems studied in this work: (a) Triple C--N bond stretching in propionitrile (CH$_3$CH$_2$CN). (b) Conformational isomerization of the [Fe(CN)$_5$(NO)]$^{2-}$ complex. The color codes are as follows: Fe (grey), N (blue), C (brown), O (red), and H (white).}
  \label{studied_systems}
\end{figure}

The WF-in-DFT embedding method has been implemented in \textsf{Psi4NumPy} quantum chemistry software~\cite{Smith2018} which was interfaced with the \textsf{MOLMPS}~\cite{Brabec2020} DMRG code. The developed method was then used to study two benchmark problems (see Figure~\ref{studied_systems}) which have a strongly correlated active part coupled to the environment, namely the triple bond stretching in propionitrile (CH$_3$CH$_2$CN) and the conformational isomerization of the model iron-nitrosyl complex [Fe(CN)$_5$(NO)]$^{2-}$\cite{Daniel2019}, which is a prototype of a transition metal complex with the non-innocent nitrosyl ligand relevant to medicinal applications~\cite{awasabisah}.
Regarding the low-level method, all the DFT calculations employed the B3LYP~\cite{Parr88_785,Becke88_3098} density functional. On the other hand, all the high-level DMRG calculations were warmed-up with the CI-DEAS procedure~\cite{Legeza2003,Szalay2015} and took advantage of the dynamical block state selection (DBSS)~\cite{legeza_2003a}, which adjusts the actual bond dimensions to fit the desired (pre-set) truncation error (TRE). 
The initial DMRG orbital orderings were optimized with the Fiedler method~\cite{Barcza2011}.
The complementary calculations listed below were carried out in the following programs: CCSD in \textsf{Psi4}~\cite{Smith2018}, CASSF/DMRG-SCF in \textsf{Orca}~\cite{orca}, adiabatic connection (AC) in \textsf{GammCor}~\cite{gammcor}, and internally contracted MRCI in \textsf{MOLPRO}~\cite{molpro}.


In our first example, we study the triple bond stretching in propionitrile (CH$_3$CH$_2$CN) molecule. The equilibrium geometry of propionitrile employed in this work is given in the Supporting Information (SI, Table~S1). For the WF-in-DFT calculations, we have employed the cc-pVDZ~\cite{Dunning1989} basis set. The active subsystem comprised the \mbox{--CN} group and the orbitals were partitioned into both subsystems by means of the SPADE procedure~\cite{Claudino2019a}. 
The stretching of the CN bond was probed by the accurate DMRG-in-B3LYP calculations with $\text{TRE} = 10^{-6}$. For comparison, we also carried out the CCSD-in-B3LYP, as well as the CCSD and DMRG calculations for the entire molecule. The frozen-core approximation was employed for the aforementioned DMRG calculations leading to the FCI space of 22 electrons in 77 orbitals and TRE was pre-set to $10^{-5}$.

\begin{figure}[!ht]
    \centering
    \includegraphics[width=0.7\textwidth]{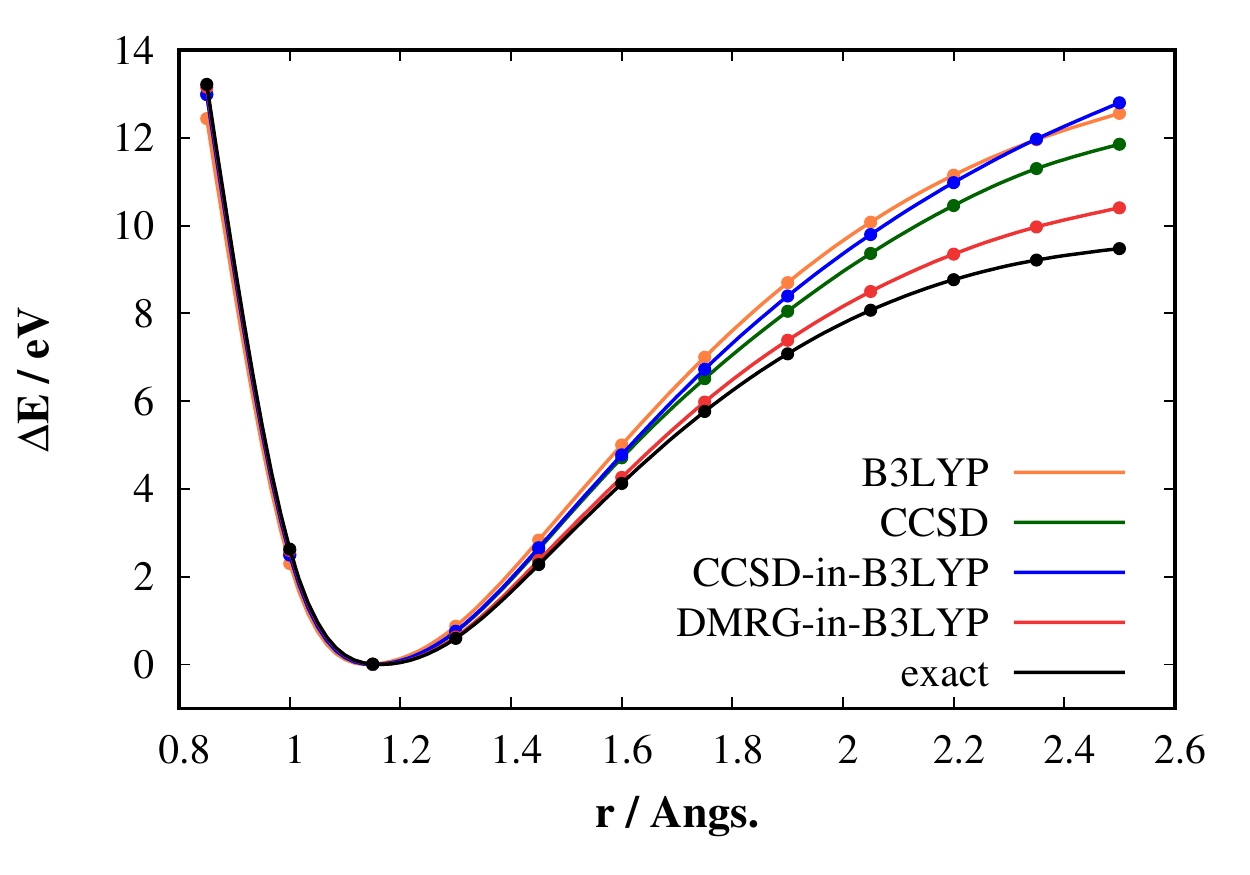}
    \caption{Comparison of the individual dissociation energy curves corresponding to the triple C-N bond stretching in CH$_3$CH$_2$CN. All calculations employ the cc-pVDZ basis set.}
    \label{prop_disoc}
\end{figure}

Figure~\ref{prop_disoc}, shows the potential energy surfaces (PES) [differences with respect to minima: $E(r_{\text{CN}}) - E_{\text{min}}$] corresponding to the triple C--N bond stretching in propionitrile. The results obtained by B3LYP, CCSD, CCSD-in-B3LYP, and DMRG-in-B3LYP are compared against the exact curve obtained by the frozen-core DMRG method. The individual absolute energies are provided in Table~S2.

As it is known, the CCSD method notoriously fails in describing correctly the triple bond breaking due to its single-determinant nature. It e.g. predicts a nonphysical bump on PES of N$_2$ molecule in the intermediate stretching region (around 2.2 \AA)~\cite{Kinoshita2005}. One can see in Figure~\ref{prop_disoc}, that the situation is unsurprisingly very similar for the triple C--N  bond stretching in CH$_3$CH$_2$CN.
The CCSD method provides much higher dissociation energies for the intermediate stretching region, than the frozen-core DMRG (at 2.5~\AA, the error is $\sim~2.4$~eV). CCSD-in-B3LYP behaves even slightly worse than CCSD itself. On the other hand, there is a huge improvement between CCSD-in-B3LYP and DMRG-in-B3LYP in description of the triple C--N bond stretching process. At 2.5~\AA, the error of DMRG-in-B3LYP with respect to DMRG is 0.9~eV, whereas for the CCSD-in-B3LYP method this error is 3.3~eV. The DMRG method as a genuine MR method is able to properly describe this process. The difference between DMRG-in-B3LYP and DMRG, which is essentially very similar to the difference between CCSD-in-B3LYP and CCSD, thus can be attributed to the lower-level (B3LYP) description of the remaining electrons plus errors of the PB-DFT embedding (density-driven errors or errors coming from the non-additivity of the exchange-correlation energy~\cite{Goodpaster2014}).

As our second example, we have studied the conformational isomerization of the model iron-nitrosyl complex [Fe(CN)$_5$(NO)]$^{2-}$. 
The B3LYP optimized geometries of the standard, flat, and reversed isomers of [Fe(CN)$_5$(NO)]$^{2-}$ (see Figure~\ref{reaction}) were taken from Ref.~\cite{Daniel2019} (also given in Table~S3-S5). For computational reasons, we used the smaller \mbox{6-31G}~\cite{Hehre1972, Rassolov1998} basis.
The active subsystem was formed by [Fe--NO]$^{3+}$ and partitioning of the orbitals into subsystems was carried out by means of the SPADE procedure~\cite{Claudino2019a}. In order to decrease the size of the virtual space, we employed the two-shell concentric localization~\cite{Claudino2019} leading to the active subsystem FCI space comprising 38 electrons in 102 orbitals. 
For comparison, we also carried out the B3LYP and CCSD calculations as well as calculations with different CAS-based MR methods. The smallest CAS(4,4) comprising the two NO $\pi^*$ orbitals together with the Fe 3d$_{xz}$ and 3d$_{yz}$ was employed for internally contracted MRCI with singles and doubles \mbox{(icMRCISD)} calculations. The larger CAS(14,15) contained the NO $\pi$ (two), $\pi^*$ (two), $\sigma$, $\sigma^*$, and Fe 3d (five), 4d (3 counterparts to the occupied 3d orbitals: 4d$_{xy}$, 4d$_{xz}$, and 4d$_{yz}$), plus one equatorial $\sigma$ orbital with the Fe 3d$_{x^2 - y^2}$ and C 2p$_{x/y}$ contributions. This CAS(14,15) was augmented with one occupied axial orbital of $\sigma$ character to form CAS(16,16). All CASSCF natural orbitals are shown in Figures~S3-S9). In the smaller CAS(14,15), we performed CASSCF computations, which were then corrected for the dynamical electron correlation by means of strongly contracted NEVPT2, the adiabatic connection (AC)~\cite{Pernal2018, Pastorczak2018}, and the linearized-AC-integrand approximation AC0~\cite{Pernal2018, Pastorczak2018}. The later two have the advantage of favourable scaling with respect to the CAS size and thus represent an ideal choice for approximate FCI solvers such as DMRG~\cite{Beran2021}.
In CAS(16,16), we performed the DMRG-SCF calculations with fixed bond dimensions equal to 2000 and subsequent AC/AC0 in order to probe the effect of the missing dynamical electron correlation.

Table~\ref{feno_occup} shows the natural orbital occupation numbers (NOONs) of the four orbitals around the Fermi level for the largest active space employed, i.e. CAS(16,16) (all occupation numbers can be found in SI). 
\begin{table}[!ht]
    \centering
    \begin{tabular}{l c c c c}\hline
         Isomer & HOMO-1 & HOMO & LUMO & LUMO+1  \\
         \hline\hline
         S & 1.82 & 1.82 & 0.21 & 0.21 \\
         F & 1.92 & 1.77 & 0.25 & 0.10 \\ 
         R & 1.72 & 1.72 & 0.32 & 0.32\\
\hline
    \end{tabular}
    \caption{DMRG-SCF(16,16) Natural Orbital Occupation Numbers for the Individual [Fe(CN)$_5$(NO)]$^{2-}$ Standard (S), Flat (F), and Reverse (R) Isomers.}
    \label{feno_occup}
\end{table}

The occupation numbers largely deviate from 2 (and 0) and confirm the non-innocent nature of the nitrosyl ligand, indicating the significant multireference character of the investigated systems. Moreover, looking at the four aforementioned orbitals (Figures~S7--S9), one can see that their electron density is mainly localized to the Fe--NO region, which corroborates the use of the WF-in-DFT embedding, in which the WF method, however, should be able to correctly describe the MR character of the \mbox{Fe--NO} moiety. The strongest MR character is observed for the reverse isomer. In this case, the weight of the HF reference in the DMRG-SCF(16,16) wave function is only 64\% and one can expect that the conventional single reference approaches might be inappropriate.

\begin{table}[!ht]
    \centering
    \begin{tabular}{c c c}\hline
         & $\Delta E_{\text{S} \rightarrow \text{F}}$\textsuperscript{a} & $\Delta E_{\text{S} \rightarrow \text{R}}$\textsuperscript{b} \\
         \hline\hline
         B3LYP & 1.77 & 1.91 \\
         CCSD & 1.72 & 2.03 \\ 
         CASSCF(14,15) & 1.63 & 1.23 \\ 
         NEVPT2(14,15) & 2.30 & 1.34 \\ 
         AC0(14,15) & 2.34 & 1.18 \\ 
         AC(14,15) & 2.20 & 1.15 \\ 
         DMRG-SCF(16,16) & 1.83 & 1.18 \\ 
         AC0(16,16) & 2.18 & 1.46 \\ 
         AC(16,16) & 2.14 & 1.38 \\ 
         icMRCISD(4,4) & 1.90 & 1.44 \\ 
         \hline
         CCSD-in-B3LYP & 1.27 & 1.85 \\
         CCSD-in-HF & 1.36 & 2.12 \\
         DMRG-in-B3LYP & 1.92 & 1.17 \\
         DMRG-in-HF & 2.01 & 1.44 \\
\hline
    \end{tabular}
    \caption{Reaction Energies in~eV Corresponding to the Conformational Isomerization of [Fe(CN)$_5$(NO)]$^{2-}$ Complex Calculated with Different Methods and 6-31G Basis Set.}
    \label{fe_no_table}

\textsuperscript{a}\small $\Delta E_{\text{S} \rightarrow \text{F}}$ denotes the energy difference between flat (F) and standard (S) isomers.\\

\textsuperscript{b}\small $\Delta E_{\text{S} \rightarrow \text{R}}$ denotes the energy difference between reverse (R) and standard (S) isomers.\\

\end{table}

\begin{figure}[!ht]
    \centering
    \includegraphics[width=0.7\textwidth]{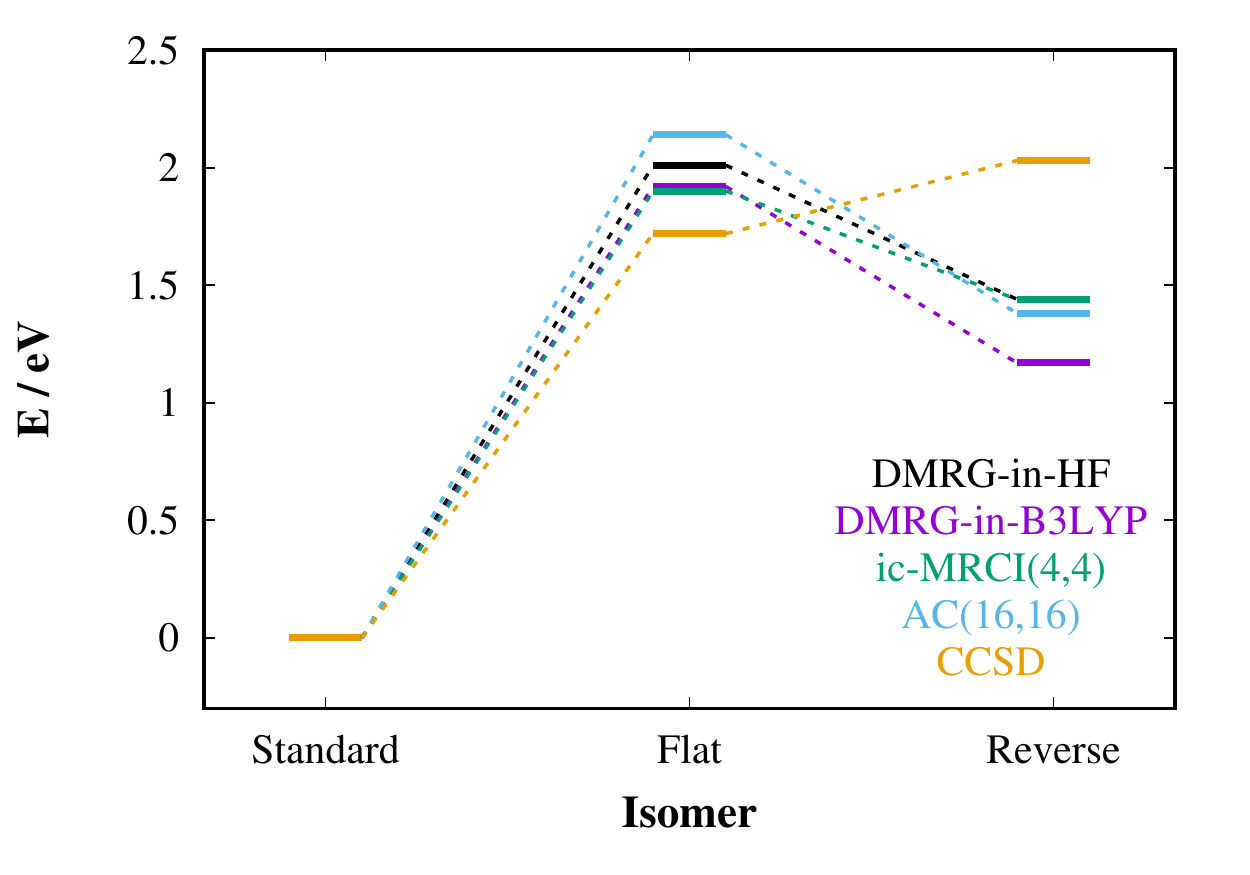}
    \caption{Graphical representation of e\-ner\-ge\-tics of [Fe(CN)$_5$(NO)]$^{2-}$ complex conformational isomerization for selected computational methods.}
    \label{feno_fig}
\end{figure}

Table~\ref{fe_no_table} shows the reaction energies of three stable isomers involved in the [Fe(CN)$_5$(NO)]$^{2-}$ complex conformational isomerization computed by various single and multi-reference methods as well as with the CCSD and DMRG methods embedded in the HF or DFT environment. The graphical summary is depicted in Figure~\ref{feno_fig}. 
Because of the significant multireference character in all three isomers, Figure~\ref{feno_fig} and Table~\ref{fe_no_table} indicate that the single reference methods (B3LYP and CCSD), in contrast to all state-of-the-art multireference approaches, incorrectly predict the reverse isomer to have the highest energy. At the CAS(14,15) level, we can observe that adding the dynamical electron correlation on top of CASSCF by means of NEVPT2 and AC0/AC results in a larger $\Delta E_{\text{S} \rightarrow \text{F}}$ by 0.6 -- 0.7~eV, whereas $\Delta E_{\text{S} \rightarrow \text{R}}$ is affected only slightly. More importantly, AC0 provides very similar energy gaps as NEVPT2 (within 0.16~eV in case of $\Delta E_{\text{S} \rightarrow \text{R}}$), as was already pointed out previously~\cite{Pastorczak2018}. The canonical AC method captures even more correlation energy than its linearized AC0 approximation and the AC(16,16) results together with the icMRCISD(4,4) results represent our best estimates of the energy gaps, in particular 1.9--2.14~eV for $\Delta E_{\text{S} \rightarrow \text{F}}$ and $\sim$1.40~eV for $\Delta E_{\text{S} \rightarrow \text{R}}$.

Looking at the results of the embedded calculations in Table~\ref{fe_no_table}, one can see that CCSD-in-HF as well as CCSD-in-B3LYP underestimate the $\Delta E_{\text{S} \rightarrow \text{F}}$ gap even more than CCSD and predict incorrectly that the flat isomer is lower in energy than the reverse one (by 0.8~eV and 0.6~eV, respectively). On contrary, the results of the DMRG embedded calculations are in a very good agreement with our best estimates of the energy gaps. Both DMRG-in-HF as well as DMRG-in-B3LYP provide $\Delta E_{\text{S} \rightarrow \text{F}}$ gaps within the margins of the MR methods, the DMRG-in-B3LYP $\Delta E_{\text{S} \rightarrow \text{R}}$ gap is slightly lower (by $\sim$0.2~eV). The DMRG-in-HF method achieves a perfect agreement of both energy gaps with our best estimates obtained by the state-of-the-art MR methods, which confirms that the Fe-NO moiety is mainly responsible for the electronic structure properties of the [Fe(CN)$_5$(NO)]$^{2-}$ complex.


In this letter, we present the projection-based DMRG-in-DFT embedding method and we test its performance on two benchmark problems, namely the triple bond stretching in CH$_3$CH$_2$CN and conformational isomerization of [Fe(CN)$_5$(NO)]$^{2-}$, a prototype of the transition metal complex containing a non-innocent ligand. Both of these systems exhibit a significant multireference character. Our numerical results indicate that the DMRG-in-DFT provides a viable way toward accurate description of molecules containing strongly correlated fragment. In case of the triple bond stretching in CH$_3$CH$_2$CN, the DMRG-in-B3LYP method substantially outperformed the single-reference CCSD and CCSD-in-B3LYP methods, whereas in case of the [Fe(CN)$_5$(NO)]$^{2-}$ complex, the DMRG-in-B3LYP and DMRG-in-HF methods provided the energy gaps between individual isomers that are in very good agreement with the state-of-the-art multireference approaches. 
This work represents the first step toward combining DMRG with PB-DFT embedding. The biggest bottleneck of this approach is the size of the virtual space which, even when it is truncated~\cite{Claudino2019}, might be too large for DMRG. It is also the reason, why we were limited to smaller basis sets. However, in case of larger basis sets, the concept of CAS can be used in which DMRG is combined with some post-DMRG method~\cite{Cheng2022} which will the subject of our following works.

\section*{Supporting Information}
Equilibrium geometry of CH$_3$CH$_2$CN; all computed absolute energies of CH$_3$CH$_2$CN for the C--N bond stretching; geometries of standard, flat, and reversed isomers of [Fe(CN)$_5$(NO)]$^{2-}$, all computed absolute energies of standard, flat, and reversed isomers of [Fe(CN)$_5$(NO)]$^{2-}$; CASSCF and DMRG-SCF natural orbitals and occupation numbers of [Fe(CN)$_5$(NO)]$^{2-}$.

\section*{Acknowledgment}
This work was supported by the the Czech Science Foundation (grant no.\ 22-04302L), the National Science Center of Poland (grant no.\ 2021/43/I/ST4/02250), the Grant Scheme of the Charles University in Prague (grant no.\ CZ.02.2.69\slash0.0\slash0.0\slash19\_073\slash0016935), the Czech Ministry of Education, Youth and Sports from the Large Infrastructures for Research, Experimental Development and Innovations
project ``IT4Innovations National Supercomputing Center-LM2015070.'', and the Center for Scalable and Predictive methods for Excitation and Correlated phenomena (SPEC), which is funded by the U.S. Department of Energy (DOE), Office of Science, Office of Basic Energy Sciences, the Division of Chemical Sciences, Geosciences, and Biosciences. The Flatiron Institute is a division of the Simons Foundation.

\bibliography{references}

\end{document}


\clearpage

\section{Propionitrile (CH$_3$CH$_2$CN) C-N bond stretching}

\begin{table}[!ht]
\centering
\begin{tabular}{c r r r r}
C   & &     -2.38207   &    -0.46087    &    0.01893 \\
N   & &     -3.18147   &   -0.80786     &   0.76930 \\
H   & &      -0.03176  &      0.93909   &     0.54131 \\
C   & &       0.02231  &      0.17867   &    -0.25262 \\
H   & &       0.75941  &      0.50795   &    -1.00054 \\
H   & &       0.38038  &     -0.75999   &     0.19640 \\
C   & &      -1.34723  &     -0.01758   &    -0.92251 \\
H   & &      -1.69162  &      0.92161   &    -1.38724 \\
H   & &      -1.28110  &     -0.76419   &    -1.73200 \\
\end{tabular}
\caption{Equilibrium geometry of CH$_3$CH$_2$CN, XYZ in \AA.}
\end{table}

\begin{table}[!ht]
    \centering
    \begin{tabular}{c c c c c c}
        $r_{\text{C-N}}$ & B3LYP & CCSD & CCSD-in-B3LYP & DMRG-in-B3LYP & DMRG(FC) \\
        \hline
        0.85 & -171.618382 & -171.106586 & -171.337520 & -171.341987 & -171.11161047 \\
        1.00 & -171.991004 & -171.492235 & -171.723038 & -171.730008 & -171.50086928 \\
        1.15 & -172.075359 & -171.584162 & -171.814725 & -171.825192 & -171.59716505 \\
        1.30 & -172.043567 & -171.556873 & -171.786935 & -171.802364 & -171.57557639 \\
        1.45 & -171.971320 & -171.487757 & -171.717006 & -171.738241 & -171.51378665 \\
        1.60 & -171.891661 & -171.411147 & -171.639154 & -171.668690 & -171.44582726 \\
        1.75 & -171.818215 & -171.344743 & -171.567481 & -171.605607 & -171.38558935 \\
        1.90 & -171.755710 & -171.288331 & -171.506139 & -171.553929 & -171.33720854 \\
        2.05 & -171.705096 & -171.239879 & -171.454591 & -171.513078 & -171.30076817 \\
        2.20 & -171.665708 & -171.199782 & -171.411130 & -171.481714 & -171.27519061 \\
        2.35 & -171.636018 & -171.168898 & -171.374676 & -171.458972 & -171.25875127 \\
        2.50 & -171.614037 & -171.148550 & -171.344320 & -171.442953 & -171.24910098 \\
    \end{tabular}
    \caption{Absolute energies of CH$_3$CH$_2$CN for a given C-N bond length (in \AA). All calculations were performed in the cc-pVDZ basis, energies are listed in a.u., and FC denotes the frozen-core approximation. DMRG(FC) calculations were performed with the DBSS procedure and TRE=$10^{-5}$.}
    \label{tab:my_label}
\end{table}

\section{[Fe(CN)$_5$(NO)]$^{2-}$ complex conformational isomerization}

\subsection{Geometries}
Source:
SI of Daniel, C.; Gourlaouen, C. Structural and Optical Properties of Metal-Nitrosyl Complexes. Molecules 2019, 24, 3638.

\begin{table}[!ht]
\centering

\begin{tabular}{c r r r r}
    Fe & & 0.00149500  & -0.00106700 & -0.09336900 \\
    C  & & -0.03222200 & 0.02103400  & 1.86410100 \\
    C  & & -1.75416100 & 0.85771400  & 0.02002800 \\
    C  & & 0.85659200  & 1.75411800  & 0.05268600 \\
    C  & & 1.75258100  & -0.85664900 & 0.09511900 \\
    C  & & -0.85785400 & -1.75326800 & 0.06284700 \\
    N  & & -0.05222500 & 0.03386700  & 3.03384600 \\
    N  & & 2.80100800  & -1.36744800 & 0.19103400 \\
    N  & & -1.37158600 & -2.80173800 & 0.14115000 \\
    N  & & 1.36881800  & 2.80371400  & 0.12573200 \\
    N  & & -2.80414500 & 1.37165900  & 0.07265600 \\
    N  & & 0.02861100  & -0.01928200 & -1.73124100 \\
    O  & & 0.04726700  & -0.03192100 & -2.87166600 \\
\end{tabular}
\caption{Geometry of the standard isomer of [Fe(CN)$_5$(NO)]$^{2-}$ complex, XYZ in \AA.}
\end{table}

\begin{table}[!ht]
\begin{tabular}{c r r r r}
    Fe  & & -0.00016900 & 0.02806300  & -0.12906000 \\
    C   & & 0.00035300  & -0.17615400 & 1.76755000 \\
    C   & & -1.38040600 & 1.40923600  & 0.04432800 \\
    C   & & 1.37930400  & 1.41018200  & 0.04440300 \\
    C   & & 1.48359600  & -1.27200700 & -0.02254300 \\
    C   & & -1.48287100 & -1.27288100 & -0.02213100 \\
    N   & & 0.00074600  & -0.27164900 & 2.93388700 \\
    N   & & 2.36724000  & -2.03798600 & 0.03986000 \\
    N   & & -2.36594400 & -2.03946500 & 0.04094100 \\
    N   & & 2.19600000  & 2.24440400  & 0.13982500 \\
    N   & & -2.19754400 & 2.24305200  & 0.13952600 \\
    N   & & 0.00022500  & -0.74637000 & -1.85226200 \\
    O   & & -0.00006700 & 0.36702500  & -2.20081400 \\
\end{tabular}
\caption{Geometry of the flat isomer of [Fe(CN)$_5$(NO)]$^{2-}$ complex, XYZ in \AA.}
\end{table}

\begin{table}[!ht]
\begin{tabular}{c r r r r}
    Fe  & & -0.00060500 & 0.00018400  & -0.05436100 \\
    C   & & 0.03567000  & -0.02895700 & 1.86383700 \\
    C   & & -1.53095900 & 1.21983800  & 0.09791900 \\
    C   & & 1.22394800  & 1.53034400  & 0.04912800 \\
    C   & & 1.53359200  & -1.22166100 & 0.00102400 \\
    C   & & -1.21846800 & -1.53465800 & 0.04870100 \\
    N   & & 0.05817700  & -0.04746800 & 3.03450200 \\
    N   & & 2.44981100  & -1.95160100 & 0.02040000 \\
    N   & & -1.94702900 & -2.45095200 & 0.09594500 \\
    N   & & 1.95614400  & 2.44363500  & 0.09789000 \\
    N   & & -2.44506900 & 1.94860300  & 0.17478400 \\
    O   & & -0.03787600 & 0.03041200  & -1.80130200 \\
    N   & & -0.06403000 & 0.05242500  & -2.92921600 \\
\end{tabular}
\caption{Geometry of the reverse isomer of [Fe(CN)$_5$(NO)]$^{2-}$ complex, XYZ in \AA.}
\end{table}

\clearpage

\subsection{Energies}

\begin{table}[!ht]
    \centering
    \begin{tabular}{c r r r}
         & standard & flat & reverse \\
         \hline
         B3LYP & -1857.204433 & -1857.139493 & -1857.134402 \\
         CCSD & -1854.298712 & -1854.235608 & -1854.224112 \\ 
         CASSCF(14,15) & -1852.95065 & -1852.890706 & -1852.905416 \\ 
         NEVPT2(14,15) & -1854.268748 & -1854.184329 & -1854.219546 \\ 
         AC0(14,15) & -1854.294758 & -1854.208607 & -1854.251568 \\ 
         AC(14,15) & -1854.089513 & -1854.008786 & -1854.047418 \\ 
         DMRG-SCF(16,16) & -1852.997355 & -1852.930239 & -1852.953831 \\ 
         AC0(16,16) & -1854.304948 & -1854.224917 & -1854.251228 \\ 
         AC(16,16) & -1854.107123 & -1854.028369 & -1854.05652 \\ 
         icMRCISD(4,4) & -1853.756956 & -1853.686976 & -1853.703916 \\ 
         \hline
         CCSD-in-B3LYP & -1856.113810 & -1856.067165 & -1856.045671 \\
         CCSD-in-HF & -1853.102533 & -1853.052473 &  -1853.024501\\
         DMRG-in-B3LYP & -1856.159181 & -1856.088551 & -1856.116160 \\
         DMRG-in-HF & -1853.146749 & -1853.072713 & -1853.093745
    \end{tabular}
    \caption{Absolute energies (in a.u.) of standard, flat, and reverse isomers of [Fe(CN)$_5$(NO)]$^{2-}$ complex in 6-31G basis.}
    \label{fe_no_absoluteenergies}
\end{table}

\subsection{Natural orbitals of [Fe(CN)$_5$(NO)]$^{2-}$ complex}
\renewcommand{\thesubfigure}{\arabic{subfigure}}
\begin{figure}[!h]
  \subfloat[$n_{\text{occup}}$ = 1.8235]{%
    \includegraphics[width=0.22\textwidth]{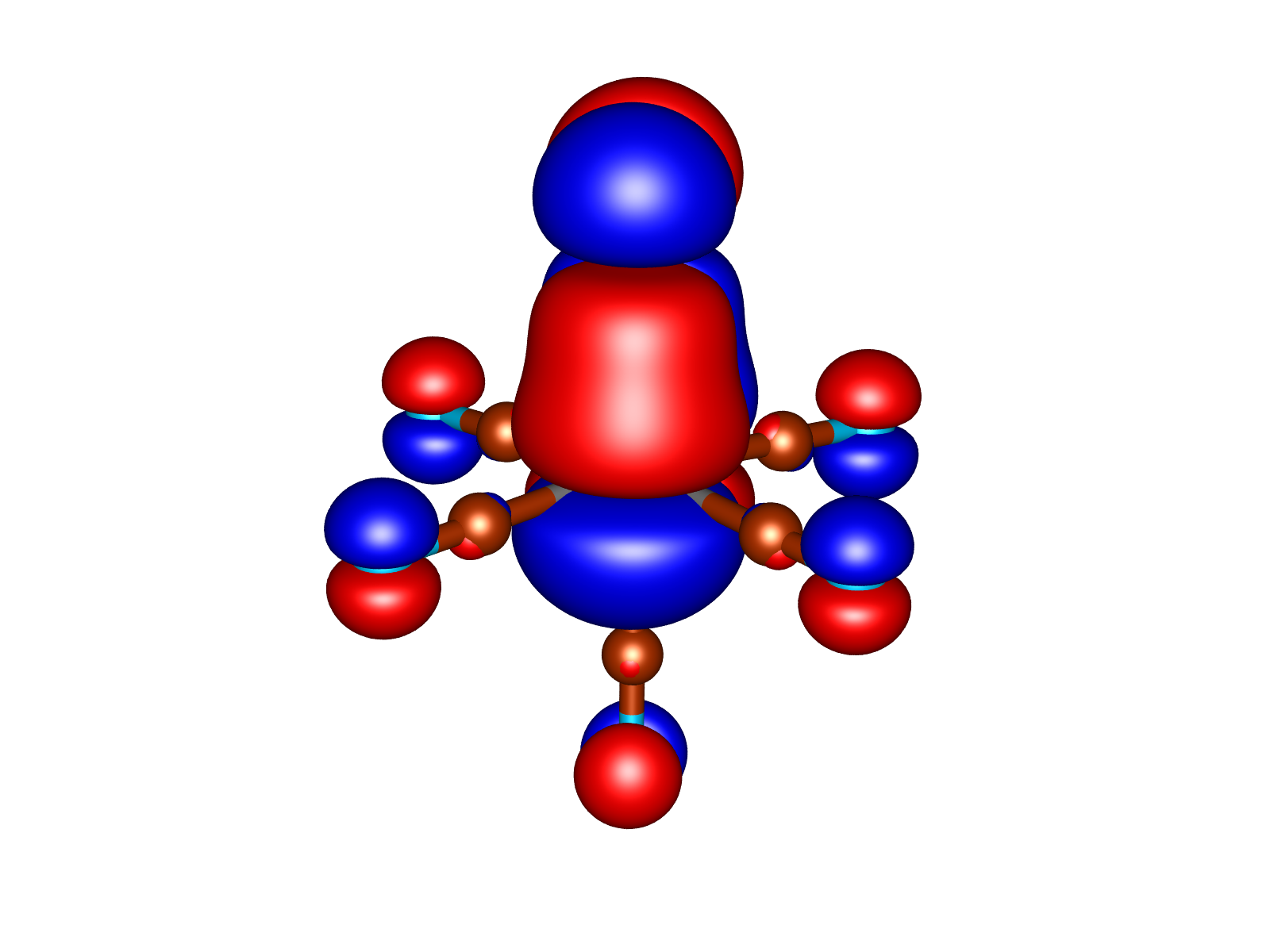}
  }
  \hfill
  \subfloat[$n_{\text{occup}}$ = 1.8231]{%
    \includegraphics[width=0.22\textwidth]{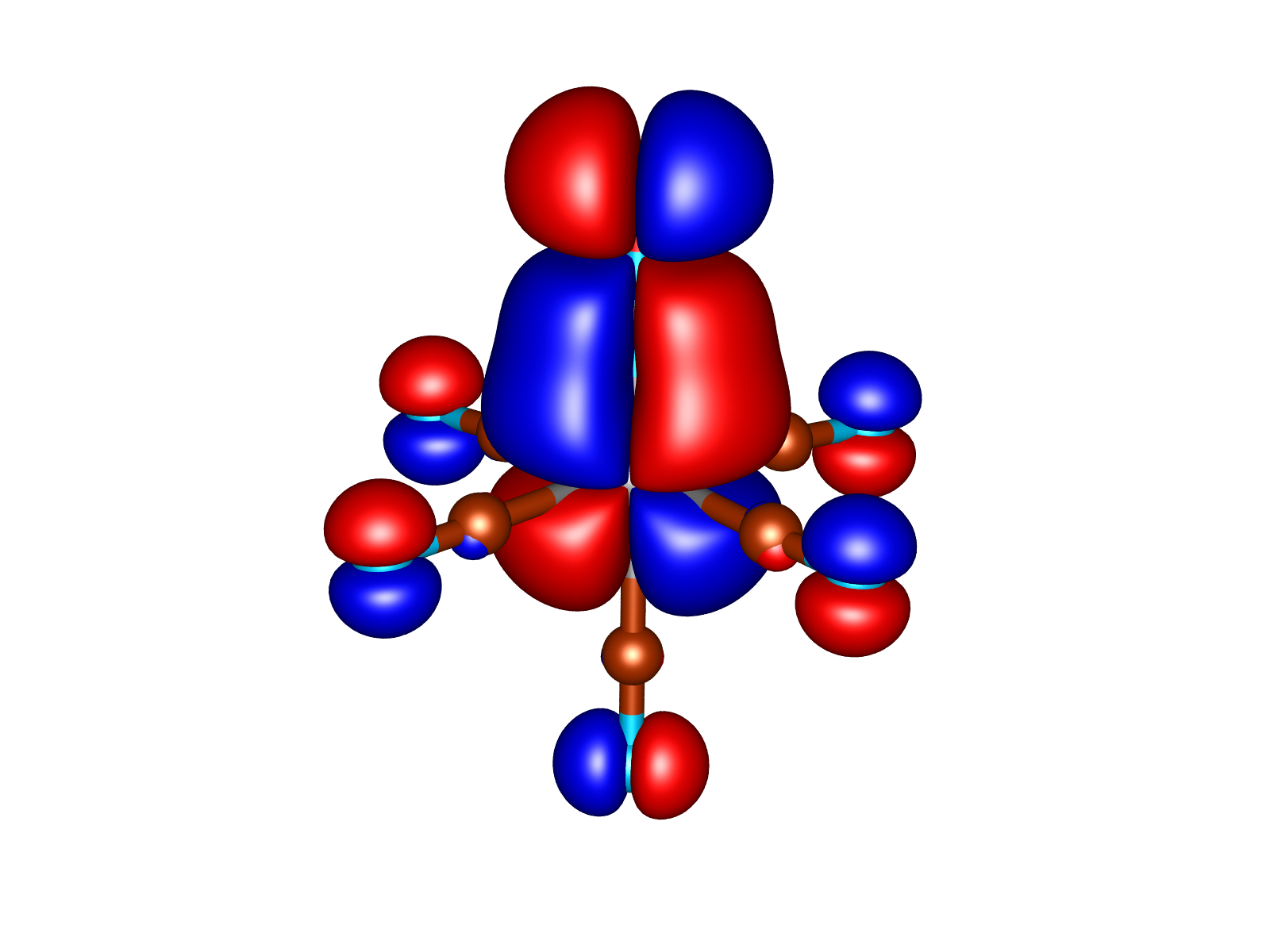}
  }
  \hfill
  \subfloat[$n_{\text{occup}}$ = 0.1765]{%
    \includegraphics[width=0.22\textwidth]{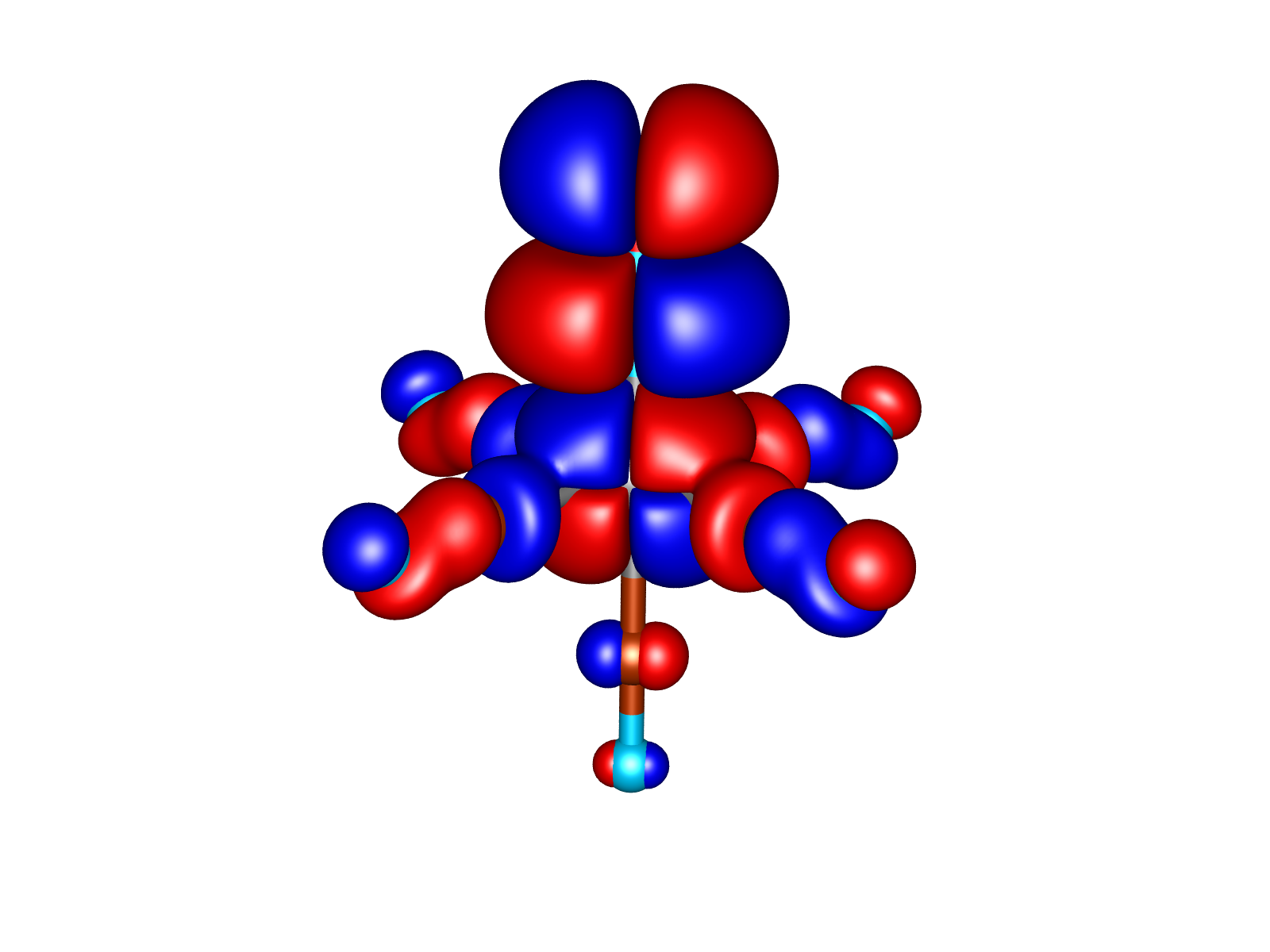}
  }
  \hfill
  \subfloat[$n_{\text{occup}}$ = 0.1765]{%
    \includegraphics[width=0.22\textwidth]{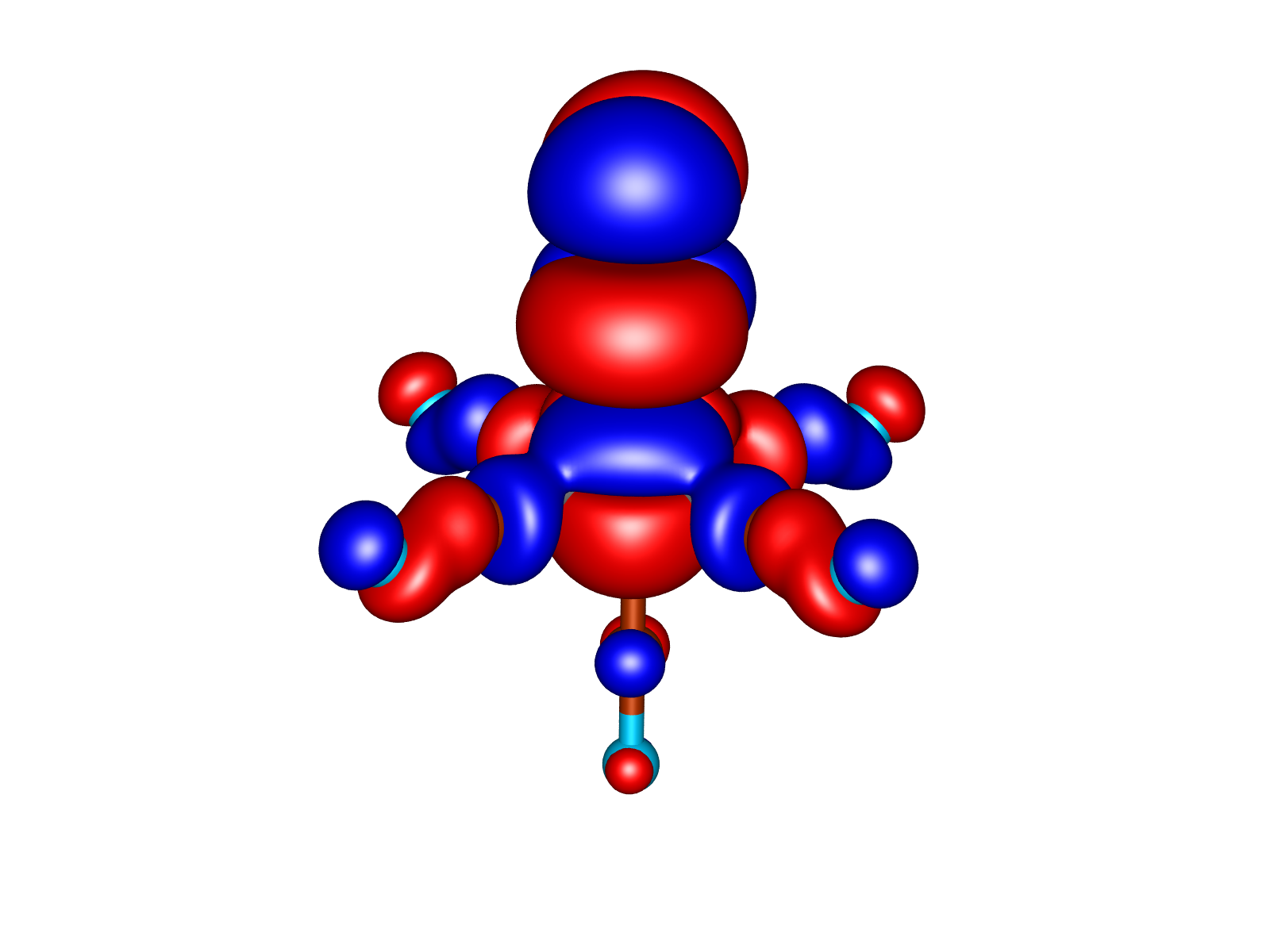}
  } \\
  \caption{Fe-NO complex, standard, CASSCF(4, 4) \label{orbs_casscf0404_1}}
\end{figure}

\renewcommand{\thesubfigure}{\arabic{subfigure}}
\begin{figure}[!h]
  \subfloat[$n_{\text{occup}}$ = 1.9024]{%
    \includegraphics[width=0.22\textwidth]{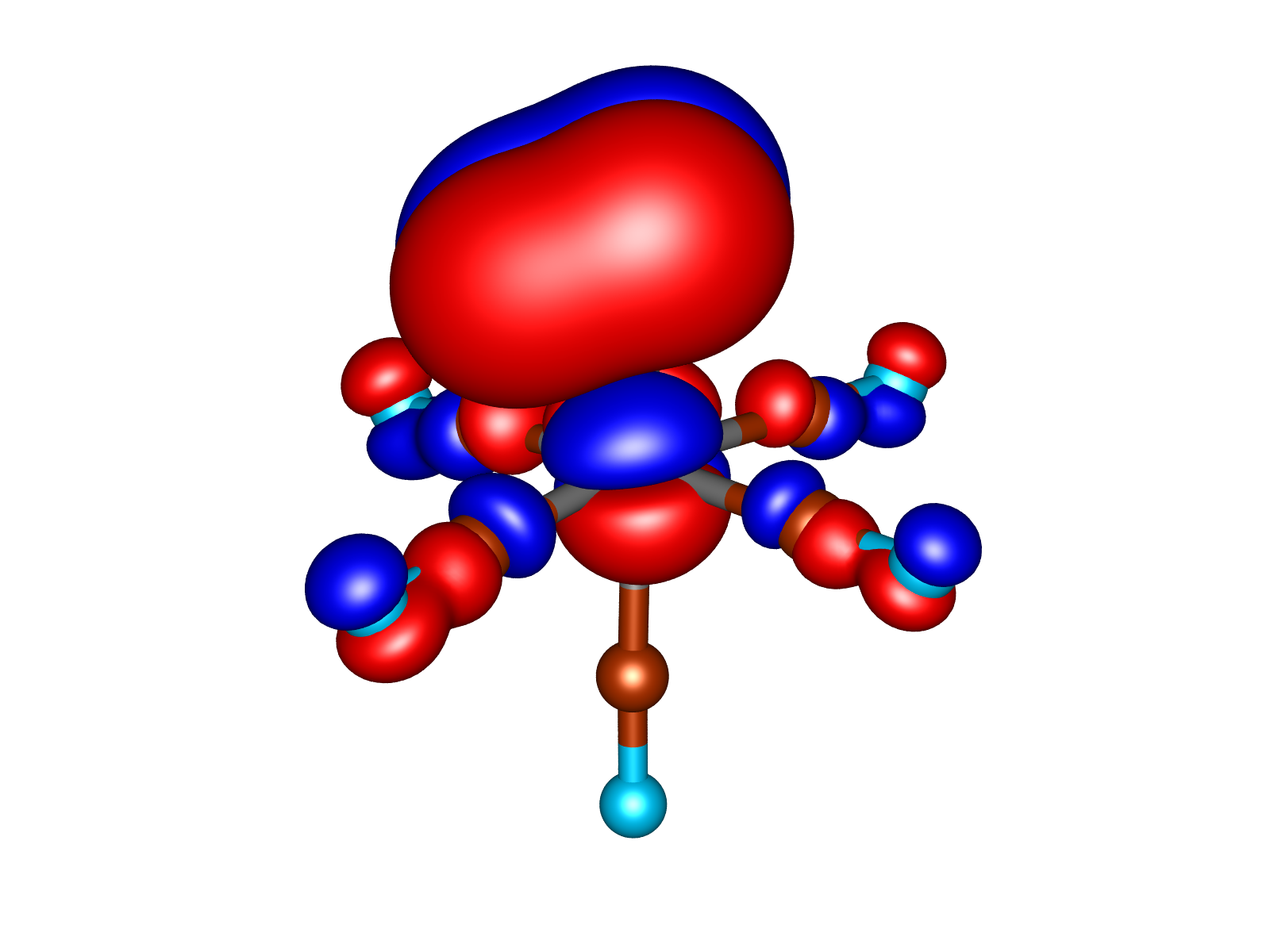}
  }
  \hfill
  \subfloat[$n_{\text{occup}}$ = 1.6449]{%
    \includegraphics[width=0.22\textwidth]{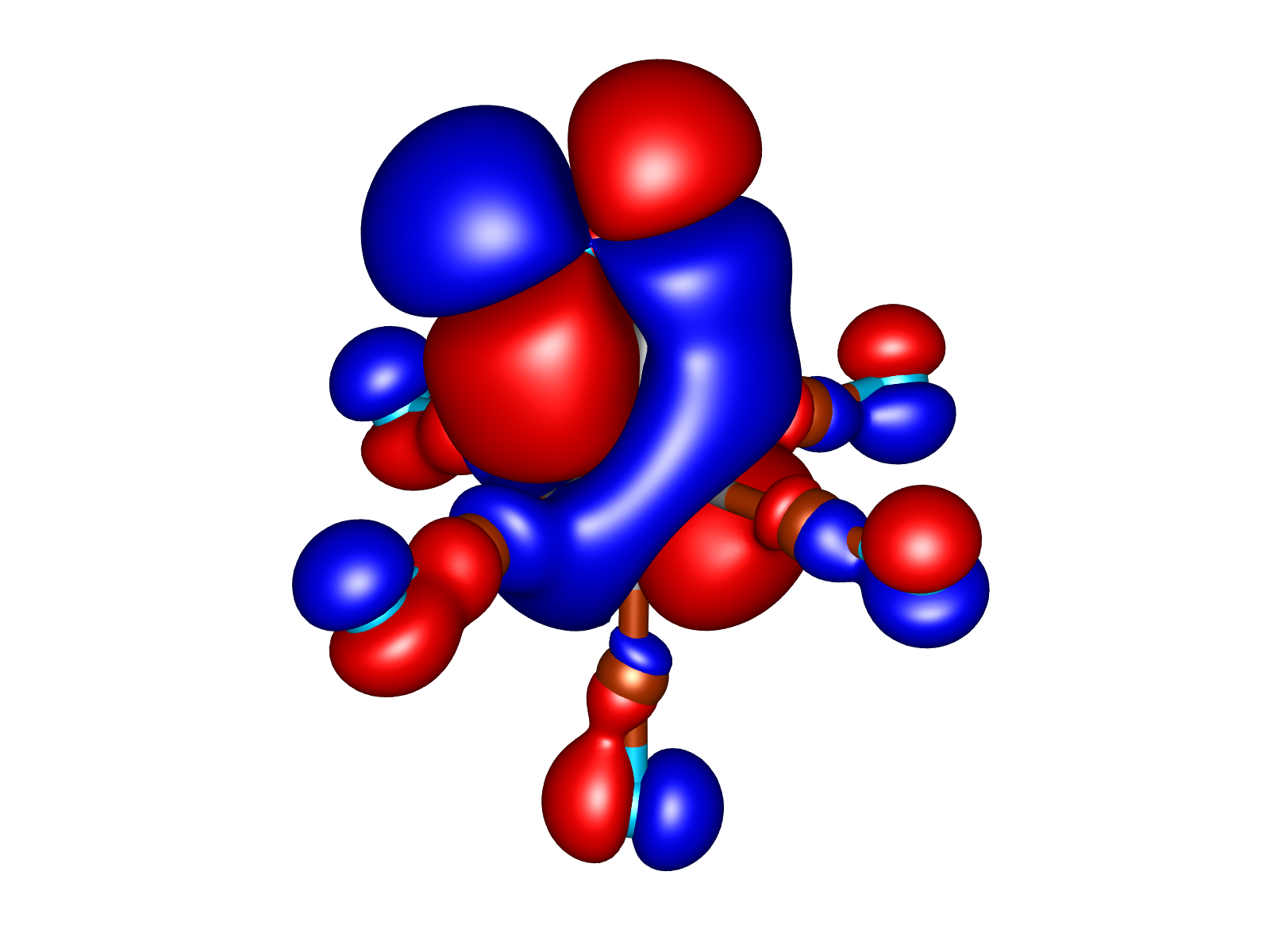}
  }
  \hfill
  \subfloat[$n_{\text{occup}}$ = 0.3550]{%
    \includegraphics[width=0.22\textwidth]{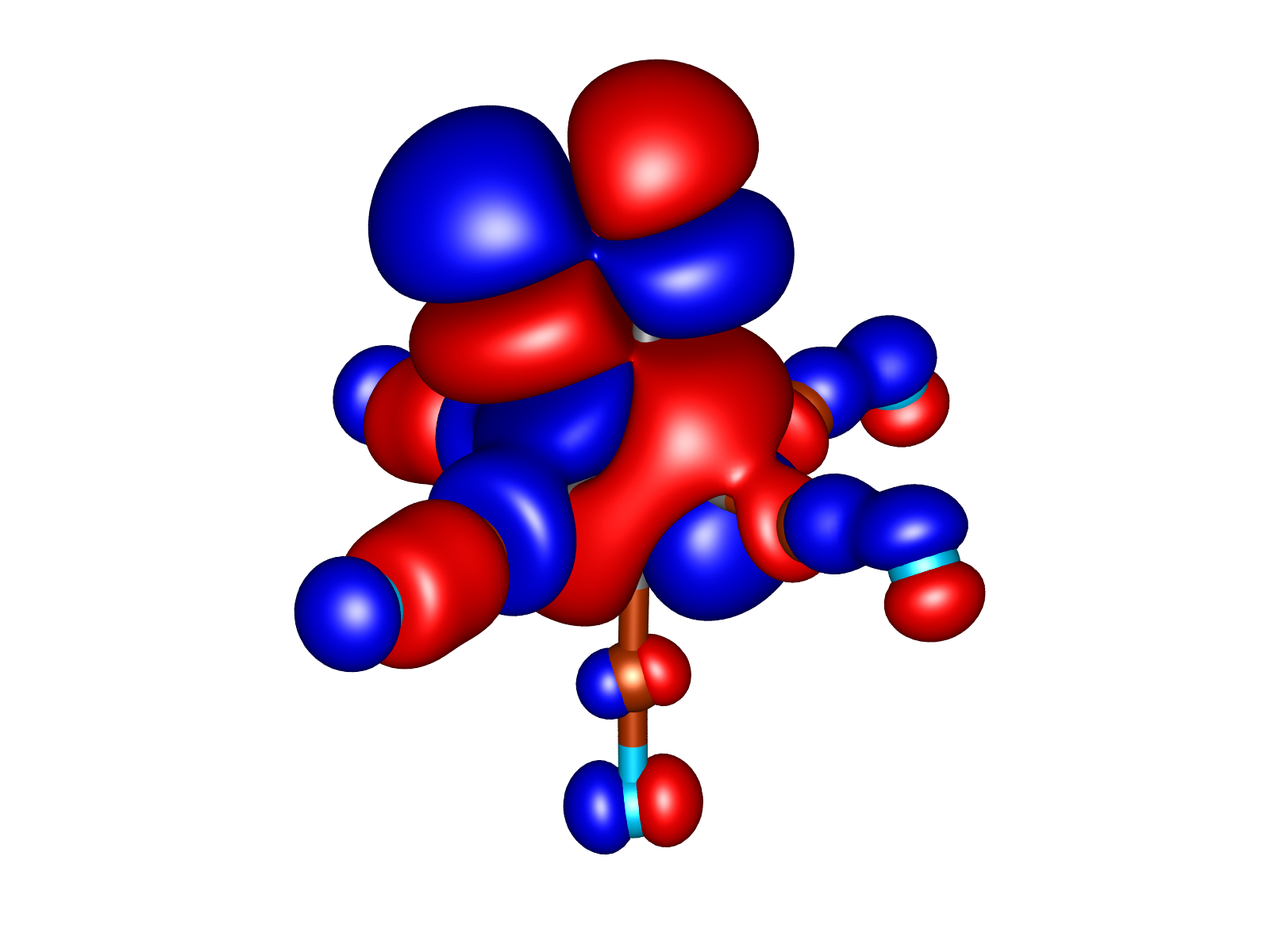}
  }
  \hfill
  \subfloat[$n_{\text{occup}}$ = 0.0977]{%
    \includegraphics[width=0.22\textwidth]{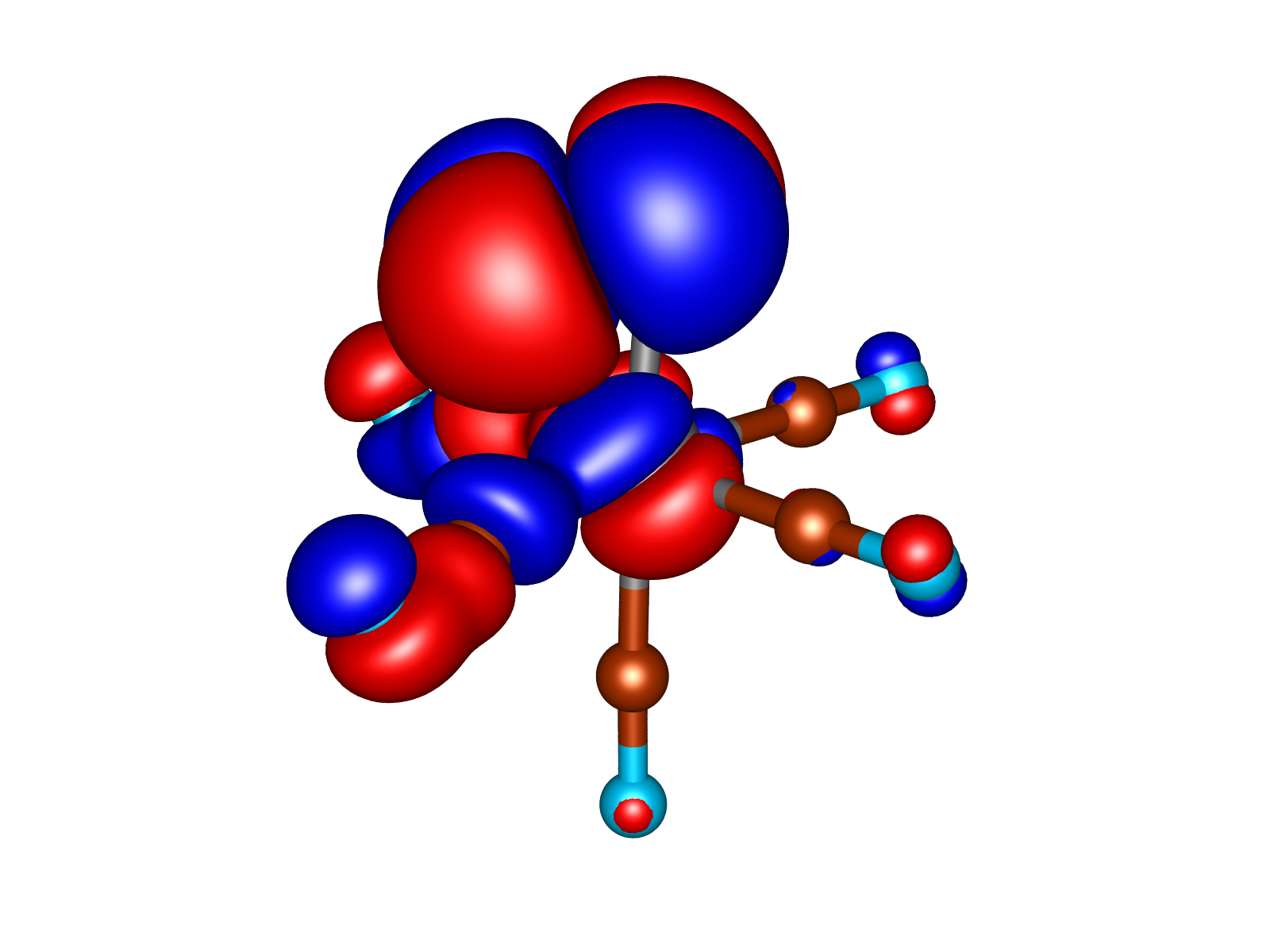}
  } \\
  \caption{Fe-NO complex, flat, CASSCF(4, 4) \label{orbs_casscf0404_3}}
\end{figure}

\renewcommand{\thesubfigure}{\arabic{subfigure}}
\begin{figure}[!h]
  \subfloat[$n_{\text{occup}}$ = 1.6793]{%
    \includegraphics[width=0.22\textwidth]{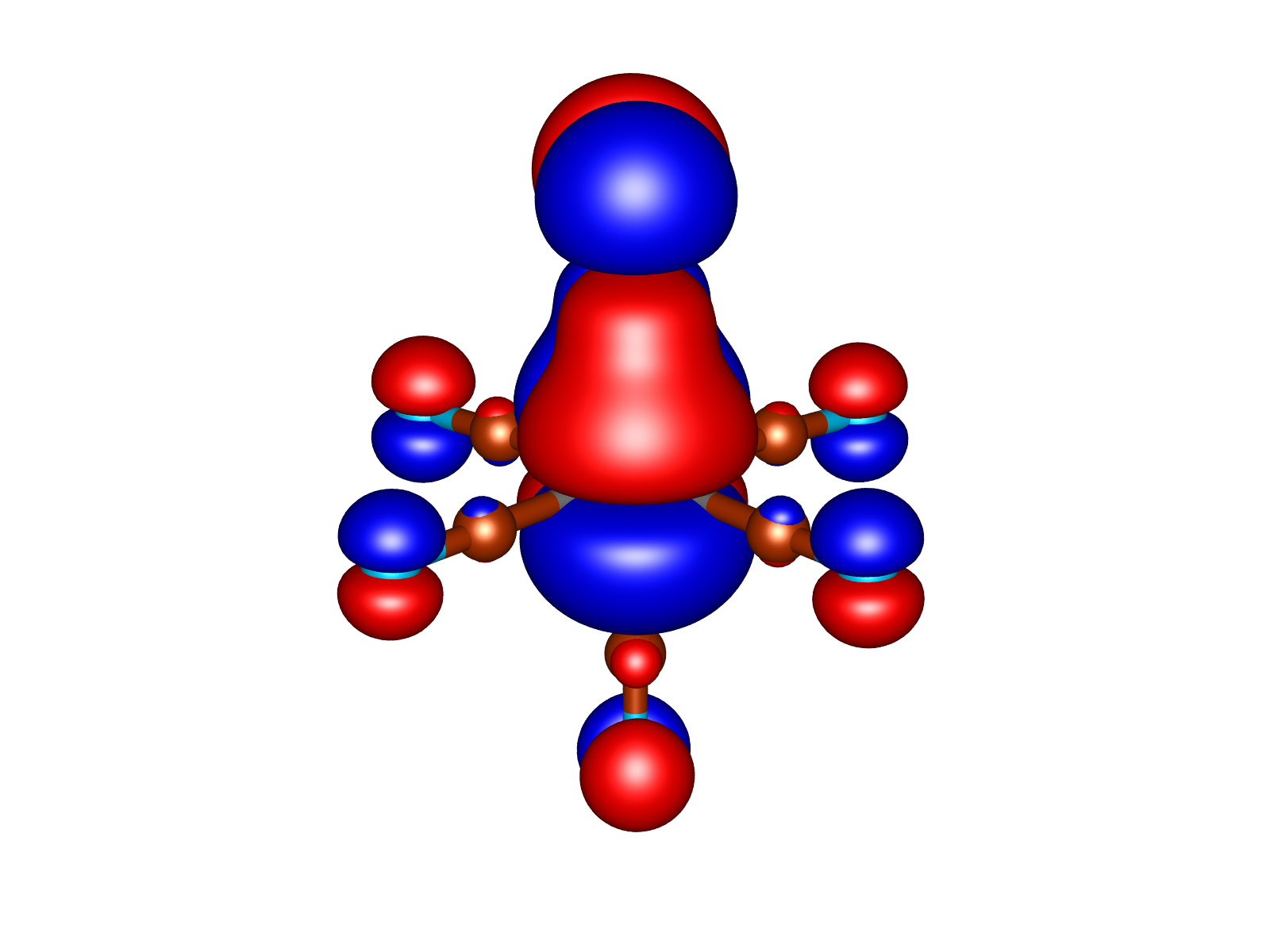}
  }
  \hfill
  \subfloat[$n_{\text{occup}}$ = 1.6679]{%
    \includegraphics[width=0.22\textwidth]{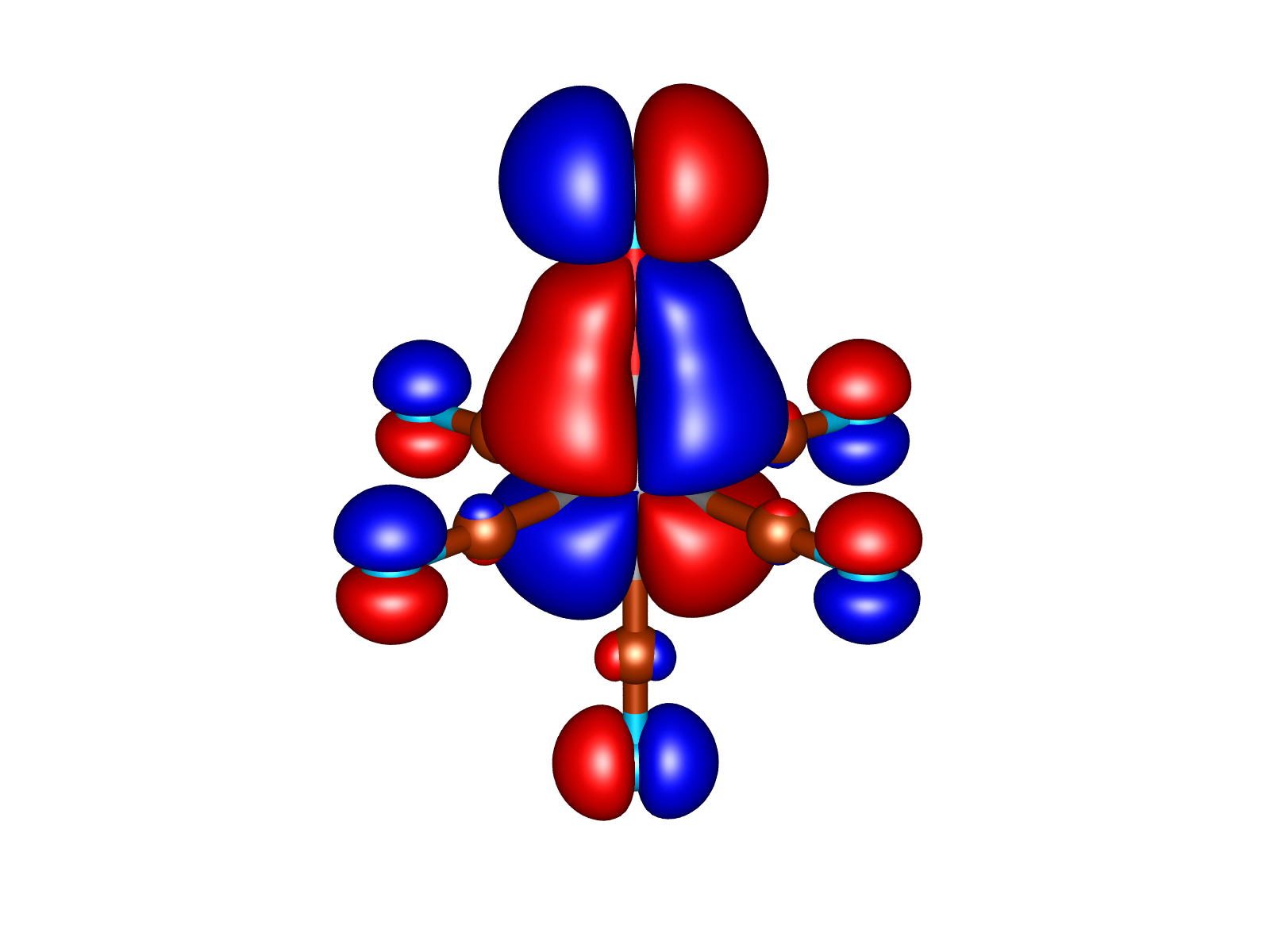}
  }
  \hfill
  \subfloat[$n_{\text{occup}}$ = 0.3321]{%
    \includegraphics[width=0.22\textwidth]{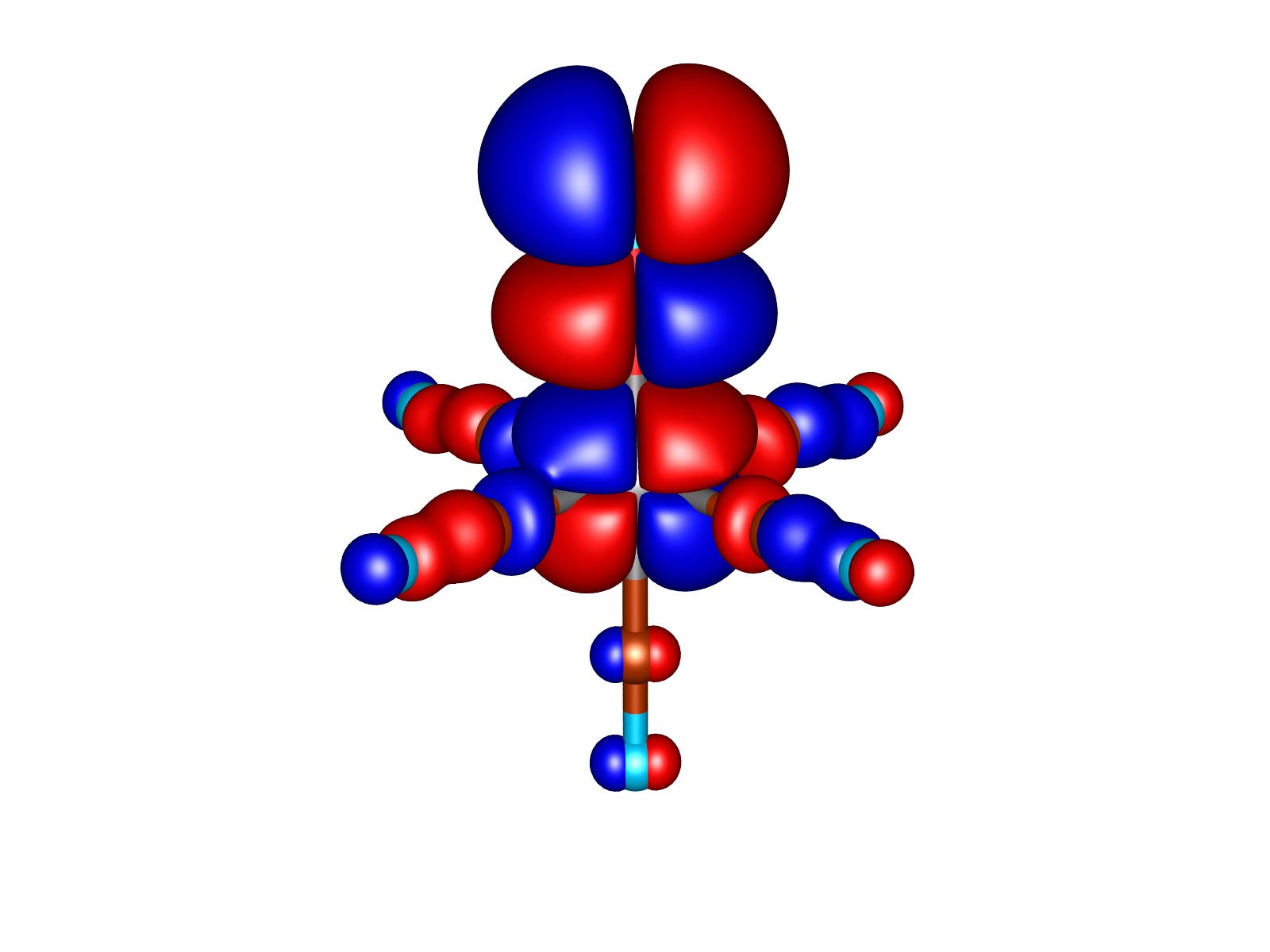}
  }
  \hfill
  \subfloat[$n_{\text{occup}}$ = 0.3207]{%
    \includegraphics[width=0.22\textwidth]{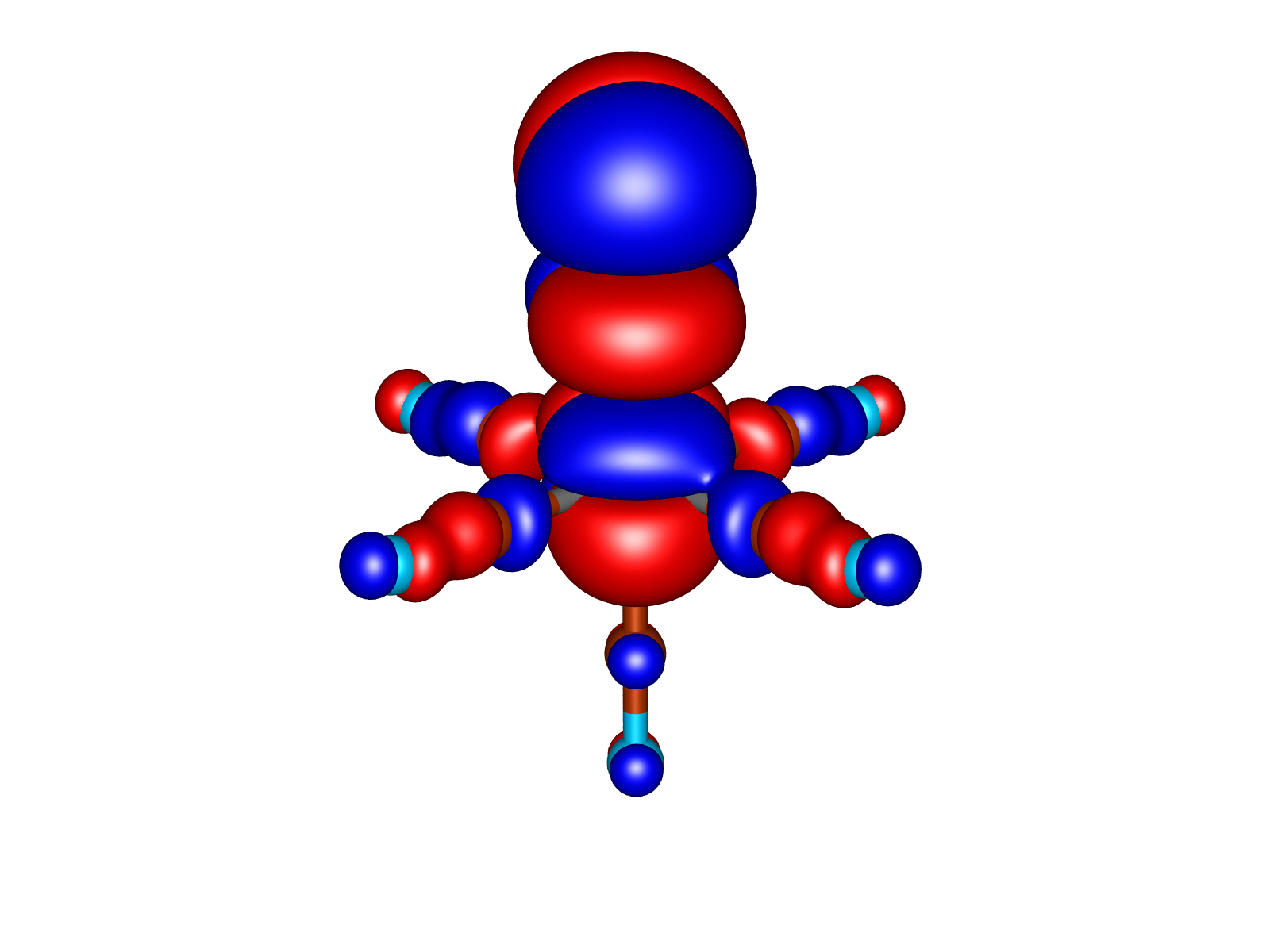}
  } \\
  \caption{Fe-NO complex, reverse, CASSCF(4, 4) \label{orbs_casscf0404_5}}
\end{figure}

\renewcommand{\thesubfigure}{\arabic{subfigure}}
\begin{figure}[!h]
  \subfloat[$n_{\text{occup}}$ = 1.9639]{%
    \includegraphics[width=0.22\textwidth]{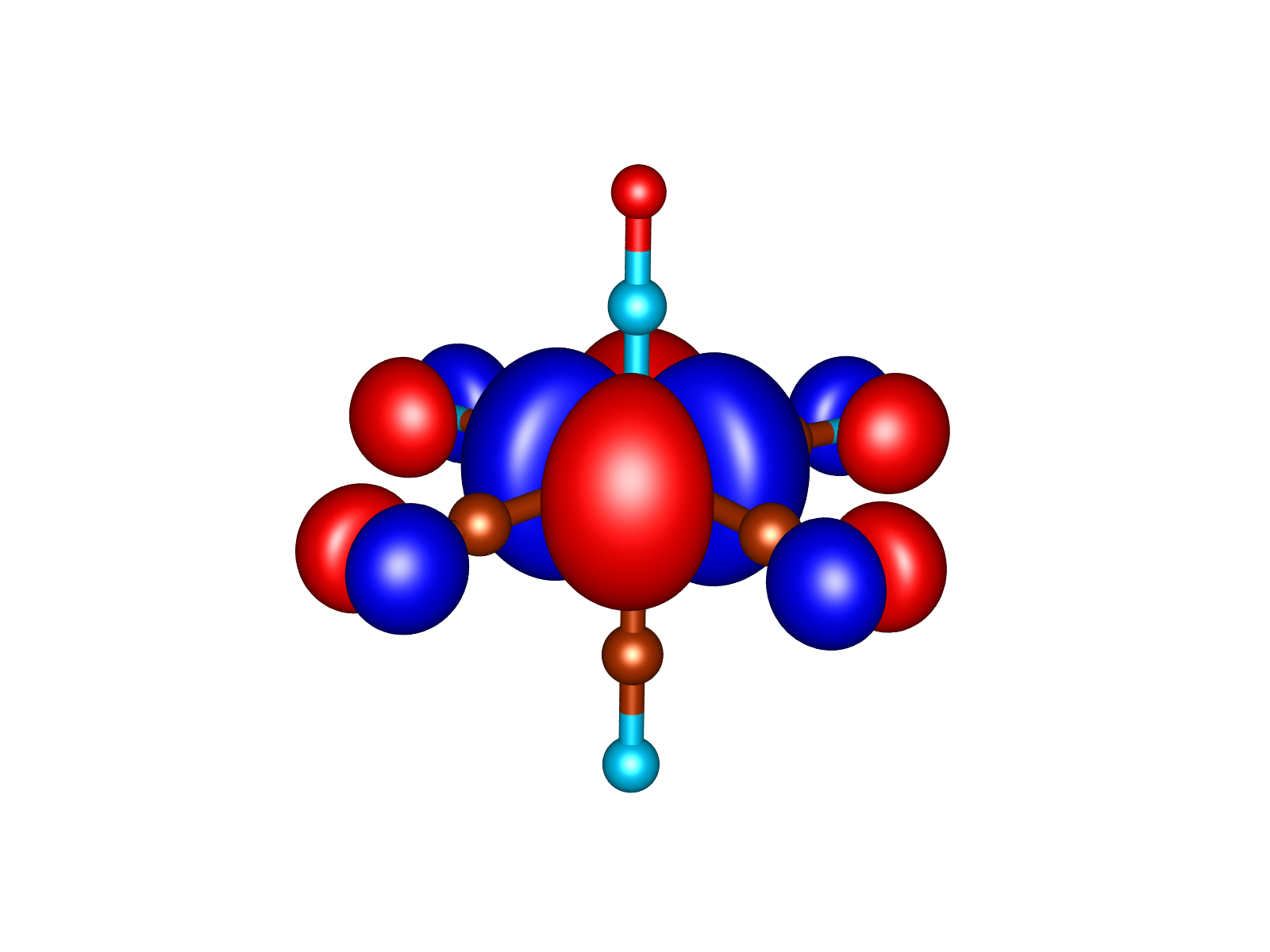}
  }
  \subfloat[$n_{\text{occup}}$ = 1.9577]{%
    \includegraphics[width=0.22\textwidth]{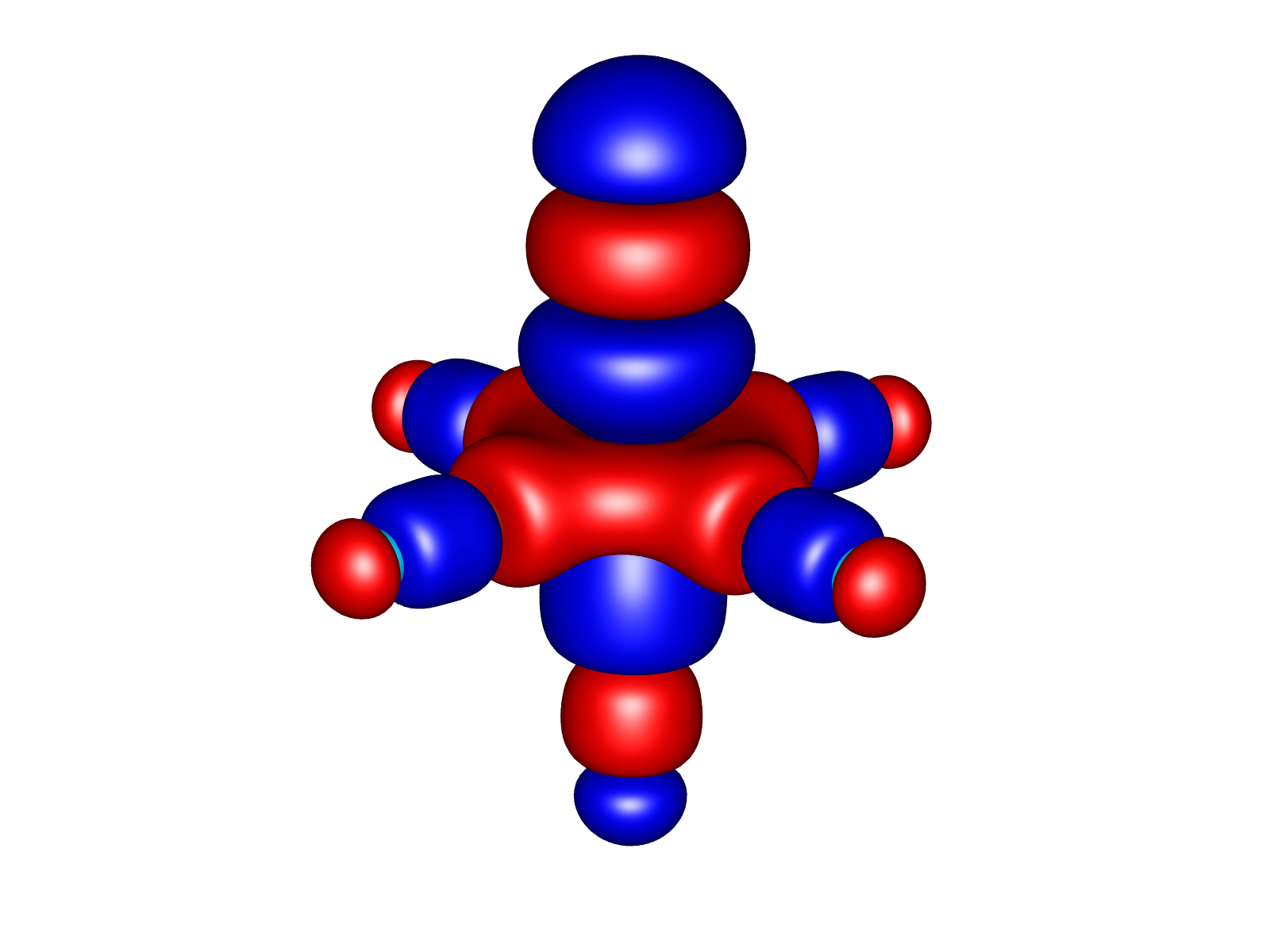}
  }
  \subfloat[$n_{\text{occup}}$ = 1.9559]{%
    \includegraphics[width=0.22\textwidth]{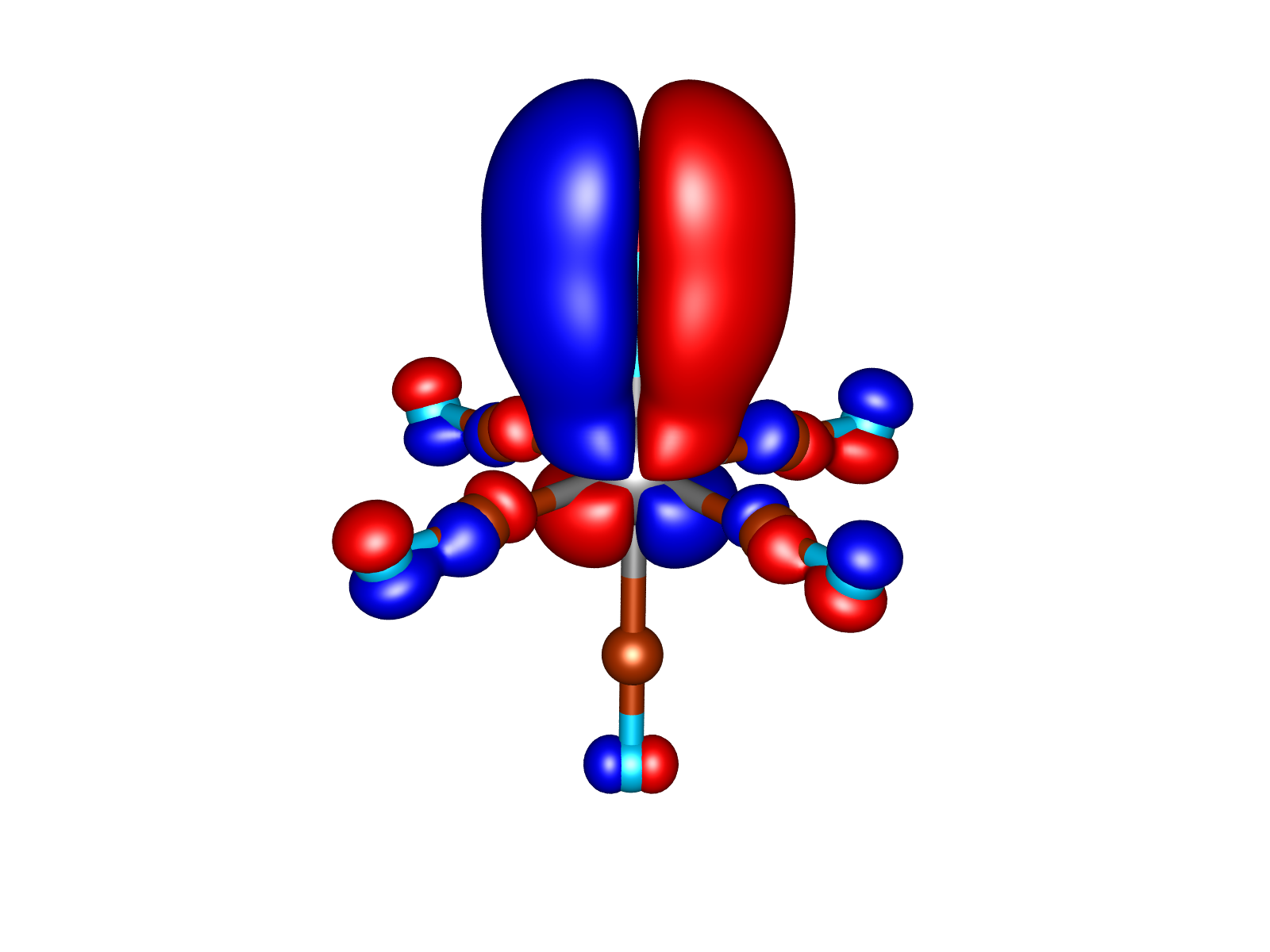}
  } \\
  \subfloat[$n_{\text{occup}}$ = 1.9559]{%
    \includegraphics[width=0.22\textwidth]{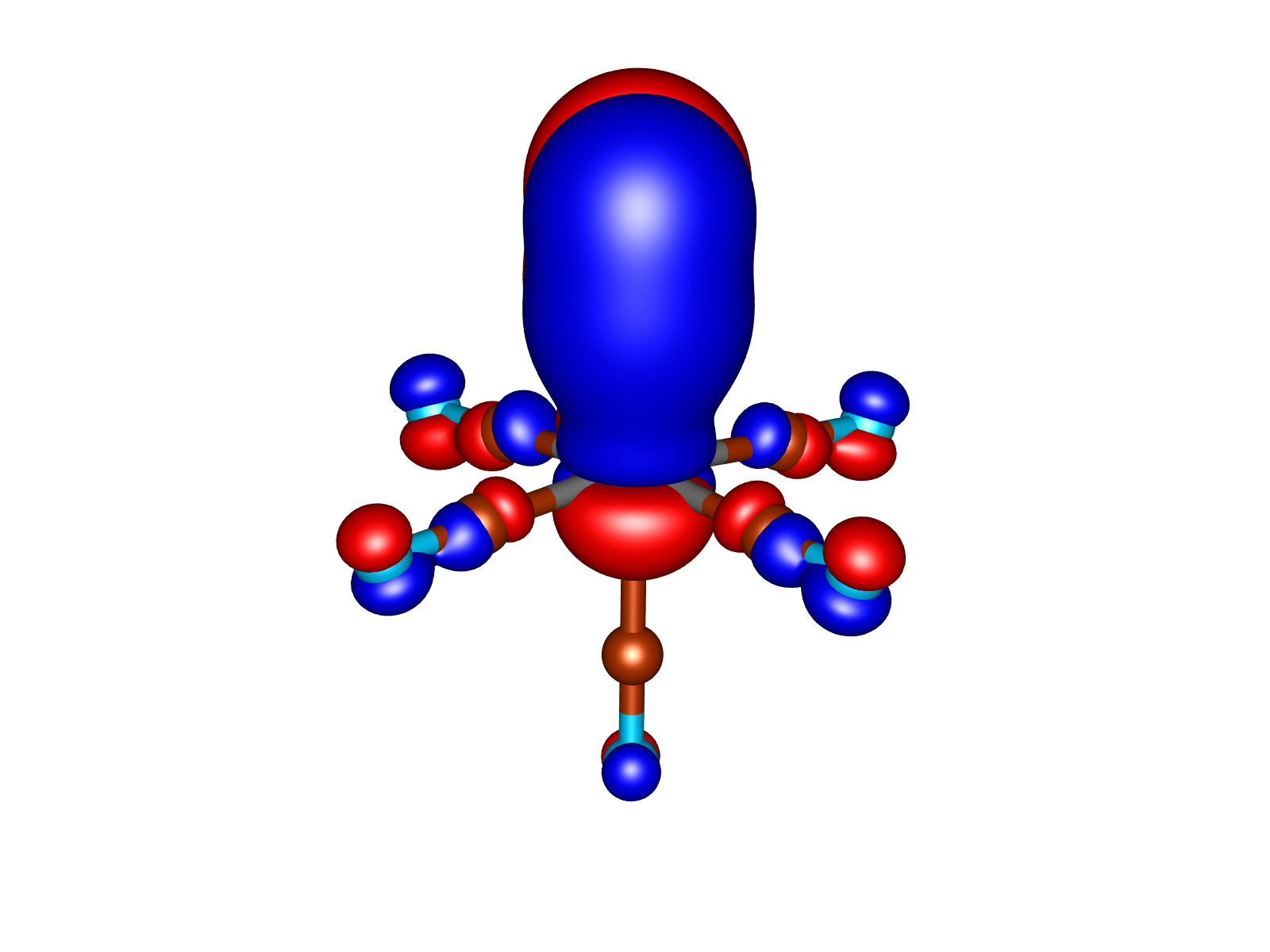}
  }
  \hfill
  \subfloat[$n_{\text{occup}}$ = 1.9539]{%
    \includegraphics[width=0.22\textwidth]{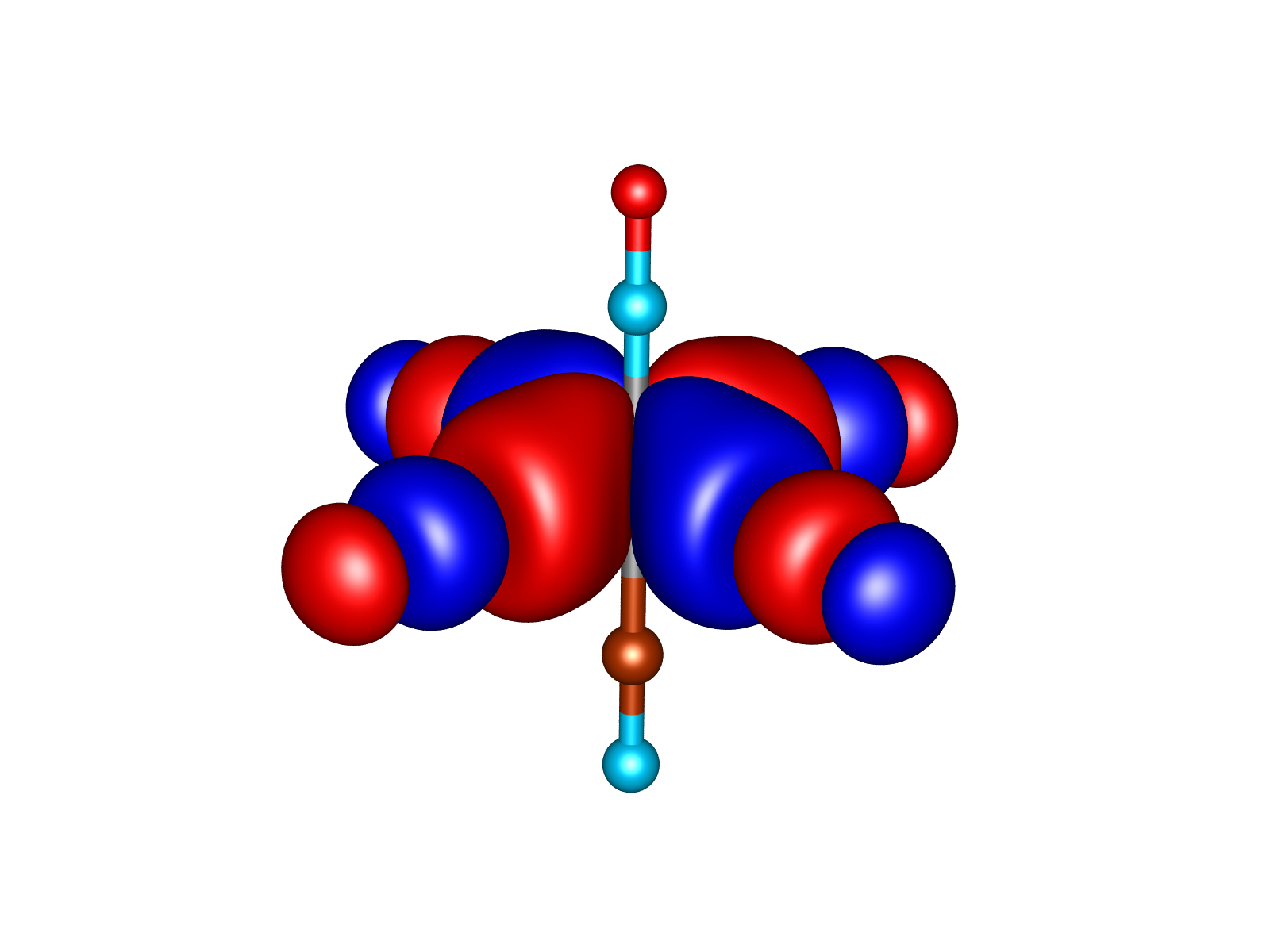}
  }
  \hfill
  \subfloat[$n_{\text{occup}}$ = 1.8246]{%
    \includegraphics[width=0.22\textwidth]{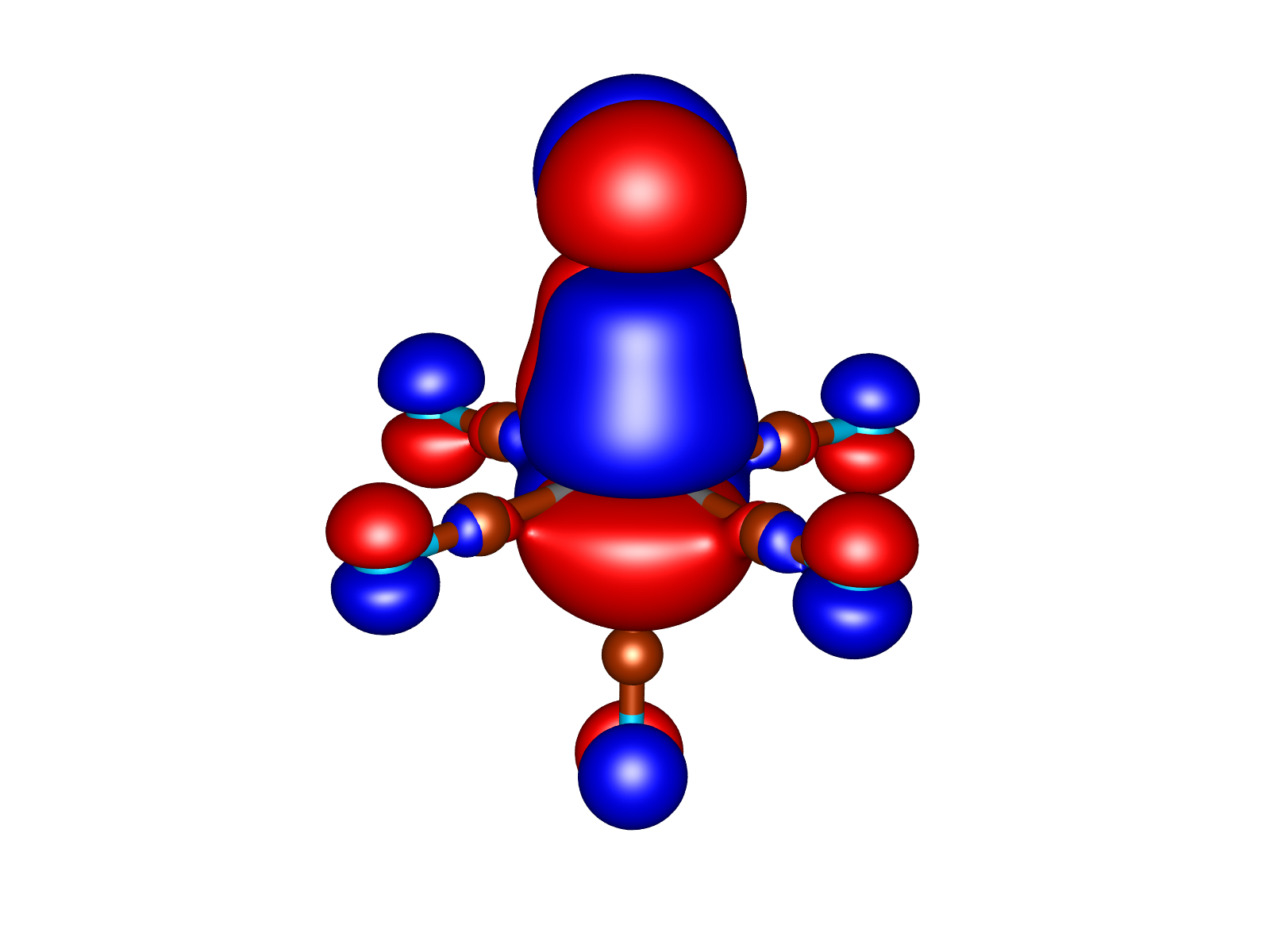}
  }
  \hfill
  \subfloat[$n_{\text{occup}}$ = 1.8246]{%
    \includegraphics[width=0.22\textwidth]{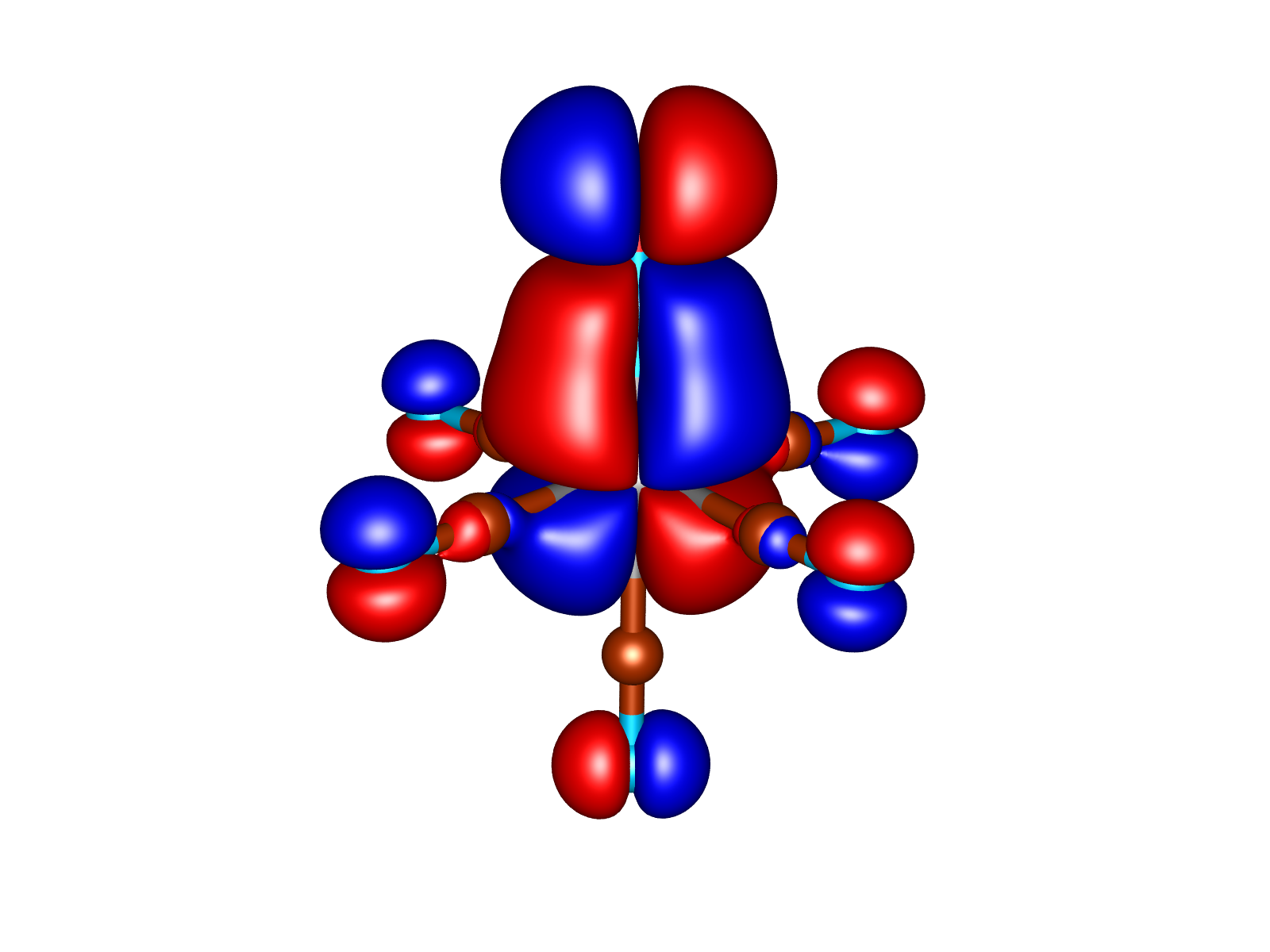}
  }
  \\
  \subfloat[$n_{\text{occup}}$ = 0.2084]{%
    \includegraphics[width=0.22\textwidth]{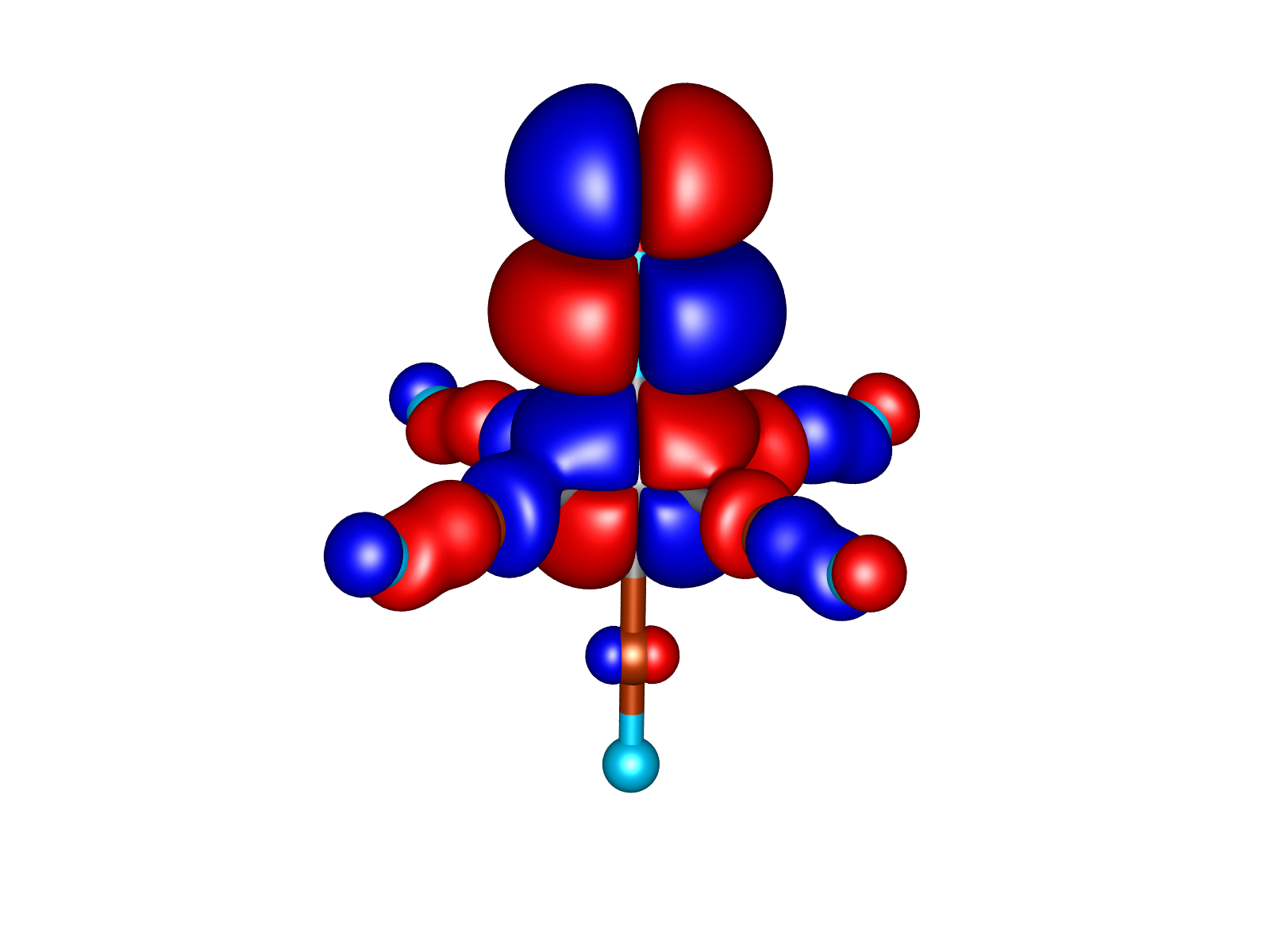}
  }
  \hfill
  \subfloat[$n_{\text{occup}}$ = 0.2084]{%
    \includegraphics[width=0.22\textwidth]{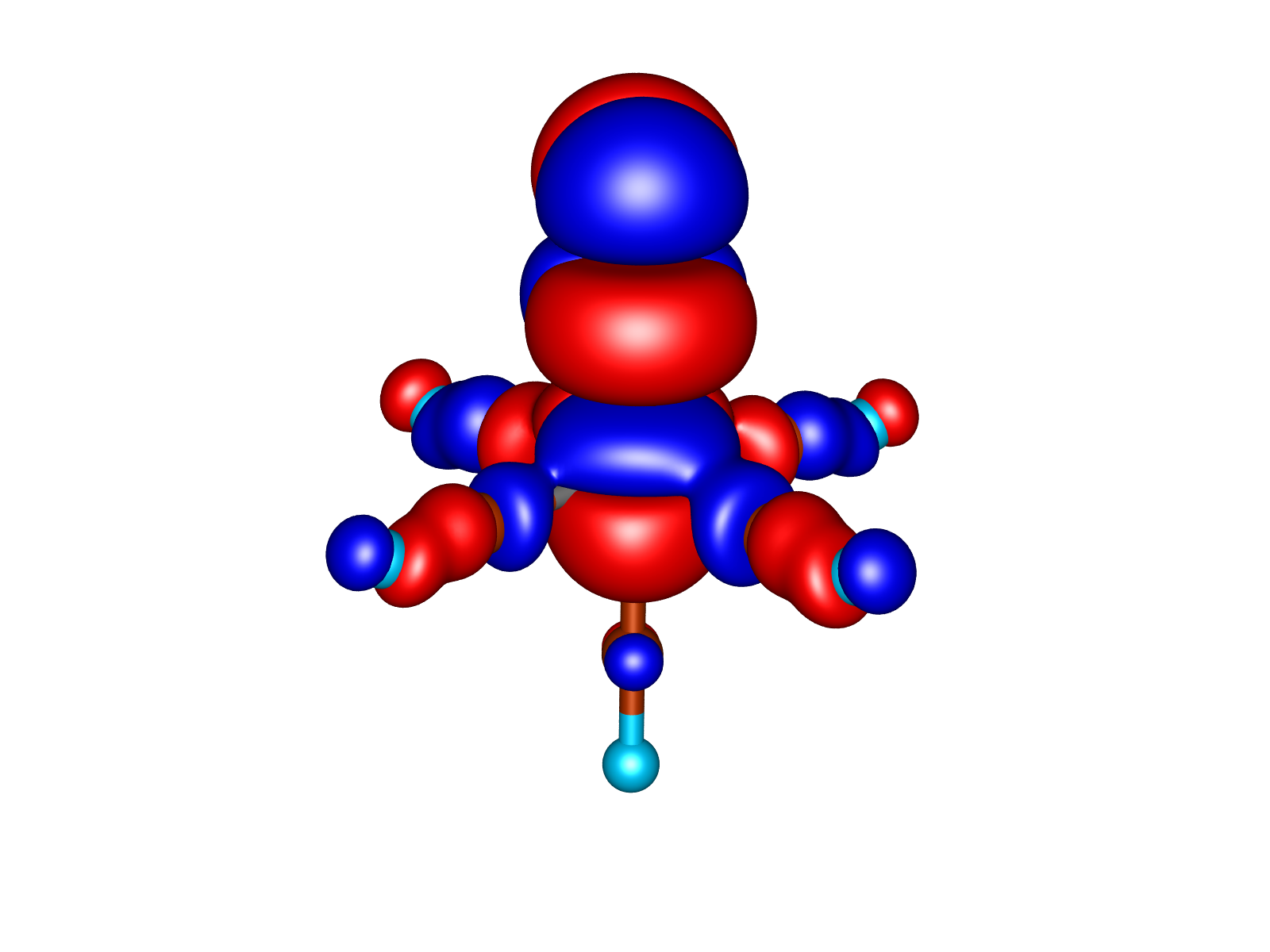}
  }
  \hfill
  \subfloat[$n_{\text{occup}}$ = 0.0510]{%
    \includegraphics[width=0.22\textwidth]{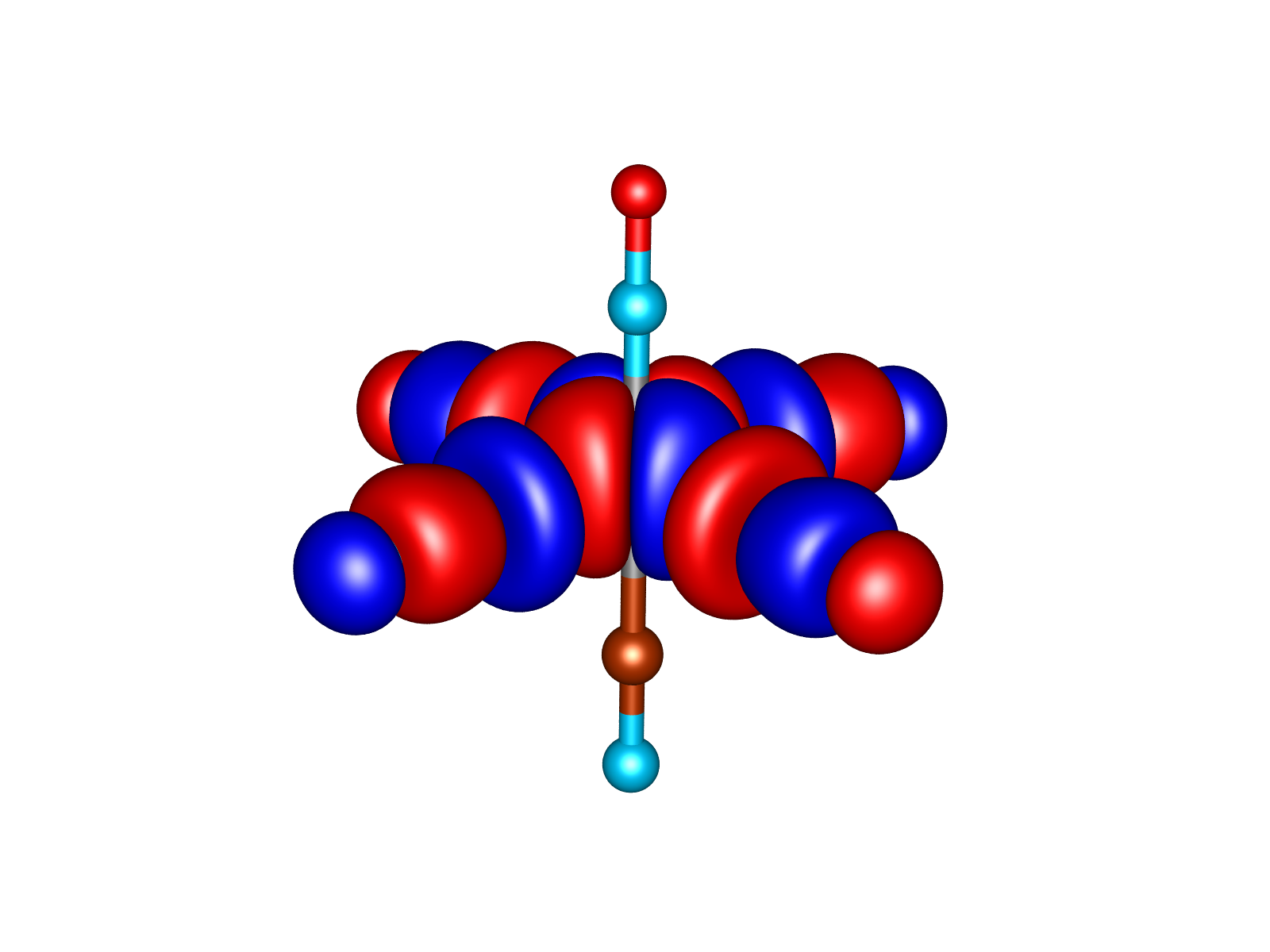}
  }
  \hfill
  \subfloat[$n_{\text{occup}}$ = 0.0404]{%
    \includegraphics[width=0.22\textwidth]{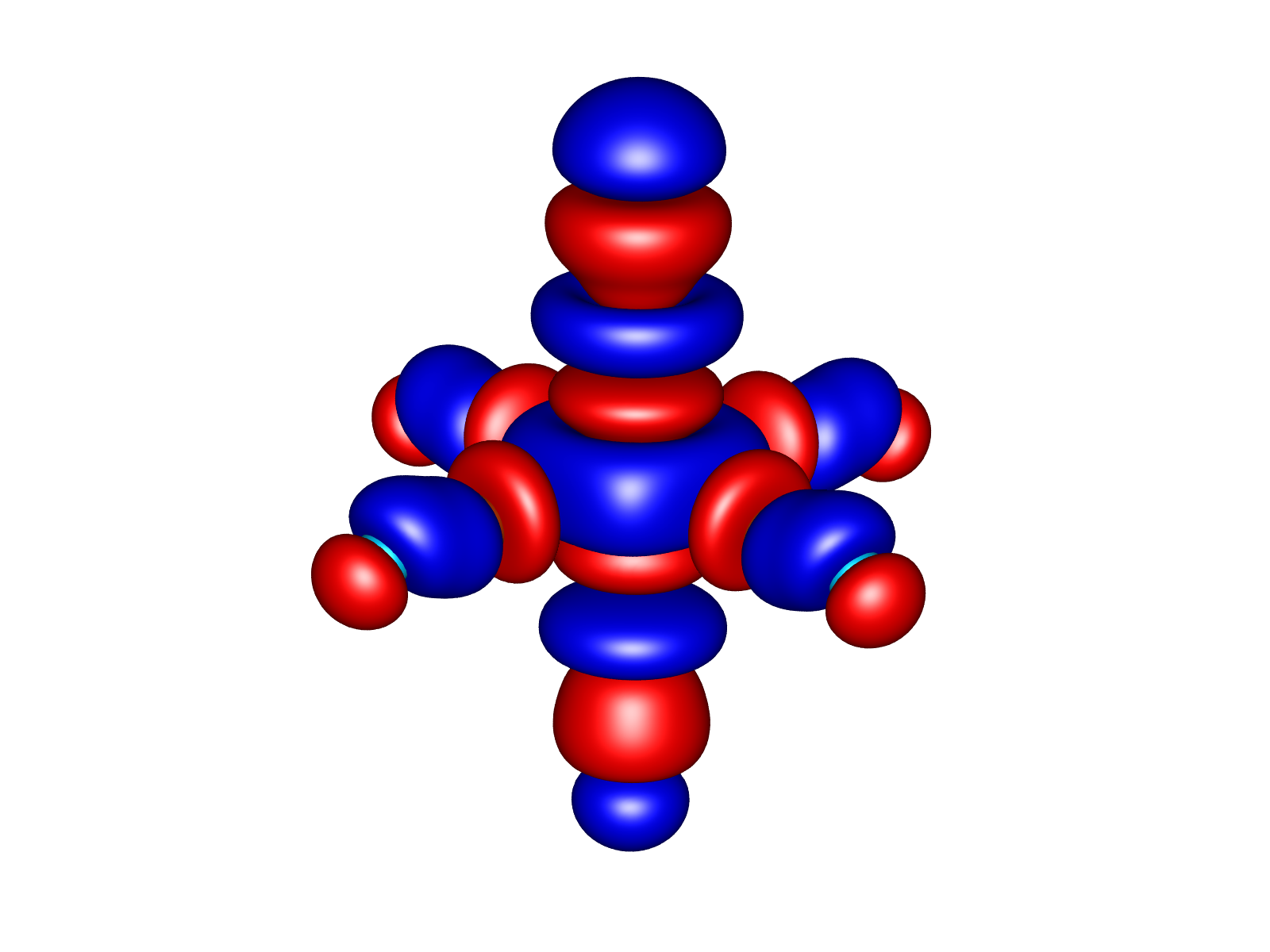}
  }
  \\
  \subfloat[$n_{\text{occup}}$ = 0.0204]{%
    \includegraphics[width=0.22\textwidth]{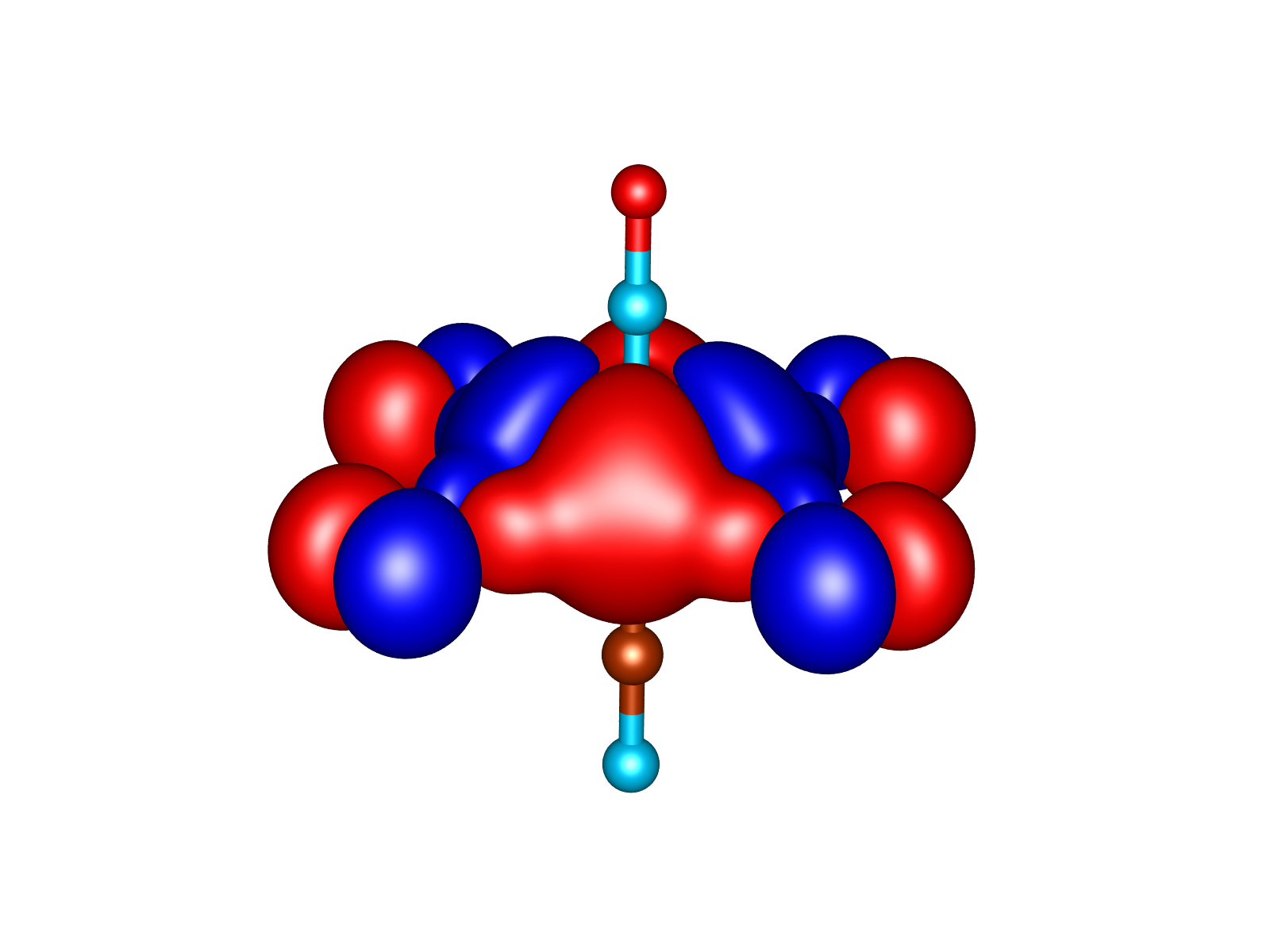}
  }
  \hfill
  \subfloat[$n_{\text{occup}}$ = 0.0156]{%
    \includegraphics[width=0.22\textwidth]{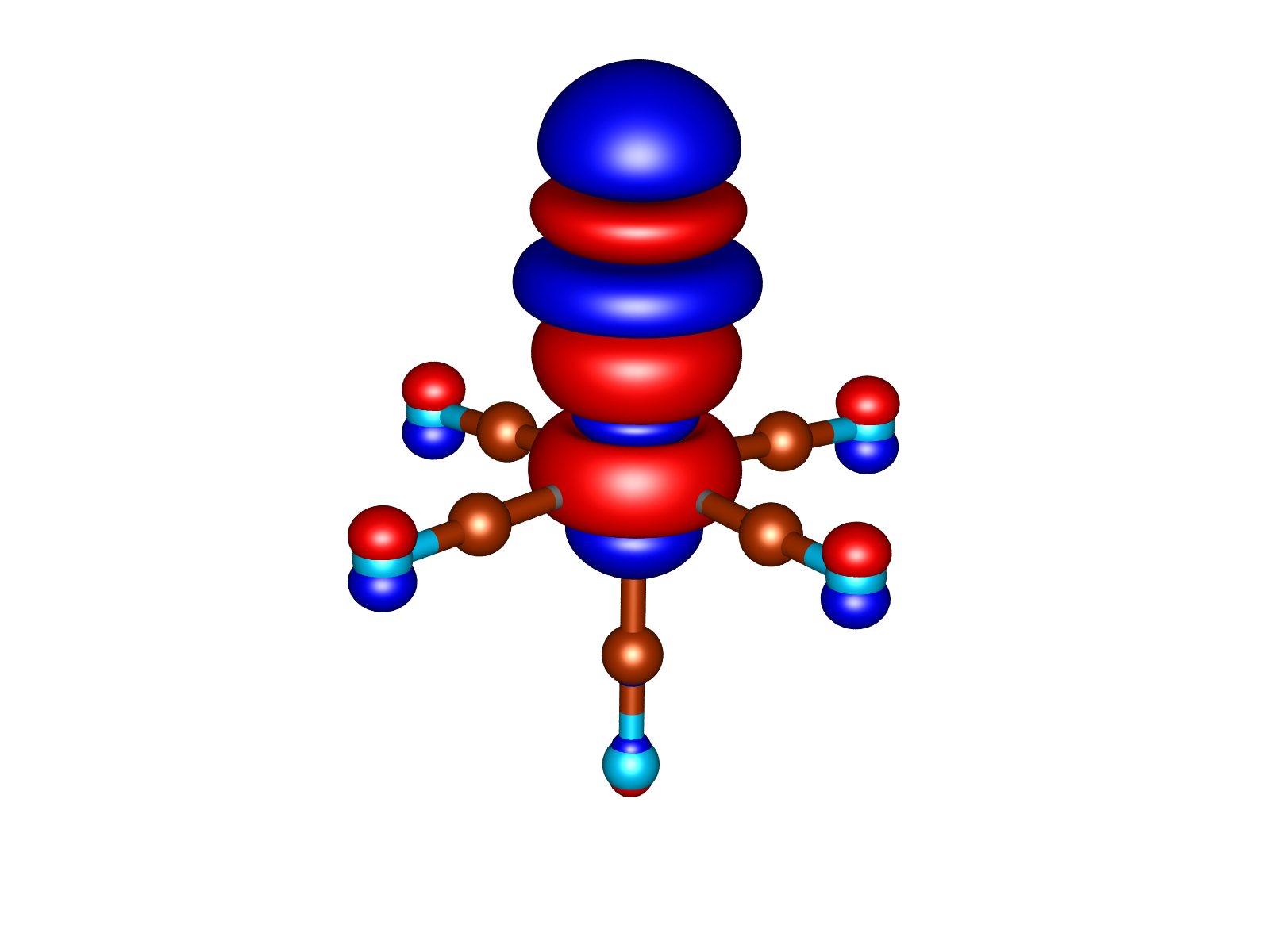}
  }
  \hfill
  \subfloat[$n_{\text{occup}}$ = 0.0095]{%
    \includegraphics[width=0.22\textwidth]{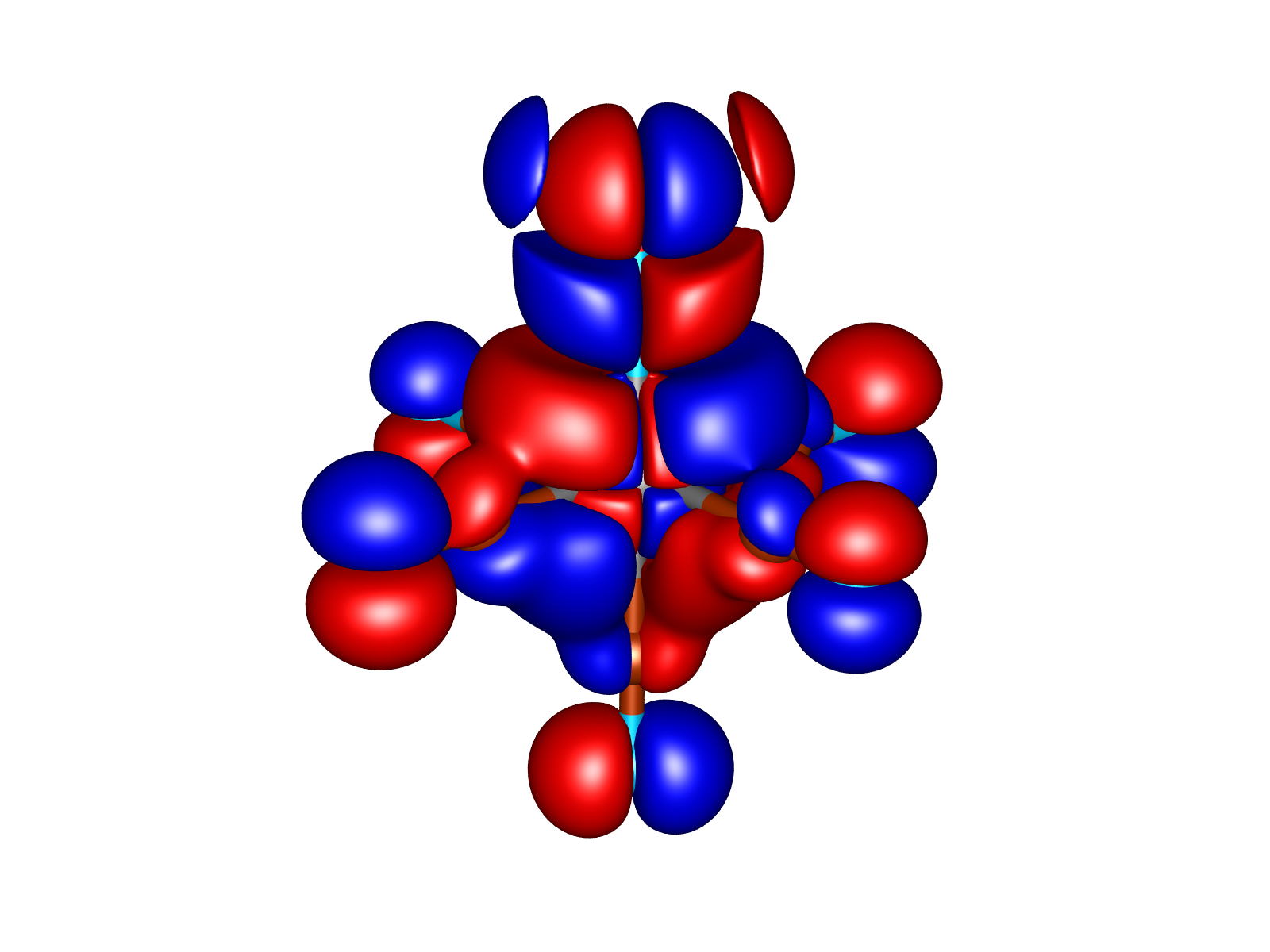}
  }
  \hfill
  \subfloat[$n_{\text{occup}}$ = 0.0095]{%
    \includegraphics[width=0.22\textwidth]{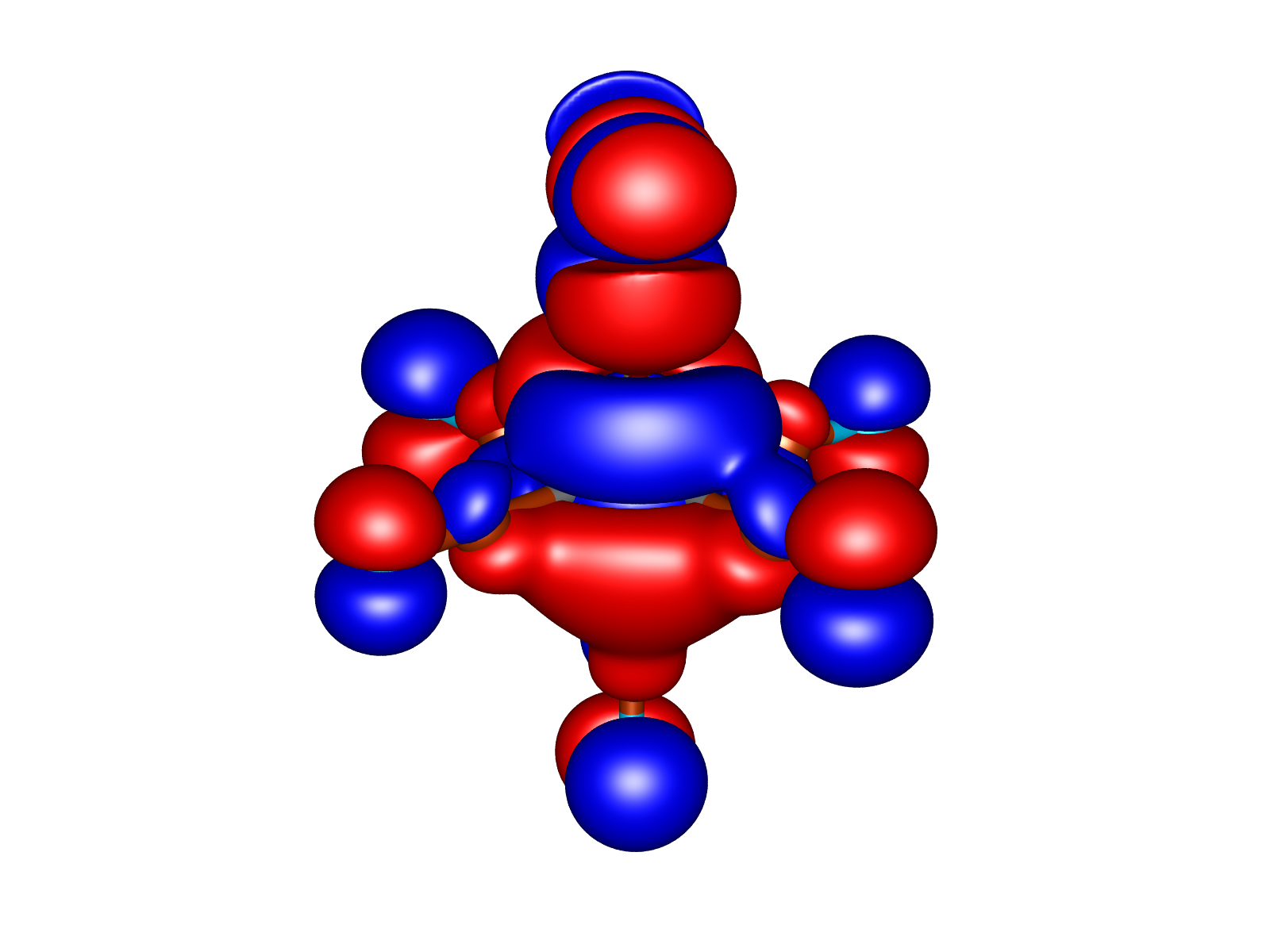}
  }
  \\
  \caption{Fe-NO complex, standard, CASSCF(14, 15) \label{orbs_cas1415_1}}
\end{figure}

\renewcommand{\thesubfigure}{\arabic{subfigure}}
\begin{figure}[!h]
  \subfloat[$n_{\text{occup}}$ = 1.9721]{%
    \includegraphics[width=0.22\textwidth]{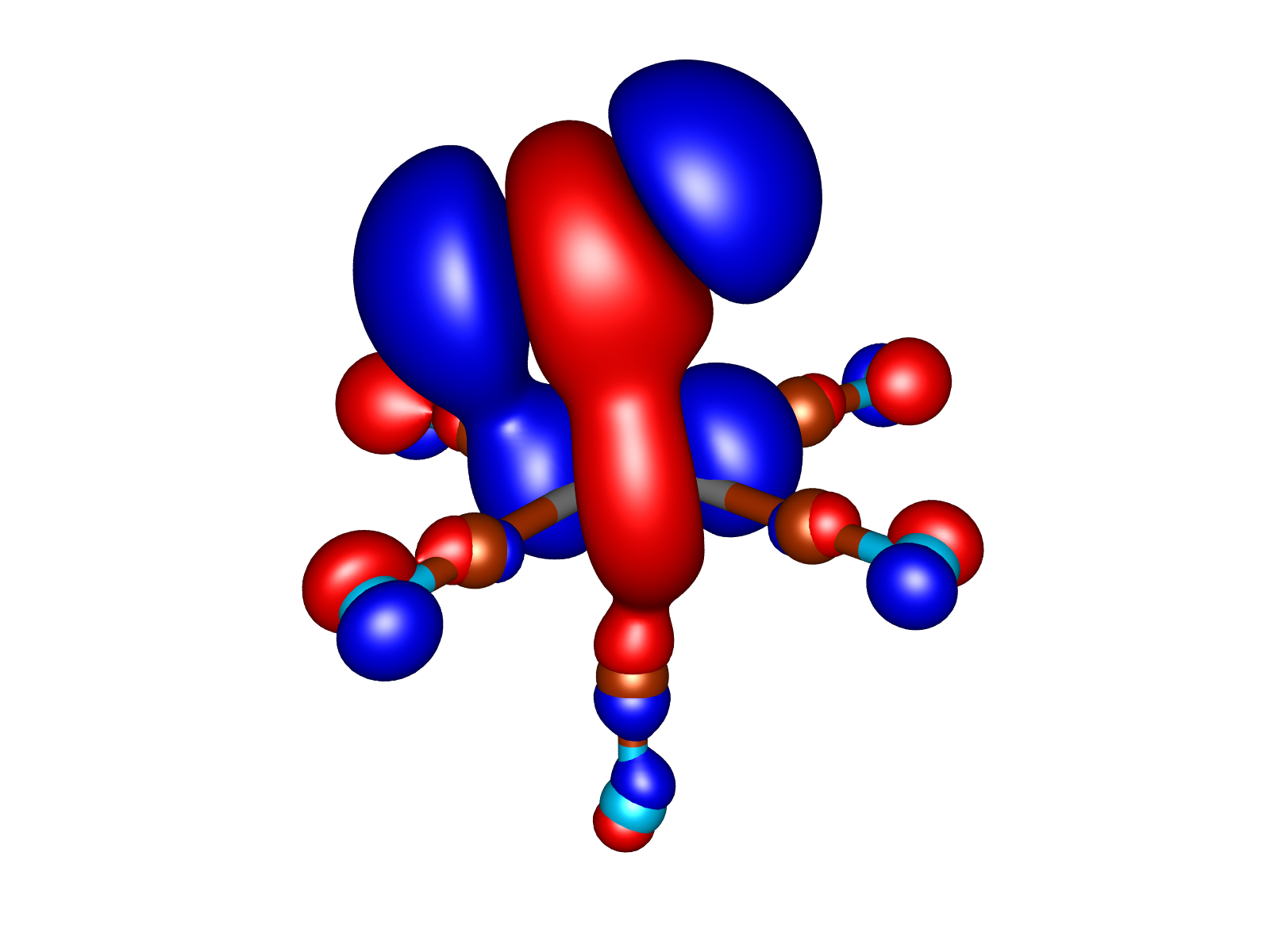}
  }
  \subfloat[$n_{\text{occup}}$ = 1.9650]{%
    \includegraphics[width=0.22\textwidth]{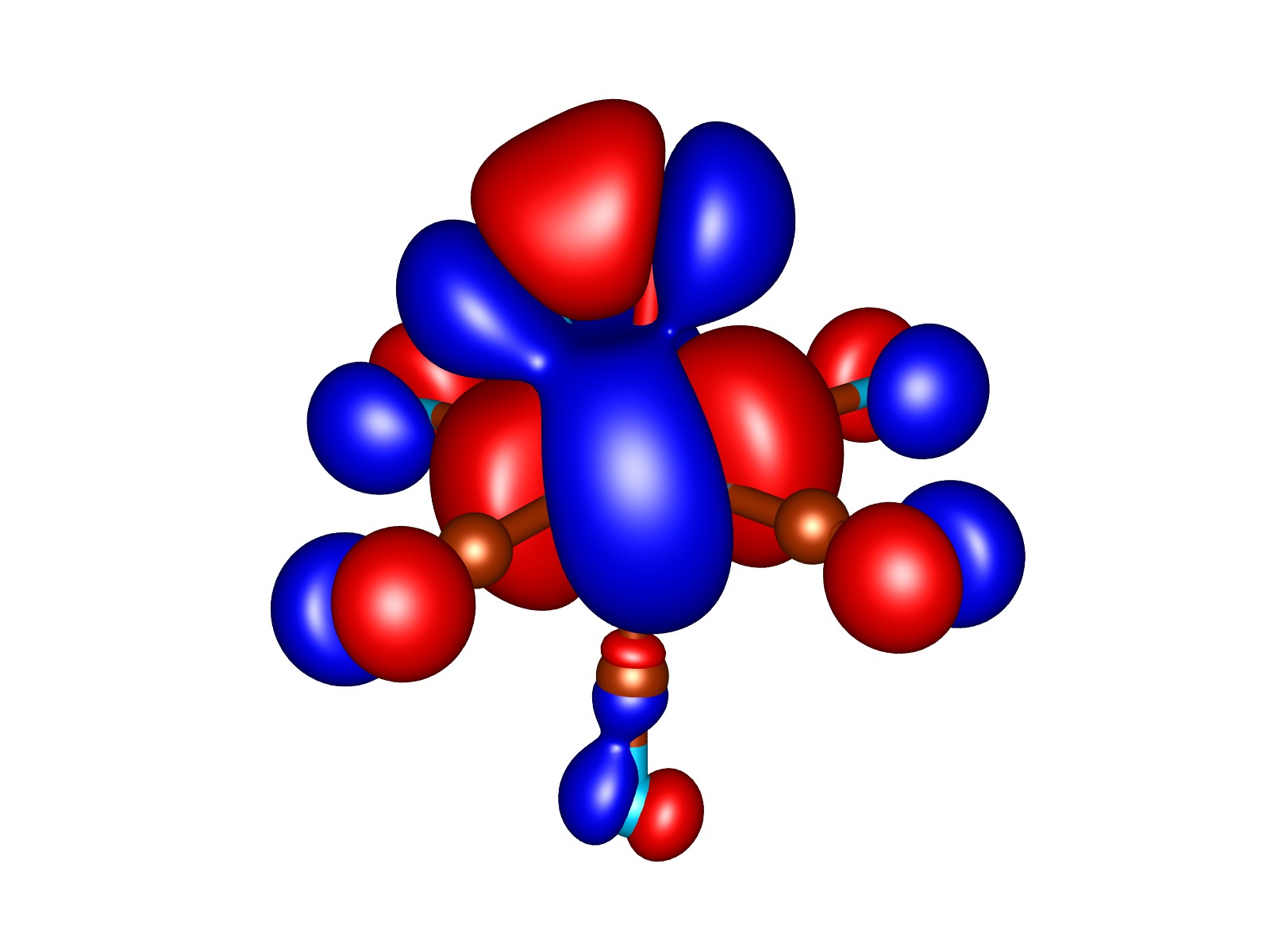}
  }
  \subfloat[$n_{\text{occup}}$ = 1.9607]{%
    \includegraphics[width=0.22\textwidth]{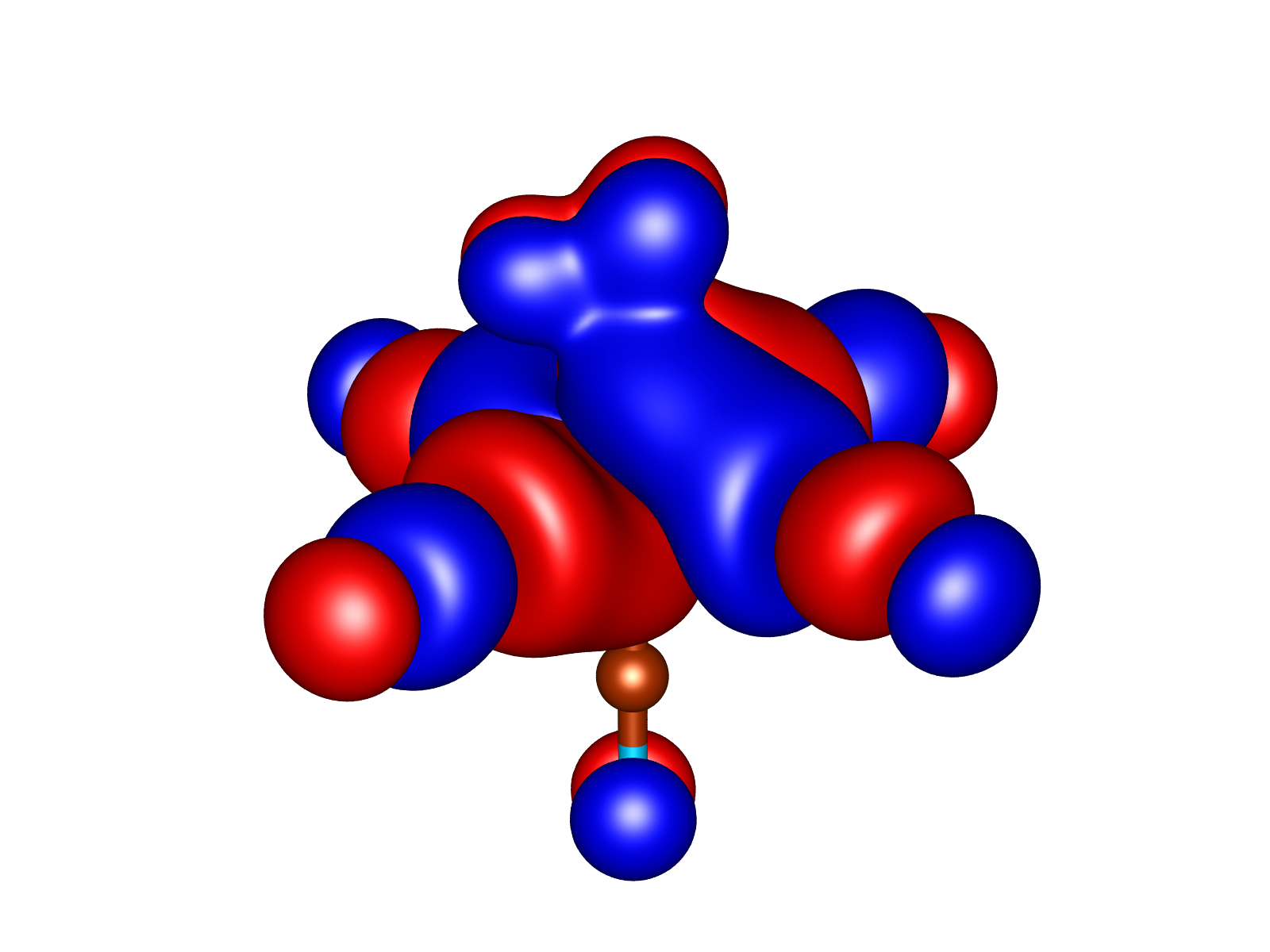}
  } \\
  \subfloat[$n_{\text{occup}}$ = 1.9535]{%
    \includegraphics[width=0.22\textwidth]{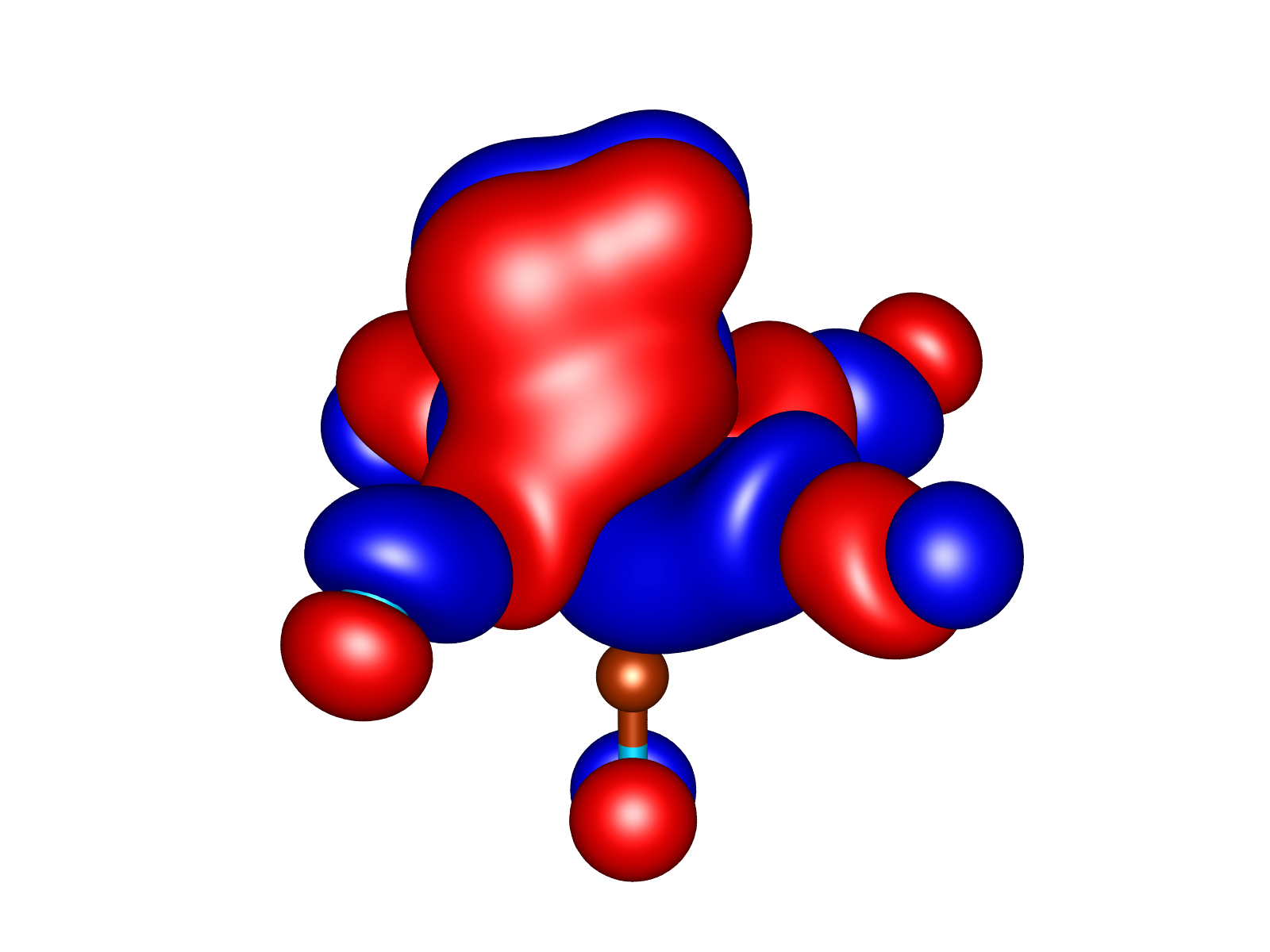}
  }
  \hfill
  \subfloat[$n_{\text{occup}}$ = 1.9529]{%
    \includegraphics[width=0.22\textwidth]{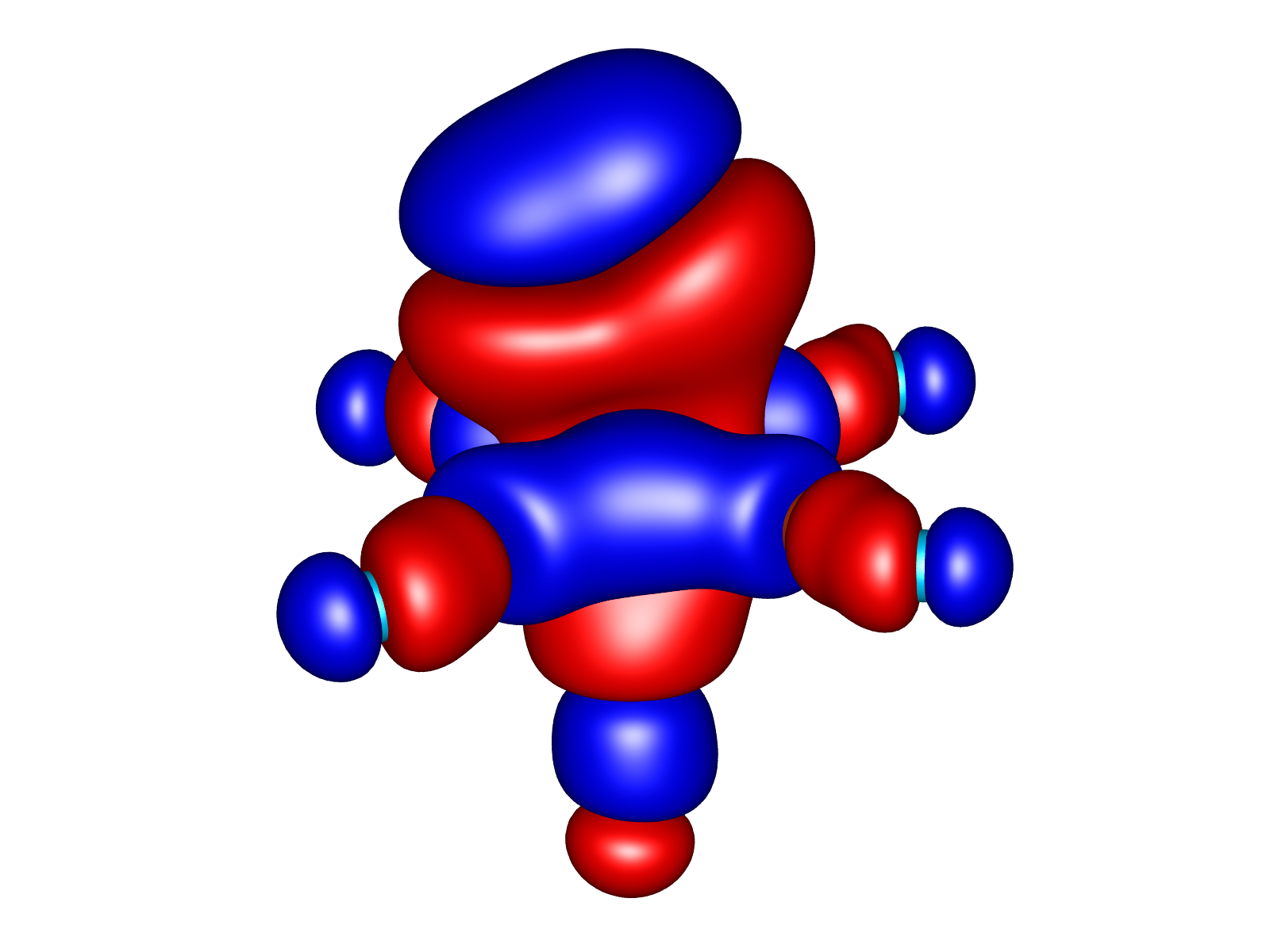}
  }
  \hfill
  \subfloat[$n_{\text{occup}}$ = 1.9269]{%
    \includegraphics[width=0.22\textwidth]{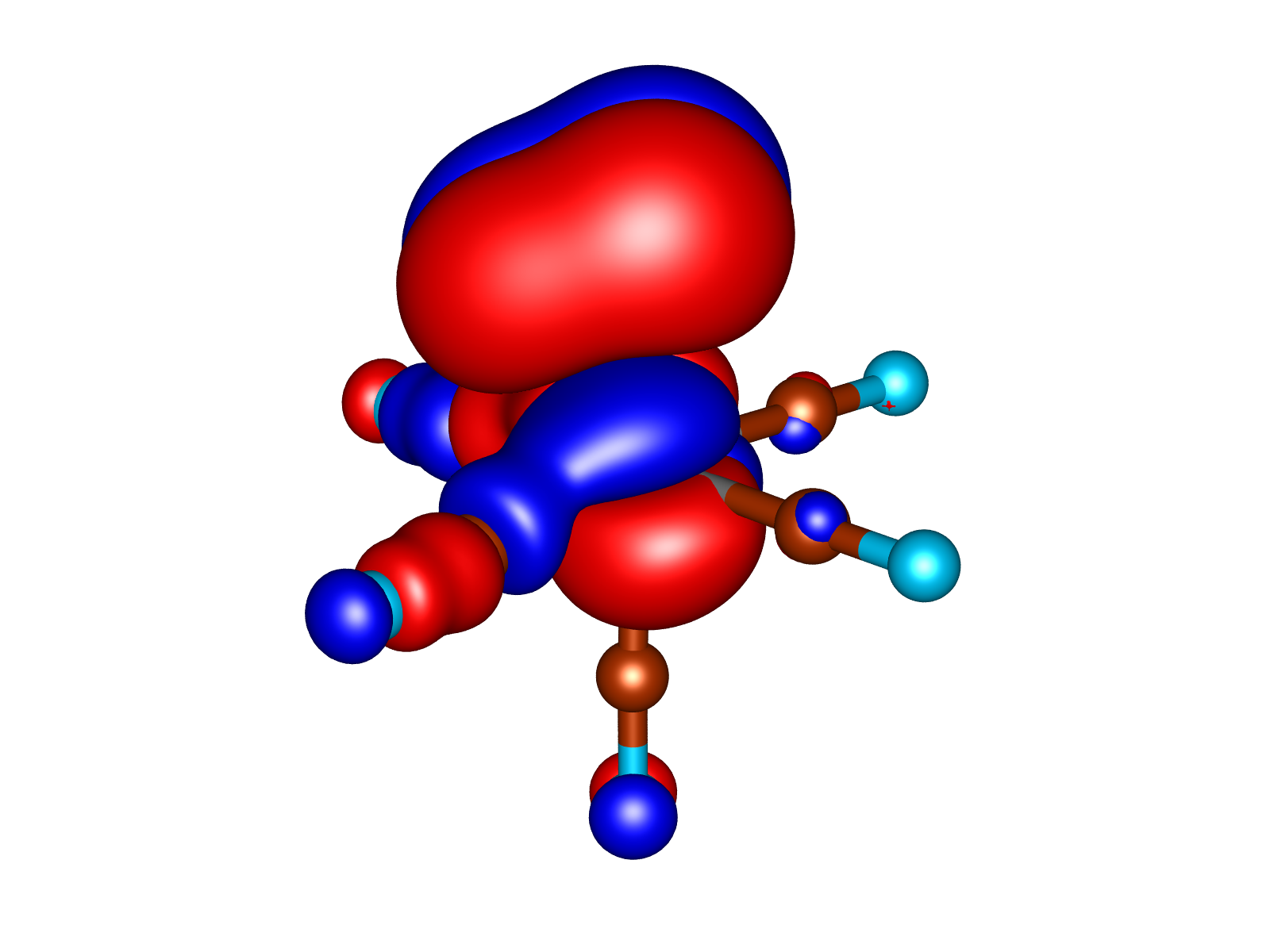}
  }
  \hfill
  \subfloat[$n_{\text{occup}}$ = 1.7694]{%
    \includegraphics[width=0.22\textwidth]{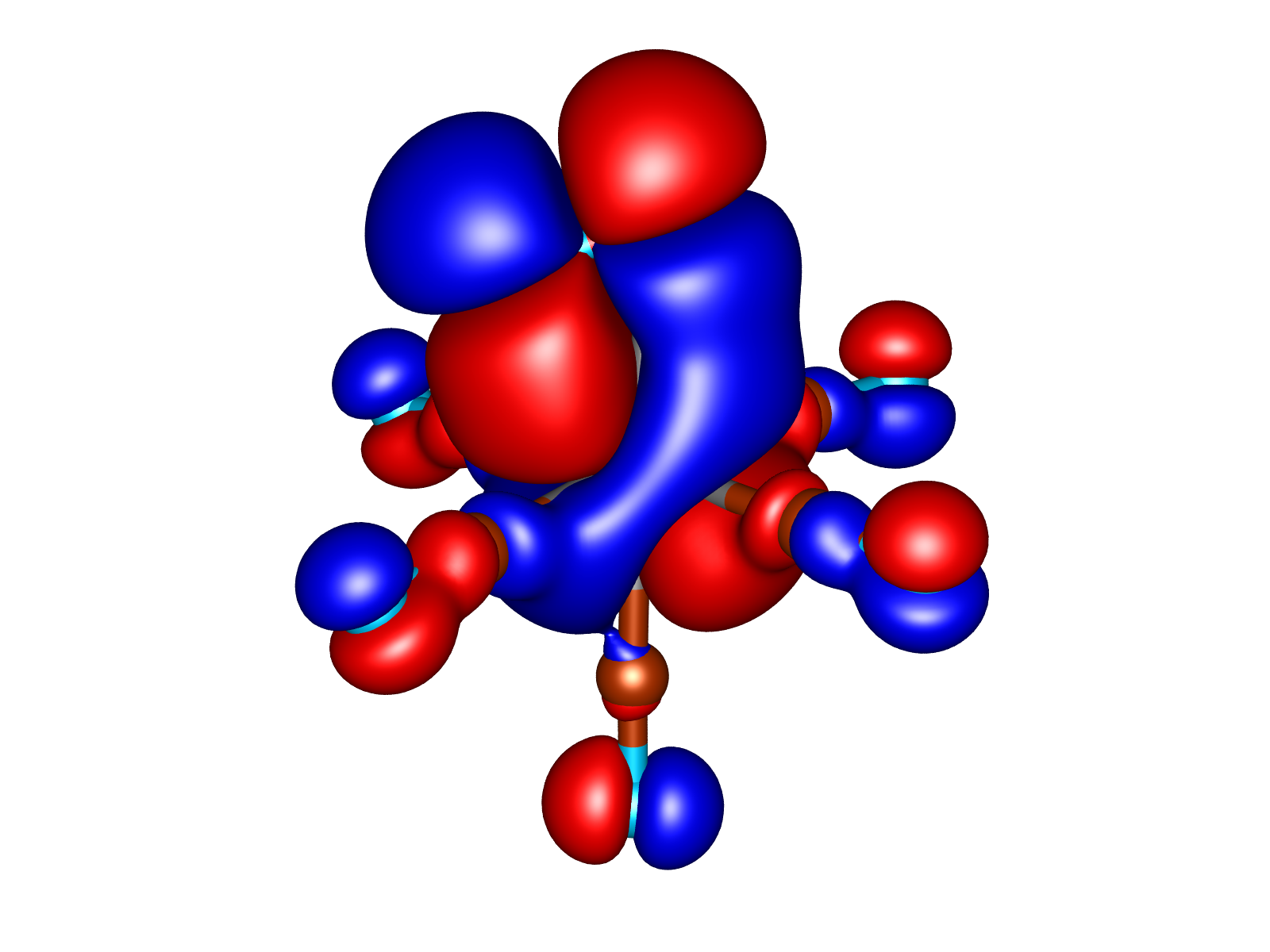}
  }
  \\
  \subfloat[$n_{\text{occup}}$ = 0.2479]{%
    \includegraphics[width=0.22\textwidth]{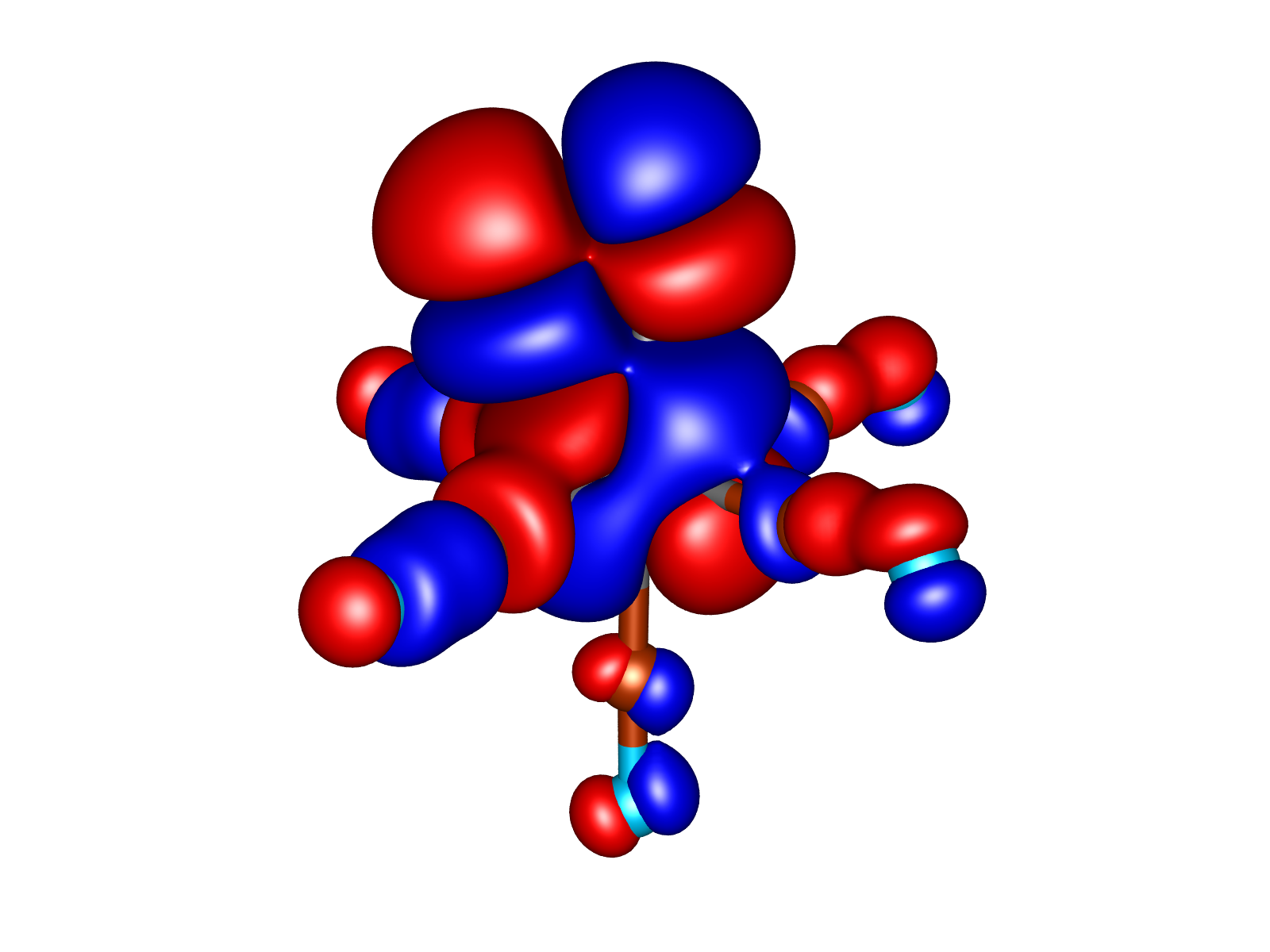}
  }
  \hfill
  \subfloat[$n_{\text{occup}}$ = 0.0939]{%
    \includegraphics[width=0.22\textwidth]{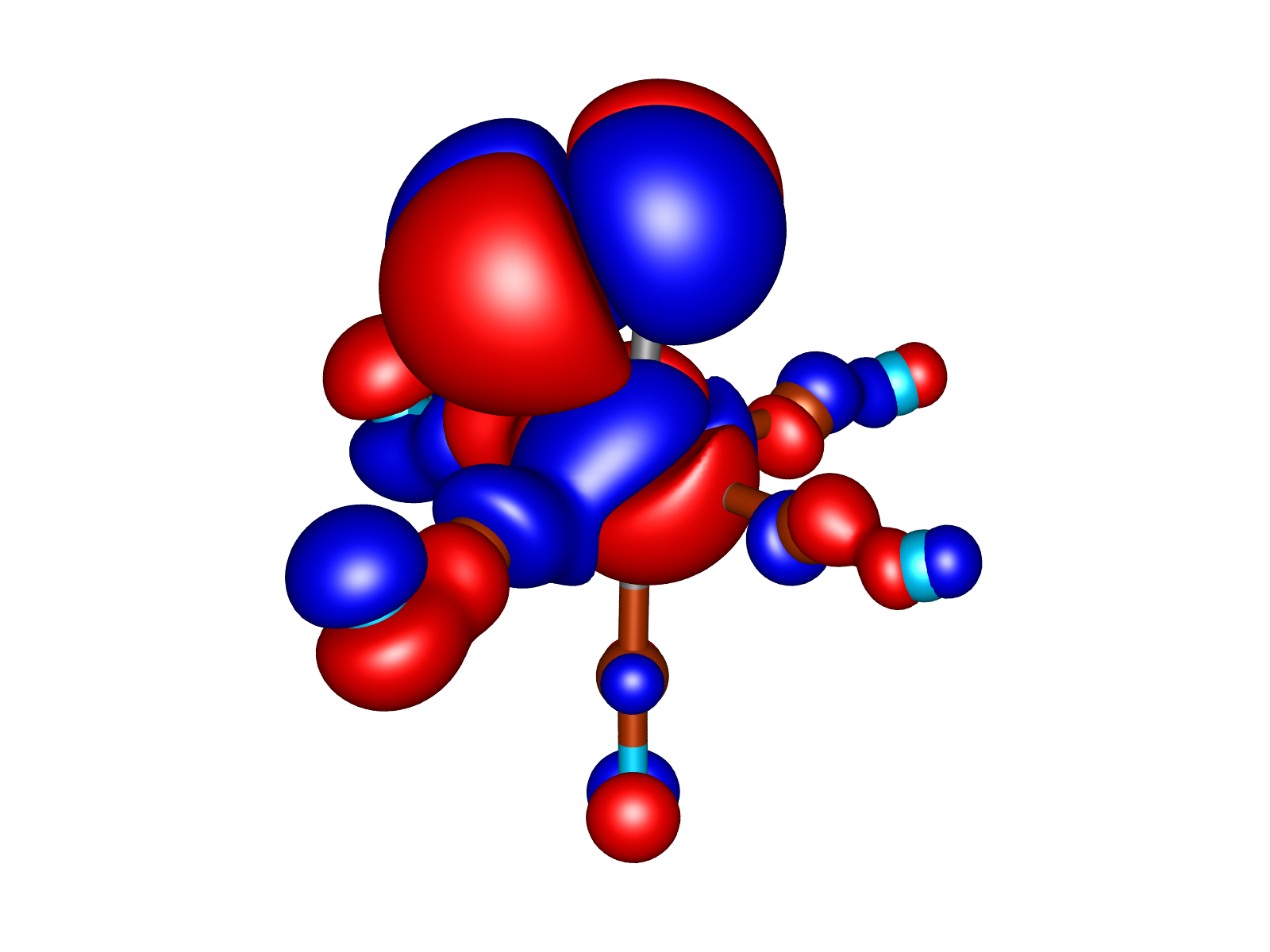}
  }
  \hfill
  \subfloat[$n_{\text{occup}}$ = 0.0468]{%
    \includegraphics[width=0.22\textwidth]{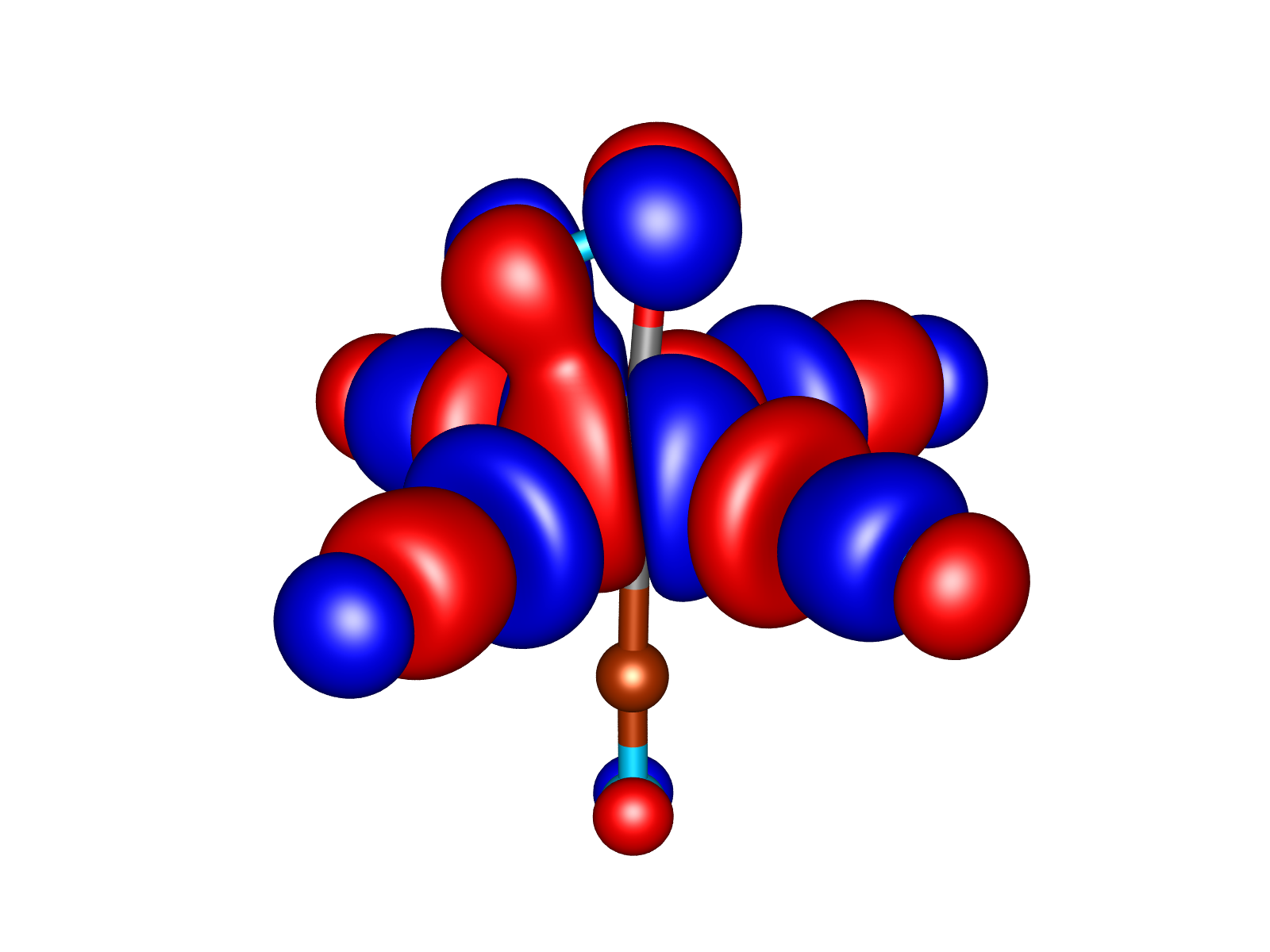}
  }
  \hfill
  \subfloat[$n_{\text{occup}}$ = 0.0402]{%
    \includegraphics[width=0.22\textwidth]{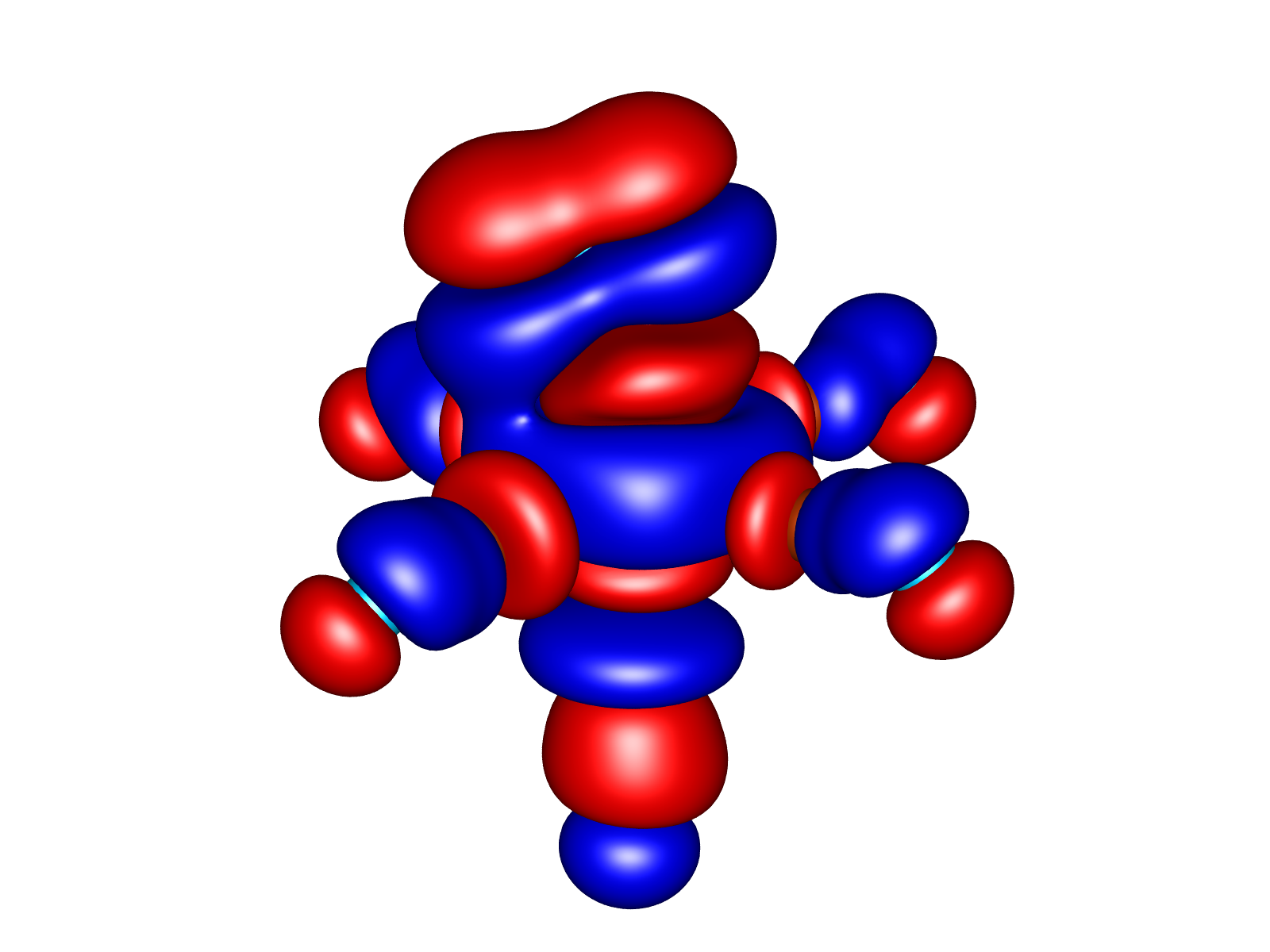}
  }
  \\
  \subfloat[$n_{\text{occup}}$ = 0.0295]{%
    \includegraphics[width=0.22\textwidth]{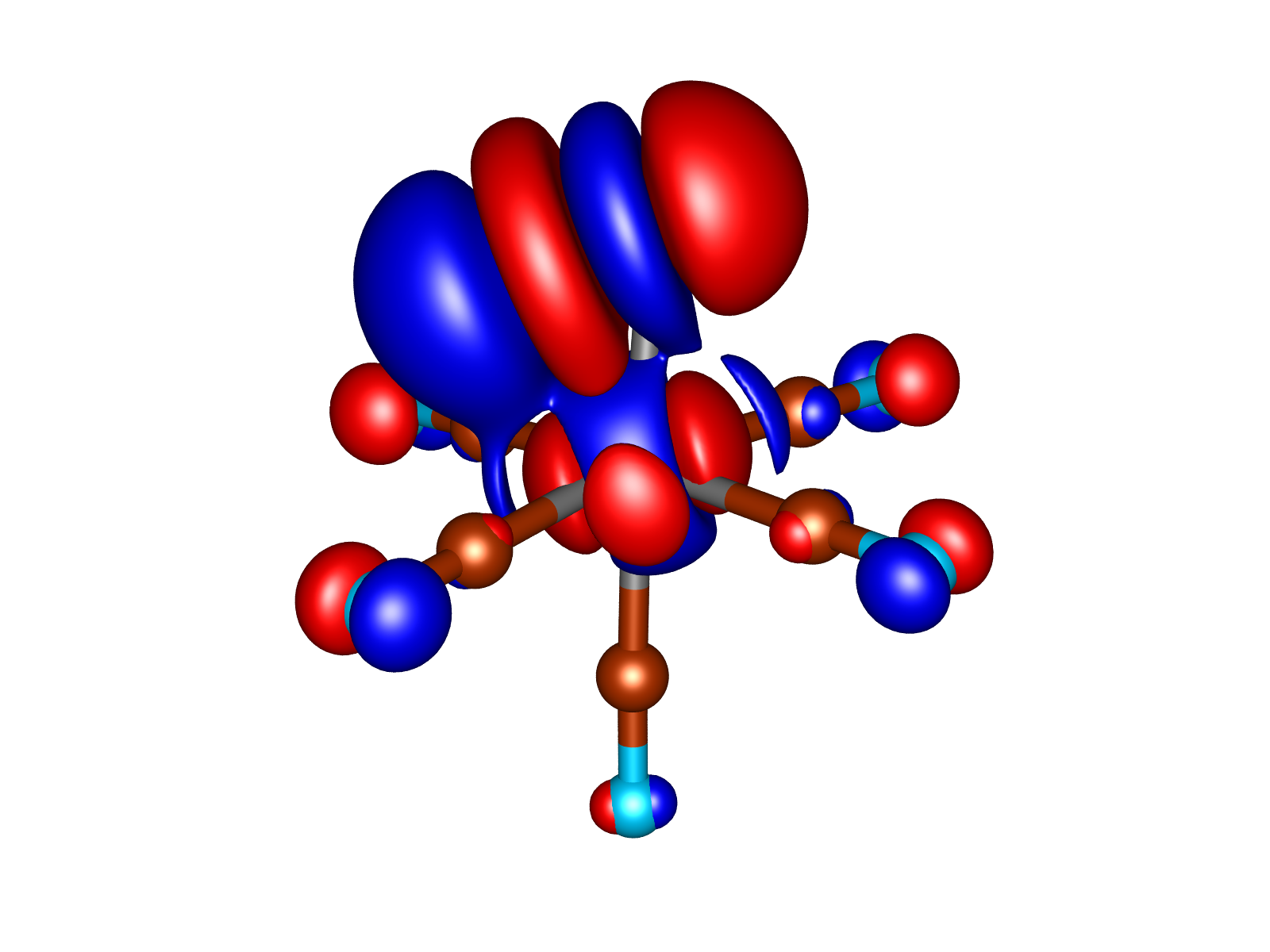}
  }
  \hfill
  \subfloat[$n_{\text{occup}}$ = 0.0192]{%
    \includegraphics[width=0.22\textwidth]{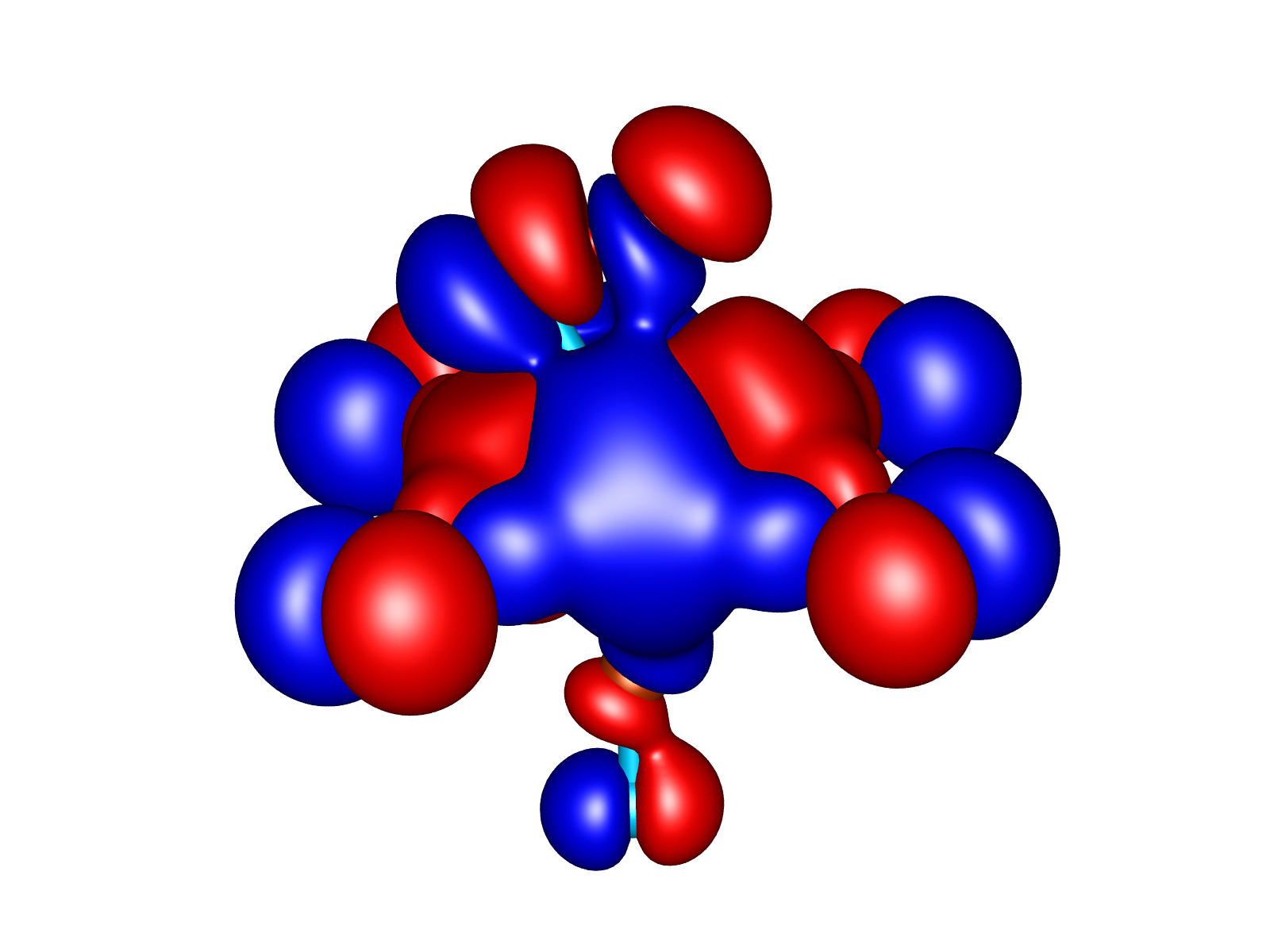}
  }
  \hfill
  \subfloat[$n_{\text{occup}}$ = 0.0140]{%
    \includegraphics[width=0.22\textwidth]{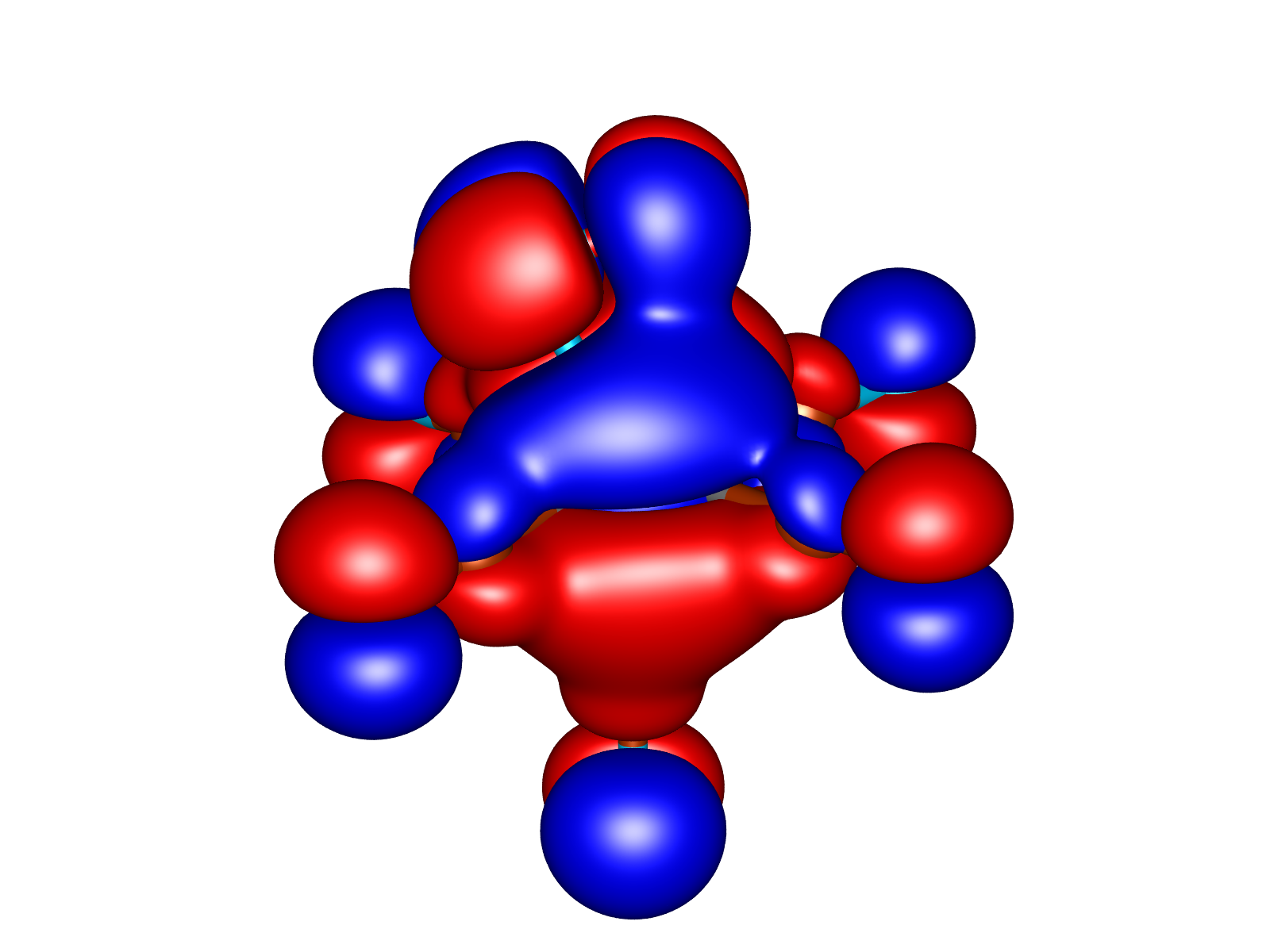}
  }
  \hfill
  \subfloat[$n_{\text{occup}}$ = 0.078]{%
    \includegraphics[width=0.22\textwidth]{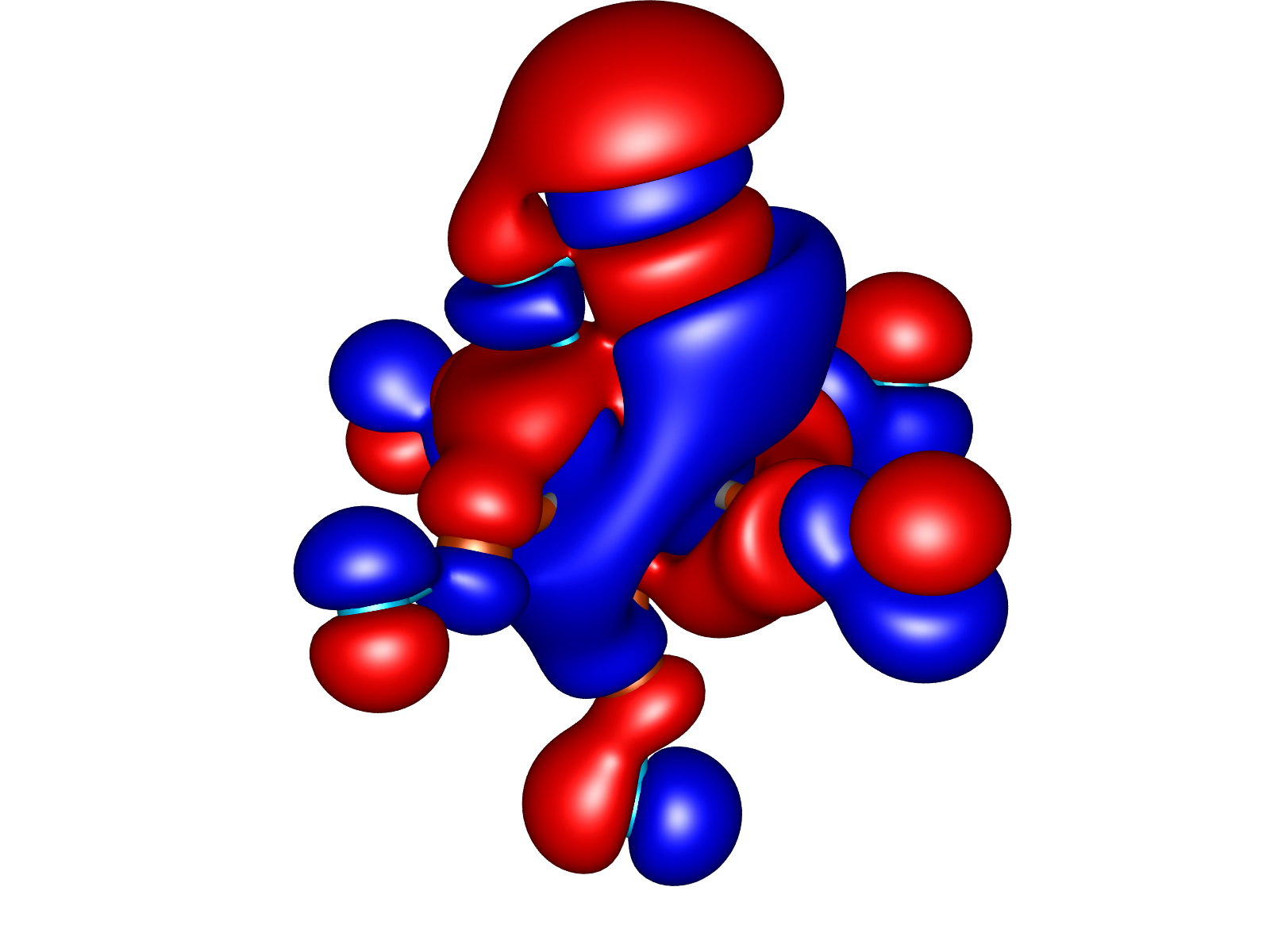}
  }
  \\
  \caption{Fe-NO complex, flat, CASSCF(14, 15) \label{orbs_cas1415_3}}
\end{figure}

\renewcommand{\thesubfigure}{\arabic{subfigure}}
\begin{figure}[!h]
  \subfloat[$n_{\text{occup}}$ = 1.9565]{%
    \includegraphics[width=0.22\textwidth]{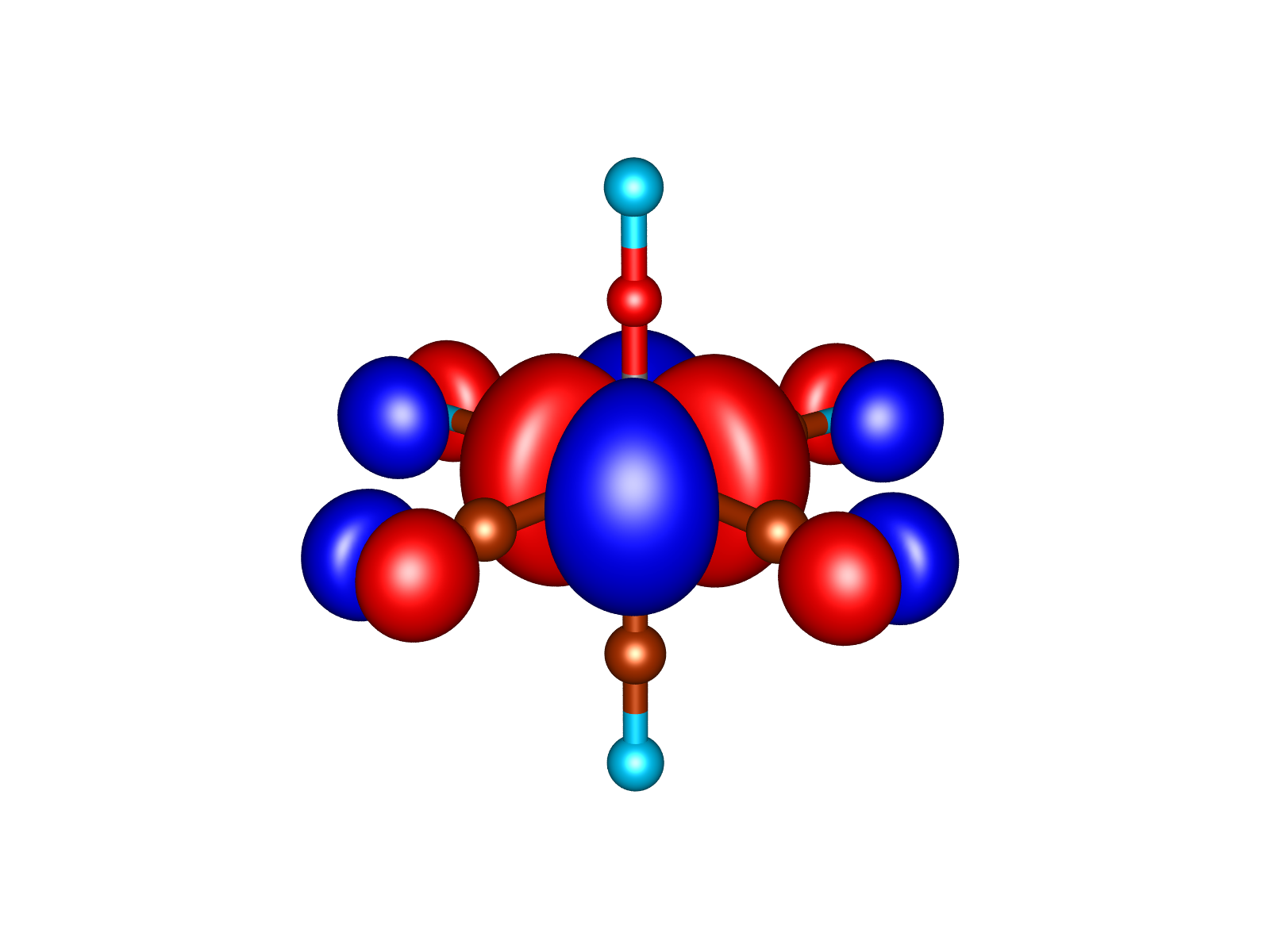}
  }
  \subfloat[$n_{\text{occup}}$ = 1.9542]{%
    \includegraphics[width=0.22\textwidth]{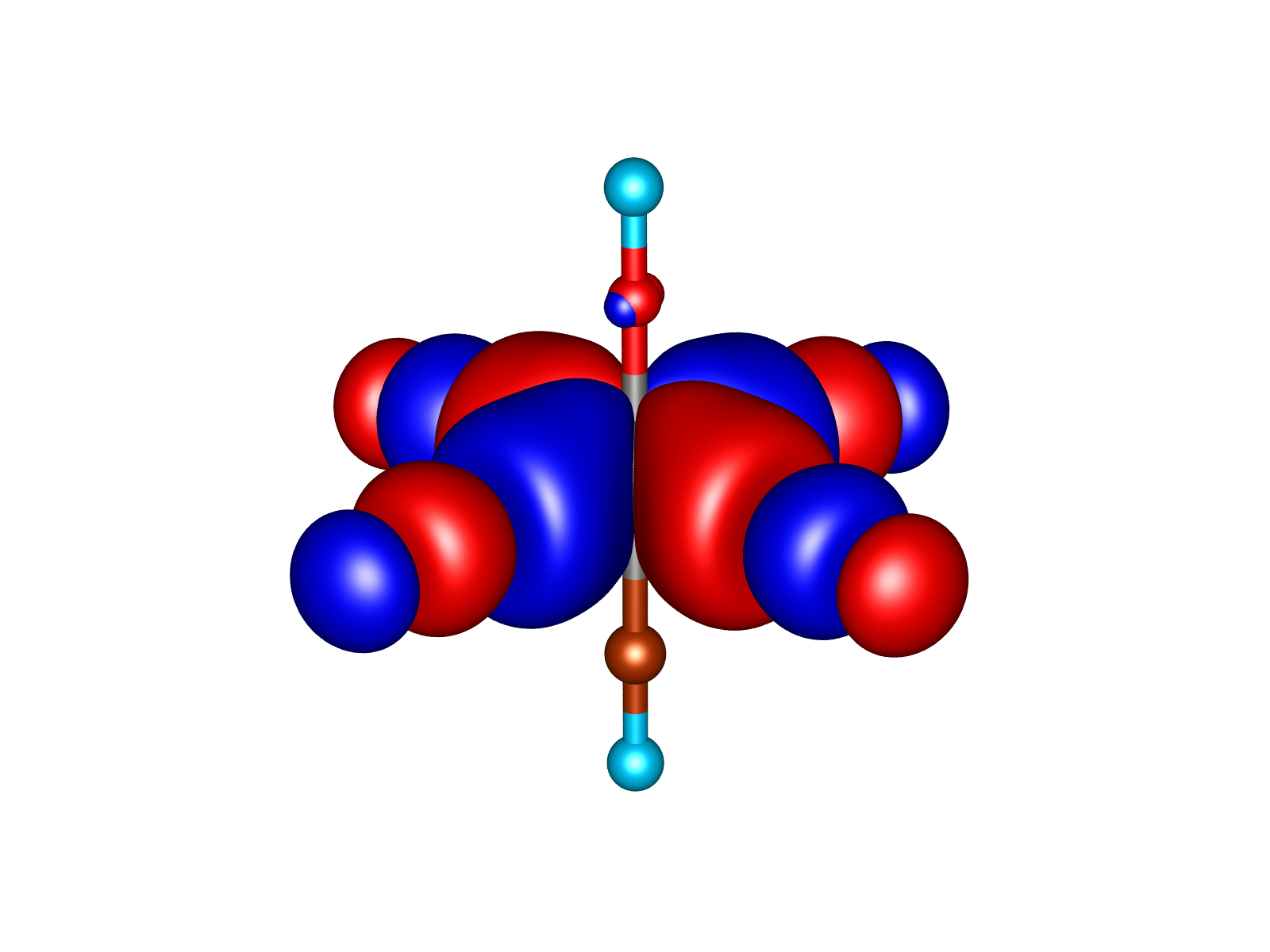}
  }
  \subfloat[$n_{\text{occup}}$ = 1.9530]{%
    \includegraphics[width=0.22\textwidth]{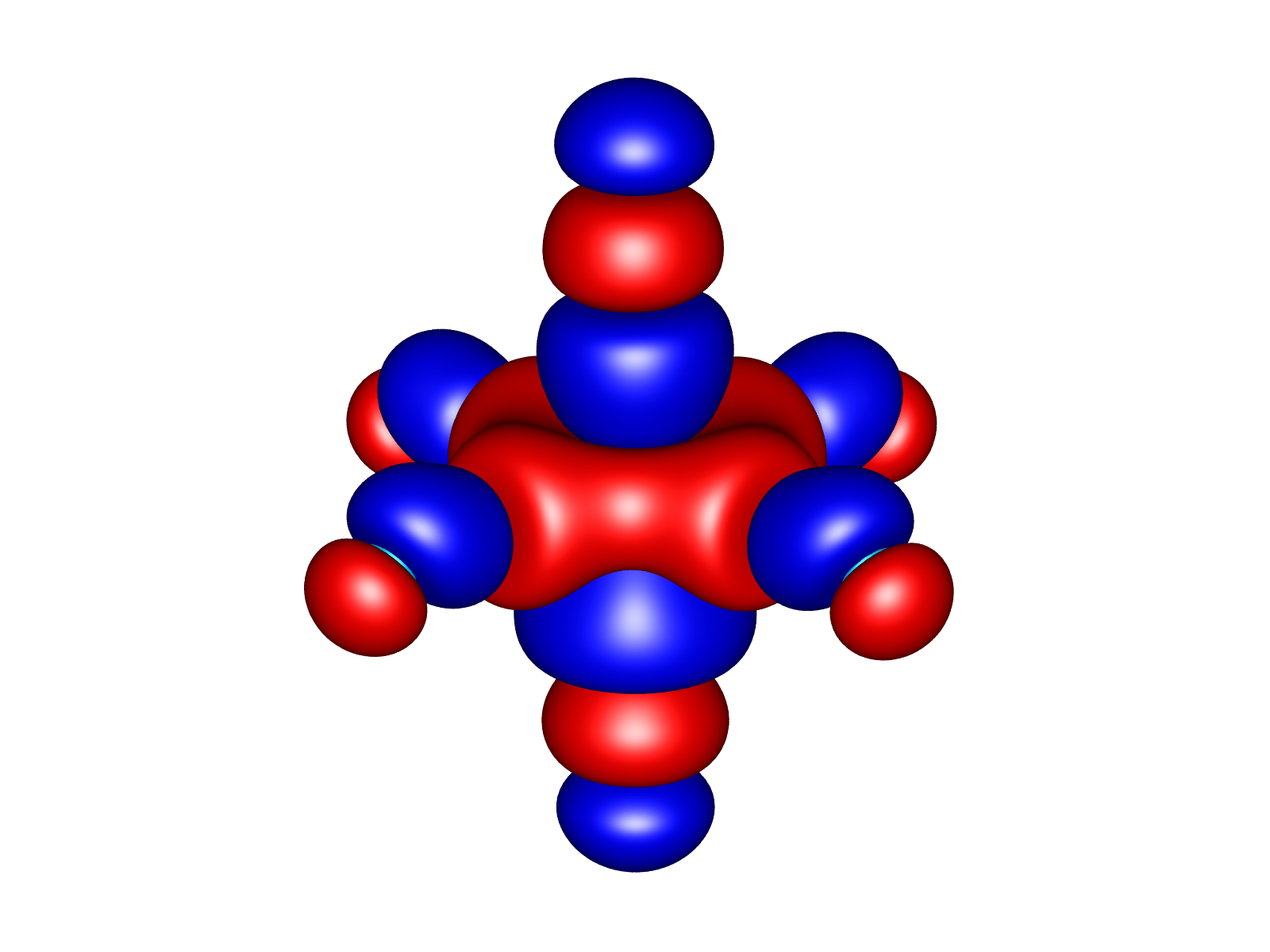}
  } \\
  \subfloat[$n_{\text{occup}}$ = 1.9484]{%
    \includegraphics[width=0.22\textwidth]{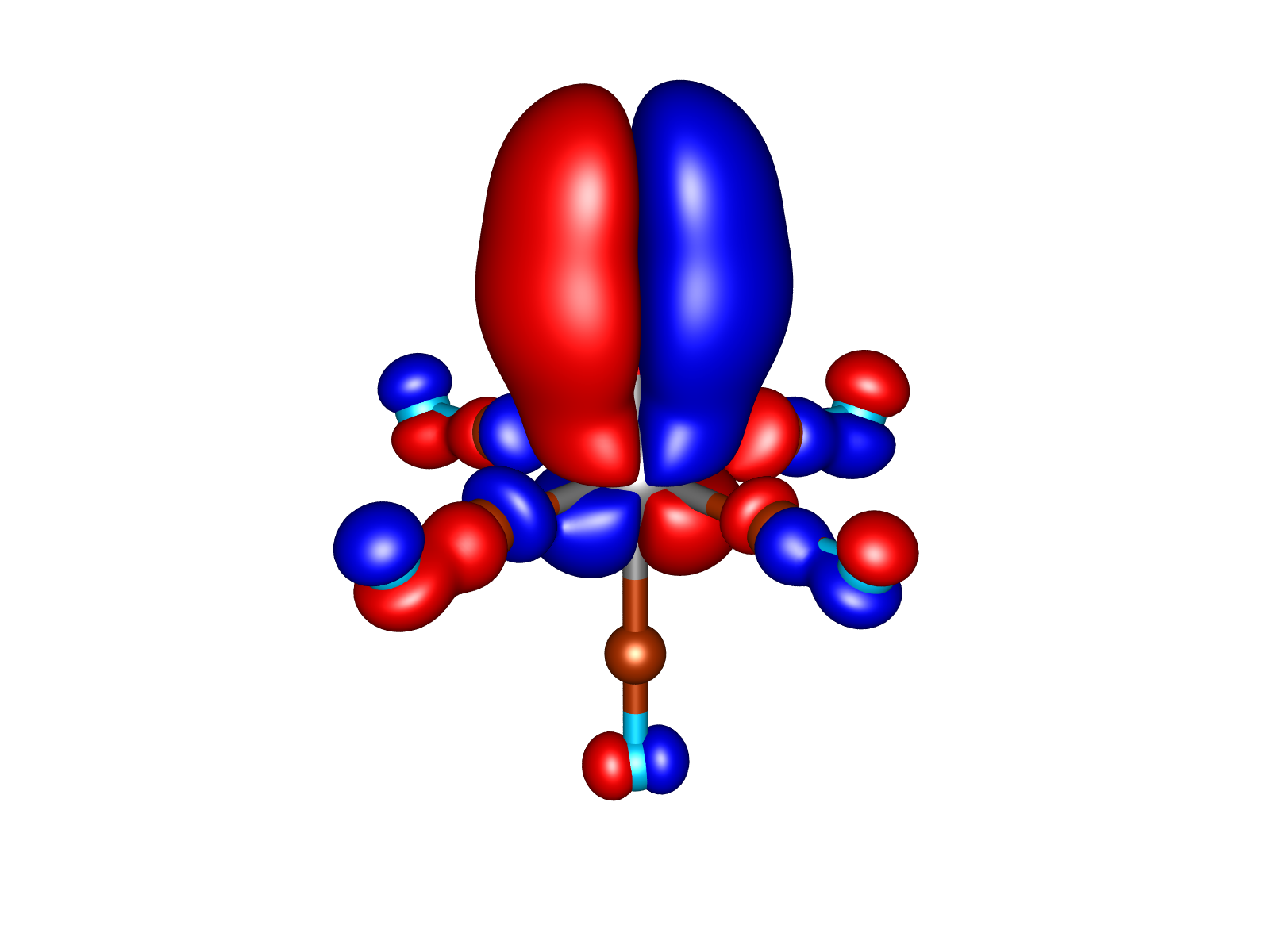}
  }
  \hfill
  \subfloat[$n_{\text{occup}}$ = 1.9483]{%
    \includegraphics[width=0.22\textwidth]{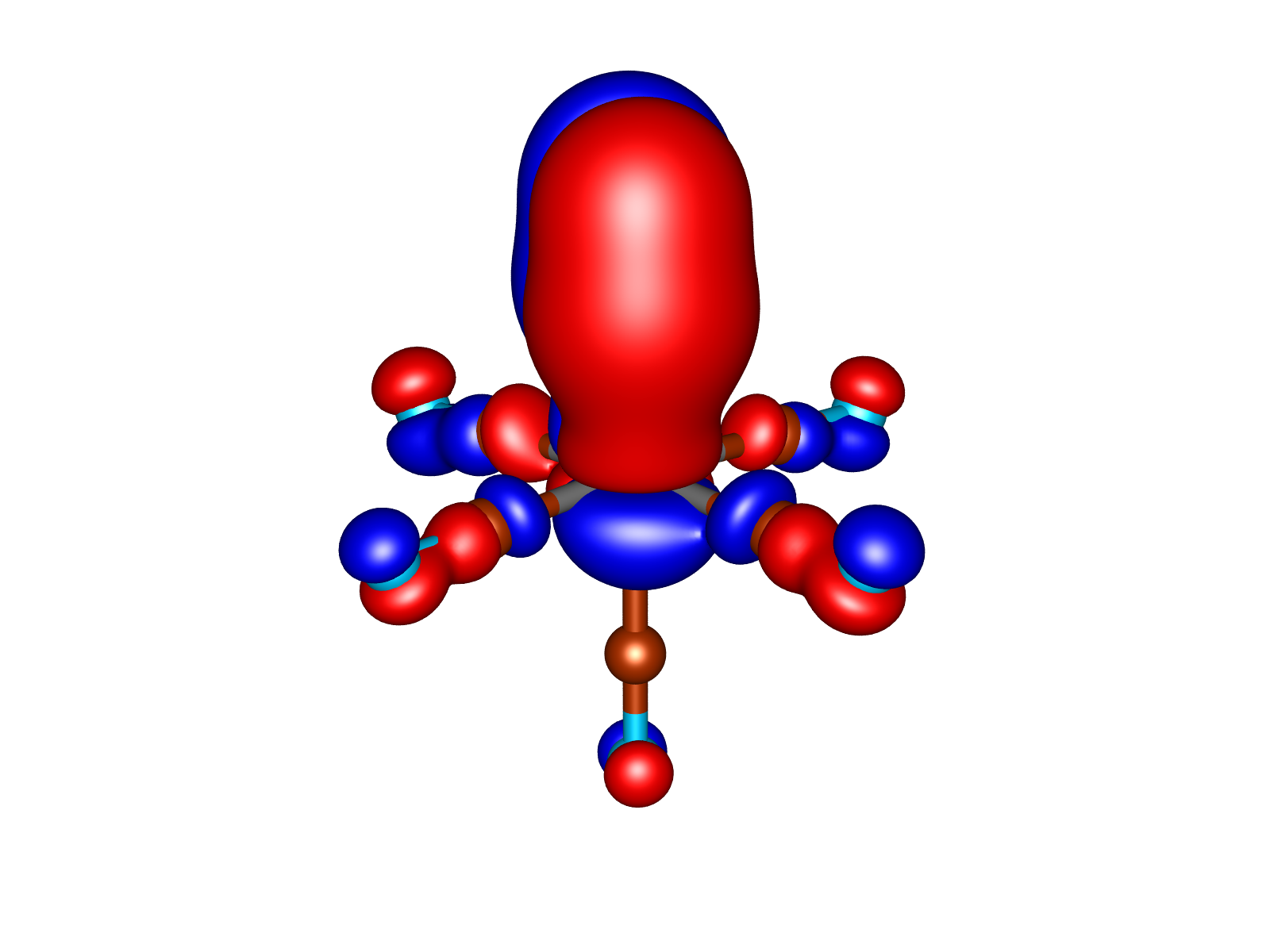}
  }
  \hfill
  \subfloat[$n_{\text{occup}}$ = 1.7036]{%
    \includegraphics[width=0.22\textwidth]{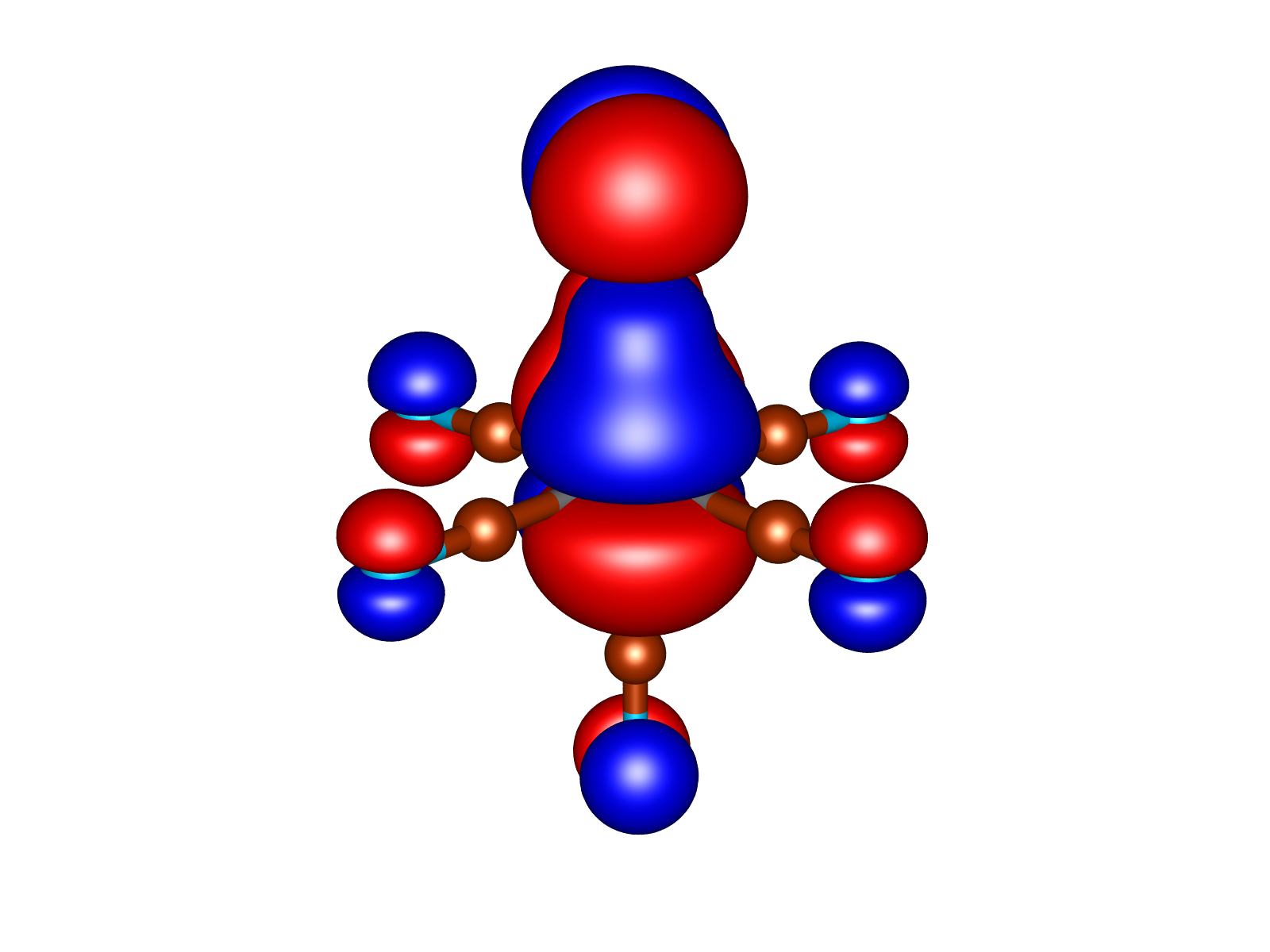}
  }
  \hfill
  \subfloat[$n_{\text{occup}}$ = 1.7035]{%
    \includegraphics[width=0.22\textwidth]{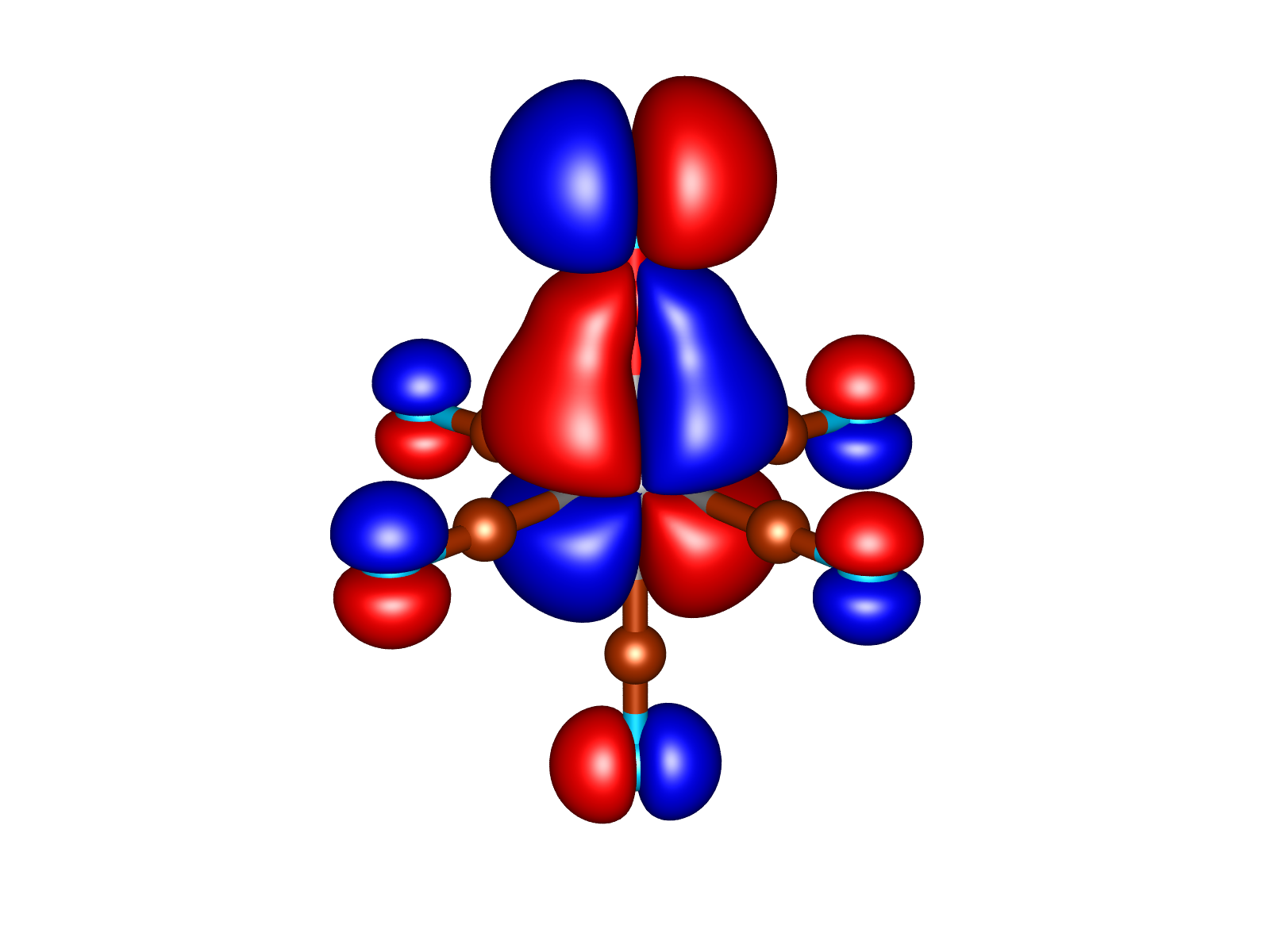}
  }
  \\
  \subfloat[$n_{\text{occup}}$ = 0.3324]{%
    \includegraphics[width=0.22\textwidth]{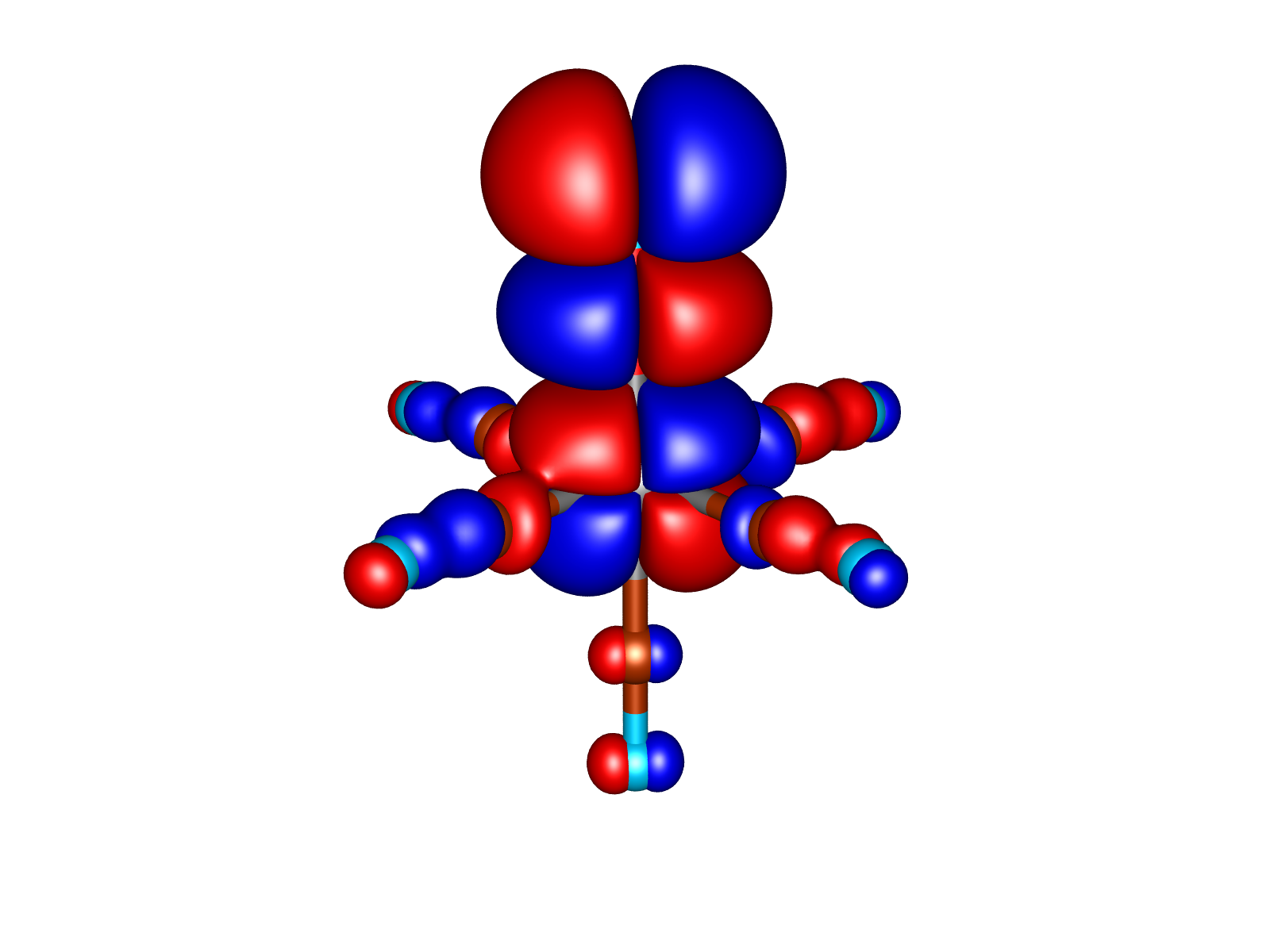}
  }
  \hfill
  \subfloat[$n_{\text{occup}}$ = 0.3323]{%
    \includegraphics[width=0.22\textwidth]{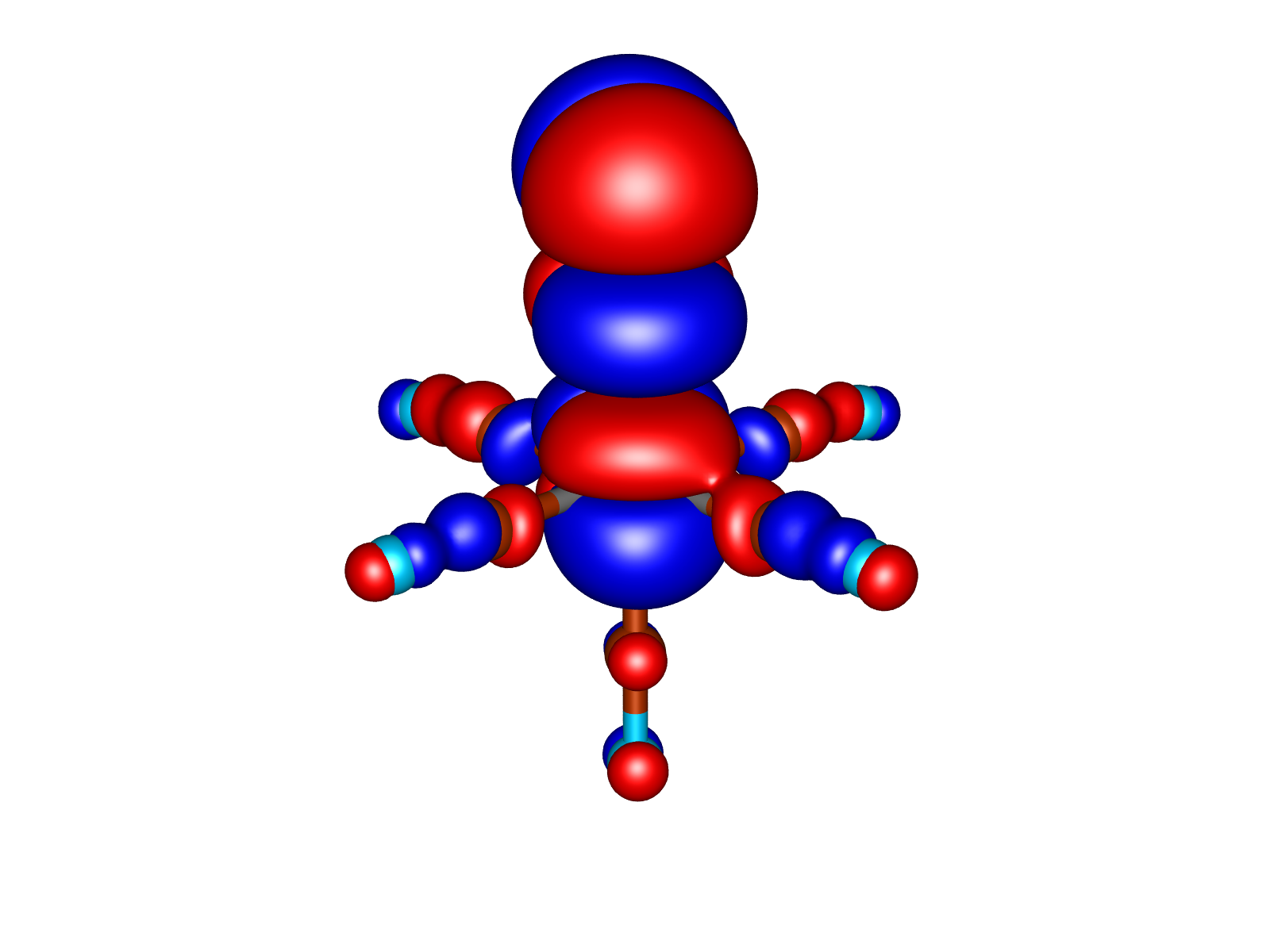}
  }
  \hfill
  \subfloat[$n_{\text{occup}}$ = 0.0639]{%
    \includegraphics[width=0.22\textwidth]{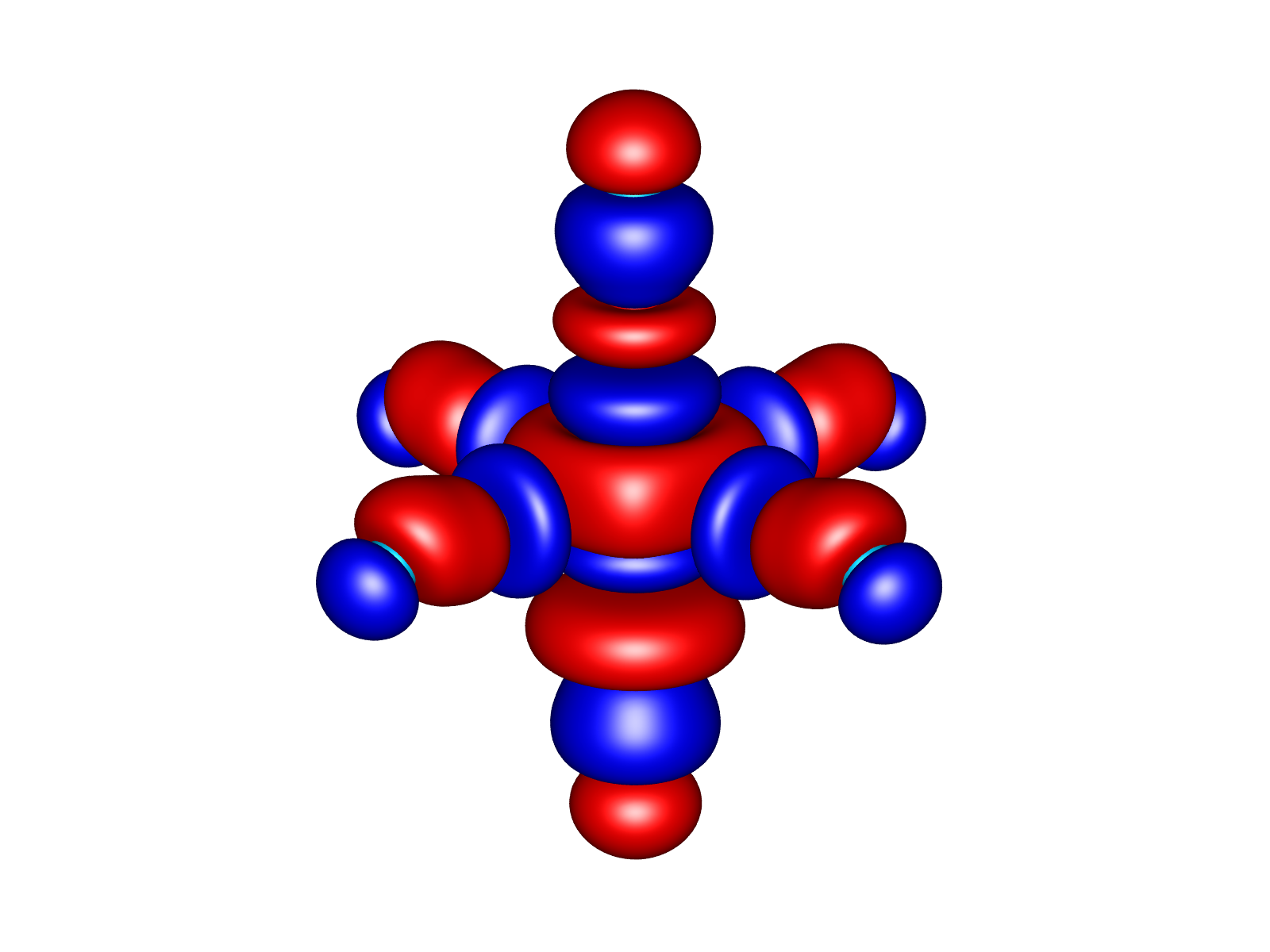}
  }
  \hfill
  \subfloat[$n_{\text{occup}}$ = 0.0533]{%
    \includegraphics[width=0.22\textwidth]{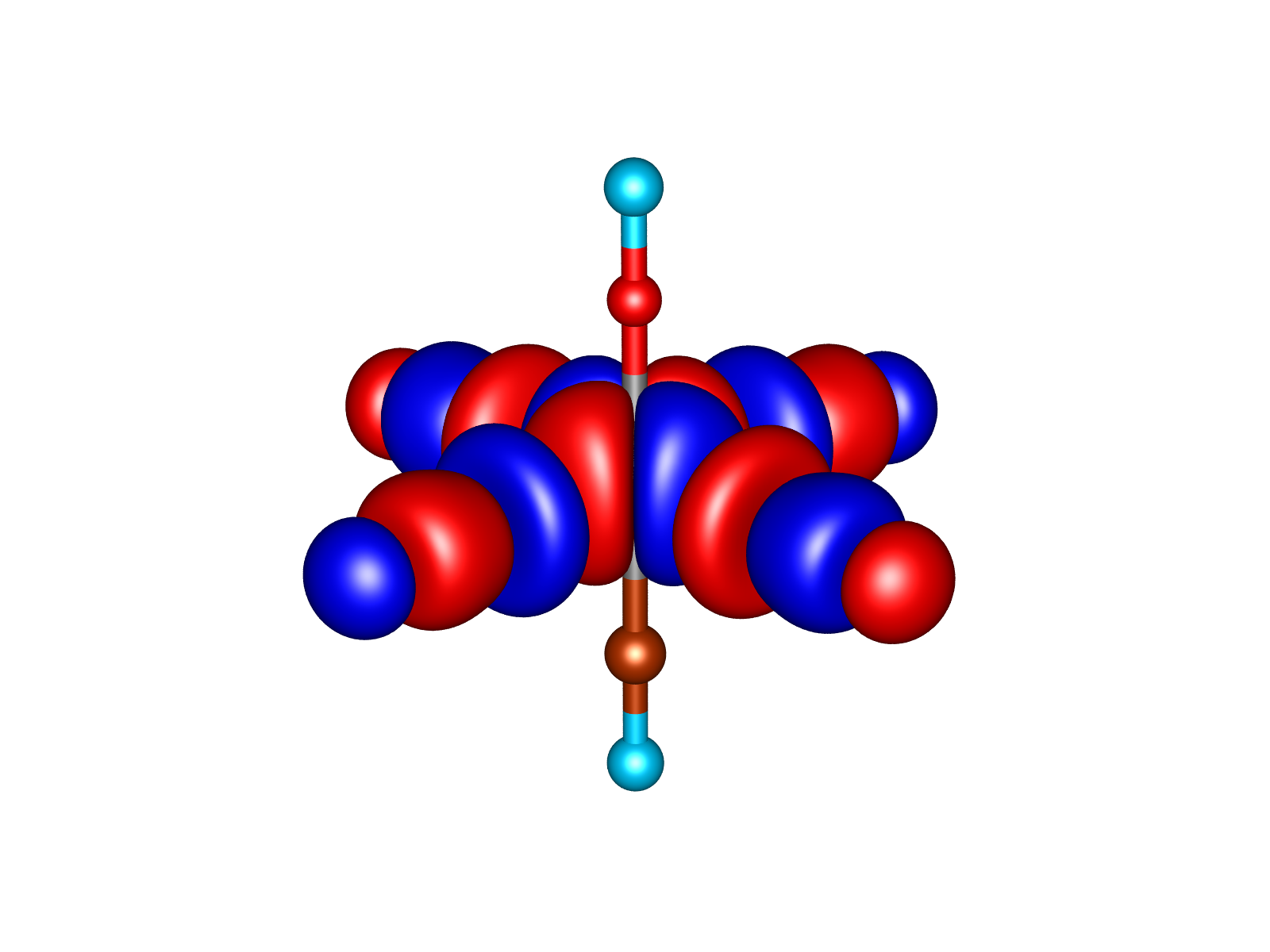}
  }
  \\
  \subfloat[$n_{\text{occup}}$ = 0.0223]{%
    \includegraphics[width=0.22\textwidth]{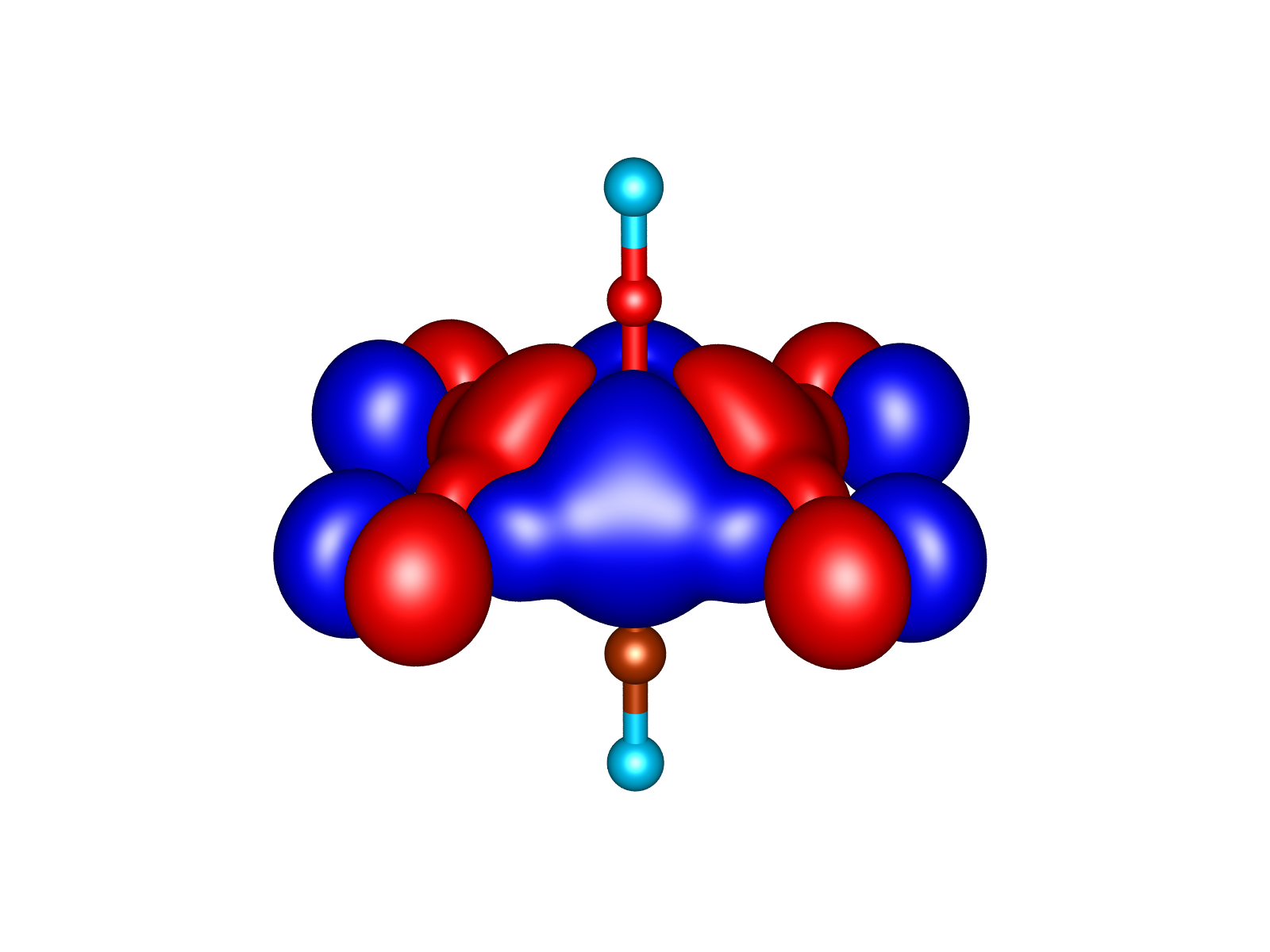}
  }
  \hfill
  \subfloat[$n_{\text{occup}}$ = 0.0131]{%
    \includegraphics[width=0.22\textwidth]{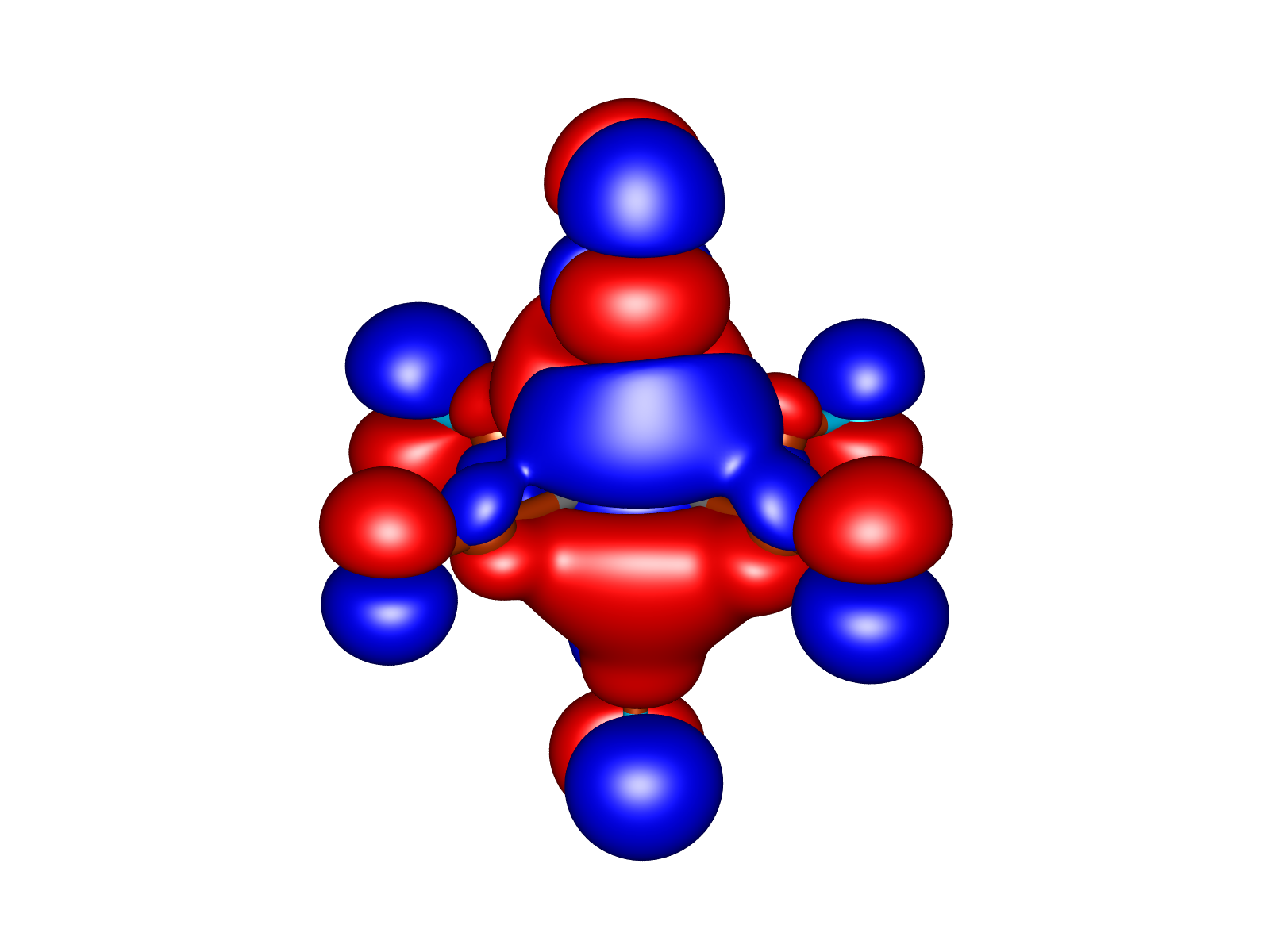}
  }
  \hfill
  \subfloat[$n_{\text{occup}}$ = 0.0131]{%
    \includegraphics[width=0.22\textwidth]{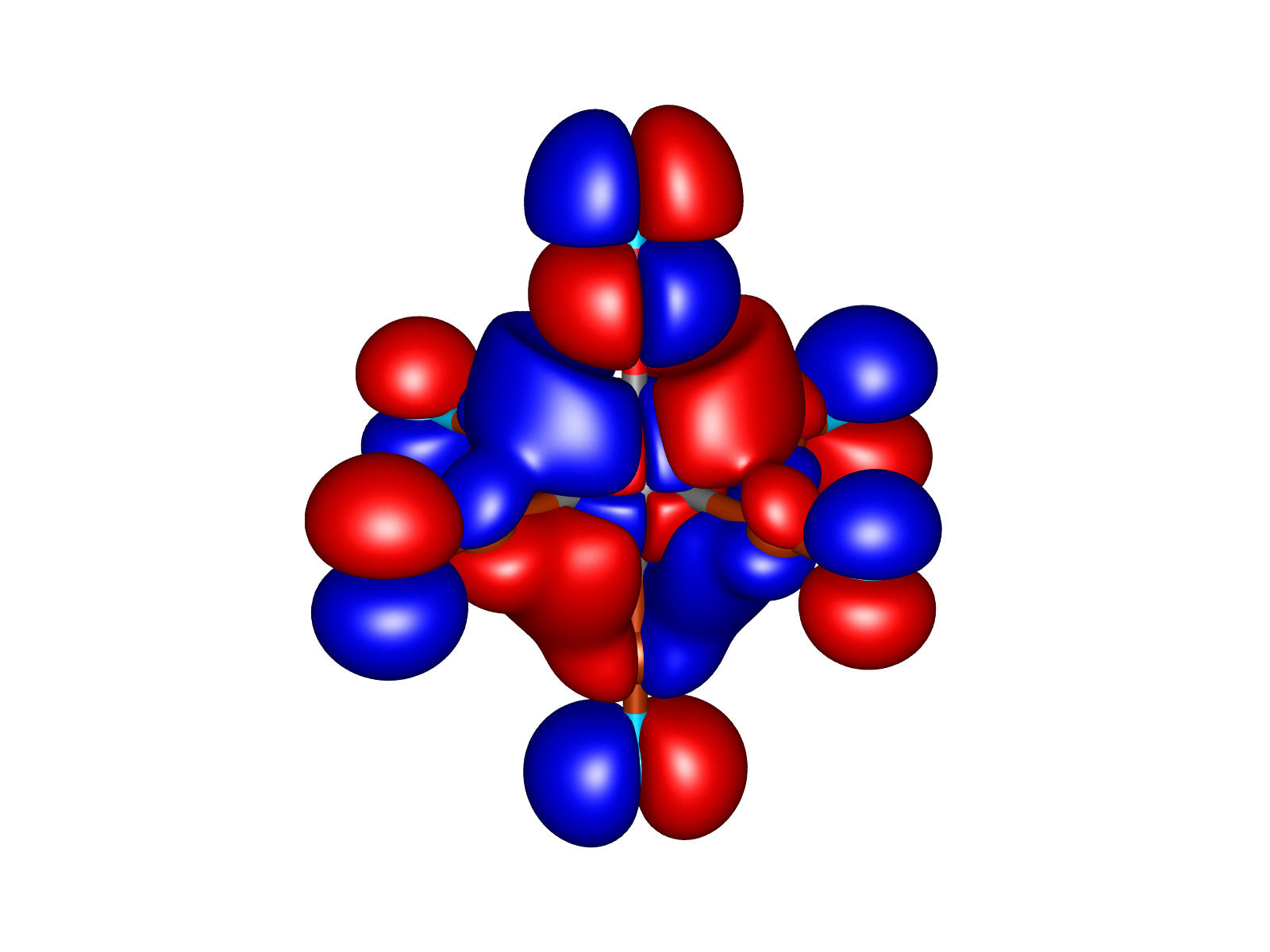}
  }
  \hfill
  \subfloat[$n_{\text{occup}}$ = 0.0020]{%
    \includegraphics[width=0.22\textwidth]{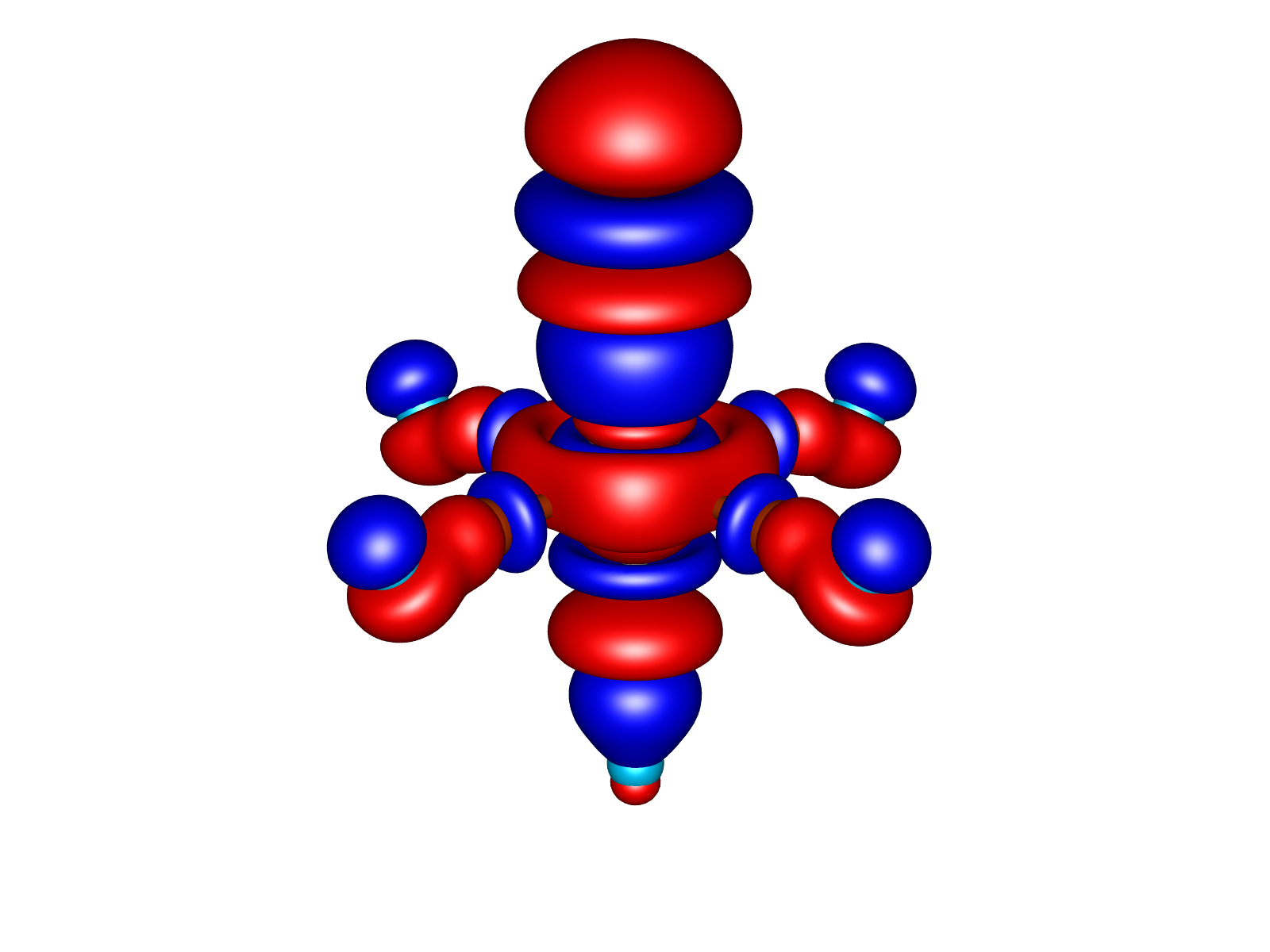}
  }
  \\
  \caption{Fe-NO complex, reverse, CASSCF(14, 15) \label{orbs_cas1415_5}}
\end{figure}

\renewcommand{\thesubfigure}{\arabic{subfigure}}
\begin{figure}[!h]
  \subfloat[$n_{\text{occup}}$ = 1.9760]{%
    \includegraphics[width=0.22\textwidth]{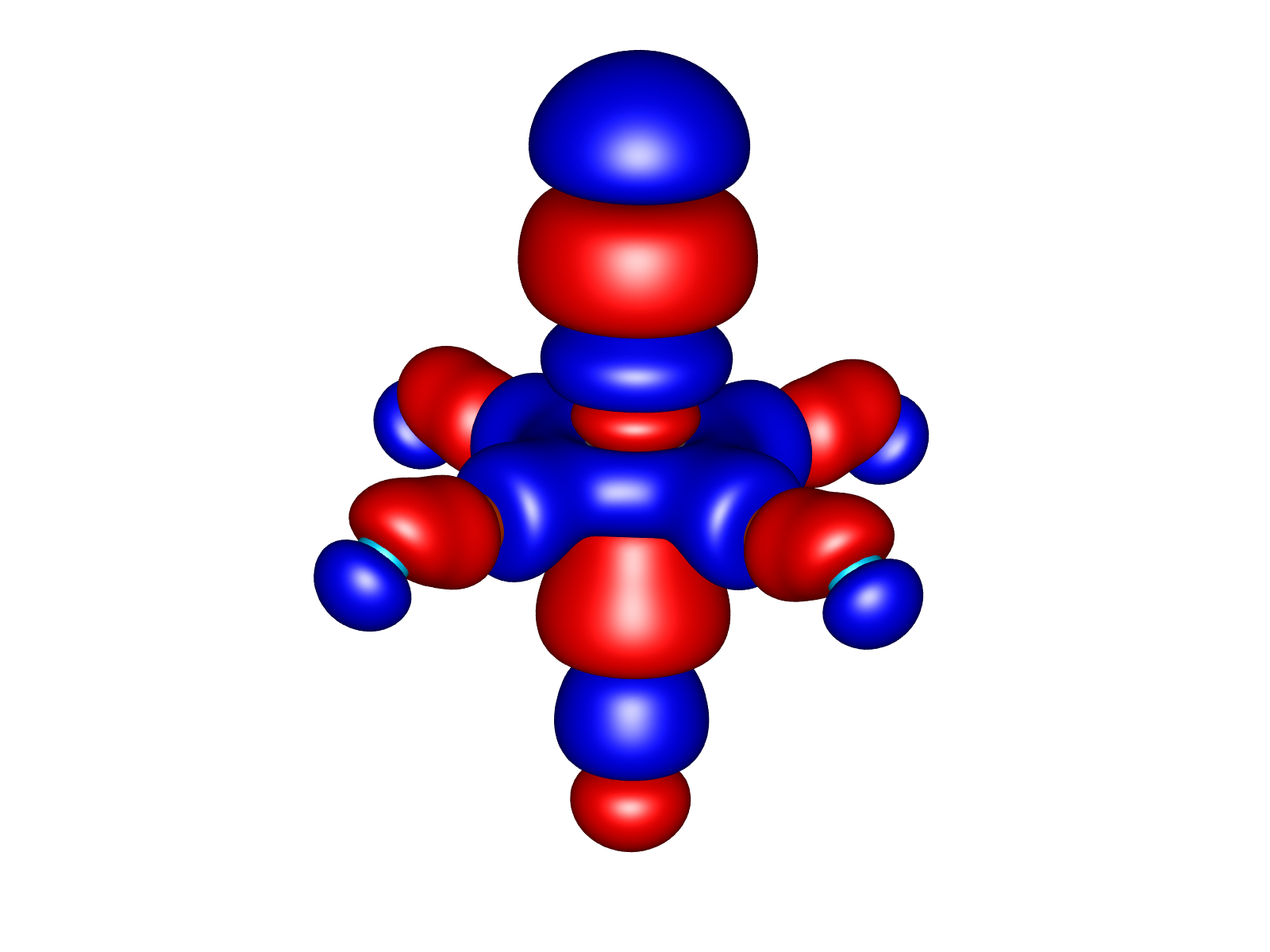}
  }
  \hfill
  \subfloat[$n_{\text{occup}}$ = 1.9624]{%
    \includegraphics[width=0.22\textwidth]{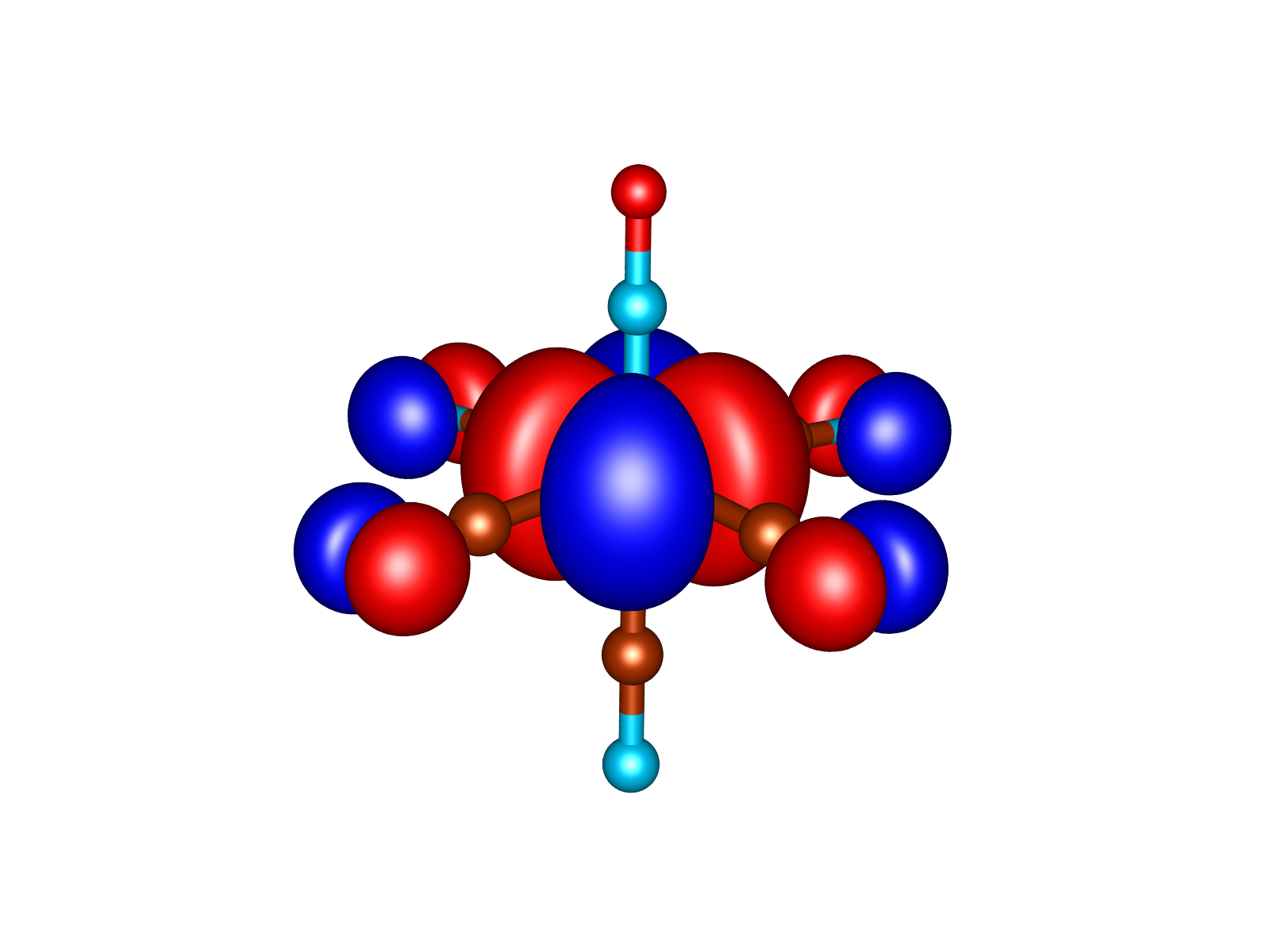}
  }
  \hfill
  \subfloat[$n_{\text{occup}}$ = 1.9573]{%
    \includegraphics[width=0.22\textwidth]{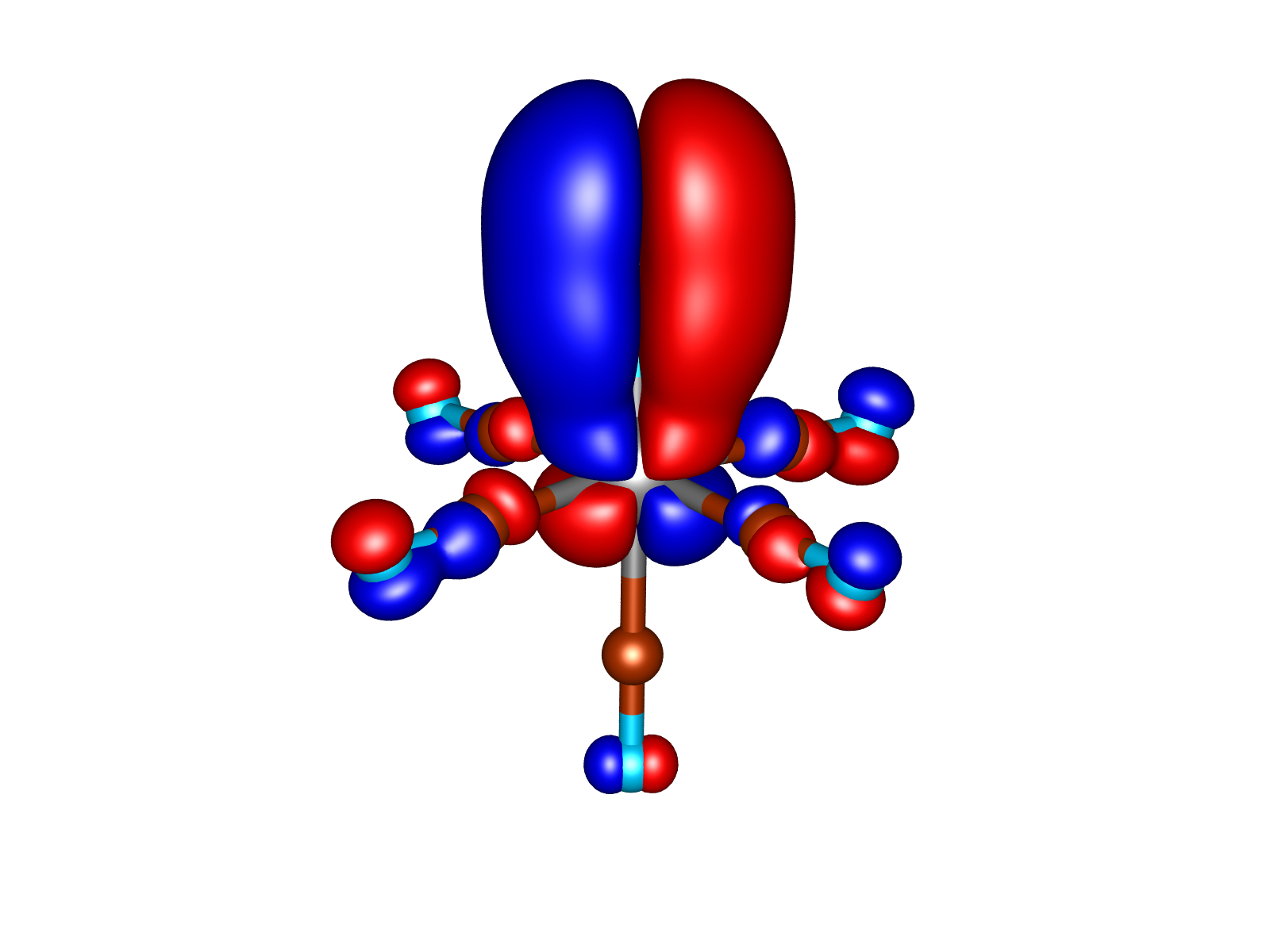}
  }
  \hfill
  \subfloat[$n_{\text{occup}}$ = 1.9573]{%
    \includegraphics[width=0.22\textwidth]{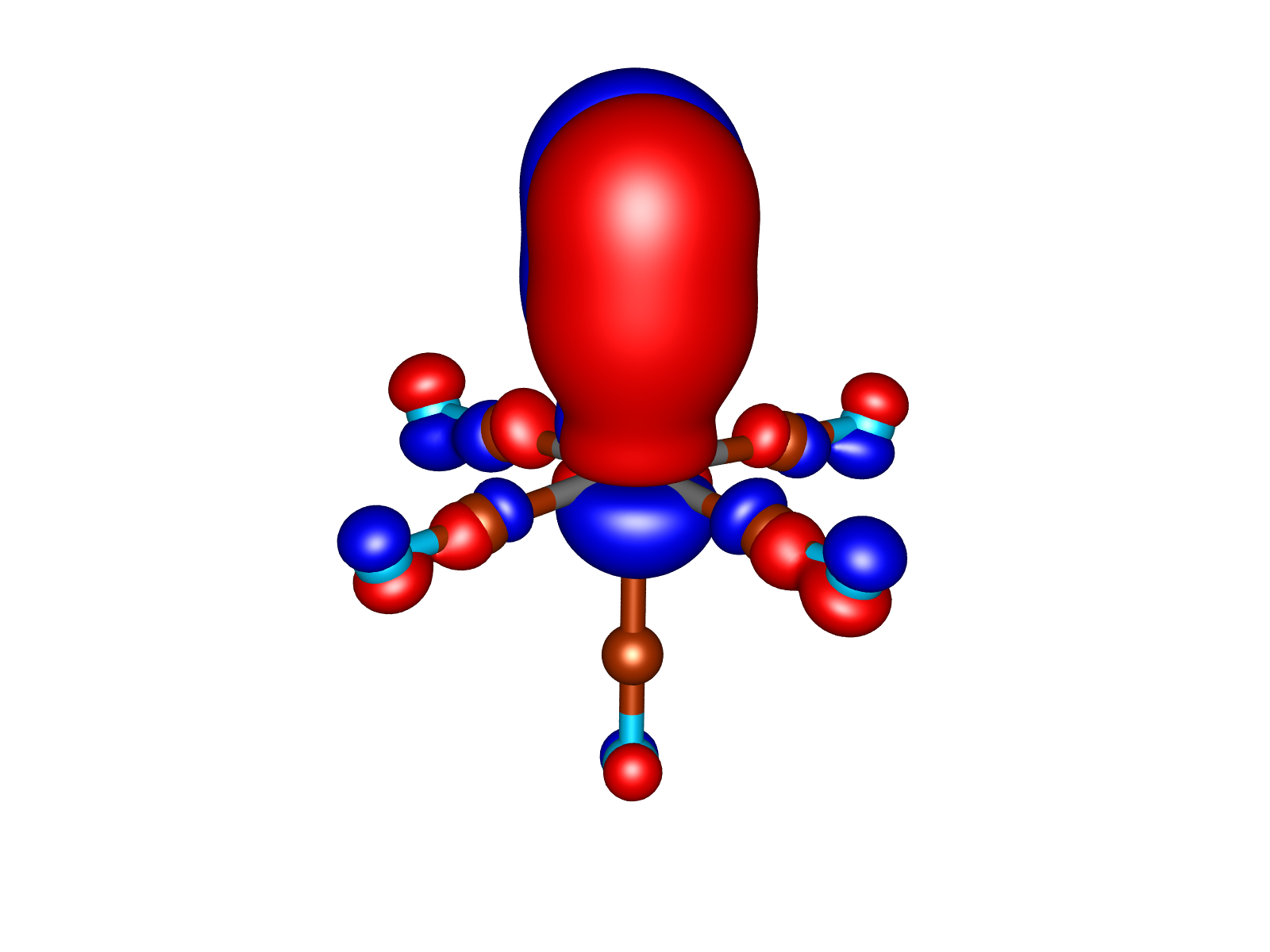}
  } \\
  \subfloat[$n_{\text{occup}}$ = 1.9517]{%
    \includegraphics[width=0.22\textwidth]{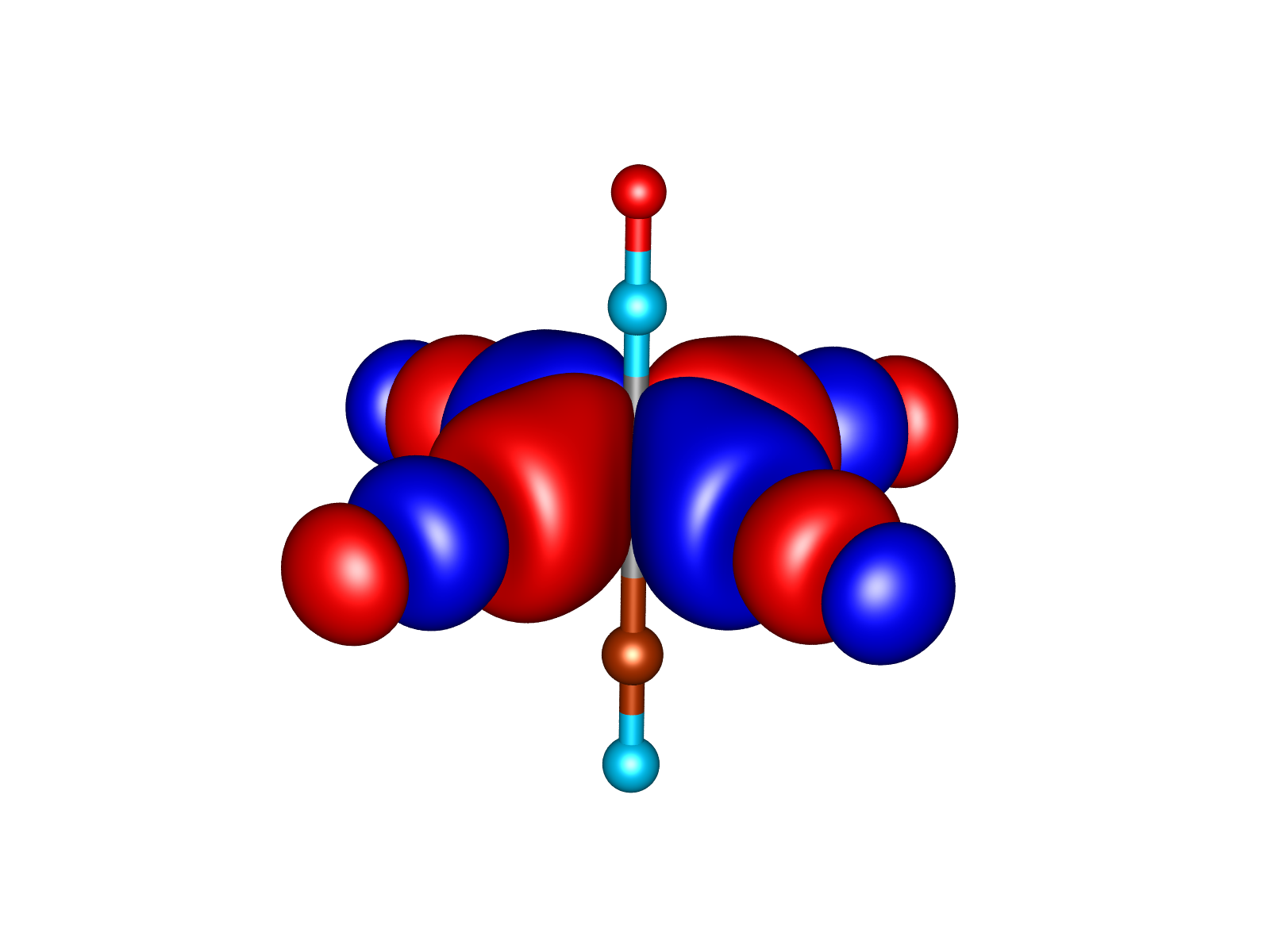}
  }
  \hfill
  \subfloat[$n_{\text{occup}}$ = 1.9502]{%
    \includegraphics[width=0.22\textwidth]{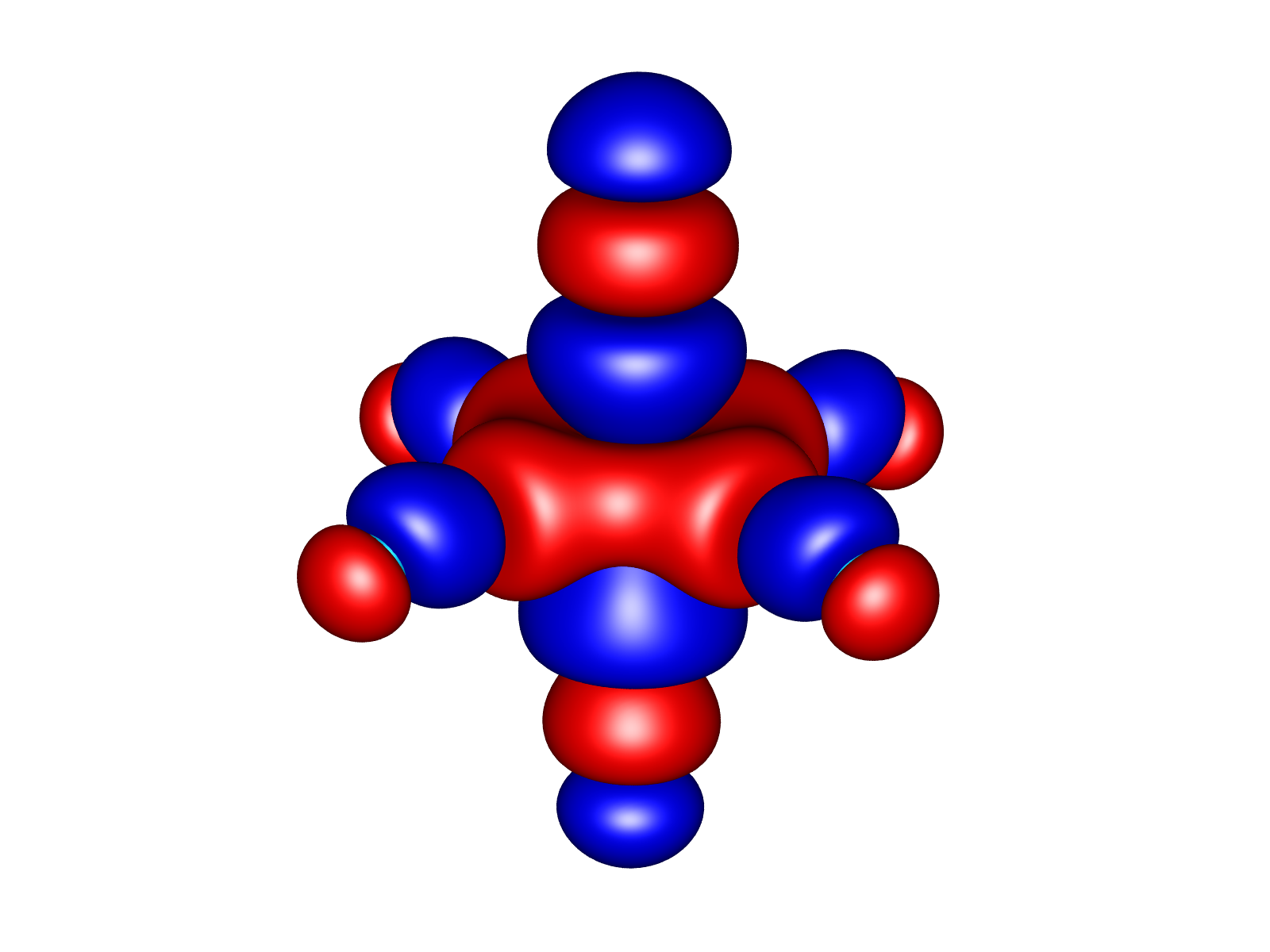}
  }
  \hfill
  \subfloat[$n_{\text{occup}}$ = 1.8244]{%
    \includegraphics[width=0.22\textwidth]{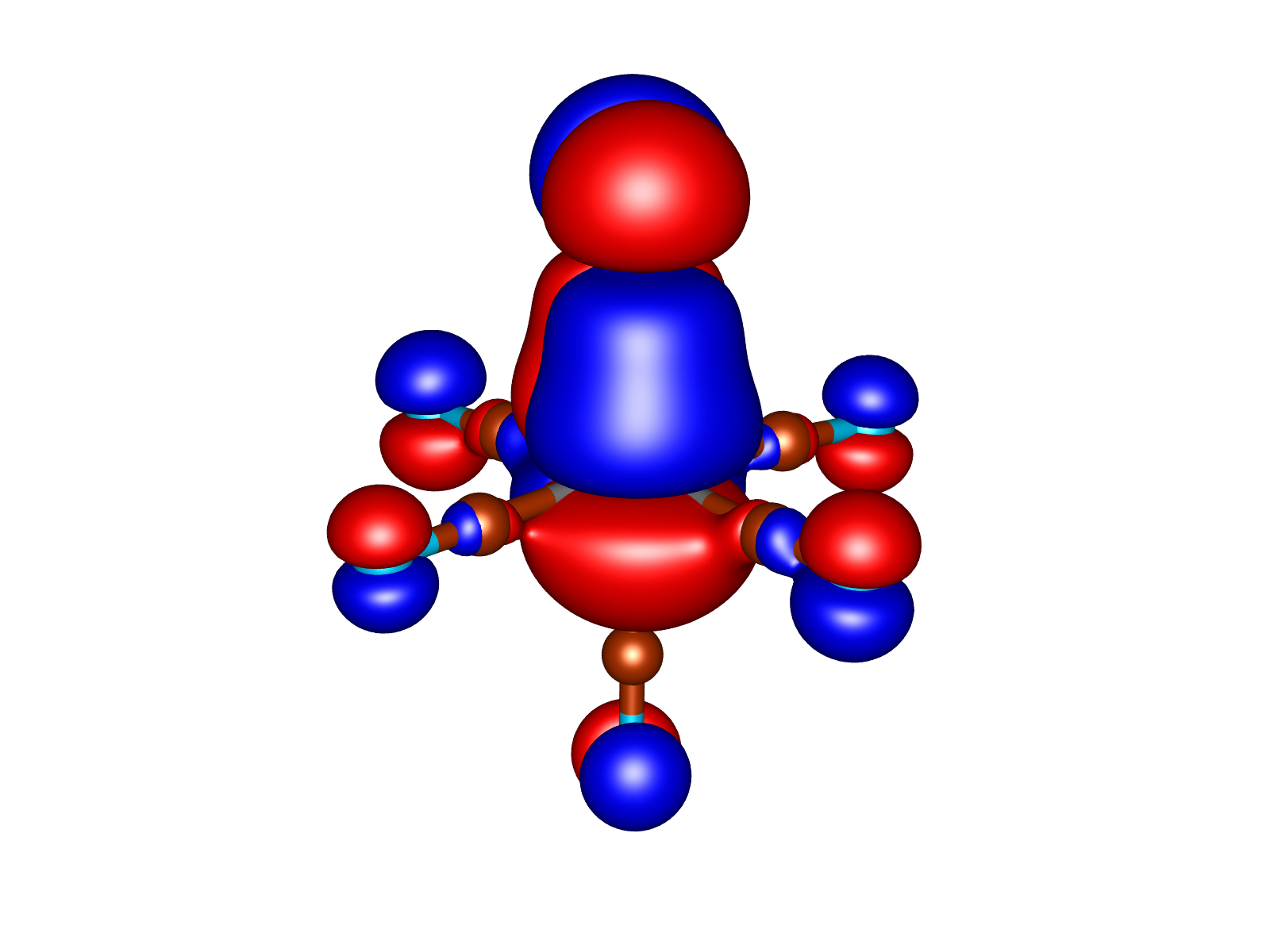}
  }
  \hfill
  \subfloat[$n_{\text{occup}}$ = 1.8244]{%
    \includegraphics[width=0.22\textwidth]{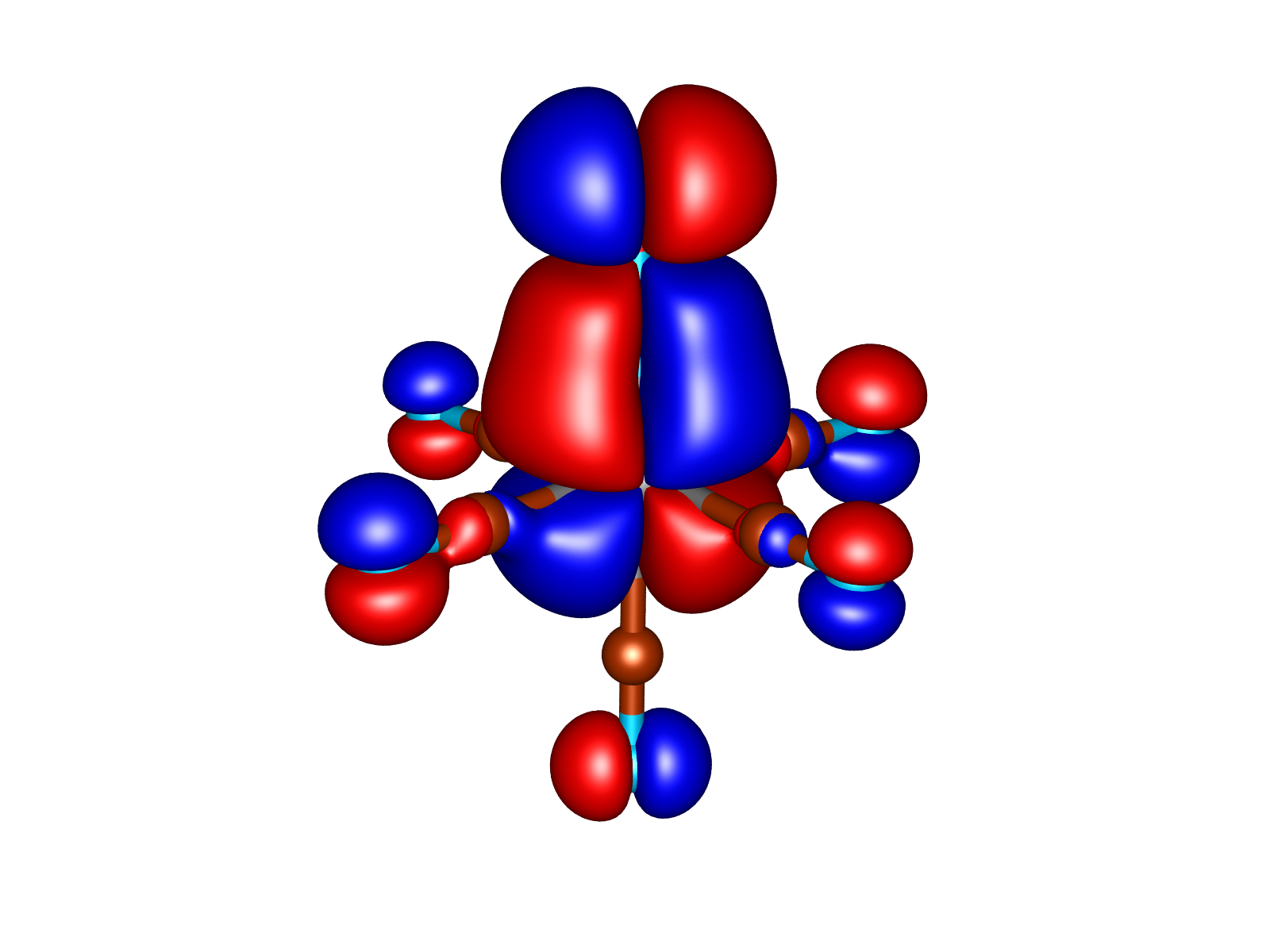}
  }
  \\
  \subfloat[$n_{\text{occup}}$ = 0.2060]{%
    \includegraphics[width=0.22\textwidth]{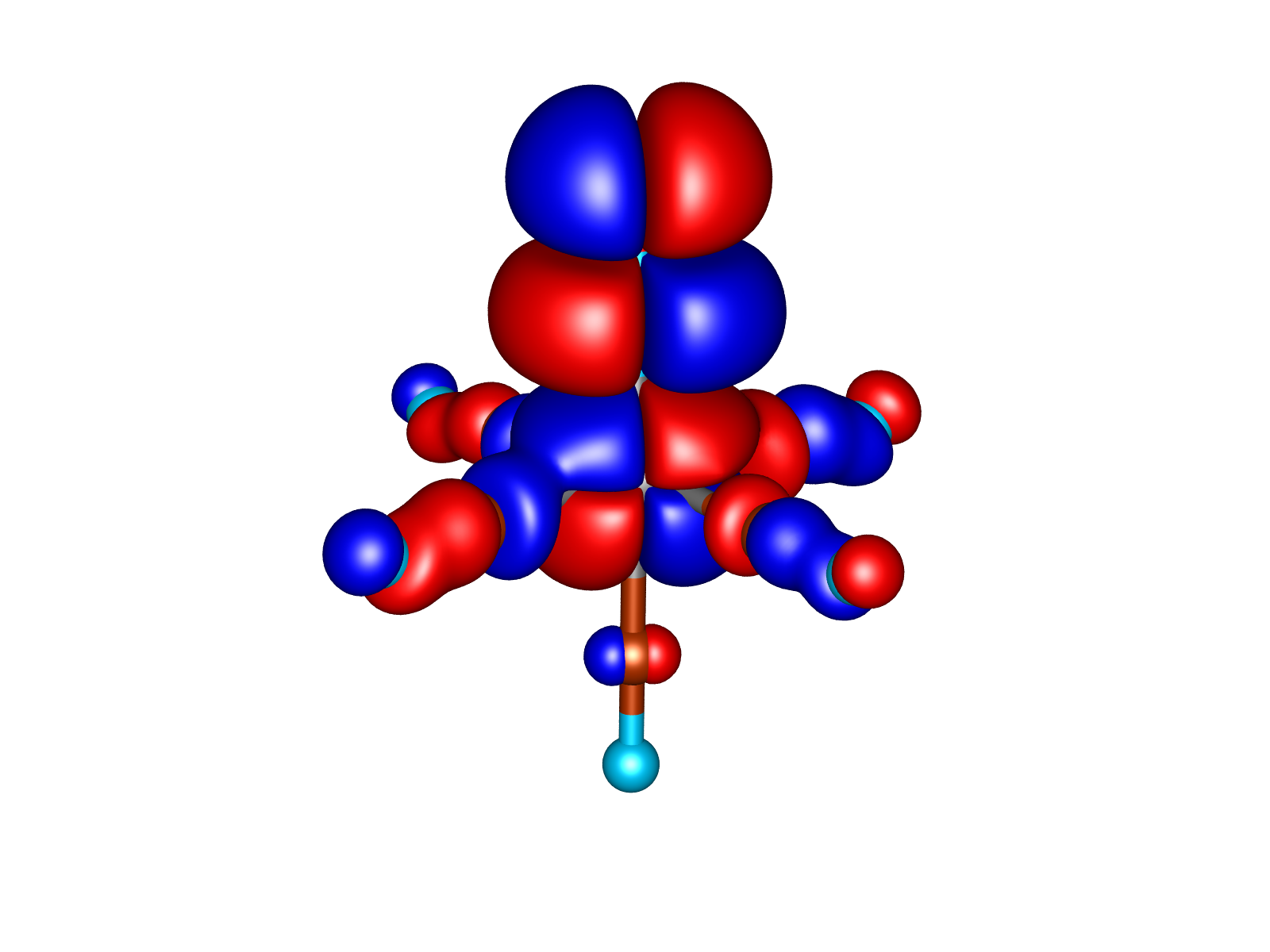}
  }
  \hfill
  \subfloat[$n_{\text{occup}}$ = 0.2060]{%
    \includegraphics[width=0.22\textwidth]{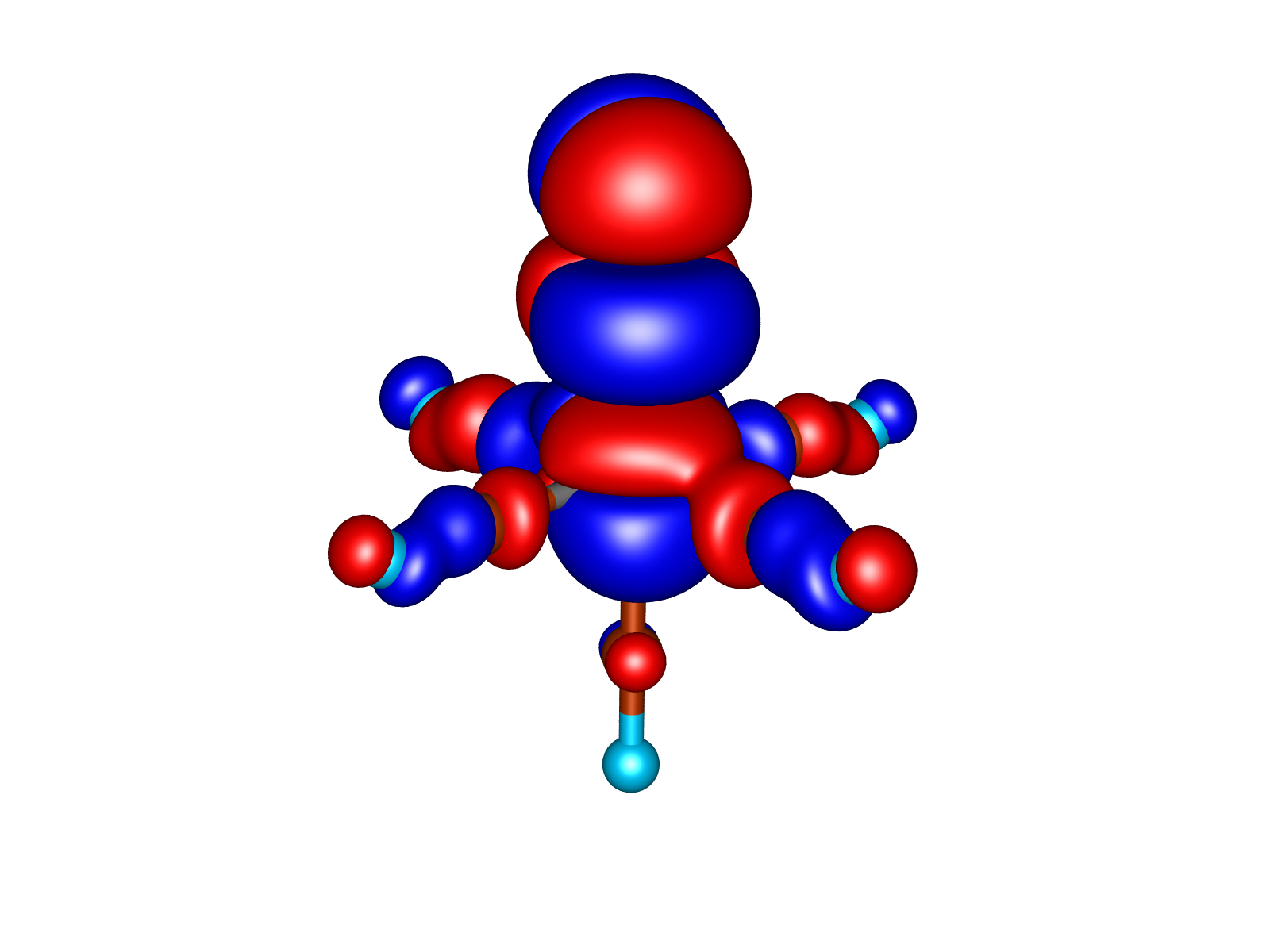}
  }
  \hfill
  \subfloat[$n_{\text{occup}}$ = 0.0610]{%
    \includegraphics[width=0.22\textwidth]{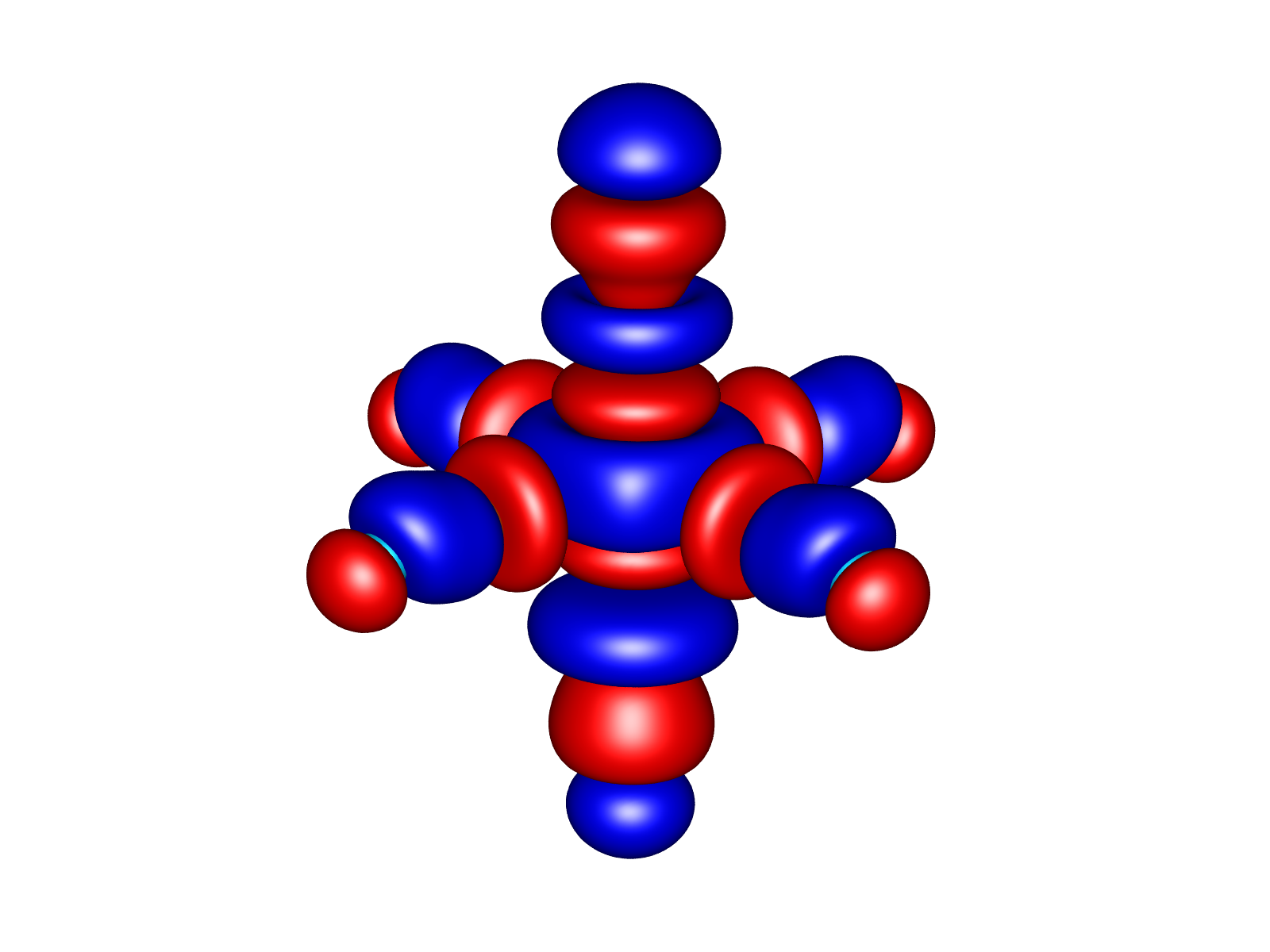}
  }
  \hfill
  \subfloat[$n_{\text{occup}}$ = 0.0537]{%
    \includegraphics[width=0.22\textwidth]{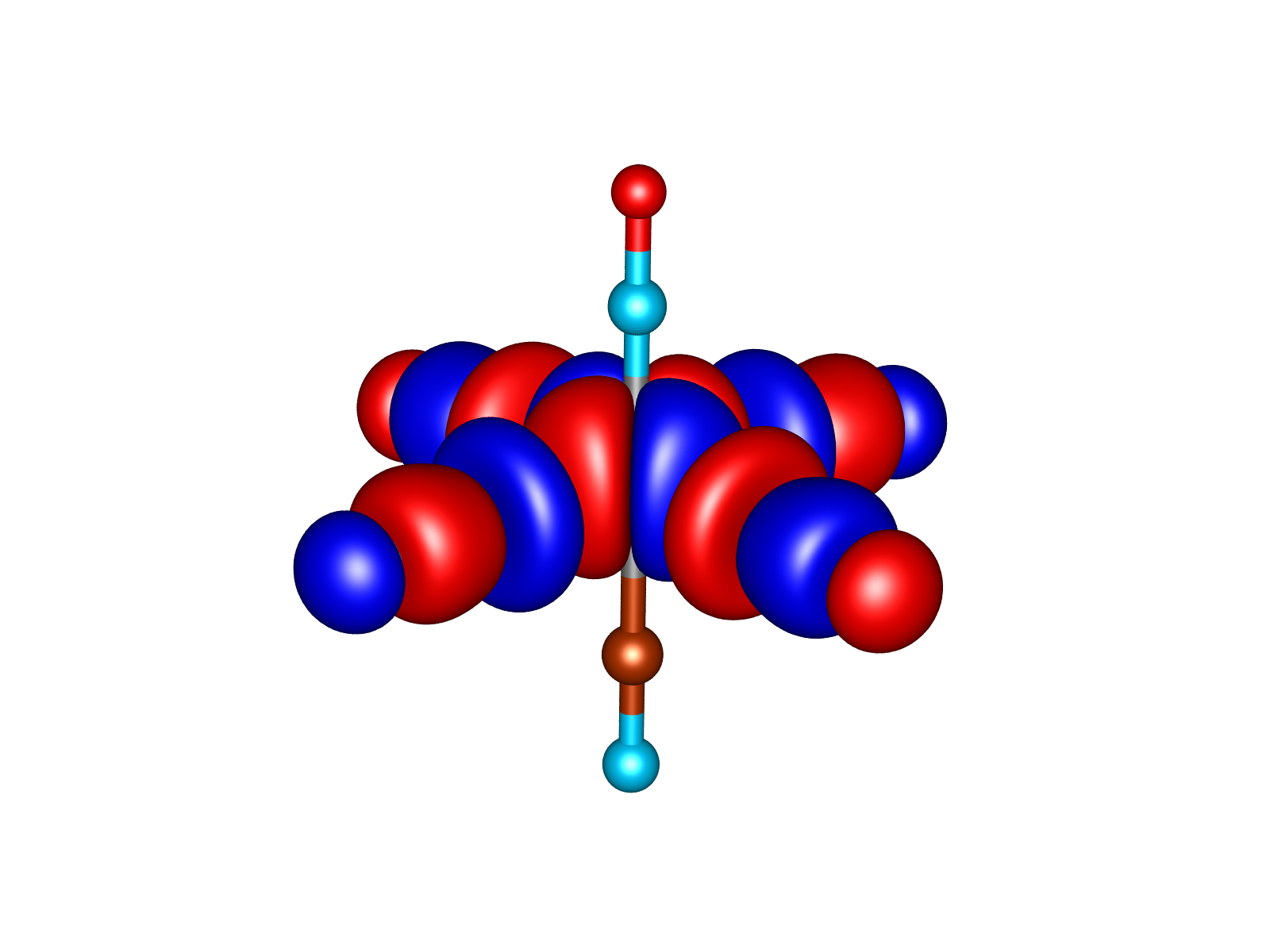}
  }
  \\
  \subfloat[$n_{\text{occup}}$ = 0.0253]{%
    \includegraphics[width=0.22\textwidth]{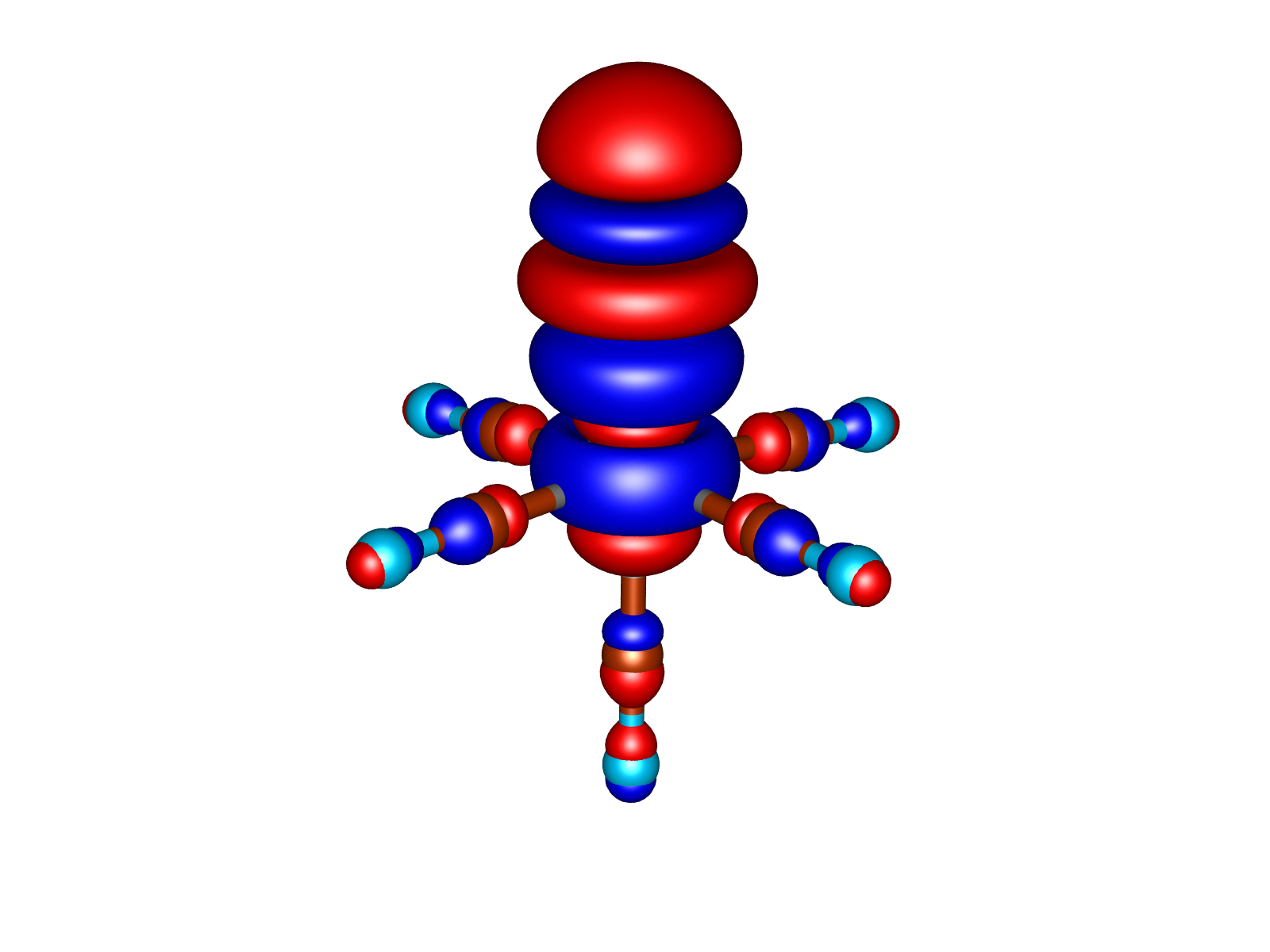}
  }
  \hfill
  \subfloat[$n_{\text{occup}}$ = 0.0227]{%
    \includegraphics[width=0.22\textwidth]{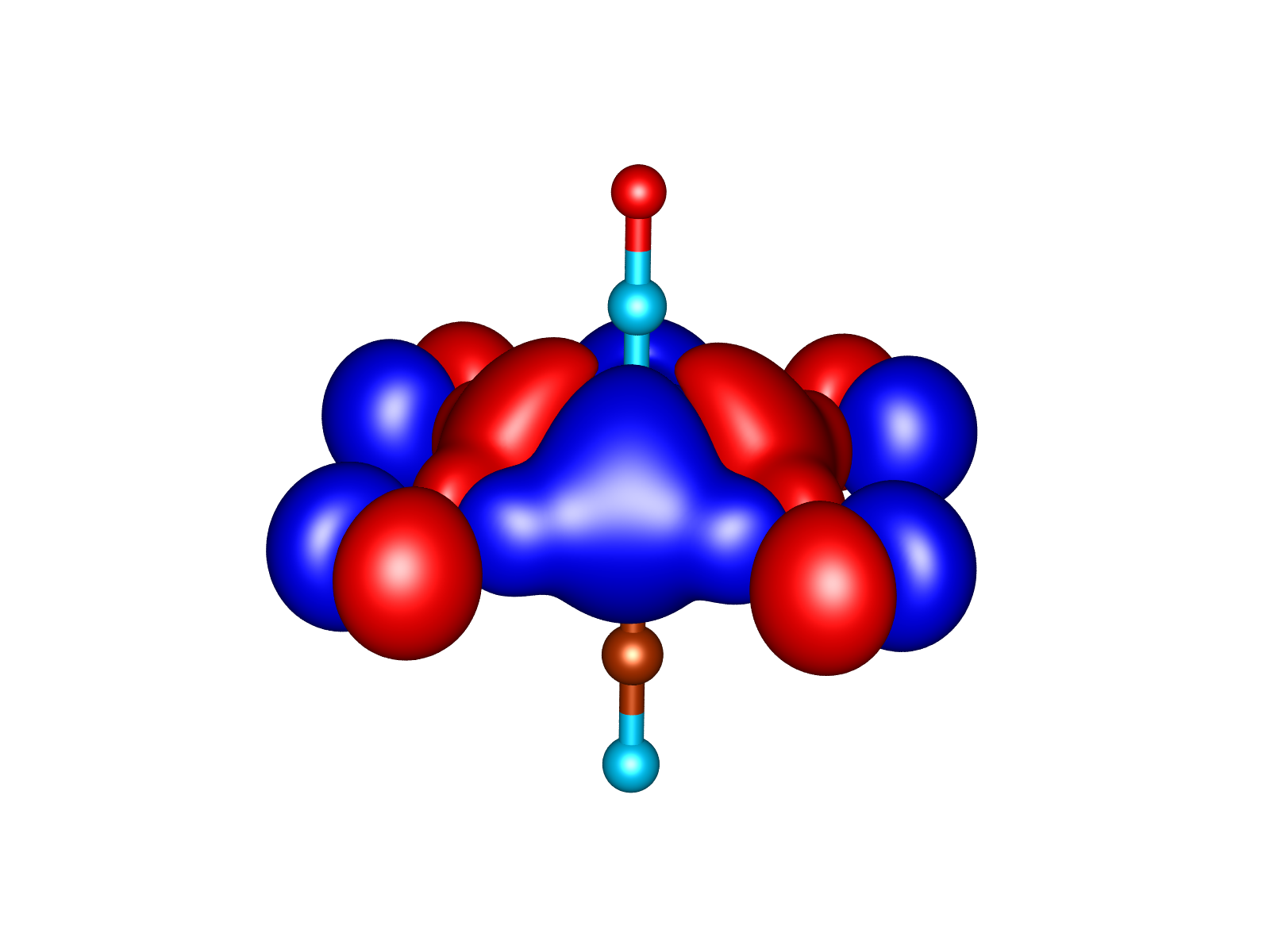}
  }
  \hfill
  \subfloat[$n_{\text{occup}}$ = 0.0109]{%
    \includegraphics[width=0.22\textwidth]{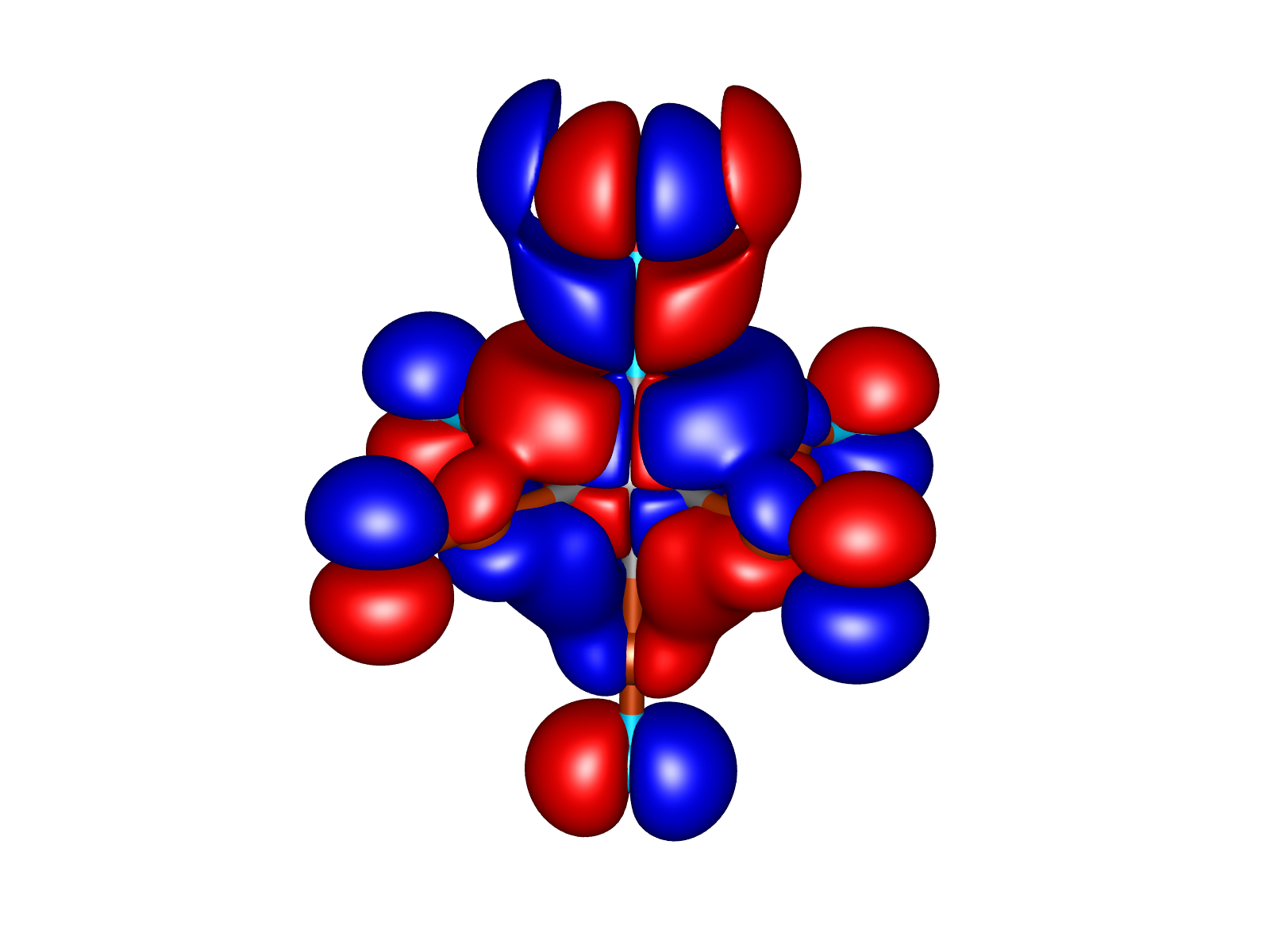}
  }
  \hfill
  \subfloat[$n_{\text{occup}}$ = 0.0109]{%
    \includegraphics[width=0.22\textwidth]{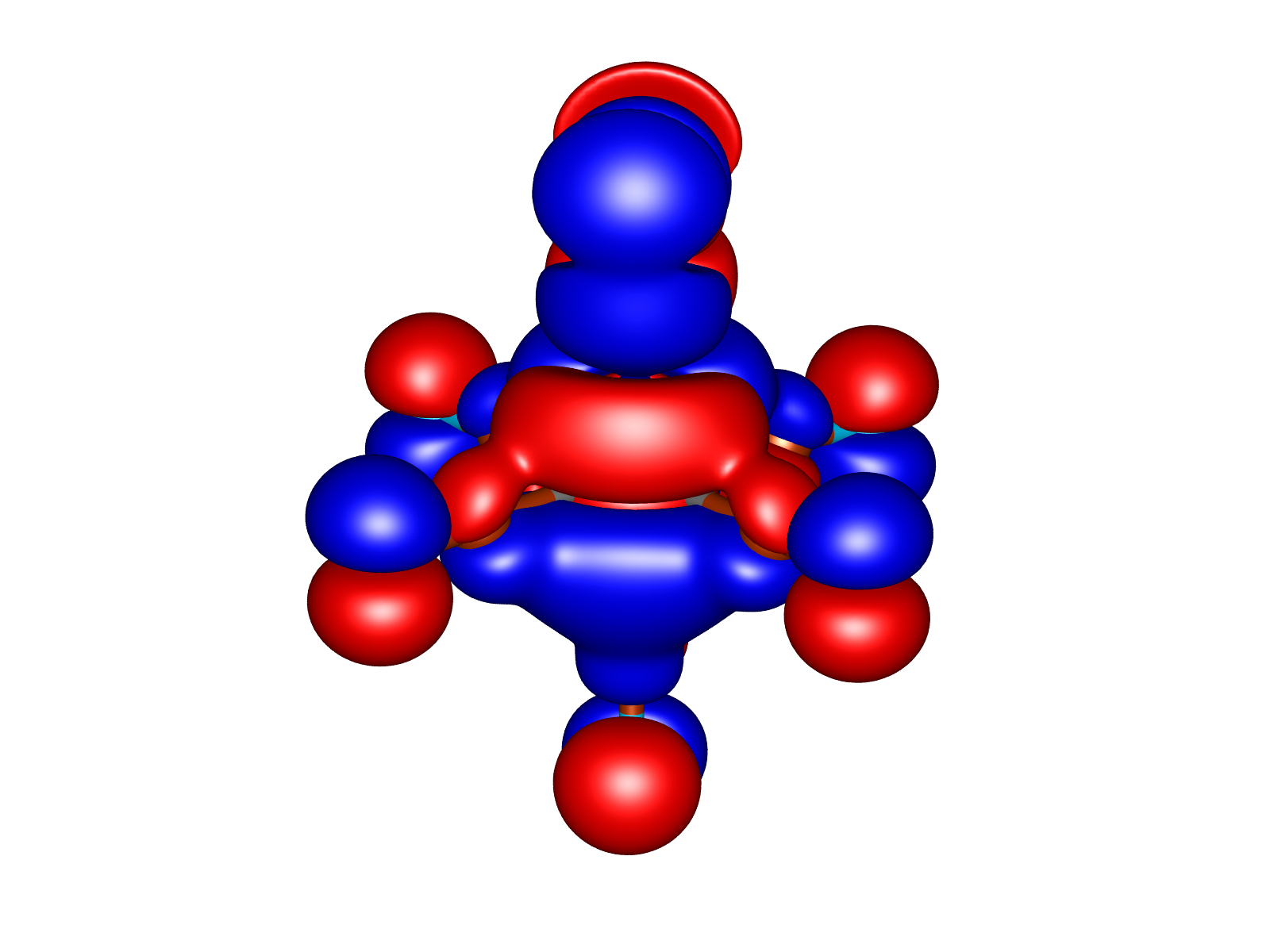}
  }
  \\
  \caption{Fe-NO complex, standard, DMRG-SCF(16, 16) \label{orbs_cas1616_1}}
\end{figure}

\renewcommand{\thesubfigure}{\arabic{subfigure}}
\begin{figure}[!h]
  \subfloat[$n_{\text{occup}}$ = 1.9720 ]{%
    \includegraphics[width=0.22\textwidth]{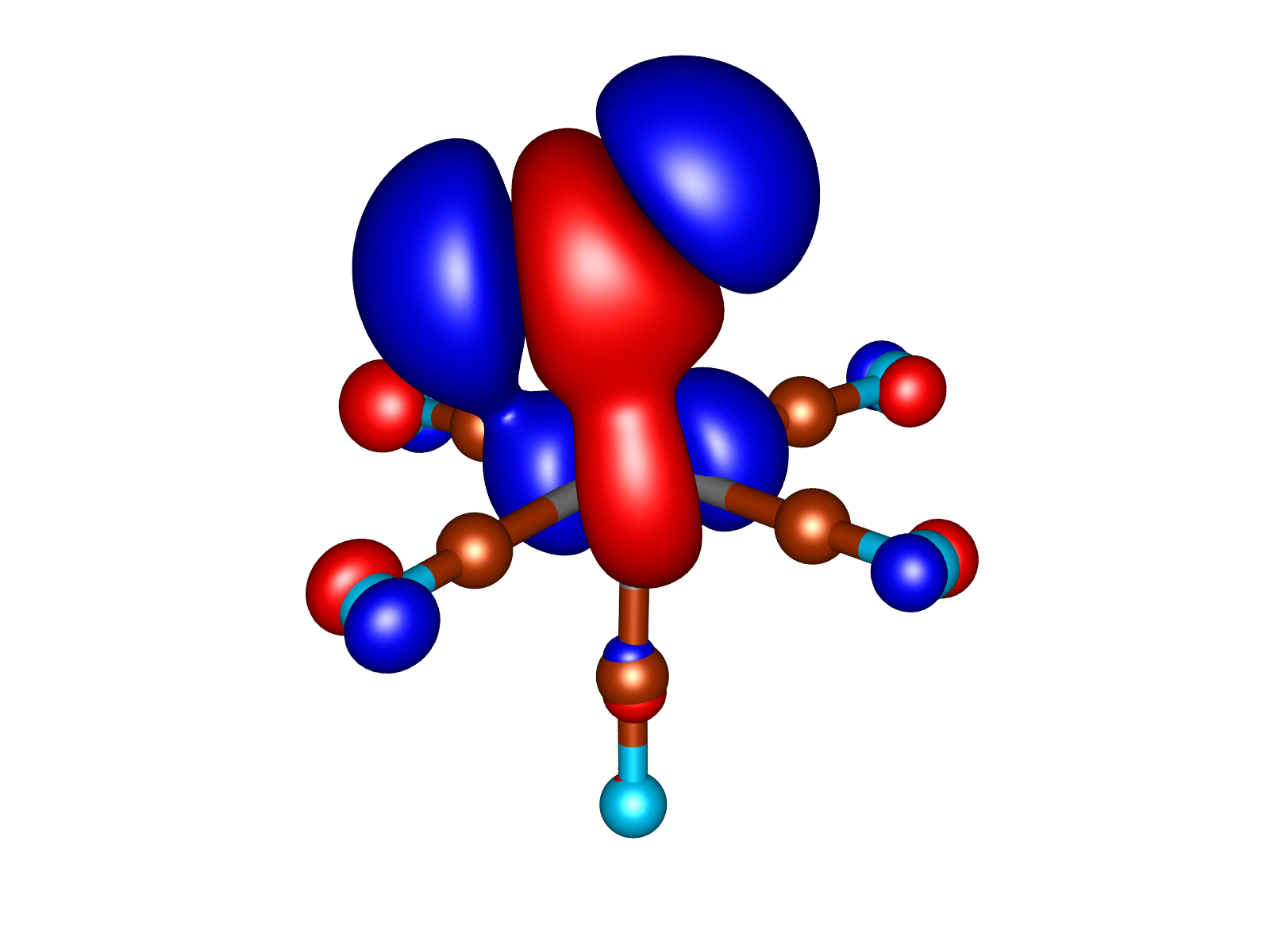}
  }
  \hfill
  \subfloat[$n_{\text{occup}}$ = 1.9680 ]{%
    \includegraphics[width=0.22\textwidth]{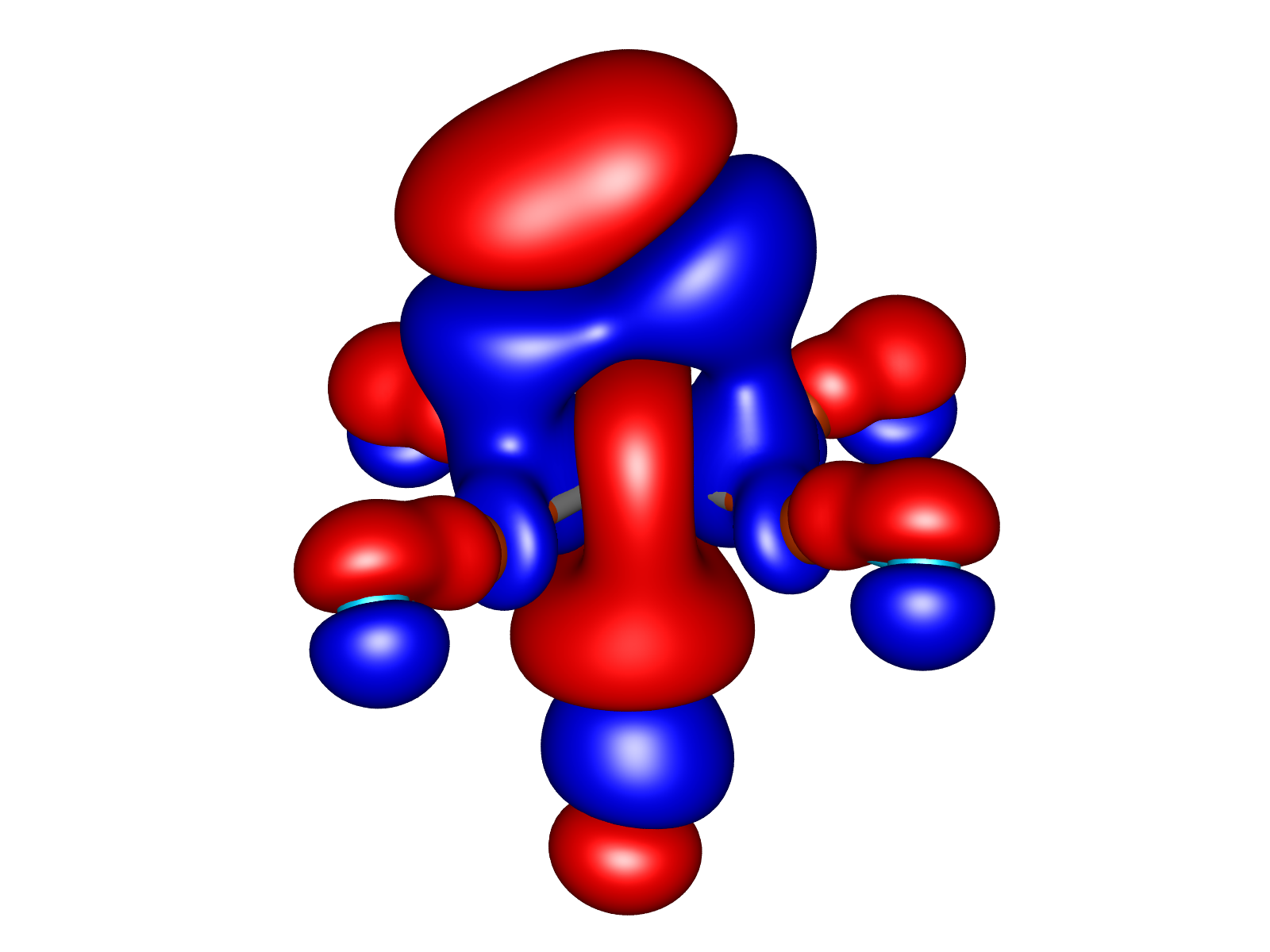}
  }
  \hfill
  \subfloat[$n_{\text{occup}}$ = 1.9638 ]{%
    \includegraphics[width=0.22\textwidth]{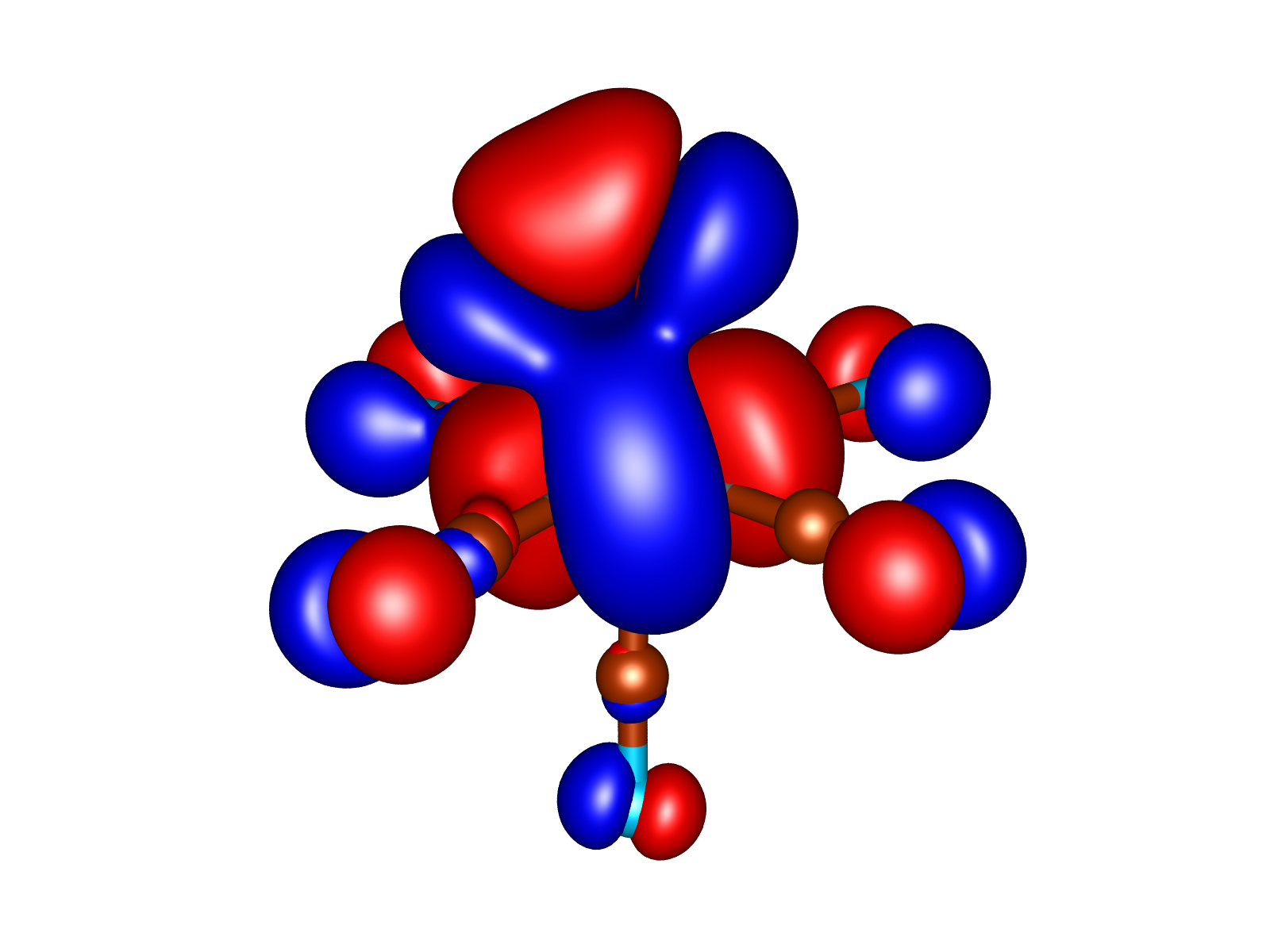}
  }
  \hfill
  \subfloat[$n_{\text{occup}}$ = 1.9578 ]{%
    \includegraphics[width=0.22\textwidth]{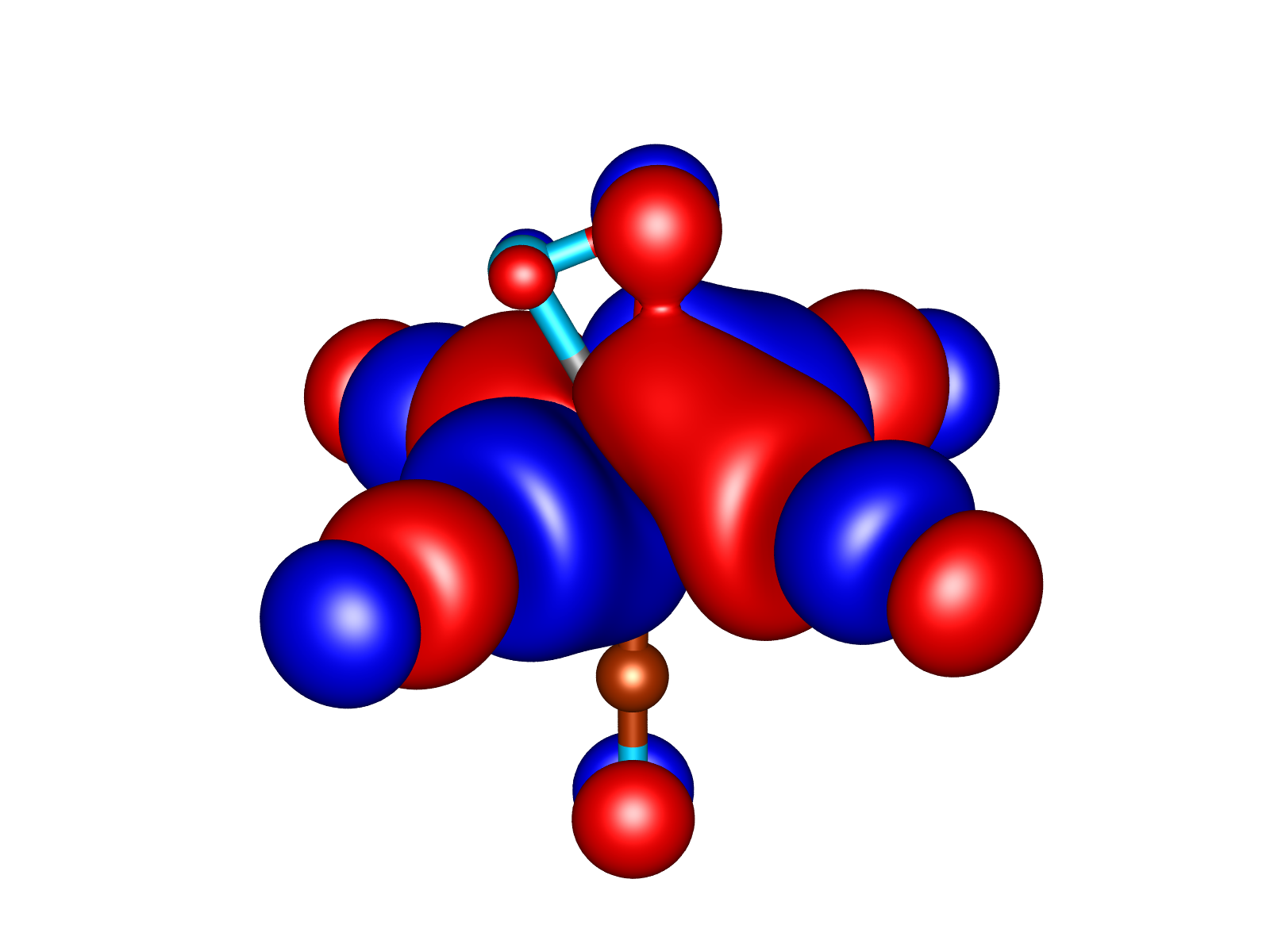}
  } \\
  \subfloat[$n_{\text{occup}}$ = 1.9506 ]{%
    \includegraphics[width=0.22\textwidth]{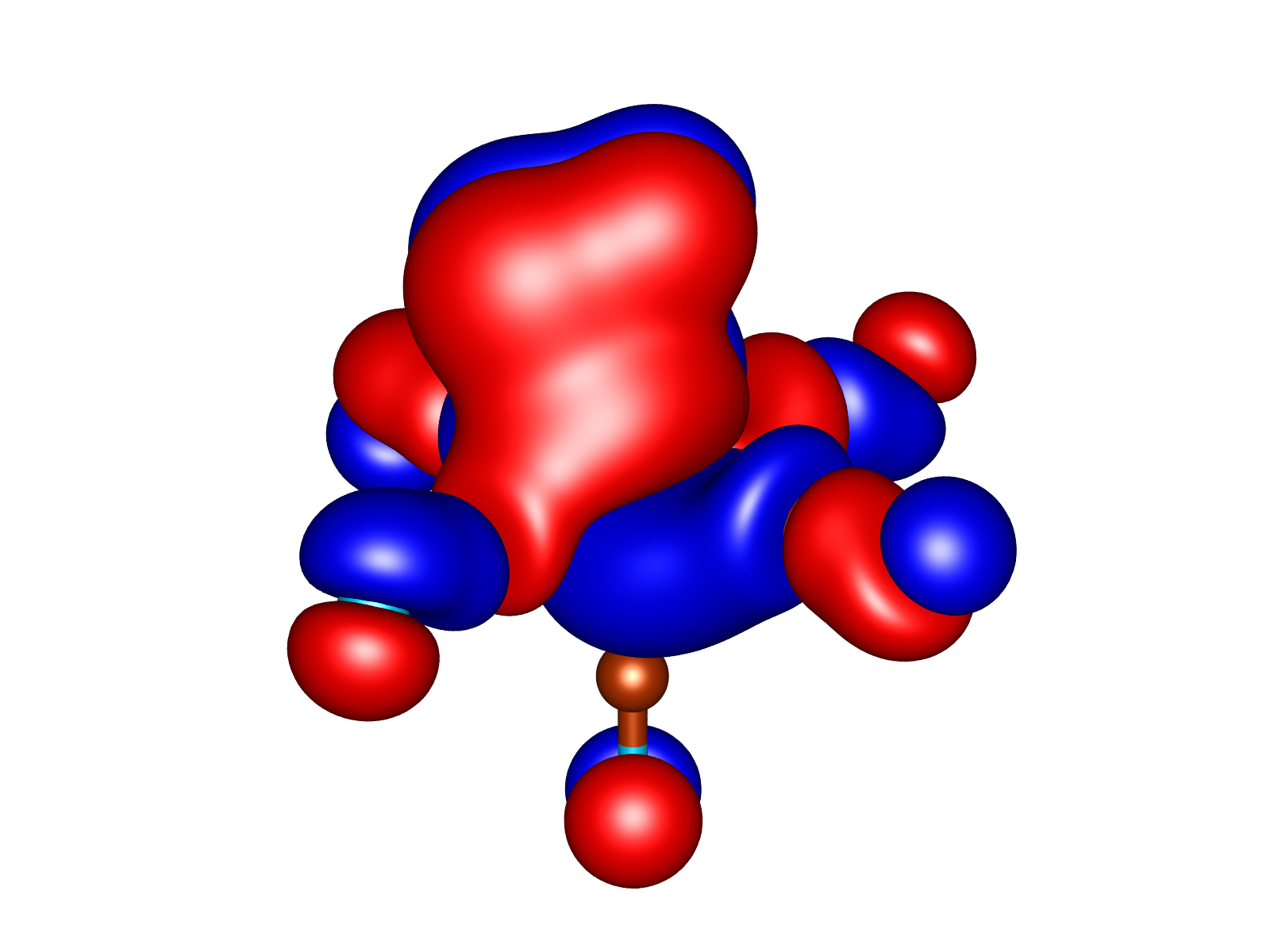}
  }
  \hfill
  \subfloat[$n_{\text{occup}}$ = 1.9419 ]{%
    \includegraphics[width=0.22\textwidth]{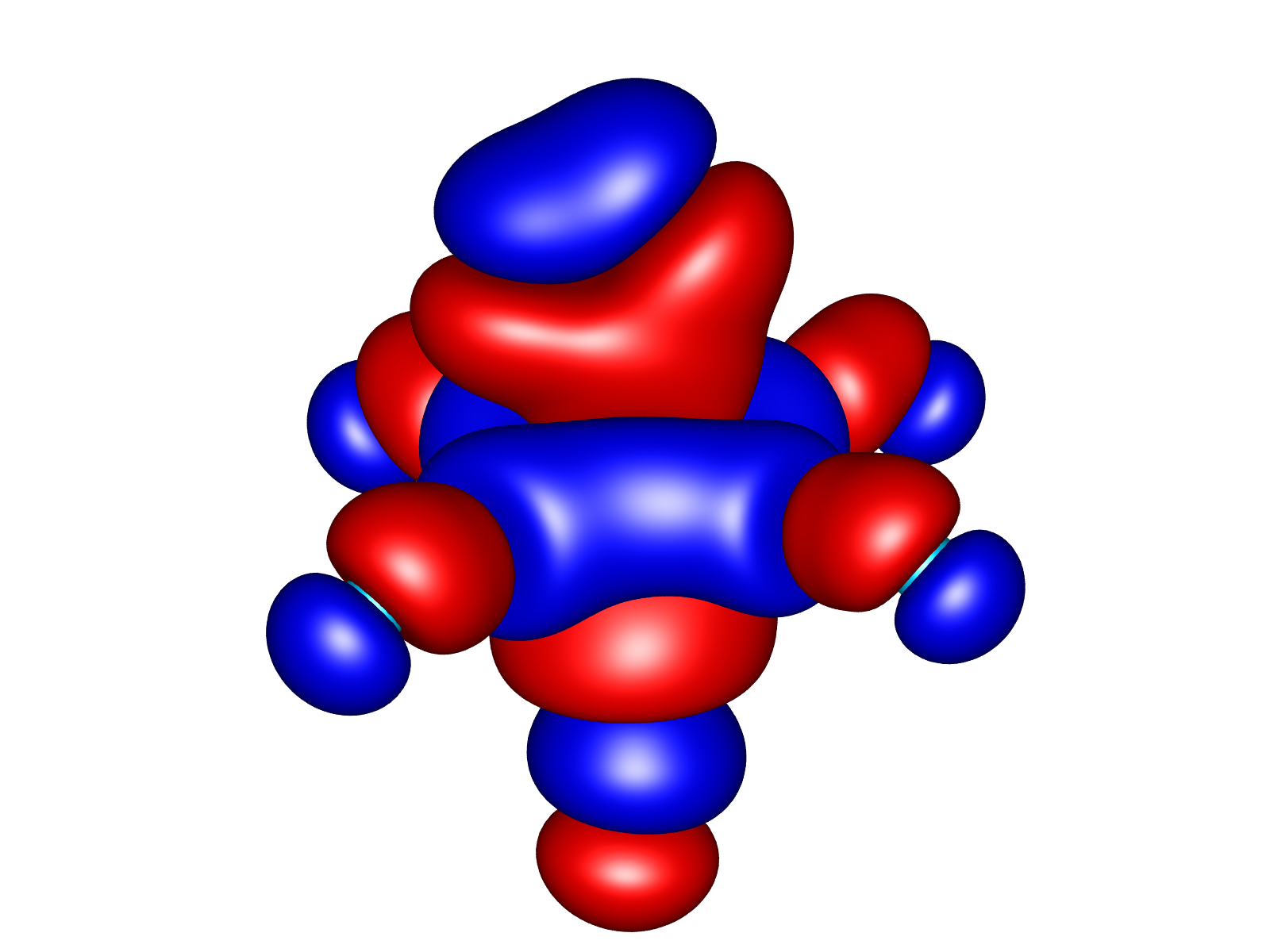}
  }
  \hfill
  \subfloat[$n_{\text{occup}}$ = 1.9265 ]{%
    \includegraphics[width=0.22\textwidth]{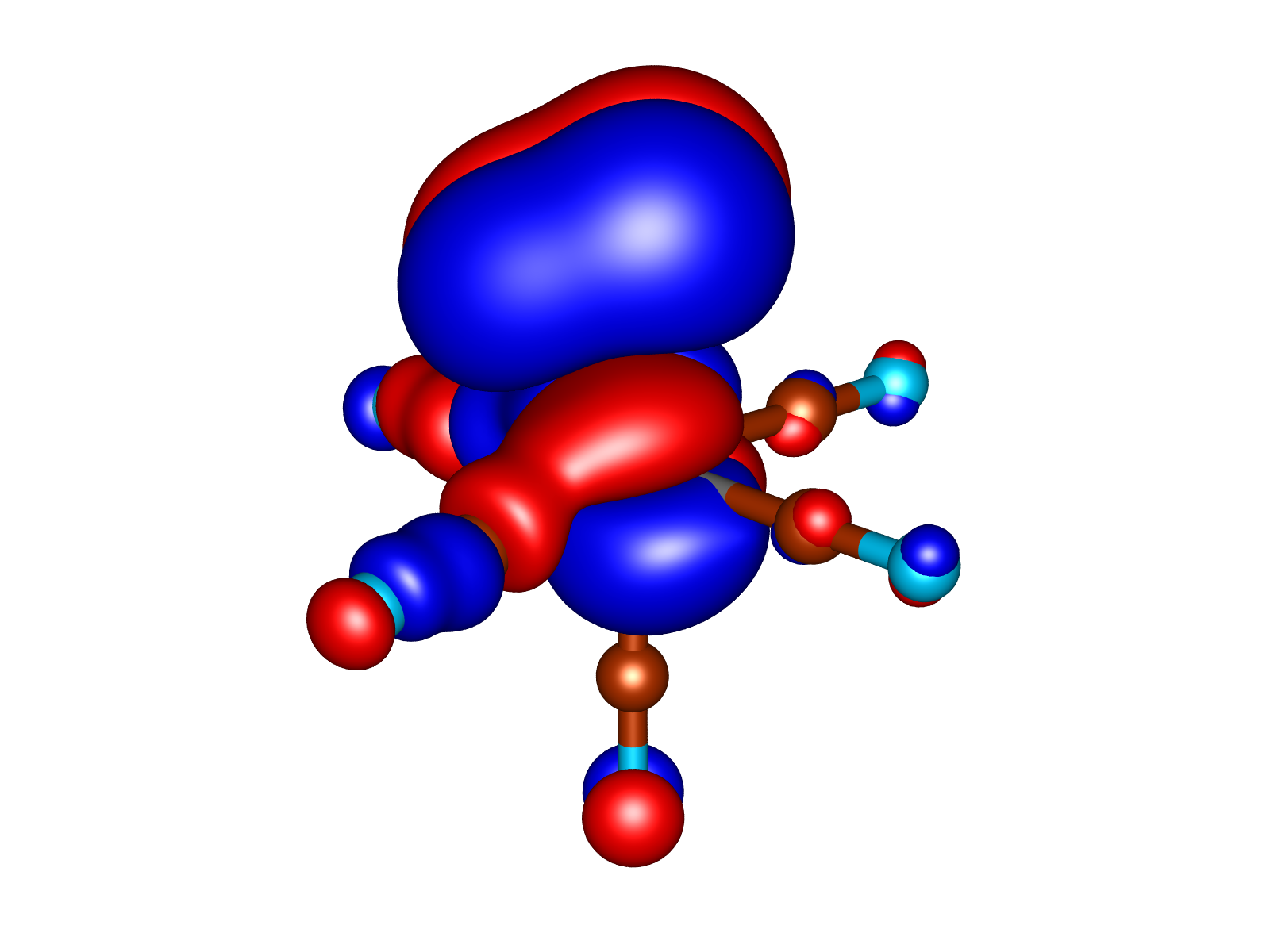}
  }
  \hfill
  \subfloat[$n_{\text{occup}}$ = 1.7734 ]{%
    \includegraphics[width=0.22\textwidth]{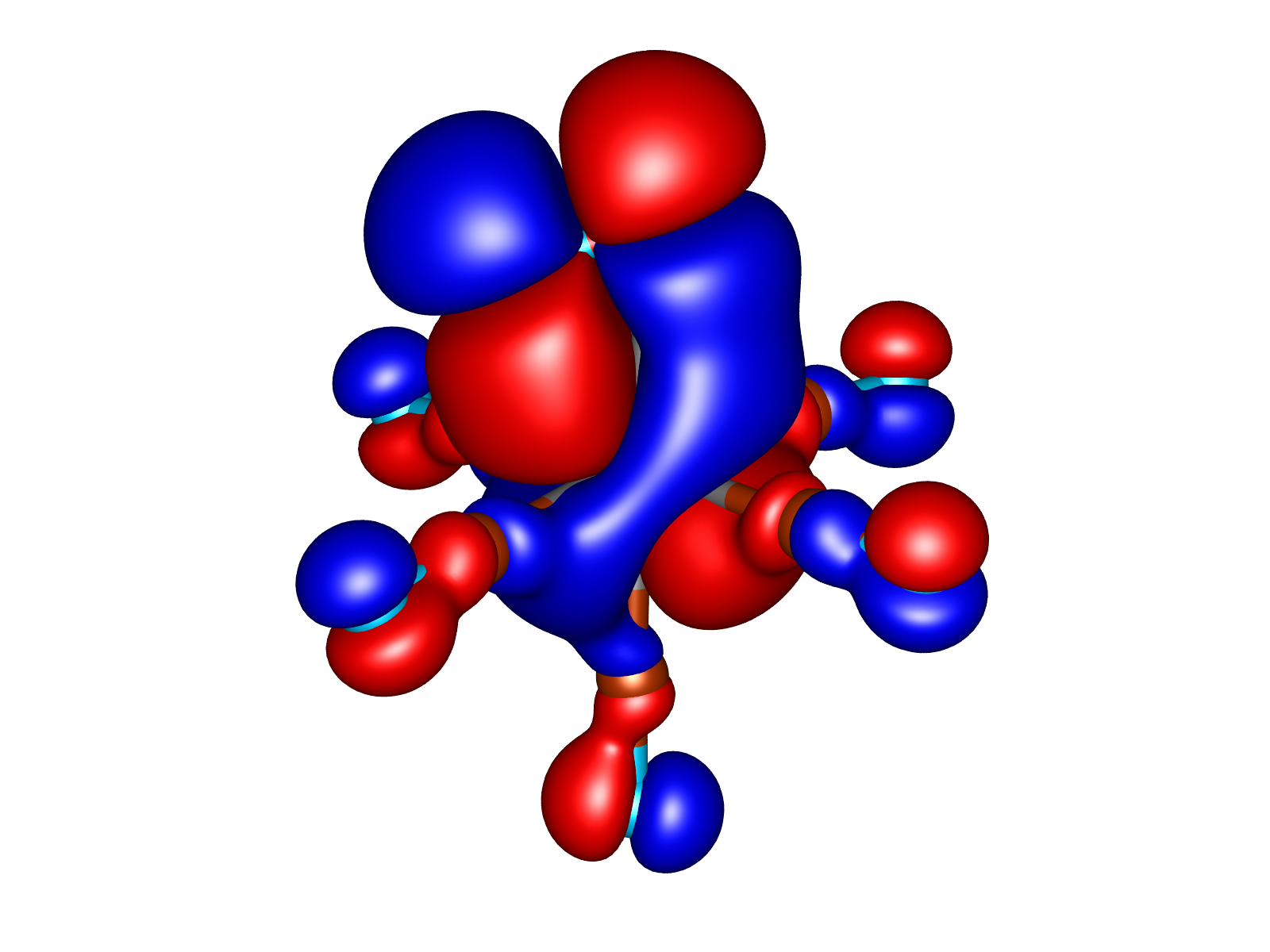}
  }
  \\
  \subfloat[$n_{\text{occup}}$ = 0.2488 ]{%
    \includegraphics[width=0.22\textwidth]{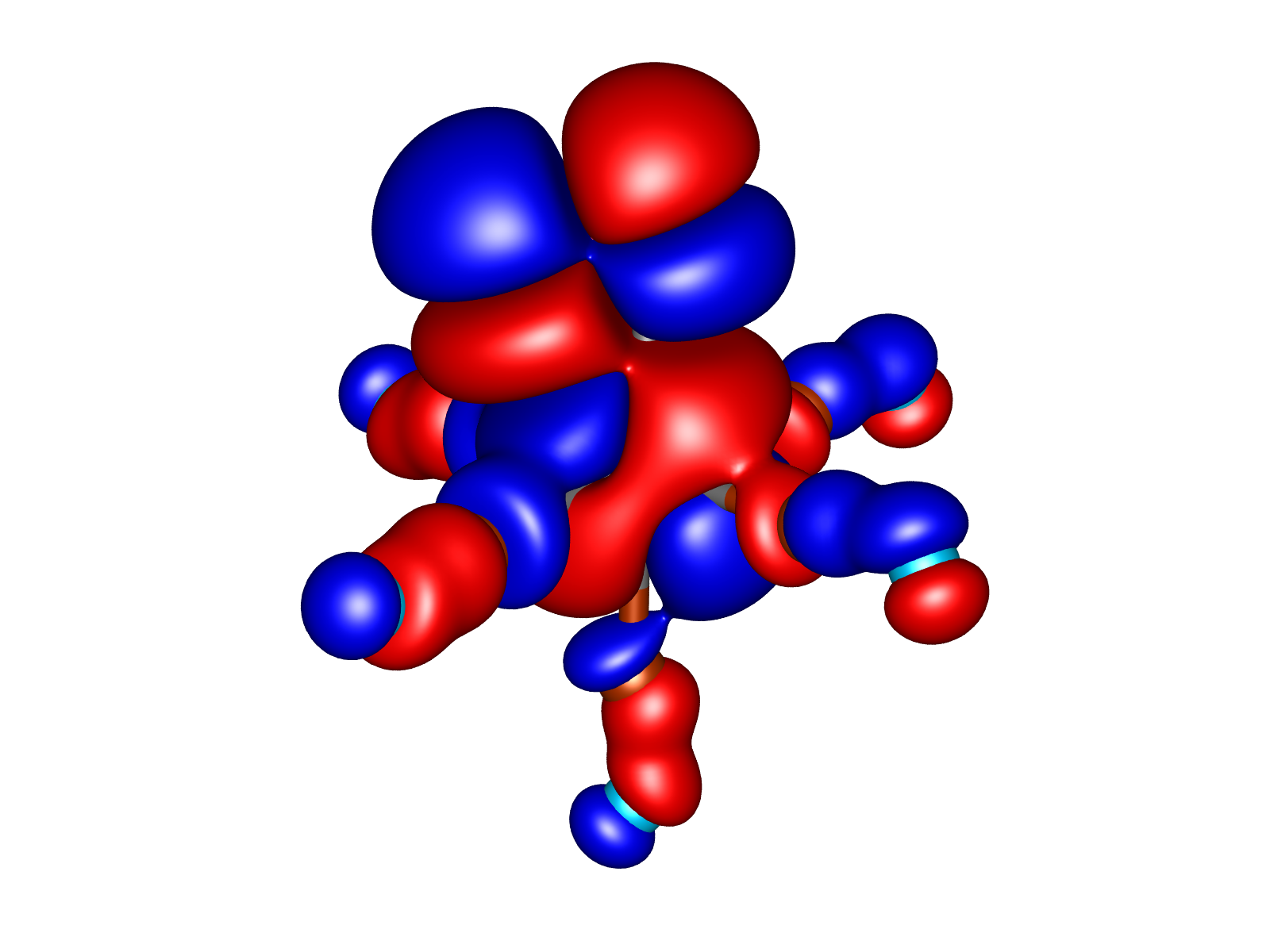}
  }
  \hfill
  \subfloat[$n_{\text{occup}}$ = 0.0966 ]{%
    \includegraphics[width=0.22\textwidth]{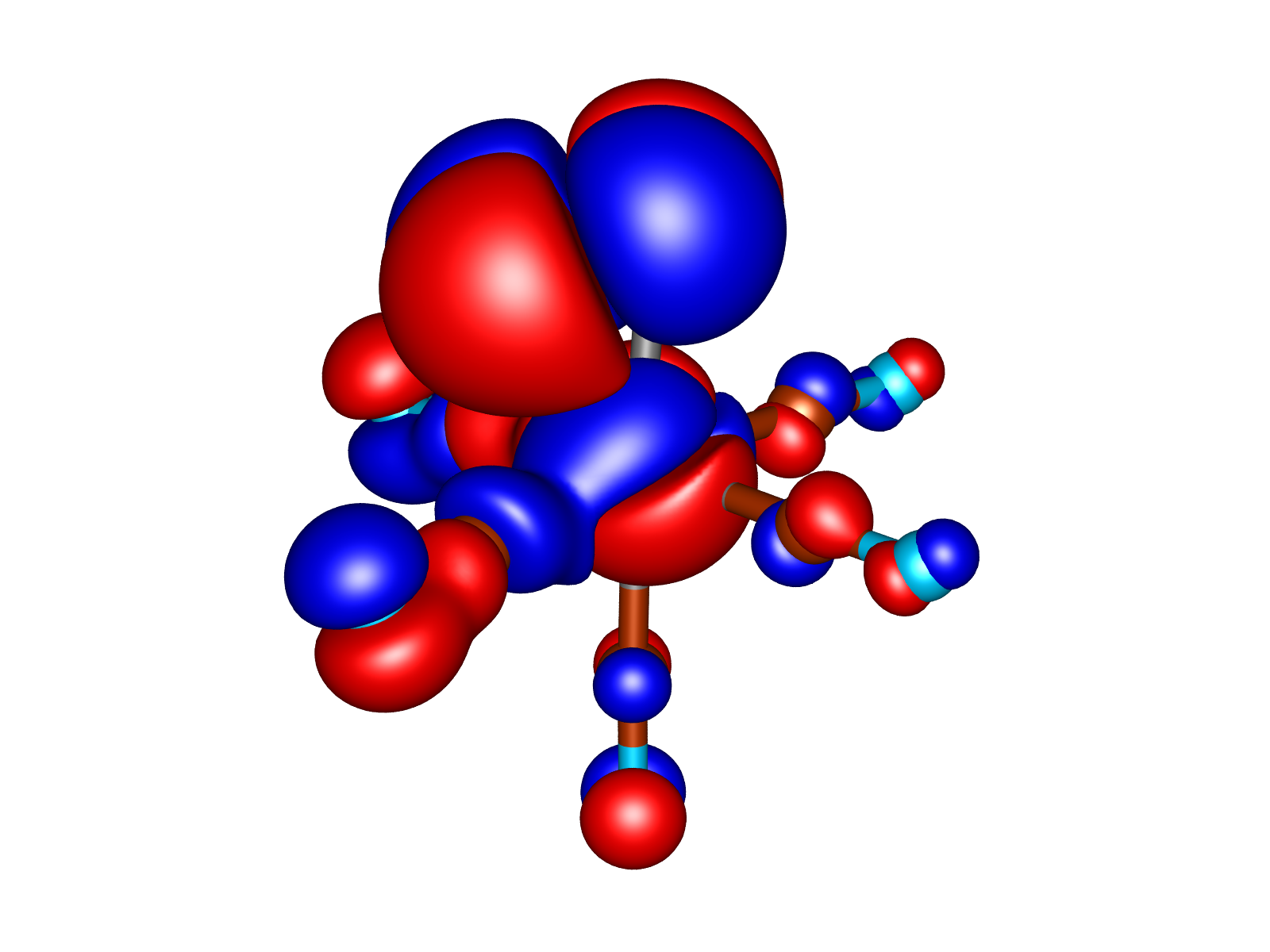}
  }
  \hfill
  \subfloat[$n_{\text{occup}}$ = 0.0701 ]{%
    \includegraphics[width=0.22\textwidth]{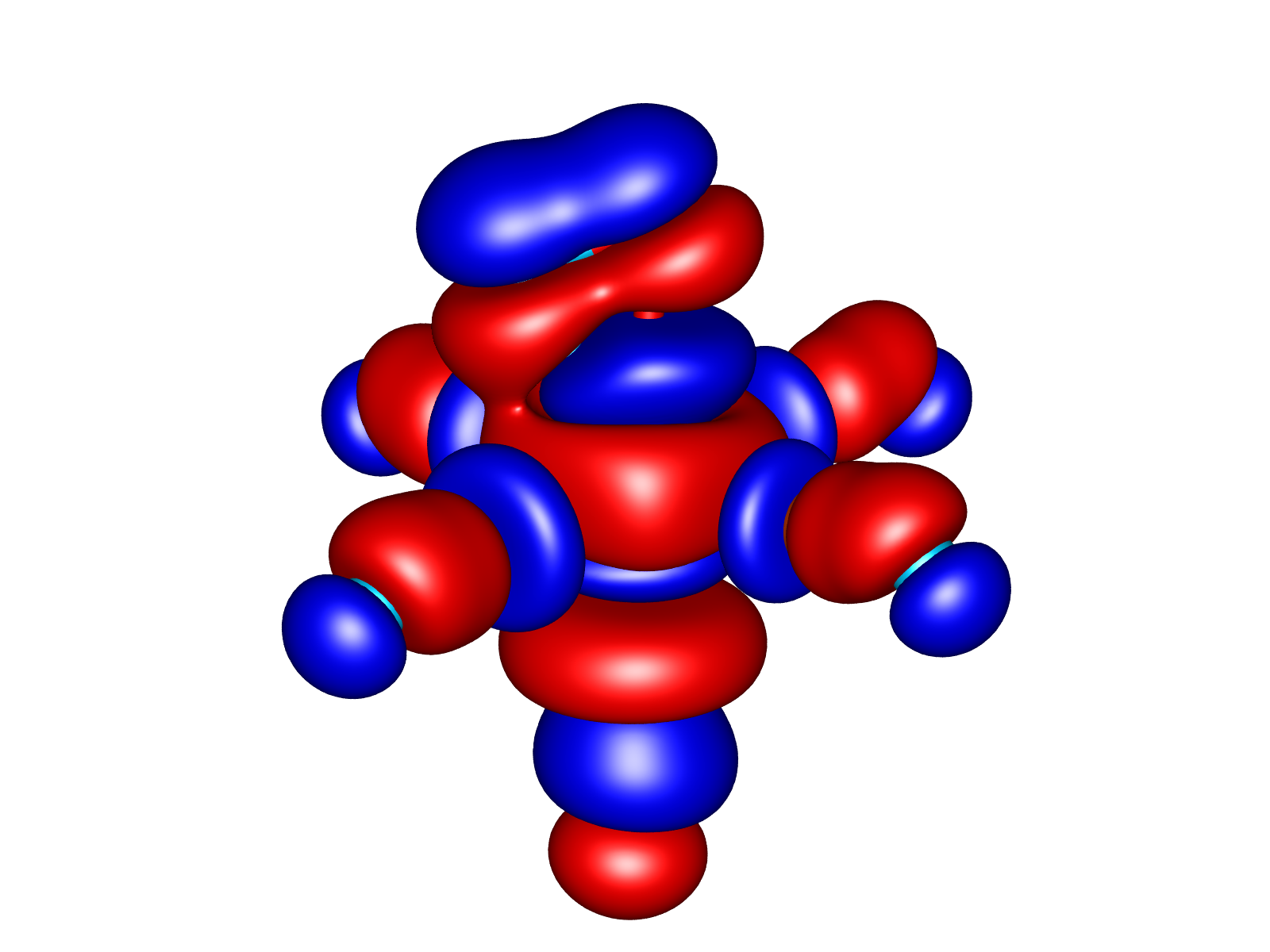}
  }
  \hfill
  \subfloat[$n_{\text{occup}}$ =  0.0498 ]{%
    \includegraphics[width=0.22\textwidth]{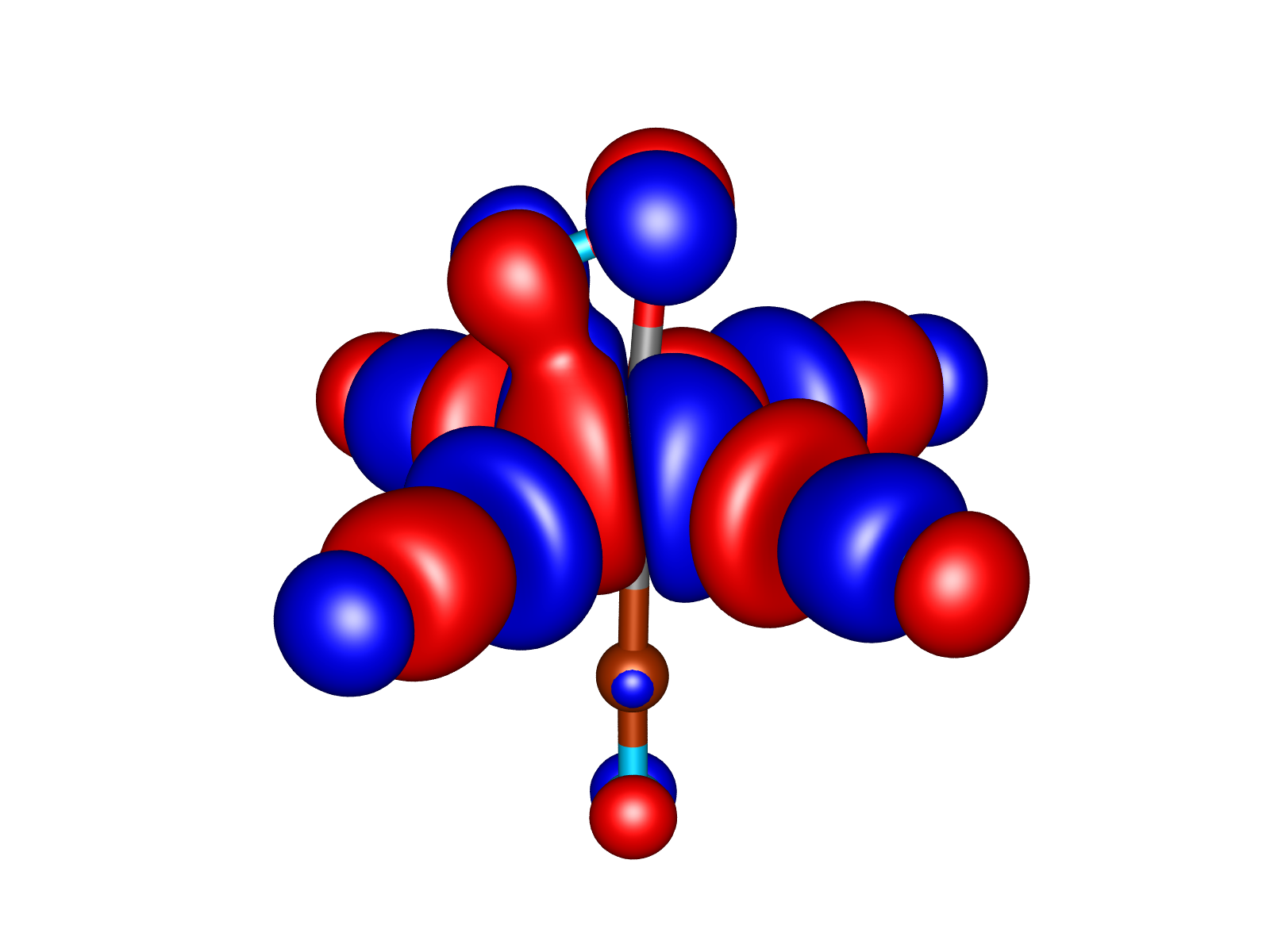}
  }
  \\
  \subfloat[$n_{\text{occup}}$ = 0.0300 ]{%
    \includegraphics[width=0.22\textwidth]{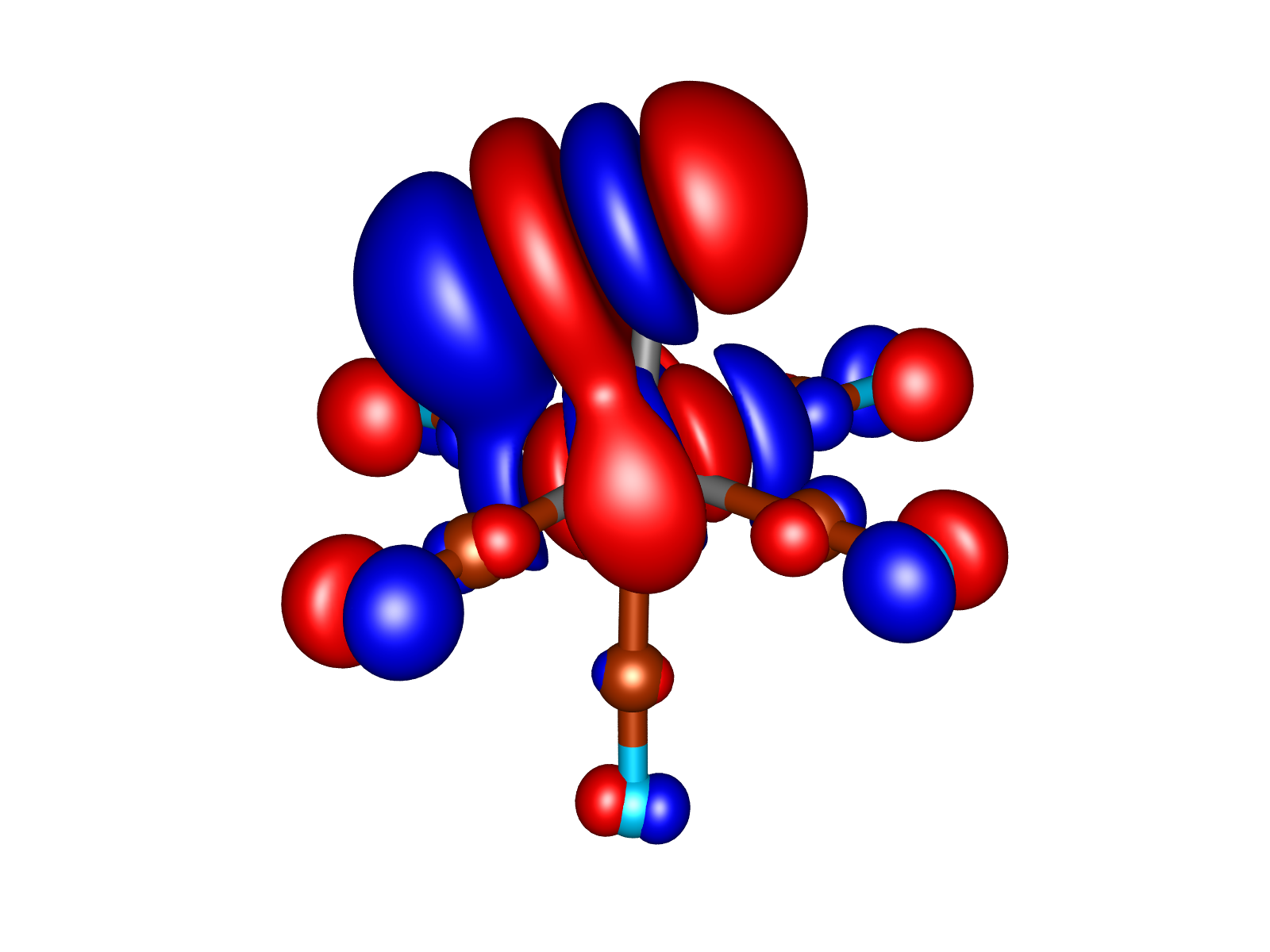}
  }
  \hfill
  \subfloat[$n_{\text{occup}}$ = 0.0227 ]{%
    \includegraphics[width=0.22\textwidth]{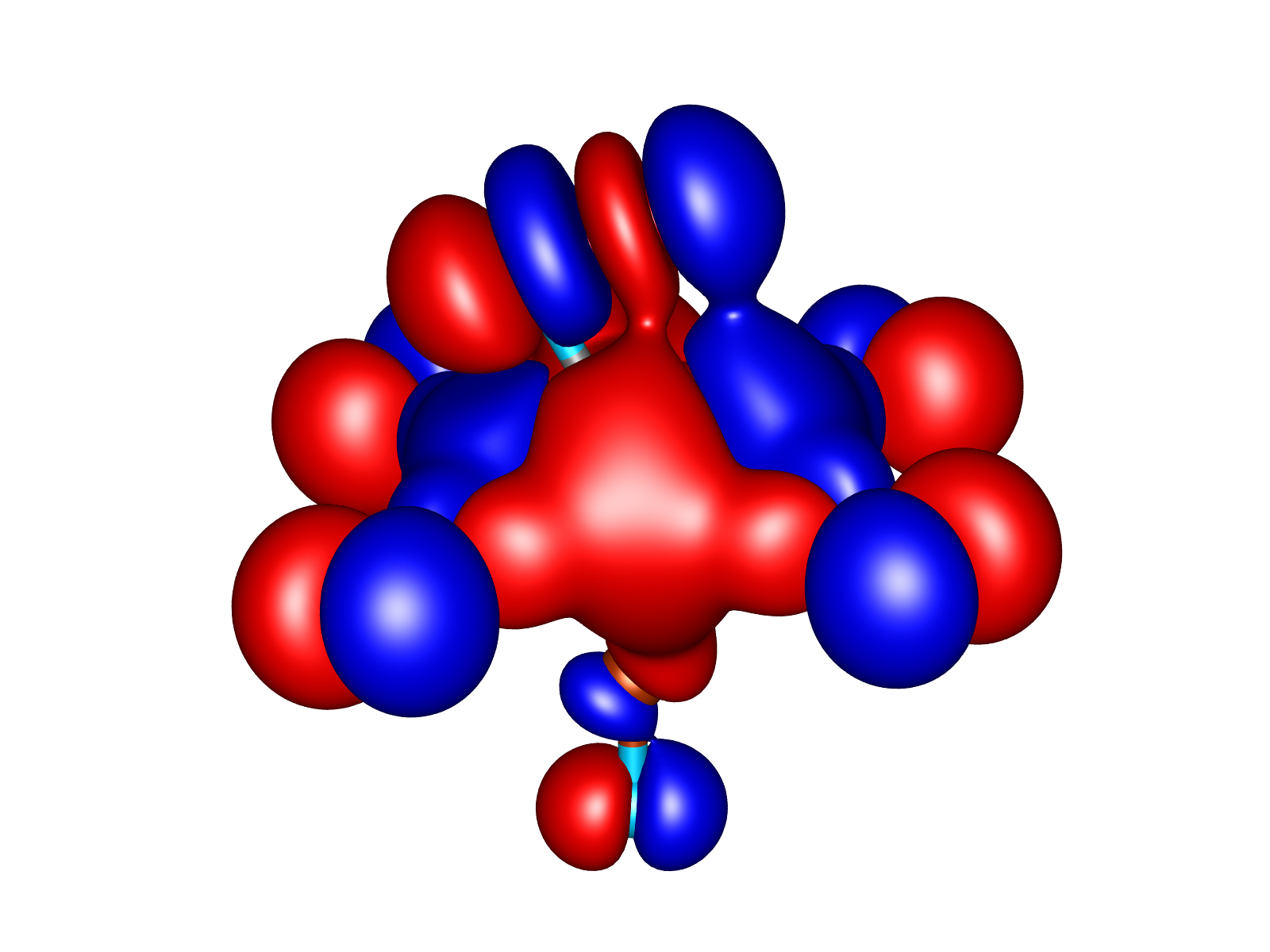}
  }
  \hfill
  \subfloat[$n_{\text{occup}}$ = 0.0170 ]{%
    \includegraphics[width=0.22\textwidth]{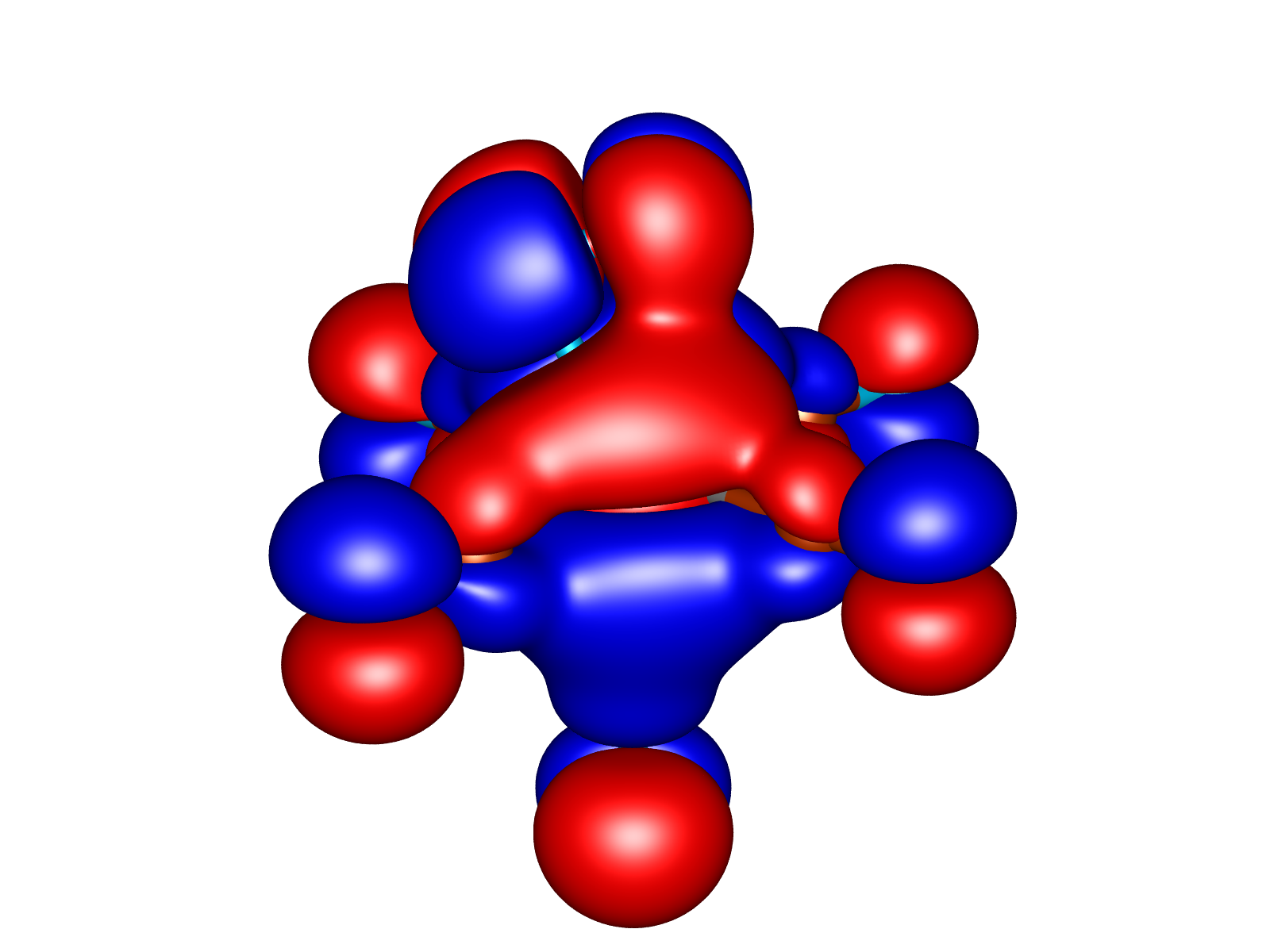}
  }
  \hfill
  \subfloat[$n_{\text{occup}}$ = 0.0112 ]{%
    \includegraphics[width=0.22\textwidth]{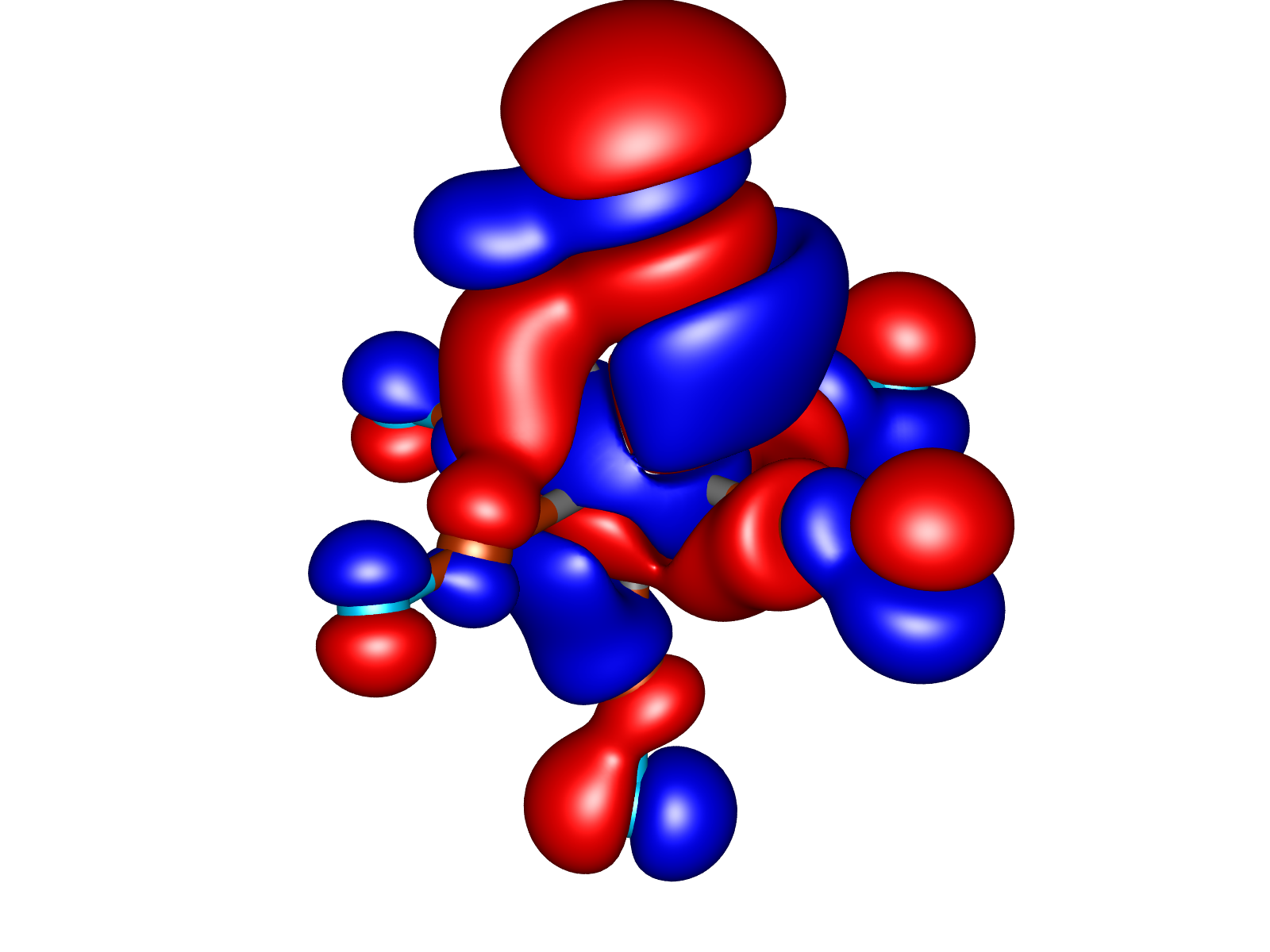}
  }
  \\
  \caption{Fe-NO complex, flat, DMRG-SCF(16, 16) \label{orbs_cas1616_3}}
\end{figure}

\renewcommand{\thesubfigure}{\arabic{subfigure}}
\begin{figure}[!h]
  \subfloat[$n_{\text{occup}}$ = 1.9768]{%
    \includegraphics[width=0.22\textwidth]{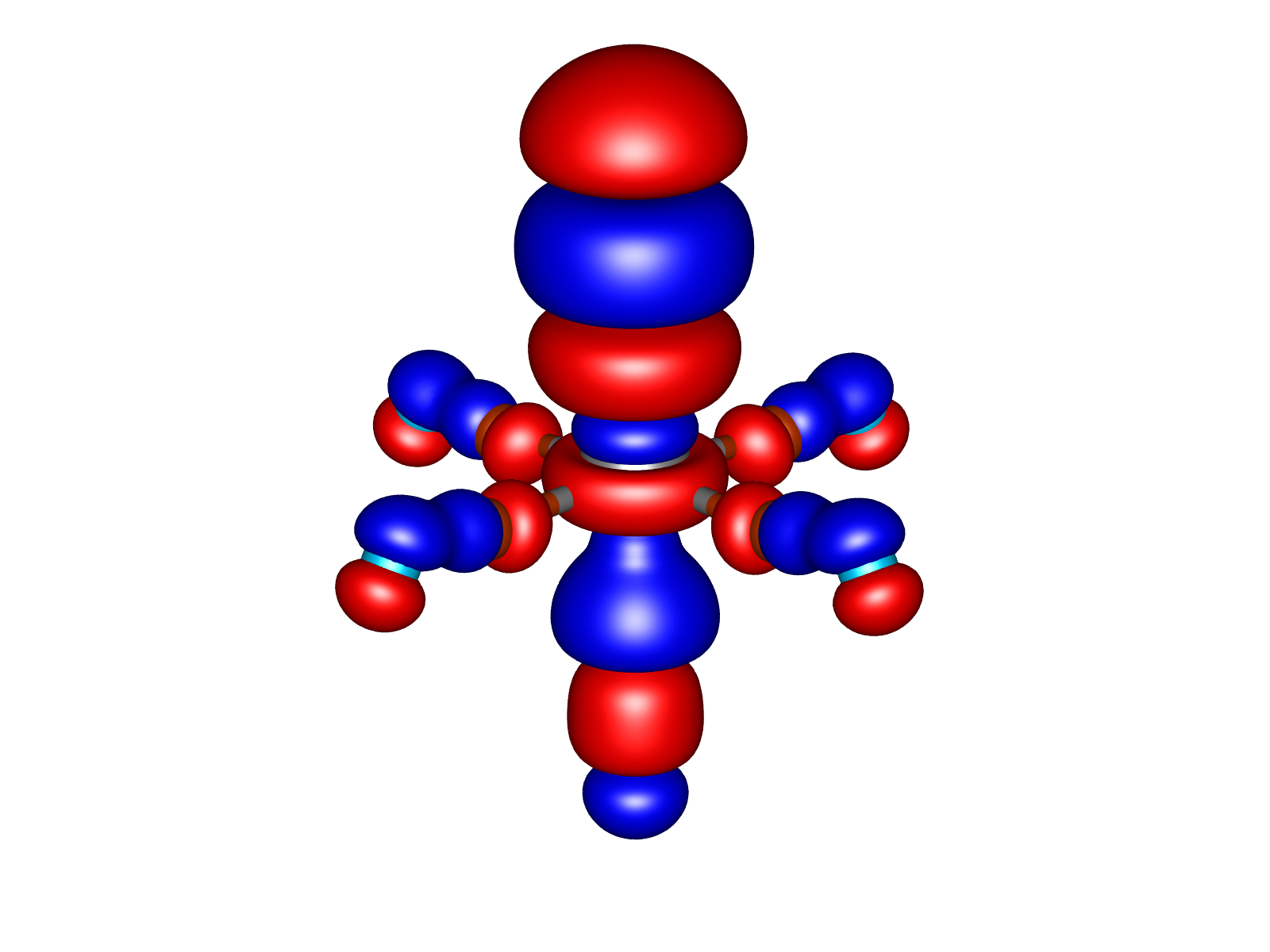}
  }
  \hfill
  \subfloat[$n_{\text{occup}}$ = 1.9572]{%
    \includegraphics[width=0.22\textwidth]{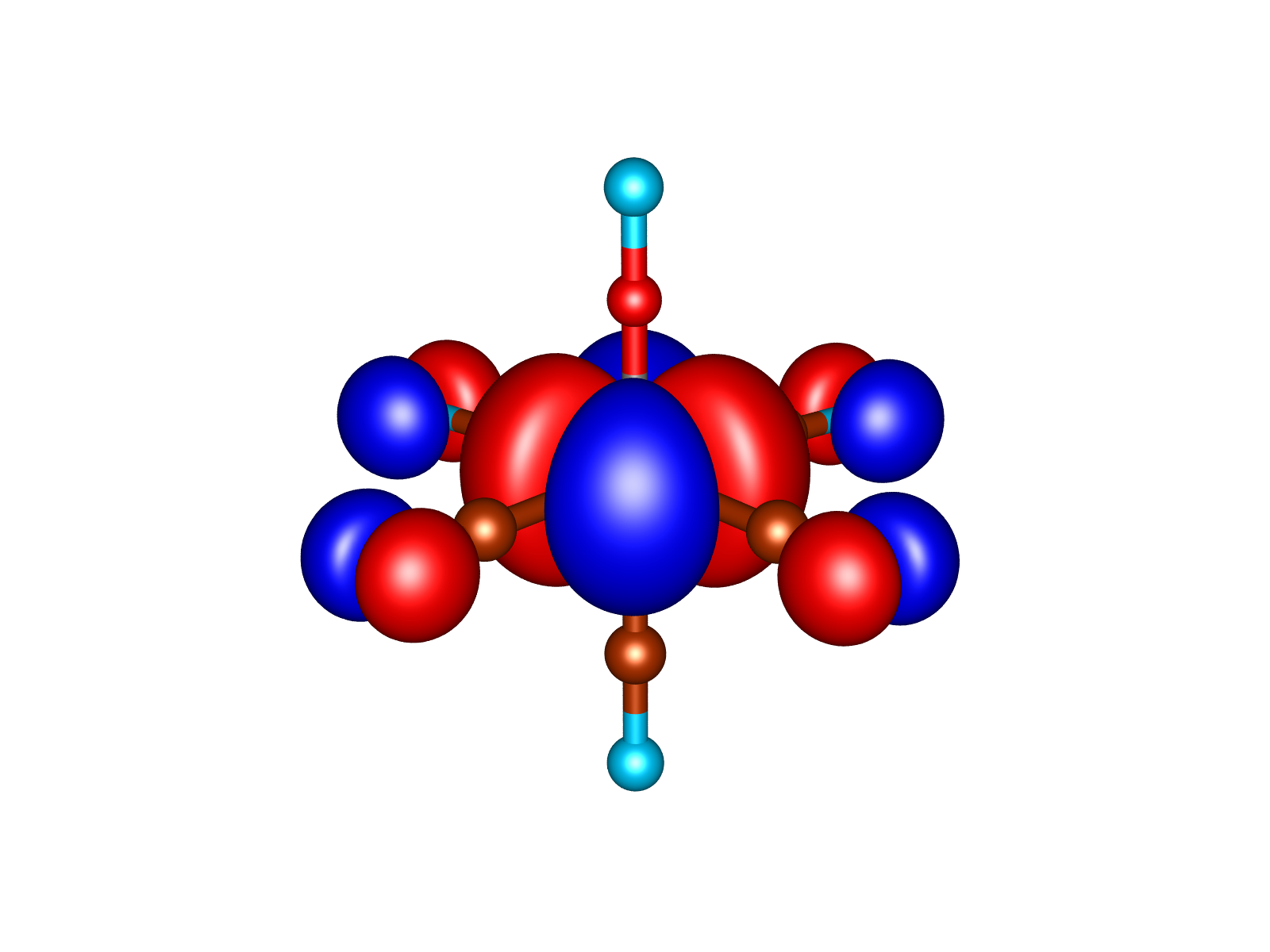}
  }
  \hfill
  \subfloat[$n_{\text{occup}}$ = 1.9538]{%
    \includegraphics[width=0.22\textwidth]{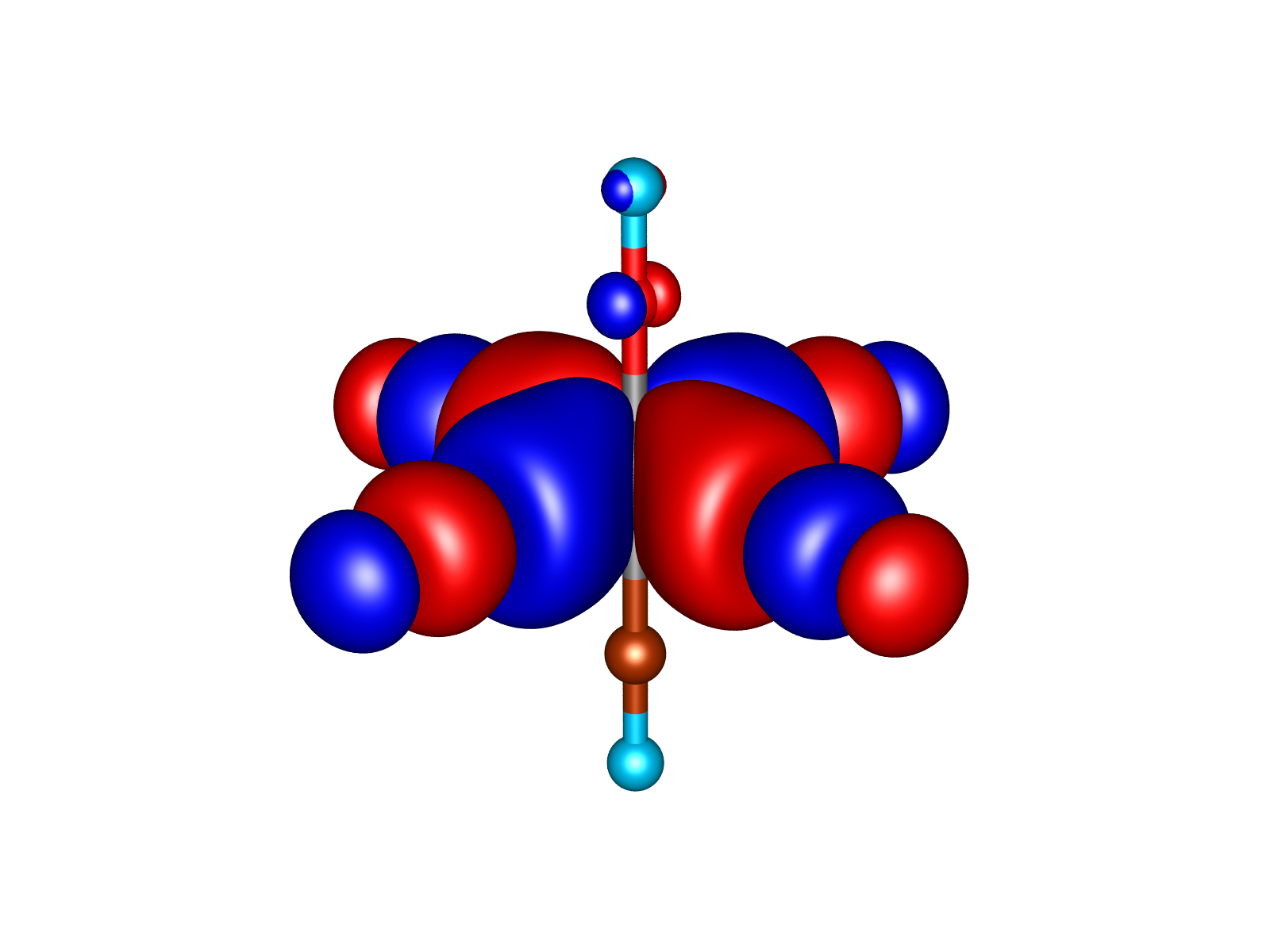}
  }
  \hfill
  \subfloat[$n_{\text{occup}}$ = 1.9518]{%
    \includegraphics[width=0.22\textwidth]{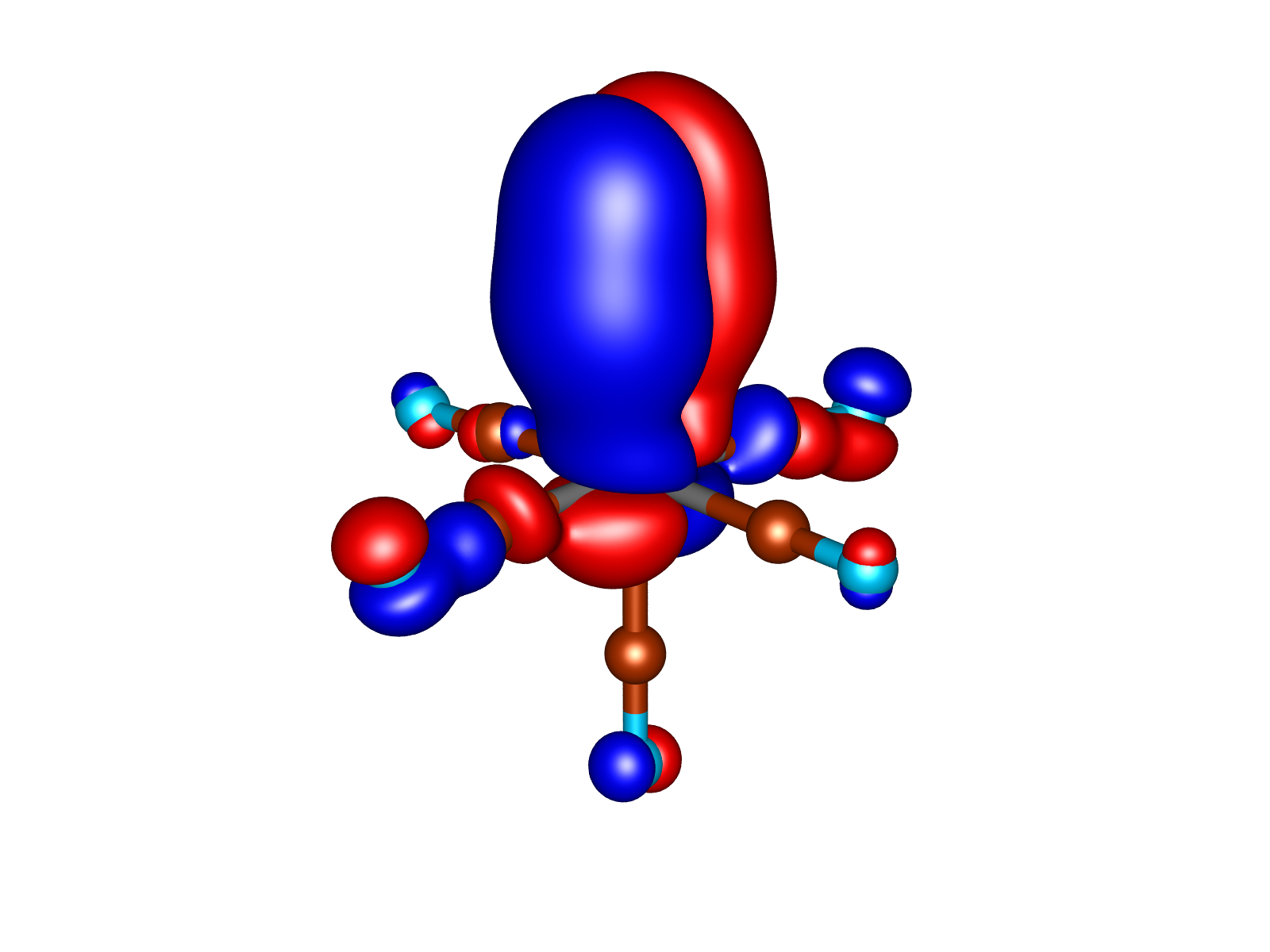}
  } \\
  \subfloat[$n_{\text{occup}}$ = 1.9518]{%
    \includegraphics[width=0.22\textwidth]{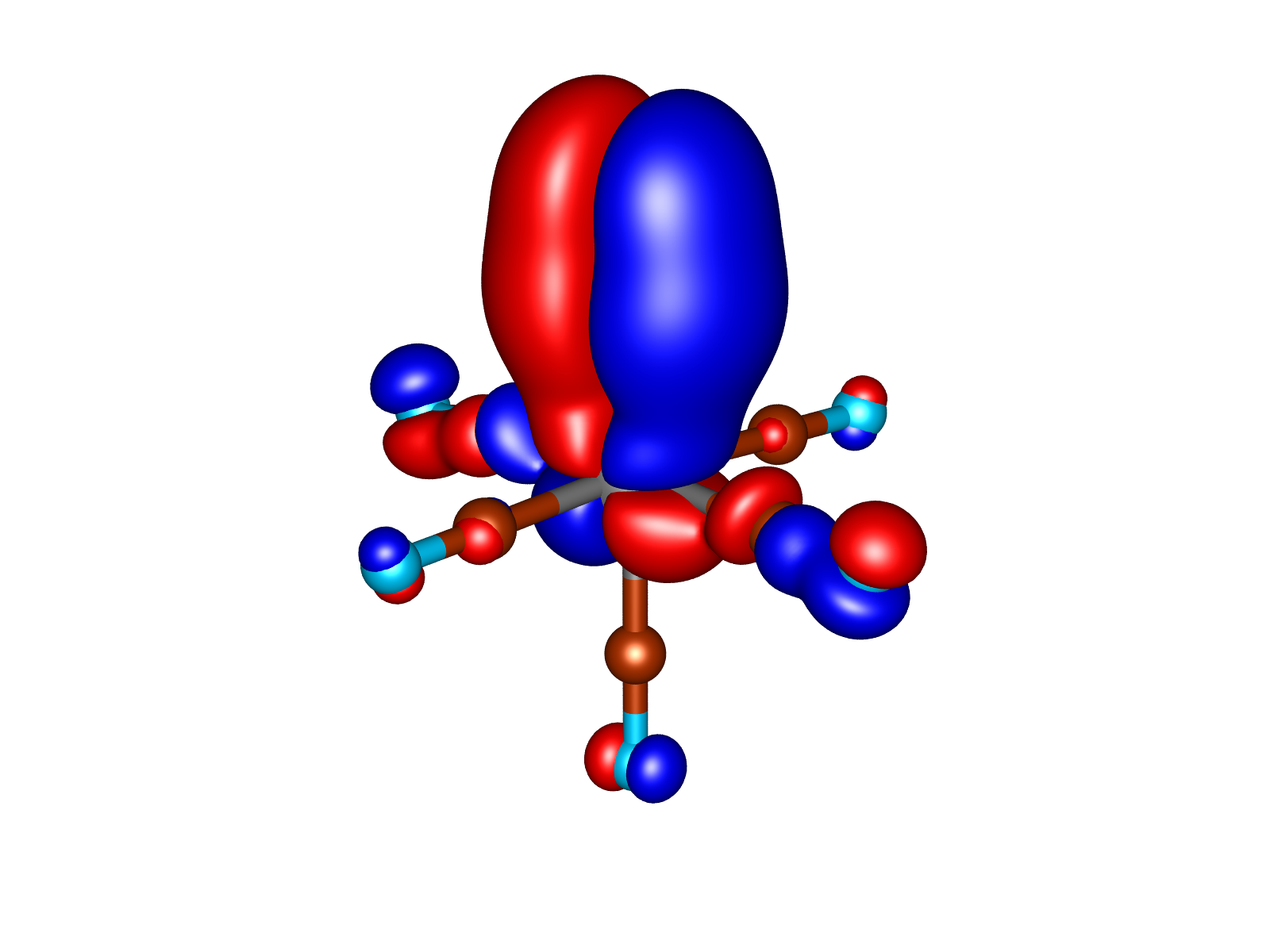}
  }
  \hfill
  \subfloat[$n_{\text{occup}}$ = 1.9488]{%
    \includegraphics[width=0.22\textwidth]{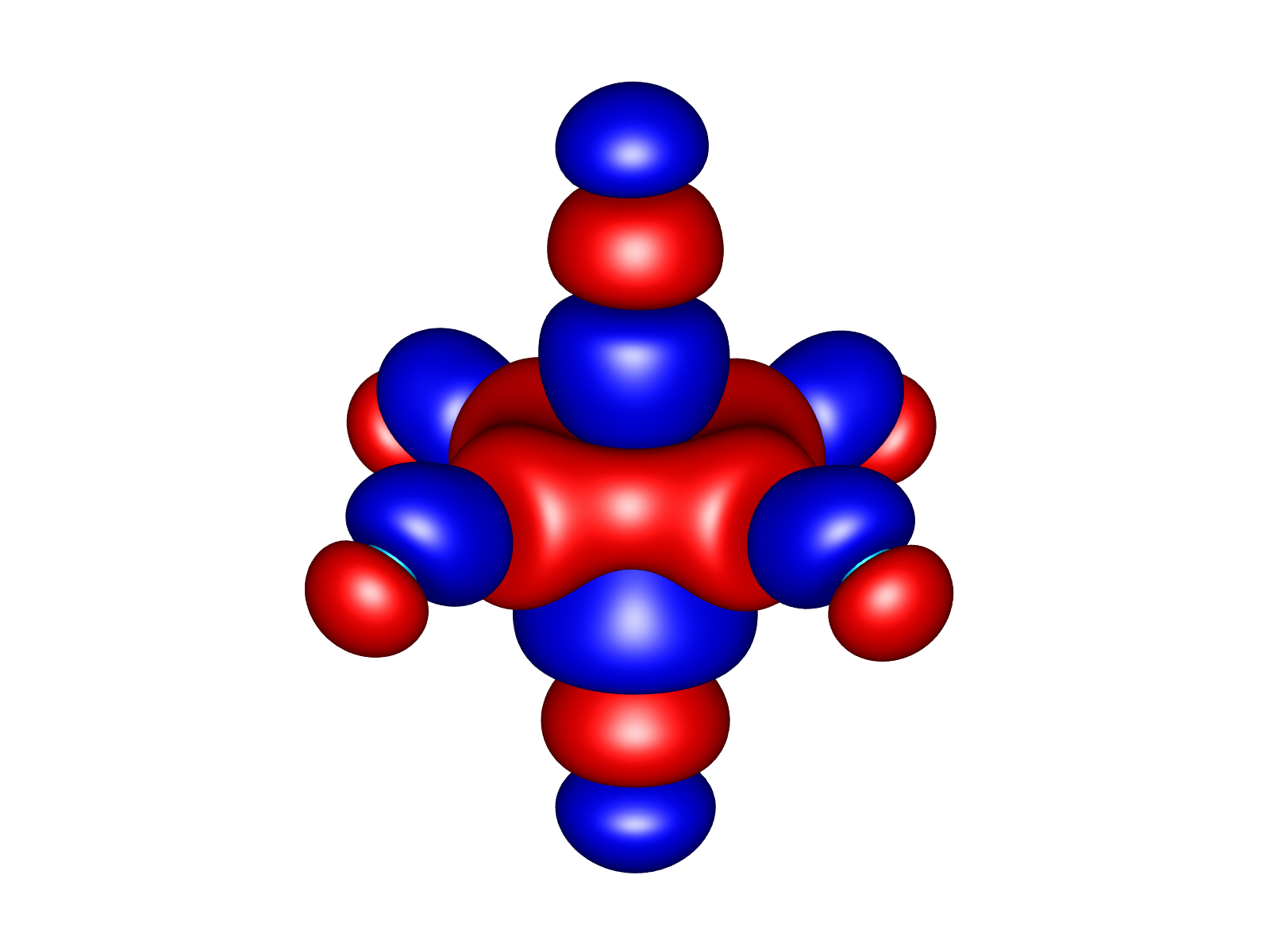}
  }
  \hfill
  \subfloat[$n_{\text{occup}}$ = 1.7169]{%
    \includegraphics[width=0.22\textwidth]{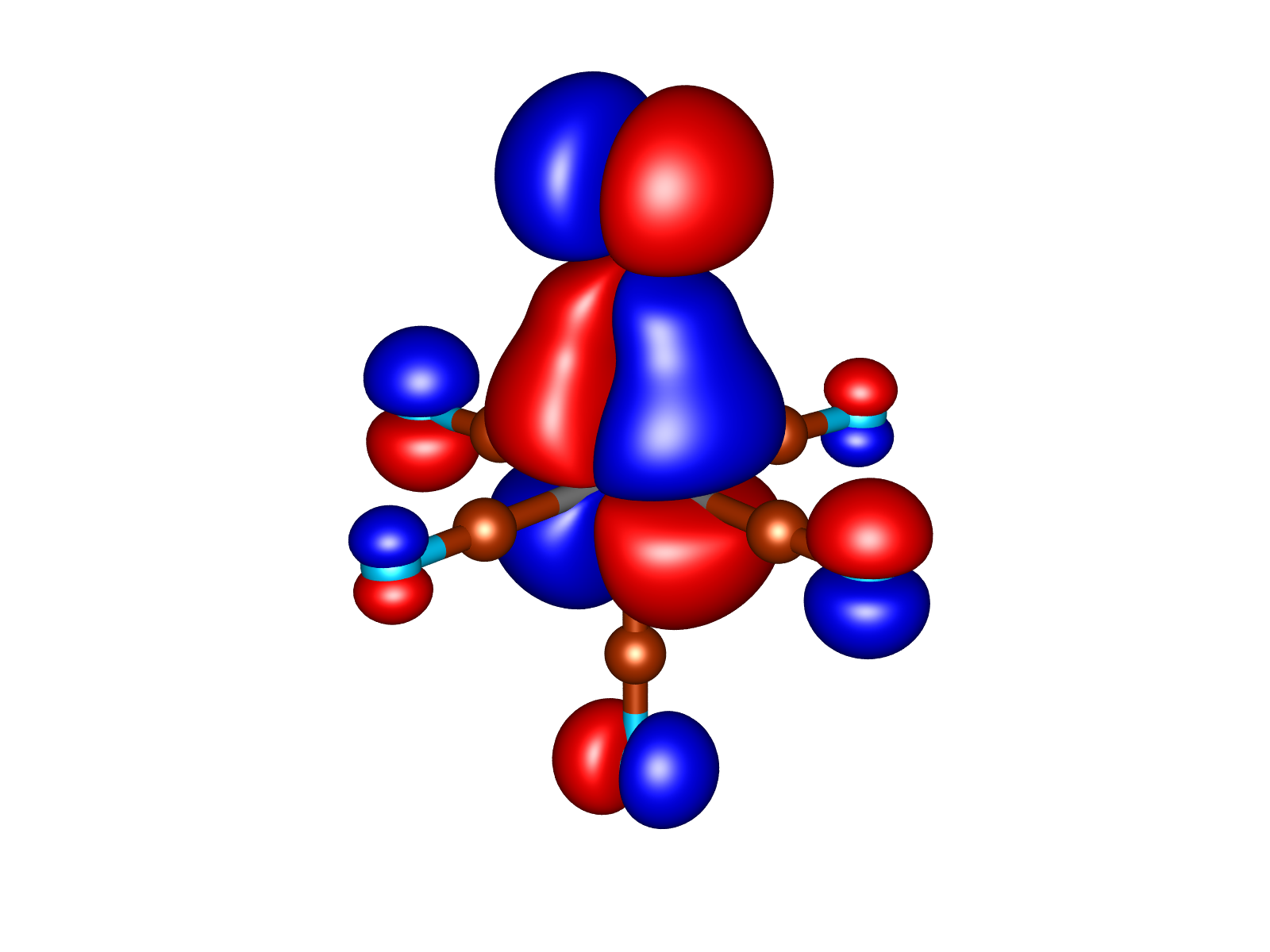}
  }
  \hfill
  \subfloat[$n_{\text{occup}}$ = 1.7167]{%
    \includegraphics[width=0.22\textwidth]{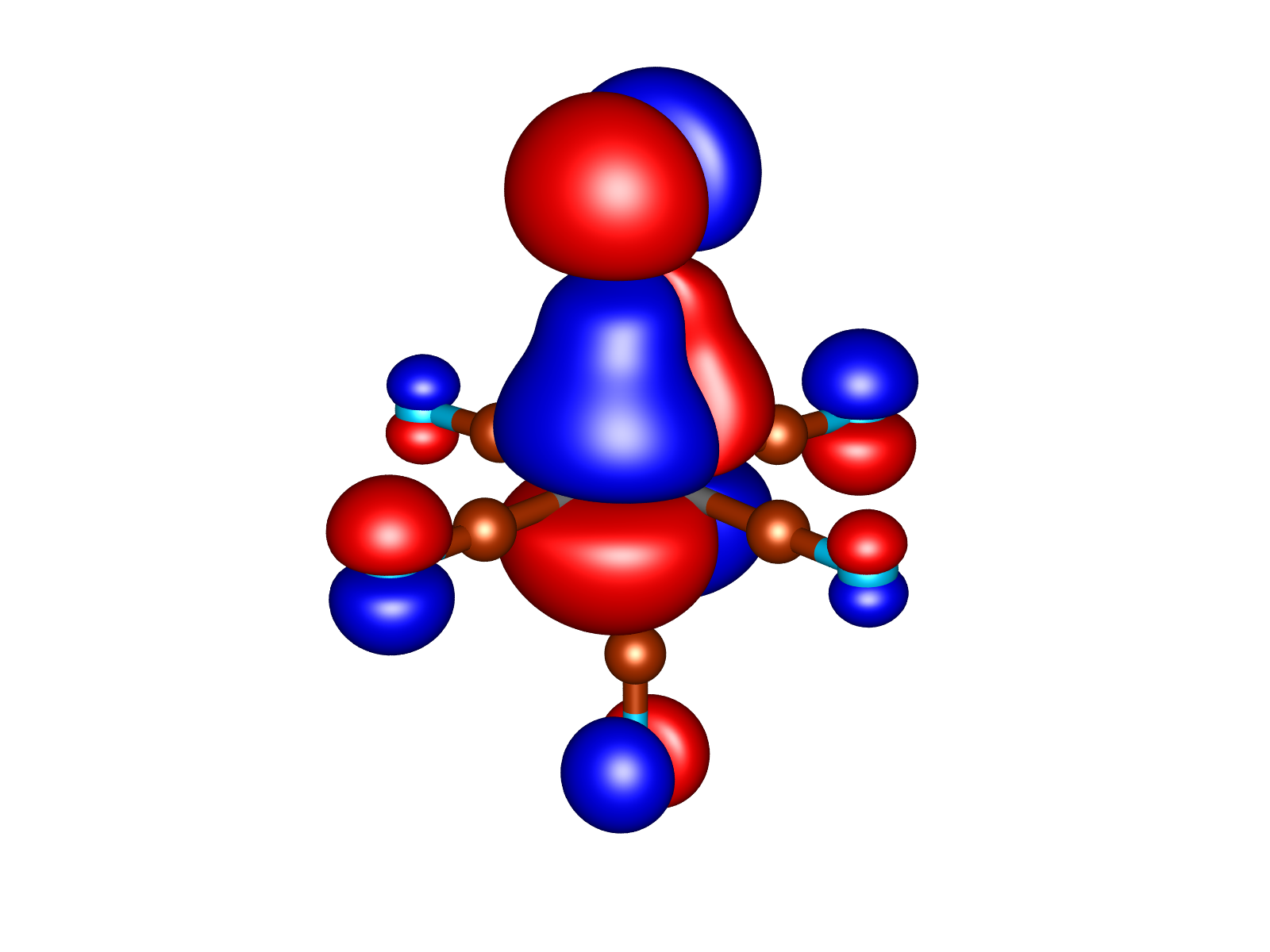}
  }
  \\
  \subfloat[$n_{\text{occup}}$ = 0.3151]{%
    \includegraphics[width=0.22\textwidth]{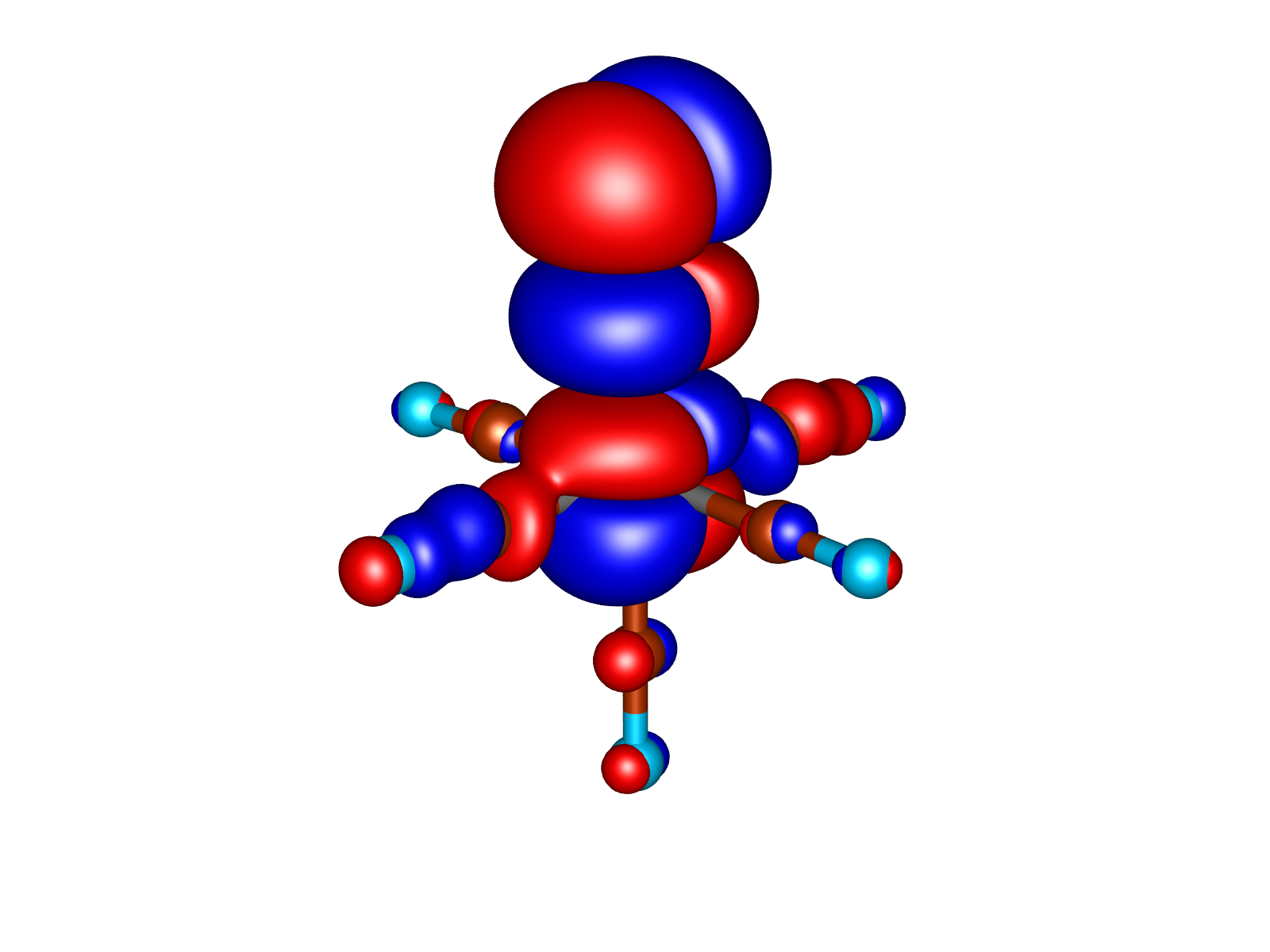}
  }
  \hfill
  \subfloat[$n_{\text{occup}}$ = 0.3149]{%
    \includegraphics[width=0.22\textwidth]{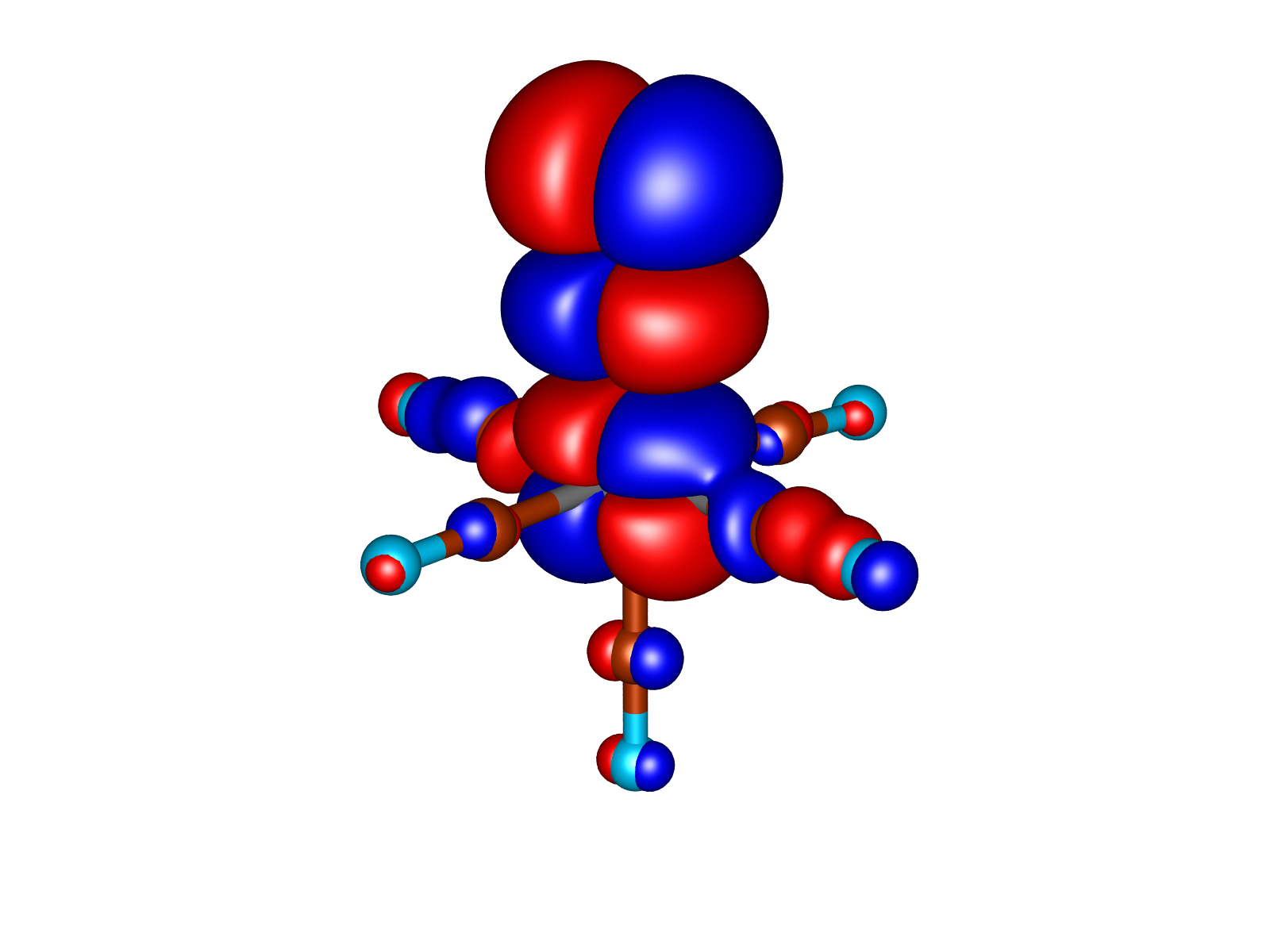}
  }
  \hfill
  \subfloat[$n_{\text{occup}}$ = 0.0682]{%
    \includegraphics[width=0.22\textwidth]{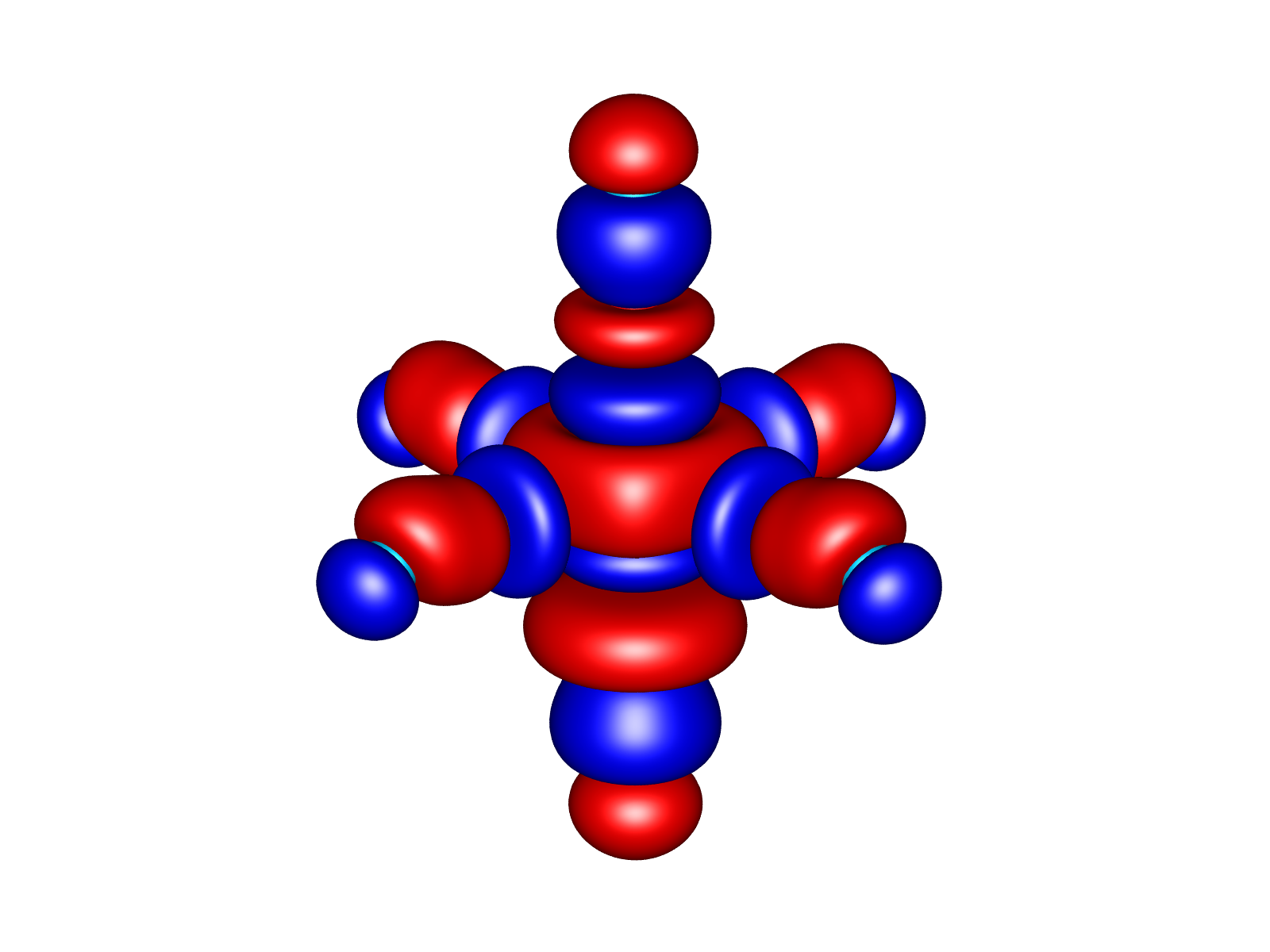}
  }
  \hfill
  \subfloat[$n_{\text{occup}}$ = 0.0540]{%
    \includegraphics[width=0.22\textwidth]{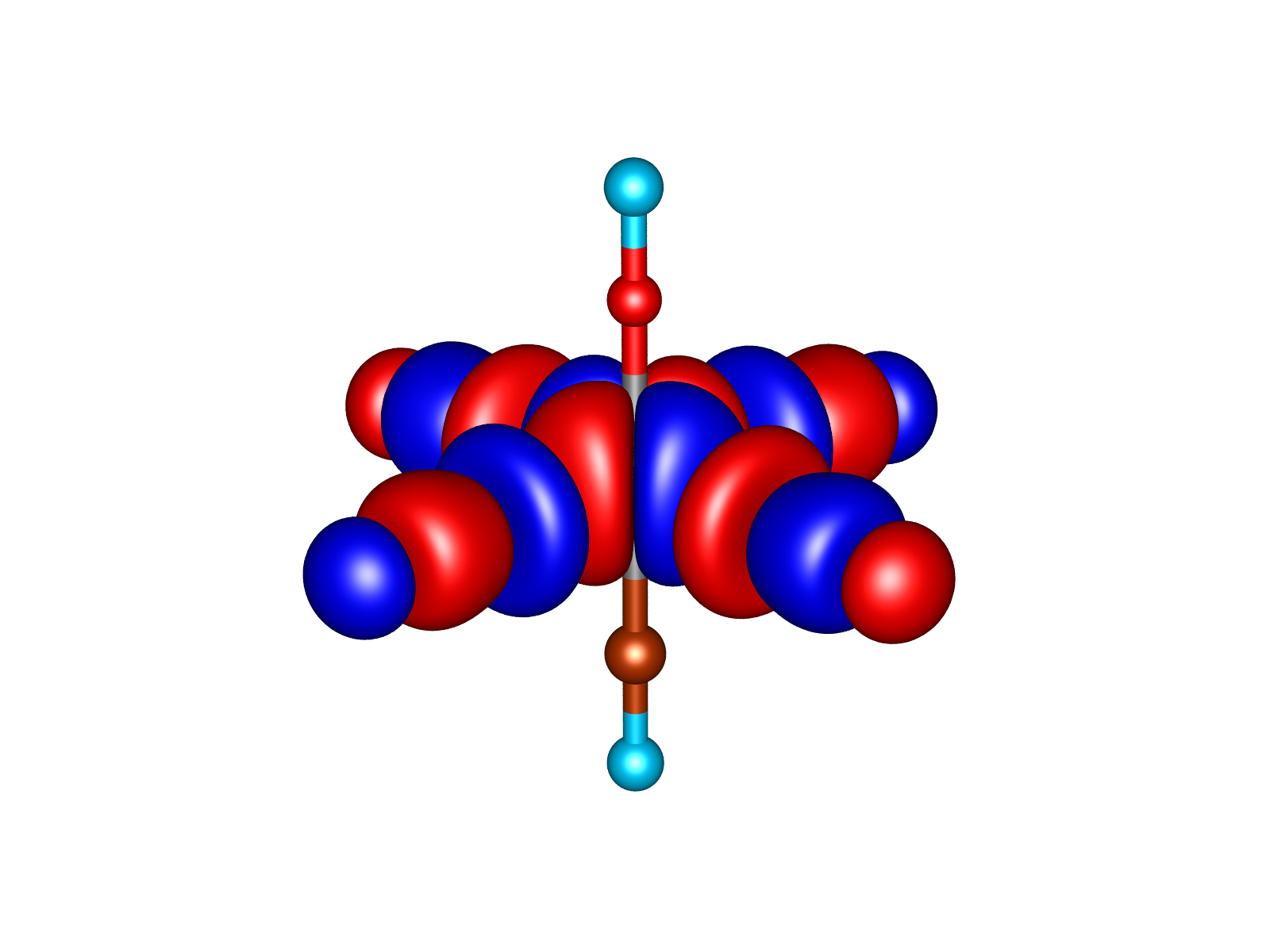}
  }
  \\
  \subfloat[$n_{\text{occup}}$ = 0.0235]{%
    \includegraphics[width=0.22\textwidth]{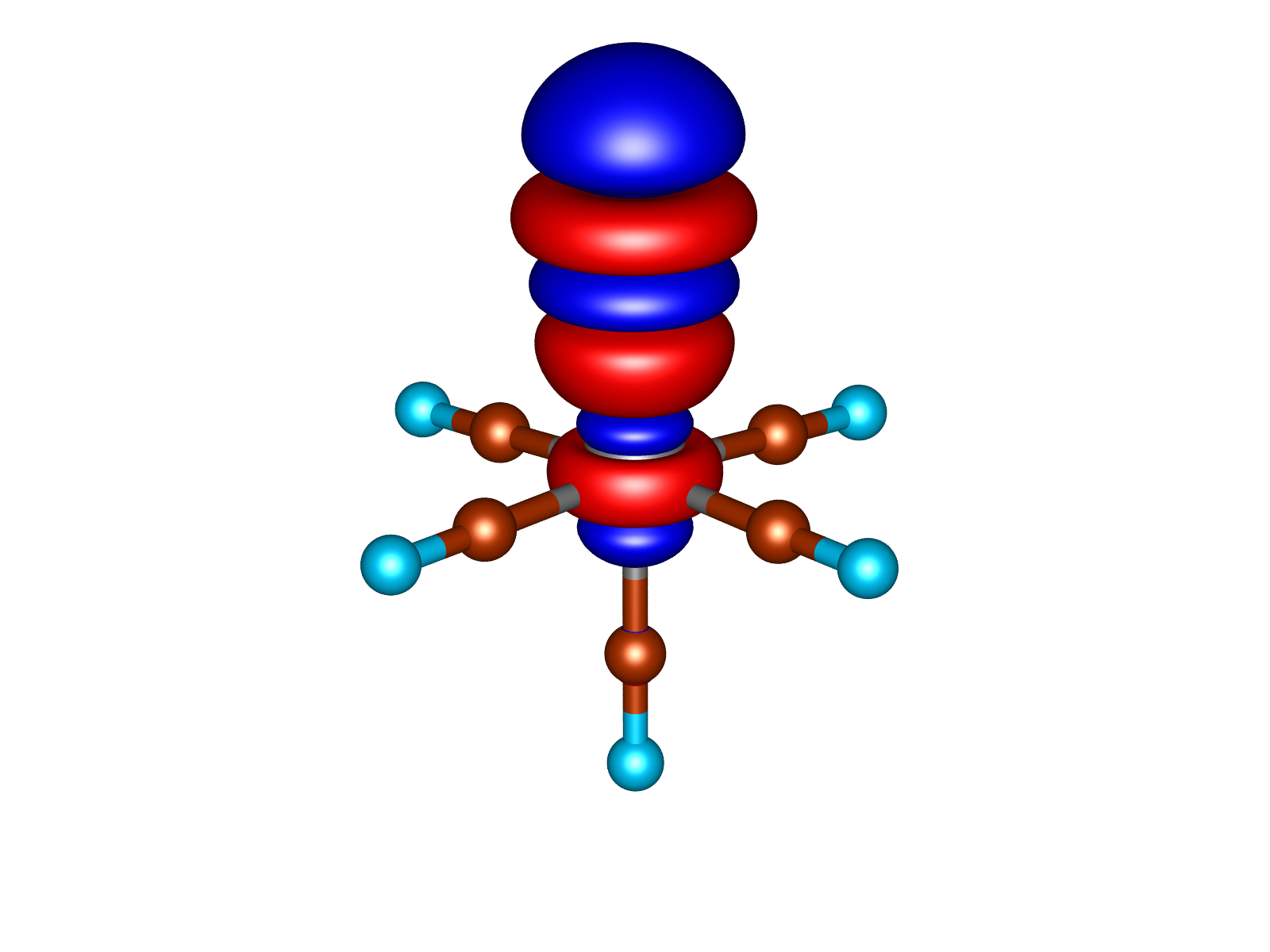}
  }
  \hfill
  \subfloat[$n_{\text{occup}}$ = 0.0230]{%
    \includegraphics[width=0.22\textwidth]{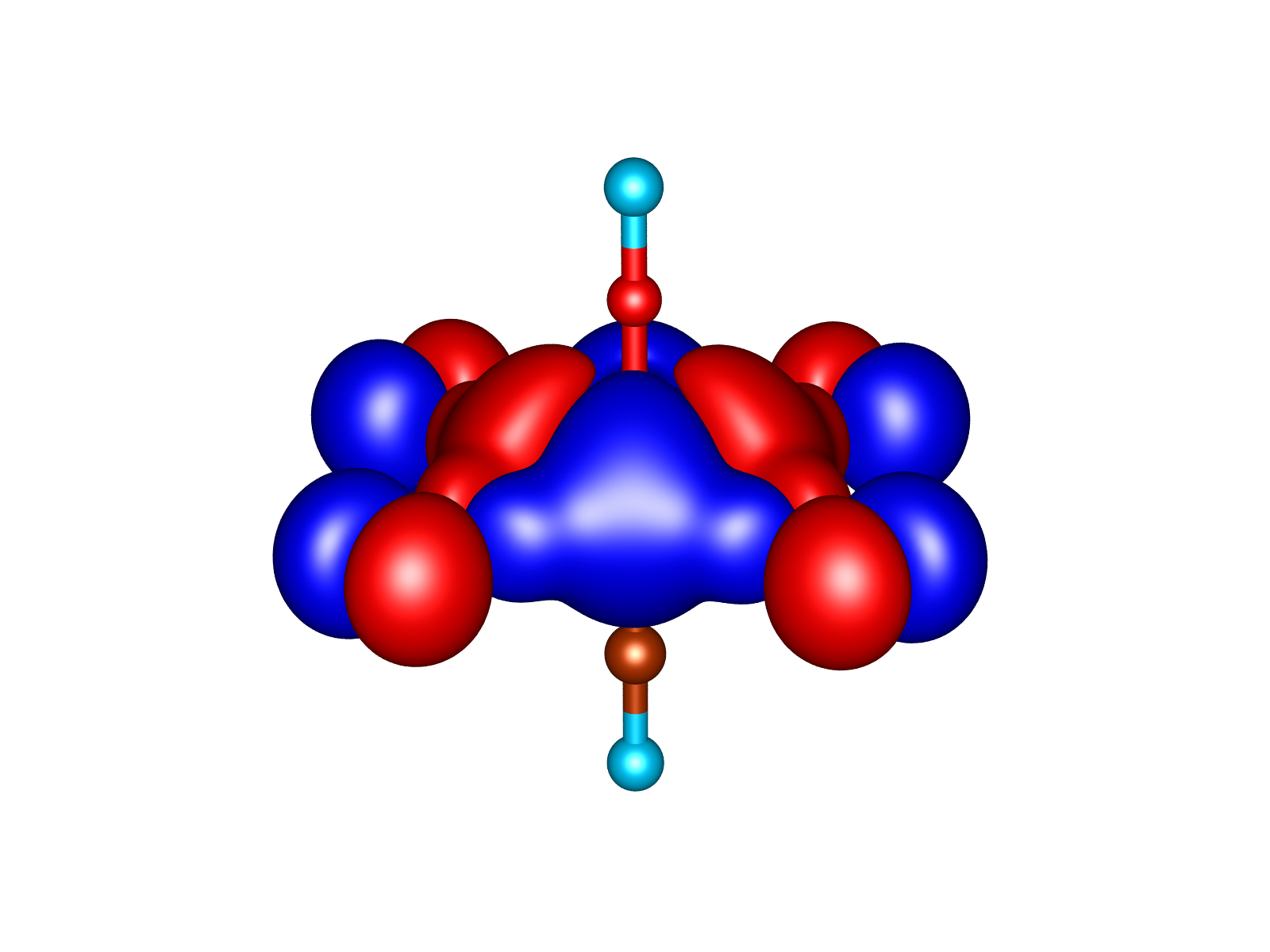}
  }
  \hfill
  \subfloat[$n_{\text{occup}}$ = 0.0138]{%
    \includegraphics[width=0.22\textwidth]{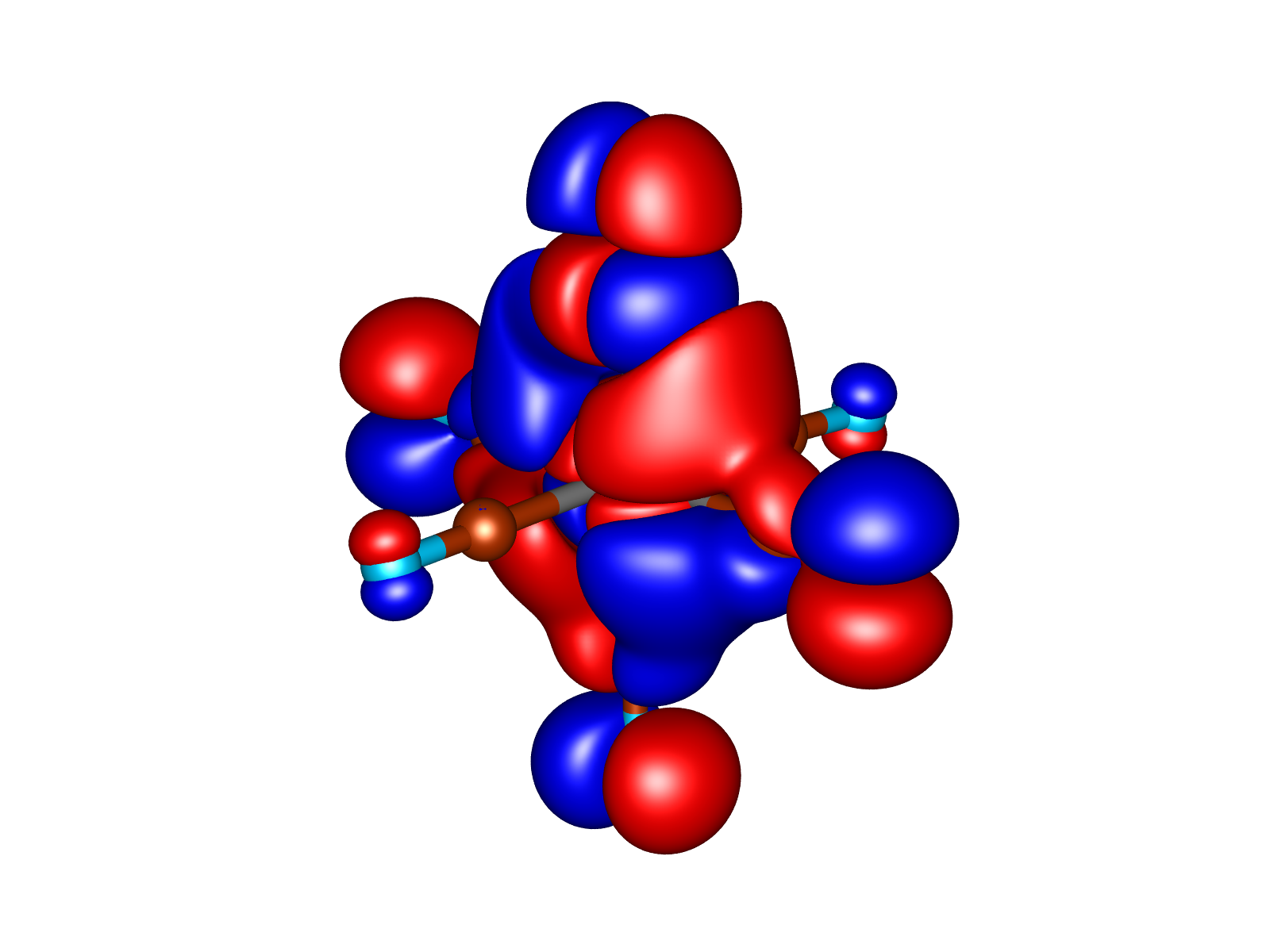}
  }
  \hfill
  \subfloat[$n_{\text{occup}}$ = 0.0138]{%
    \includegraphics[width=0.22\textwidth]{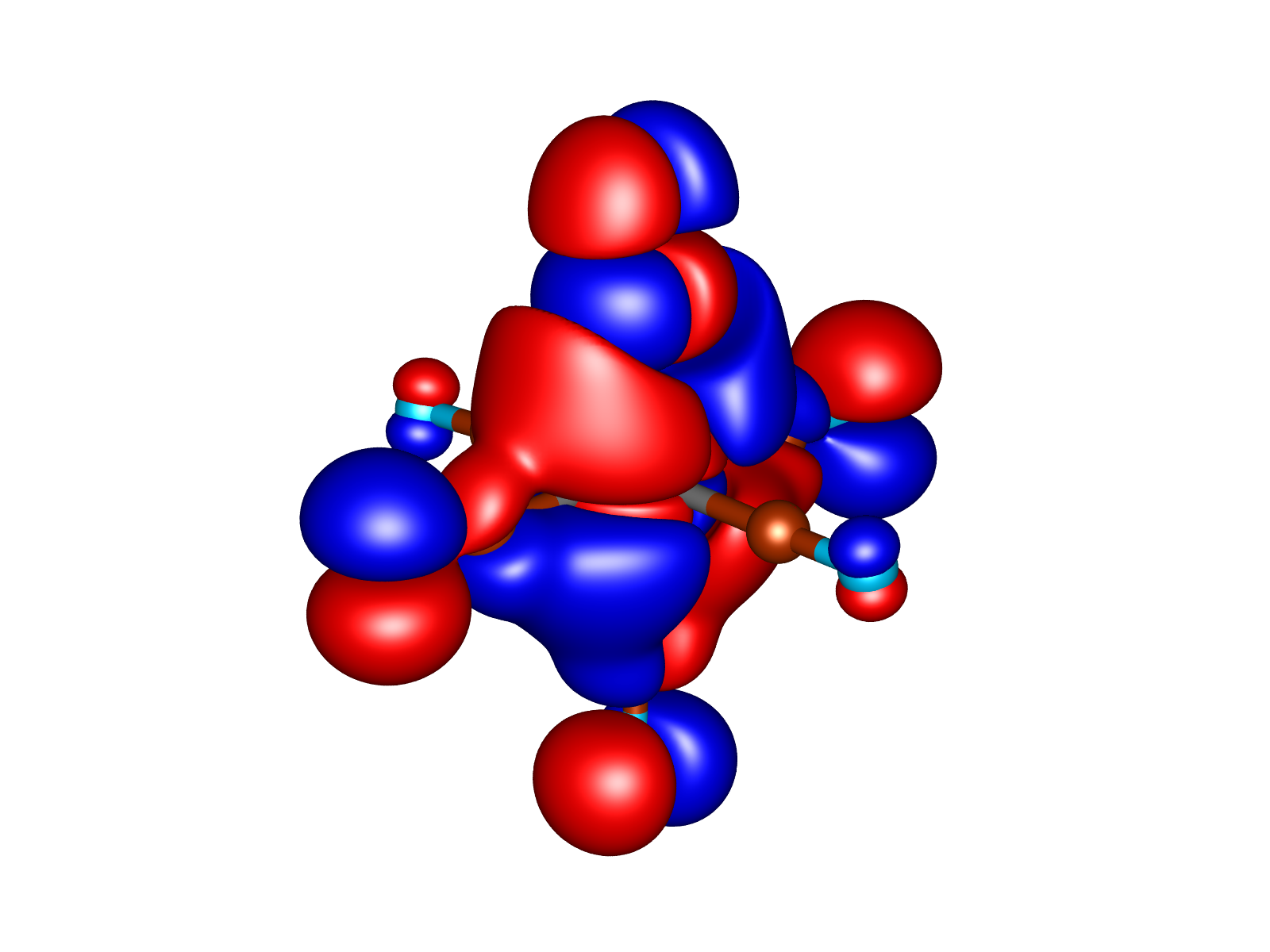}
  }
  \\
  \caption{Fe-NO complex, reverse, DMRG-SCF(16, 16) \label{orbs_cas1616_5}}
\end{figure}